\documentclass[aps,pra,reprint, amsmath, amssymb,superscriptaddress]{revtex4-1}
 
\usepackage{bm}
\usepackage[retainorgcmds]{IEEEtrantools}
\usepackage{graphicx}
\usepackage{mathrsfs}
\usepackage{amsmath}
\usepackage{amsfonts}
\usepackage{amssymb}
\usepackage{color}
\usepackage{times,txfonts}
\usepackage{nicefrac}
\usepackage{ragged2e}
\usepackage{tikz}
\usepackage{bbm}

\usepackage[colorlinks=true,linkcolor=blue,urlcolor=blue,citecolor=blue,pdfusetitle]{hyperref}

\usepackage{blindtext}

\usepackage[framemethod=TikZ]{mdframed}
\usepackage[T1]{fontenc}
\mdfdefinestyle{MyFrame}{%
    linecolor=black,
    outerlinewidth=1pt,
    roundcorner=0pt,
    innertopmargin=0.77\baselineskip,
    innerbottommargin=0.75\baselineskip,
    innerrightmargin=14pt,
    innerleftmargin=14pt,
    backgroundcolor=gray!20!white}
\mdfdefinestyle{MyFrameTitled}{%
    linecolor=black,
    outerlinewidth=1pt,
    roundcorner=0pt,
    innertopmargin=0.77\baselineskip,
    innerbottommargin=0.75\baselineskip,
    innerrightmargin=14pt,
    innerleftmargin=14pt,
    backgroundcolor=gray!20!white,
     frametitlebackgroundcolor   =black!15,
    frametitlerule =true,
}

\newcommand{\genK}{K}    
\newcommand{\genJ}{\mathcal{J}}    
\newcommand{\nops}{\mathsf{r}}    
\newcommand{\trans}{^{\text{T}}}
\newcommand{\MP}{^{\text{MP}}}
\newcommand{\tr}{\mathrm{tr}}
\newcommand{\rhoss}{\rho_{\rm ss}}
\newcommand{\rhossV}{|\rho_{\rm ss}\rrangle}
\newcommand{\idV}{\llangle \id |}
\newcommand{\Hnh}{H_{\rm eff}}
\newcommand{\id}{\mathbbm{1}}
\newcommand{\ket}[1]{\vert #1 \rangle}
\newcommand{\bra}[1]{\langle #1 \vert}
\newcommand{\gtwo}{g^{(2)}}
\newcommand{\gone}{g^{(1)}}
\newcommand{\upket}{\ket{\!\!\uparrow}}
\newcommand{\dwket}{\ket{\!\!\downarrow}}
\newcommand{\upbra}{\bra{\uparrow\!\!}}
\newcommand{\dwbra}{\bra{\downarrow\!\!}}

\usepackage{amssymb}
\usepackage{pifont}
\newcommand{\rev}[1]{{\color{black}#1}}

\makeatletter
\newsavebox{\@brx}
\newcommand{\llangle}[1][]{\savebox{\@brx}{\(\m@th{#1\langle}\)}%
  \mathopen{\copy\@brx\kern-0.5\wd\@brx\usebox{\@brx}}}
\newcommand{\rrangle}[1][]{\savebox{\@brx}{\(\m@th{#1\rangle}\)}%
  \mathclose{\copy\@brx\kern-0.5\wd\@brx\usebox{\@brx}}}
\makeatother

\usepackage[framemethod=TikZ]{mdframed}
\mdfdefinestyle{MyFrame}{%
    linecolor=black,
    outerlinewidth=1pt,
    roundcorner=0pt,
    innertopmargin=0.77\baselineskip,
    innerbottommargin=0.75\baselineskip,
    innerrightmargin=14pt,
    innerleftmargin=14pt,
    backgroundcolor=gray!20!white}
\mdfdefinestyle{MyFrameTitled}{%
    linecolor=black,
    outerlinewidth=1pt,
    roundcorner=0pt,
    innertopmargin=0.77\baselineskip,
    innerbottommargin=0.75\baselineskip,
    innerrightmargin=14pt,
    innerleftmargin=14pt,
    backgroundcolor=gray!20!white,
     frametitlebackgroundcolor   =black!15,
    frametitlerule =true,
}

\begin{document}

\title{Current fluctuations in open quantum systems: \\ Bridging the gap between quantum continuous measurements and full counting statistics
}

\date{\today}

\author{Gabriel T. Landi}
\email{gabriel.landi@rochester.edu}
\affiliation{Department of Physics and Astronomy, University of Rochester, Rochester, New York 14627, USA}

\author{Michael J. Kewming}
\email{kewmingm@tcd.ie}
\affiliation{School of Physics, Trinity College Dublin, College Green, Dublin 2, D02K8N4, Ireland}

\author{Mark T. Mitchison}
\email{mark.mitchison@tcd.ie}
\affiliation{School of Physics, Trinity College Dublin, College Green, Dublin 2, D02K8N4, Ireland}
\affiliation{Trinity Quantum Alliance, Unit 16, Trinity Technology and Enterprise Centre, Pearse Street, Dublin 2, D02YN67, Ireland}

\author{Patrick P. Potts}
\email{patrick.potts@unibas.ch}
\affiliation{Department of Physics and Swiss Nanoscience Institute, University of Basel, Klingelbergstrasse 82, 4056 Basel, Switzerland}

\begin{abstract}

Continuously measured quantum systems are characterized by an output current, in the form of a stochastic and correlated time series, which conveys crucial information about the underlying quantum system. 
The many tools used to describe current fluctuations are scattered across different communities: quantum opticians often use stochastic master equations, while a prevalent approach in condensed matter physics is provided by full counting statistics. 
These, however, are simply different sides of the same coin. 
Our goal with this tutorial is to 
provide a unified toolbox for describing current fluctuations.
This not only provides novel insights, by bringing together different fields in physics, but also yields various analytical and numerical tools for computing quantities of interest. 
We illustrate our results with various pedagogical examples, and connect them with topical fields of research, such as waiting-time statistics, quantum metrology, thermodynamic uncertainty relations, quantum point contacts and Maxwell's demons. 

\end{abstract}

\maketitle{}

\tableofcontents

\section{Introduction}

\begin{figure*}
    \centering
    \includegraphics[width=0.9\textwidth]{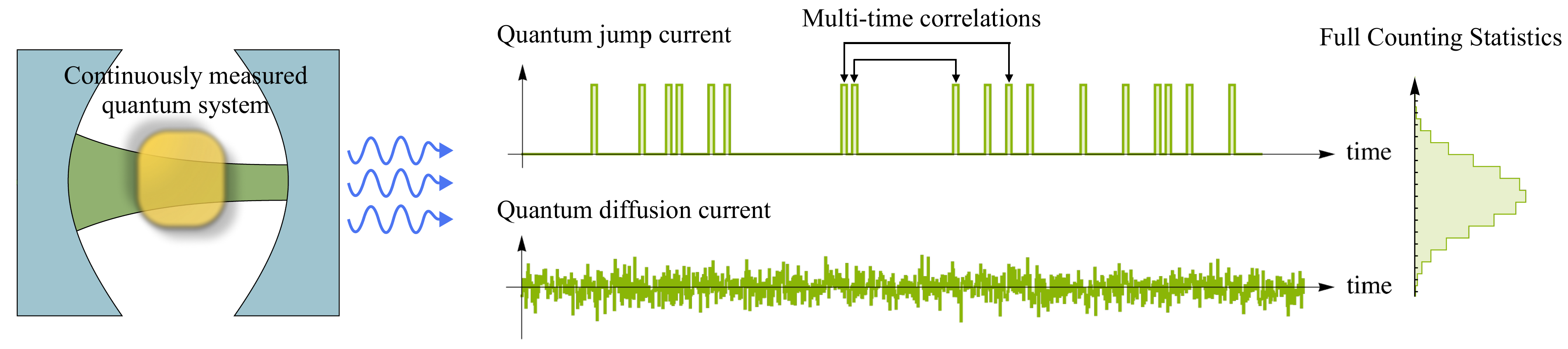}
    \caption{
    \rev{Continuously measured quantum systems and their output currents. 
    Left: paradigmatic example of photo-detector collecting the photons that leak out of an optical cavity. Middle:}
    the measurement gives rise to a stochastic output current; i.e., a correlated time-series which contains information about the underlying quantum system. 
    These currents usually come in two types: quantum jumps, which involve discrete clicks like in photo detection; and quantum diffusion, where the current is a continuous noisy signal. 
    In this tutorial we provide the tools and the intuition to analyze the statistical properties of these currents, such as their multi-time correlations. We also show how this is connected to the framework of Full Counting Statistics \rev{(right)}, which analyzes  the properties of the integrated current (called the net charge). 
    }
    \label{fig:schematics}
\end{figure*}

The importance of quantum measurements to modern quantum science cannot be overstated.
Not only do they relate to the foundations of quantum theory, but they are also the doorway to the quantum realm, since in experiments we only observe quantum systems indirectly through these measurements. 
One particularly important scenario is that of continuous measurements, where information about the system is available in the form of an \emph{output current}, i.e., a classical time series which is fundamentally stochastic in nature (Fig.~\ref{fig:schematics}). 
Owing to the recent advances in quantum science and technology, there is growing interest in understanding the properties of these currents in more detail. 
In particular, their stochastic nature means that it is vital to also address their \emph{fluctuations}.
On the one hand, fluctuations represent a fundamental limitation to the measurement precision. 
But on the other, they also encode important information.
This principle extends throughout all of quantum physics, from optical coherence functions ($g^{(1)},g^{(2)},\ldots$) ~\cite{Nagourney1986,Sauter1986,Bergquist1986,Plenio_1998,Doherty_1999,Wiseman_2009,Jacobs_2014} to electric charge transport in mesoscopic conductors~\cite{buttiker_1990,buttiker_1992,landauer_1998,blanter_2001,nazarov_book,birk_1995,reznikov_1995,gustavsson_2006, braggio2006,flindt_2009,Klich2009,Esposito_2009,kambly2011,Stegmann2017} and gravitational-wave detectors using radiation squeezing~\cite{Caves_1981,Grote_2013}. 

Current fluctuations carry information about the system which is not available in the average current $J := E\big( I(t)\big)$, where $I(t)$ denotes the classical stochastic output and $E(\cdot)$ denotes an expectation value \rev{over different realizations of the experiment}. 
For instance, the two-point correlation function 
\begin{equation}\label{two_point_function}
F(\tau) = E\Big(I(0) I(\tau)\Big) - J^2, 
\end{equation}
contains information on dynamical processes such as Rabi oscillations and relaxation rates. 
In photodetection, it is associated with the Glauber coherence functions $g^{(1)}$ and $g^{(2)}$~\cite{Glauber_1963}.
Current fluctuations also play a key role in the stable operation of mesoscopic and microscopic devices. 
Quantum dot systems, for example, can operate as autonomous thermoelectric engines~\cite{Benenti_2017}, whose output is an electric current. This current fluctuates due to the microscopic nature of the system. It thus becomes crucial to develop strategies for keeping these fluctuations in check, so that they do not compromise the device's operation~\cite{Pietzonka_2018,Guarnieri_2019,Saryal_2021}. 

The first motivation of this tutorial is to provide the intuition, as well as the tools, to compute current fluctuations from models based on quantum master equations. 
The results should therefore be useful for both theoreticians aiming to uncover fundamental features of quantum fluctuations, as well as experimentalists interested in modeling their findings with the tools of open quantum systems. 

Our second motivation is to bridge a gap that exists between approaches used by quantum opticians and by condensed matter physicists, who study the same kinds of problems but with different languages and different tools. 
For example, quantum opticians 
often employ stochastic master equations (SMEs)~\cite{Wiseman_2009,Jacobs_2014} and the quantum regression theorem~\cite{Lax1963,gardiner_book}.
A typical object of study is the current power spectrum
\begin{equation}\label{power_spectrum}
    S(\omega) = \int_{-\infty}^{\infty}e^{-i\omega \tau} F(\tau)d\tau\,.
\end{equation}
Condensed matter physicists, on the other hand, often employ a tool box called Full Counting Statistics (FCS), which describes the statistics of the stochastic charge (integrated current) $N(t) = \int_0^t dt'~I(t')$.
This involves concepts such as tilted Liouvillians and generalized quantum master equations (gQME)~\cite{levitov_charge_1993,levitov_1996,nazarov_2003,nazarov_book,Esposito_2009, Flindt_2010, Schaller2014}.
Among the quantities studied in FCS, the most popular one is 
\begin{equation}\label{D_variance}
    D = \lim\limits_{t\to \infty}~\frac{d}{dt} {\rm Var}(N(t)),
\end{equation}
called the noise, the diffusion coefficient, or the scaled variance. 
It is often not appreciated, however, that the noise is just the zero-frequency component of the power spectrum
\begin{equation}
    D = S(0). 
\end{equation}
This is but one simple example of how connected the different fields are. 
In fact, several such connections exist, for instance between SMEs and gQMEs.
Our goal is that, by bridging the gap between these fields, one would benefit from a much broader set of tools, which will  likely lead to new insights.

New researchers working with QMEs will often compute the average current, which is rather straightforward~\cite{Landi_Nonequilibrium_2022}. 
\rev{But when it comes to extracting current fluctuations, the procedures are not very well disseminated in the literature. 
Different communities often use different techniques and the relations between them are not clear.}
Our aim here is to show that making the step from master equations to current fluctuations is actually straightforward, and there exist simple and efficient methods for computing said quantities. 
This tutorial provides a pedagogical overview of several efficient numerical techniques that can be used to characterize current fluctuations.
\rev{A set of numerical tools, including the notebooks used to generate all figures of this tutorial, can be found on~\footnote{The files can be accessed in \href{https://github.com/mikeal888/Tutorial_Paper}{https://github.com/mikeal888/Tutorial\_Paper}.}.
It contains code in both Mathematica, using the Melt Library~\footnote{\href{https://melt1.notion.site/}{https://melt1.notion.site/}}, and in Python, using QuTIP~\cite{Johansson2013}.}

\begin{figure}
    \centering
    \includegraphics[width=0.5\textwidth]{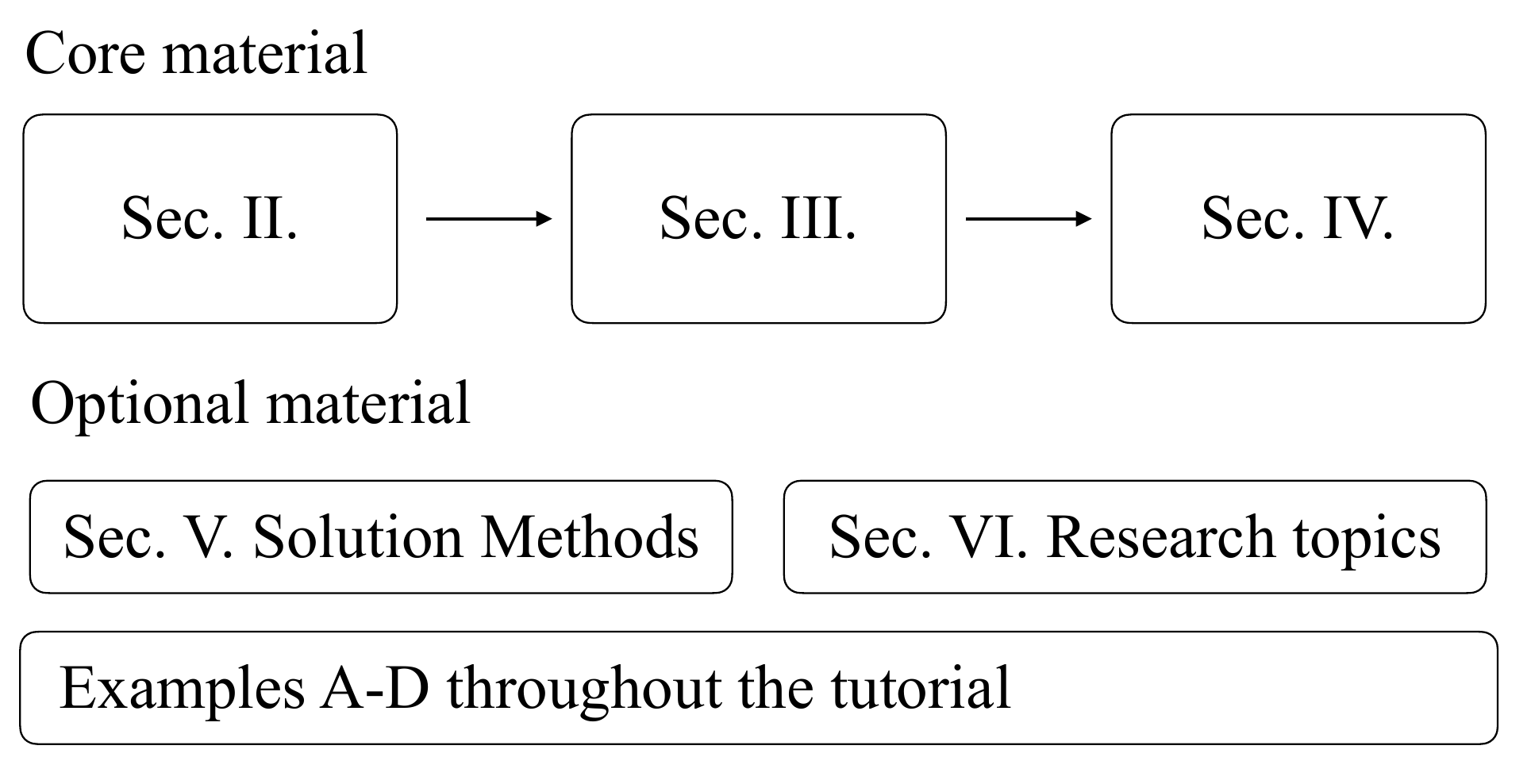}
    \caption{Structure of the tutorial, divided into core and optional material. 
    }
    \label{fig:reads}
\end{figure}

\begin{figure*}[!t]
    \centering
    \includegraphics[width=\textwidth]{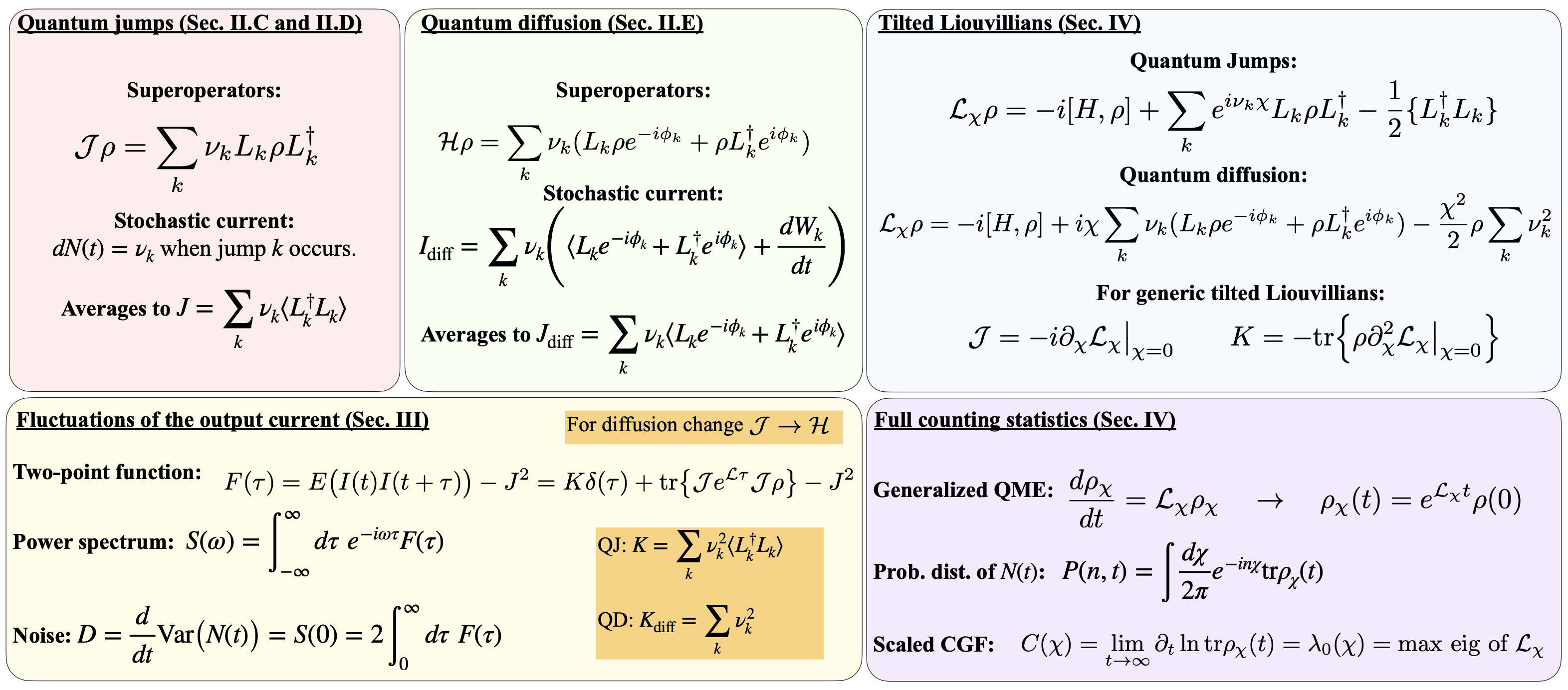}
    \caption{
    Overview of the main results in this tutorial, with the corresponding sections. 
    }
    \label{fig:overview}
\end{figure*}

The basic structure of the tutorial is shown in Fig.~\ref{fig:reads}, and a bird's-eye view of the main formulas is given in Fig.~\ref{fig:overview}. 
The main results are structured in Secs.~\ref{sec:currents}-\ref{sec:FCS}. 
In each subsection, the concepts are illustrated with 4 main examples (Sec.~\ref{sec:examples} and Fig.~\ref{fig:examples}), covering 
qubits, fermionic transport and quantum optics.
On a first pass, the reader may skip these examples, or consult them at will, without compromising the structure of the core material. 
Optional material includes Sec.~\ref{sec:methods}, which contains solution methods for computing the relevant quantities in practice, and Sec.~\ref{sec:topical} which discusses the connection with various topical fields of research. 

\begin{table*}
    \centering
    \caption{Main results for the four basic examples treated in this tutorial and which figures/equations they correspond to. 
    See also Fig.~\ref{fig:examples} for a schematic of the four examples. }
    \begin{tabular}{c|c|c|c|c}
    & Ex. A & Ex. B & Ex. C & Ex. D\\
    \hline
    Name & 
    Driven qubit
    & 
    Quantum dot
    & 
    Dephased qubit
    & 
    Parametric oscillator
    \\[0.1cm]
    Definition & Eqs.~\eqref{ExampleA_H},~\eqref{ExampleA_M} &  
    Eqs.~\eqref{ExampleB_H},~\eqref{ExampleB_M} & 
    Eqs.~\eqref{ExampleA_H},~\eqref{ExampleC_M} &
    Eqs.~\eqref{ExampleD_H},~\eqref{ExampleD_M}
    \\[0.1cm]
    Process type 
    &
    Jump
    & 
    Jump
    & 
    Diffusion
    & 
    Both
    \\[0.1cm]
    Quantum trajectories    & Fig.~\ref{fig:ExampleA_trajs} & Fig.~\ref{fig:ExampleB_trajs} & Fig.~\ref{fig:quantum_diffusion} &
    Figs.~\ref{fig:ExampleD_photodetection},~\ref{fig:ExampleD_hom}
    \\[0.1cm]
    Average current    
    & Eqs.~\eqref{ExampleA_J}, \eqref{ExampleA_Jss} 
    & Eq.~\eqref{ExampleB_Jss} 
    & $J_{\rm diff} = 0$ 
    & Sec.~\ref{sec:parametric_oscillator}
    \\[0.1cm]    
    $F(\tau)$, $S(\omega)$ or $D$ 
    &
    Fig.~\ref{fig:ExampleA_FS},  Eq.~\eqref{example_A_power_spectrum} 
    & Eq.~\eqref{ExampleB_D}
    & Figs.~\ref{fig:quantum_diffusion_power}, ~\ref{fig:ExampleC_D} 
    &
    Sec.~\ref{sec:parametric_oscillator}
    \\[0.1cm]
    $P(n,t)$ or $C(\chi)$
    & Fig.~\ref{fig:fcs_prob}
    & Secs.~\ref{sec:scgfexb},\ref{sec:saddlexB}
    & Fig.~\ref{fig:fcs_prob_diffusion_example}
    & Fig.~\ref{fig:Pnt_exampleD}
    \end{tabular}
    \label{tab:examples}
\end{table*}

\section{\label{sec:currents}Quantum master equations and output currents}

\subsection{Basic setting}

In this tutorial we will focus on systems  evolving according to quantum master equations (QMEs) of the form 
\begin{equation}
\label{M}
    \frac{d\rho}{dt} = \mathcal{L}\rho = -i[H,\rho] + \sum\limits_{k=1}^\nops \mathcal{D}[L_k]\rho.
\end{equation}
The superoperator $\mathcal{L}$ is called the Liouvillian. 
The first term is the unitary dynamics where
$H$ is the system Hamiltonian. The second term encompasses the dissipation, with  $\mathcal{D}[L]=L \rho L^\dagger - \frac{1}{2} \{ L^\dagger L, \rho\}$ being a Lindblad dissipator. 
\rev{The set of operators $\{L_k\}$, which depend on the problem at hand, are called \emph{jump} operators. 
Below we refer to each $k$ value as a \emph{jump channel}. 
That is, we might say ``$L_k$ is the jump operator of channel $k$.'' The meaning of different ``channels'' will become clear once we discuss examples in Sec.~\ref{sec:examples}. 
}
The solution of Eq.~\eqref{M} is~\footnote{All results can be generalized to time-dependent Liouvillians, e.g. when the Hamiltonian is time-dependent. One must simply replace $e^{\mathcal{L}t}$ with a time-ordered exponential $\mathcal{T} e^{\int_0^t dt' \mathcal{L}_{t'}}$, where $\mathcal{T}$ is the time-ordering operator.}
\begin{equation}\label{M_sol}
    \rho(t) = e^{\mathcal{L} t} \rho_0.
\end{equation}
Often (but not always~\cite{Buca_2012}), for any initial state, this solution will evolve towards a unique steady-state $\rhoss$, which is the solution of 
\begin{equation}\label{steady_state}
    \mathcal{L} \rhoss = 0.
\end{equation}
We will henceforth assume this is the case, unless stated otherwise.
Equation~\eqref{M} encompasses a myriad of interesting problems, across various quantum platforms. 
We assume the reader has some familiarity with equations of this form. Throughout the tutorial, we set $\hbar = k_B = 1$.

\subsection{Examples}
\label{sec:examples}
\begin{figure}
    \centering
    \includegraphics[width=\columnwidth]{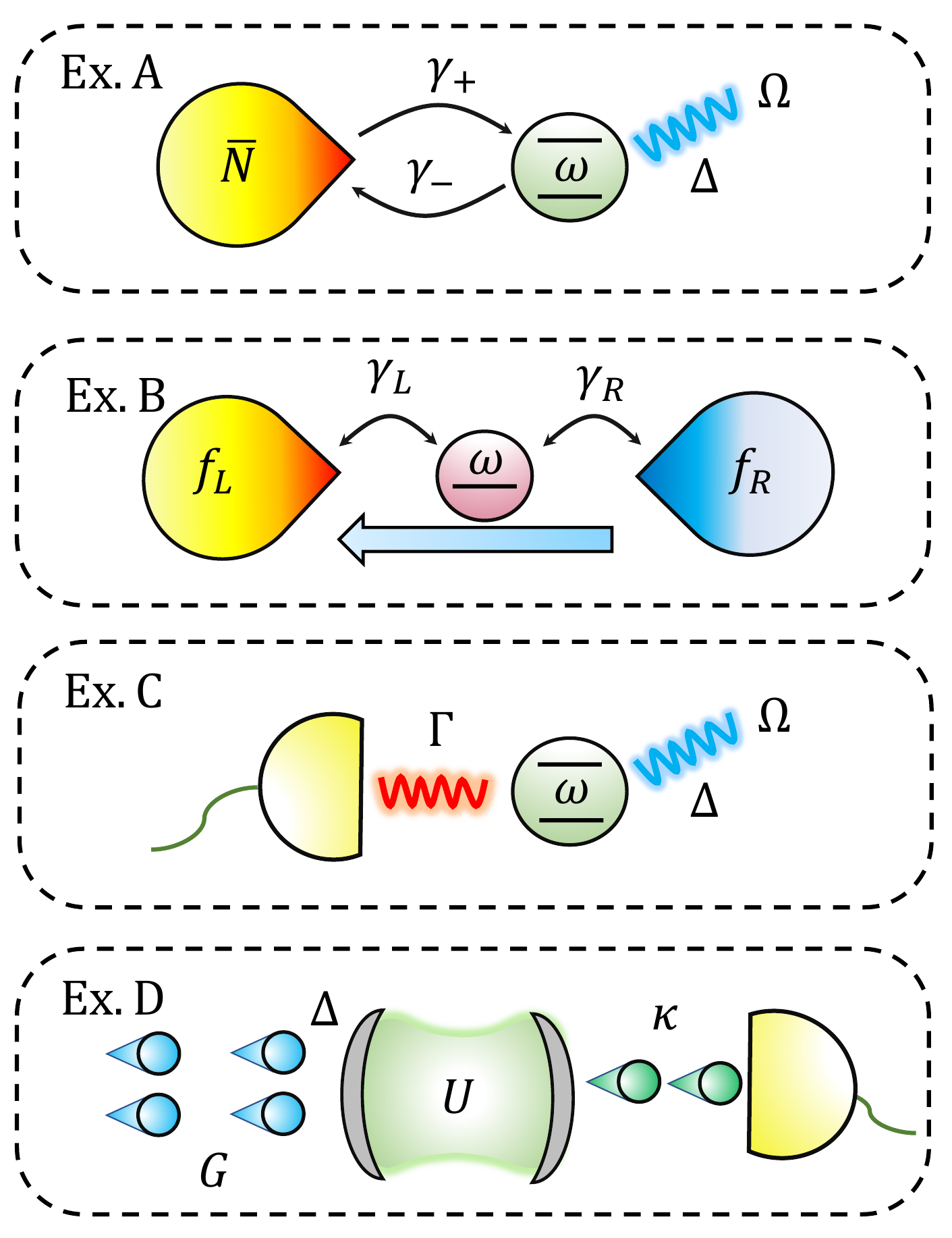}
    \caption{Depictions of the 4 basic examples discussed throughout the tutorial. See Table~\ref{tab:examples} for more details.}
    \label{fig:examples}
\end{figure}

We now list four examples that will be used throughout the tutorial, labeled A, B, C and D. 
These examples have been chosen specifically for their pedagogical value and their ubiquitous implementation, both experimentally and theoretically.
The examples are depicted pictorially in Fig.~\ref{fig:examples}, and a summary of the main results for each one (discussed throughout the review) is given in Table~\ref{tab:examples}.

\subsubsection{Example A. Coherently driven qubit coupled to a thermal reservoir}
\label{ssec:exampleA}

The first example is a single qubit  driven off-resonantly at a detuning $\Delta$ and  Rabi frequency $\Omega$. The Hamiltonian is
\begin{equation}
\label{ExampleA_H}
    H = \frac{\Delta}{2} \sigma_z + \Omega \sigma_x 
    = \Delta \sigma_+\sigma_- + \Omega(\sigma_+ + \sigma_-) - \Delta/2
    ,
\end{equation}
where $\sigma_{x,y,z,+,-}$ are Pauli matrices.
We work in the computational basis, $|0\rangle = \dwket$
and $|1\rangle = \upket$ (eigenstates of $\sigma_z$). 
The above model describes various physical scenarios. One is a two-level atom driven by a strong coherent laser field.
The Schr\"odinger picture Hamiltonian is $H(t) = \omega \sigma_+ \sigma_- + \Omega (\sigma_+ e^{-i \omega_d t} + \sigma_- e^{i \omega_d t})$, where $\omega$ is the atom's transition frequency and $\omega_d$ is the laser drive frequency. Moving to a rotating frame with $e^{i \omega_d t \sigma_+ \sigma_-}$ removes the time dependence and detunes the frequency to $\Delta = \omega-\omega_d$ yielding Eq.~(\ref{ExampleA_H}).

We assume that the qubit is coupled to a thermal bath, described by the QME
\begin{equation}\label{ExampleA_M}
    \frac{d\rho}{dt} = - i [H,\rho] + \gamma (\bar{N}+1) \mathcal{D}[\sigma_-]\rho + \gamma 
    \bar{N}\mathcal{D}[\sigma_+]\rho,
\end{equation}
where $\gamma>0$ is the coupling strength and $\bar{N} = (e^{\beta \omega}-1)^{-1}$ is the Bose-Einstein distribution, with $\beta = 1/T$.
The two dissipators describe the emission and absorption of thermal photons.
Notice that even at zero temperature ($\bar{N}=0$) the first dissipator does not vanish: this corresponds to spontaneous emission.
If $\Omega = 0$ the steady-state is diagonal in the $\sigma_z$ basis, with $\langle \sigma_z \rangle = -1/(2\bar{N}+1)$.
When $\Omega\neq 0$, the steady-state will involve both populations and coherences in the computational basis. 

\rev{The master equation~\eqref{ExampleA_M}, as well as the ones for the other examples below, are usually a good approximation when the coupling constant $\gamma$ to the reservoir is small. 
Strong system-environment coupling leads to various complications and corresponds to a vast field of research, with different approaches having been developed in the literature. Recent overviews can be found in~\cite{Landi_2022,Trushechkin_2022}.
Here we only mention that many of these approaches still involve appropriately modified master equations. 
Hence, the overarching ideas of this tutorial continue to apply. 
}

\subsubsection{Example B. Quantum dot coupled to two fermionic leads}
\label{ssec:exampleB}
The second example consists of a single quantum dot, described by fermionic operators $c, c^\dagger$  satisfying the anti-commutation relation $\{c,c^\dagger\} = 1$ and $c^2 = 0$.
The Hamiltonian is
\begin{equation}
\label{ExampleB_H}
    H = \omega c^{\dagger}c\,,
\end{equation}
where $\omega$ is the energy of the dot.
For fermions, terms such as $c + c^\dagger$ are not allowed due to parity superselection rules~\cite{Wick1952}. 
We assume that this dot is connected to two fermionic reservoirs,  each at a temperature $T_\alpha$ and chemical potential $\mu_\alpha$, with $\alpha = L, R$. 
The corresponding master equation is 
\begin{equation}
    \label{ExampleB_M}
    \frac{d\rho}{dt} = - i [H,\rho] + \sum_{\alpha = L, R}\gamma_-^\alpha \mathcal{D}[c]\rho + \gamma_+^\alpha \mathcal{D}[c^{\dagger}]\rho,
\end{equation}
where 
\begin{equation}\label{ExampleB_parametrization}
    \gamma_-^\alpha = \gamma_\alpha (1-f_\alpha), 
    \qquad 
    \gamma_+^\alpha = \gamma_\alpha f_\alpha,
\end{equation}
with $\gamma_\alpha$ being the coupling strength and $f_\alpha = (e^{\beta_\alpha(\omega - \mu_\alpha)} +1)^{-1}$ the Fermi-Dirac distribution of reservoir $\alpha = L, R$. 
Notice that since the two baths involve the same jump operators, we could also group them and define effective rates $\gamma_\pm = \gamma_\pm^{R} + \gamma_\pm^{R}$ \cite{goan_2001}.
But since they represent different physical processes, it will be important to keep them separate. 
We also note that, mathematically speaking, this model is equivalent to Example A with $\Omega = 0$ [Eq.~\eqref{ExampleA_H}], provided we map $\sigma_- \to c$ and relate $\bar{N} = f/(1-2f)$ (the strengths $\gamma_\alpha$ also change).

\subsubsection{Example C. Coherently driven qubit subject to dephasing}
\label{ssec:exampleC}

The third example is a single qubit undergoing pure dephasing.
The Hamiltonian is  identical to Eq.~(\ref{ExampleA_H}), but the master equation changes to
\begin{equation}\label{ExampleC_M}
    \frac{d\rho}{dt} = - i [H,\rho] + \Gamma \mathcal{D}[\sigma_z]\rho,
\end{equation}
where $\Gamma$ is the rate of dephasing. 
While the two dissipators in Eq.~\eqref{ExampleA_M} involve jumps between the qubit's computational basis, the dephasing only causes decoherence in this basis. 
This arises, for instance,  from fast and random fluctuations in the energy difference between the computational basis states \cite{Kiely2021}. 
For example, in quantum dots (with $\sigma_z \to 2 c^\dagger c - 1$) 
the electron couples via the Coulomb interaction to any nearby charges, whose fluctuations lead to dephasing. 
The dissipator in this case could be equivalently written as $2\Gamma \mathcal{D}[c^\dagger c]$.
As will be discussed, dephasing also appears whenever an observable, in this case $\sigma_z$, is being continuously monitored.

For $\Omega\neq 0$, the  steady-state of~\eqref{ExampleC_M}  is unique and given by the maximally mixed state $\rho = \id_2/2$. 
For $\Omega = 0$, any state diagonal in the computational basis is a valid steady state (that is, the steady state is no longer unique).

\subsubsection{Example D. Parametrically driven Kerr oscillator}
\label{ssec:exampleD}

The last example is a parametrically driven optical cavity subject to single-photon losses and a Kerr non-linearity \cite{roberts_driven-dissipative_2019}. 
This system can be modeled by a single bosonic mode, with frequency $\omega_c$ and annihilation operator $a$. 
The Hamiltonian in a frame rotating at twice the pump frequency, $\omega_p$,  reads
\begin{equation}\label{ExampleD_H}
    H = \Delta a^\dagger a + \frac{1}{2}\big(Ga^{\dagger 2} + G^* a^2 \big) + \frac{U}{2} a^\dagger a^\dagger a a,
\end{equation}
where $G$ is the 'two-photon' pumping strength, $\Delta = \omega_c - \omega_p/2$ is the detuning and $U$ is the Kerr non-linearity. 
The system is also subject to photon losses from the cavity, which we model by the master equation
\begin{equation}\label{ExampleD_M}
    \frac{d\rho}{dt} = -i [H,\rho] + \kappa \mathcal{D}[a]\rho,
\end{equation}
with $\kappa$ being the photon loss rate.
Sometimes, it is also necessary to include two-photon losses,  described by the dissipator $\mathcal{D}[a^{2}]$, but  we assume  these can be neglected for simplicity. 

\begin{figure}
    \centering
    \includegraphics[width=\columnwidth]{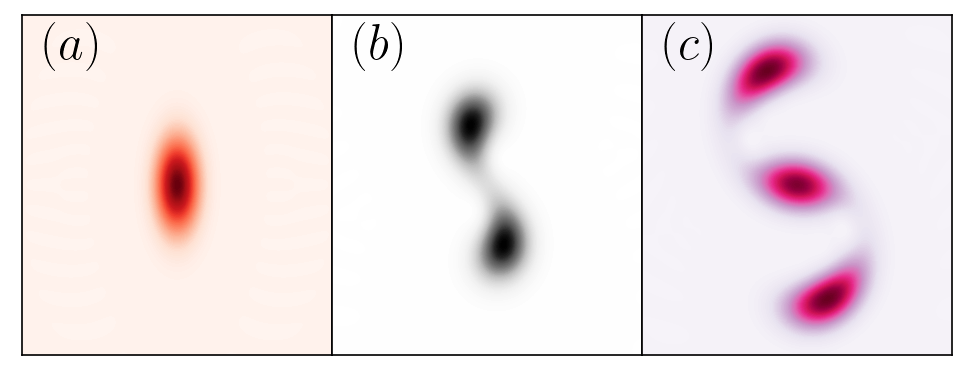}
    \caption{Steady-state Wigner function 
    \rev{$W(x,p)=\frac{1}{\pi} \int_{-\infty}^\infty dy~\langle x-y| \rhoss|x+y\rangle e^{2i py}$ (where $|x\rangle$ are the eigenstates of position) 
    }
    for Example D \rev{(Sec.~\ref{ssec:exampleD})}. The horizontal and vertical axes refer to the field quadratures (position and momentum), not shown for clarity. 
    Parameters: (a) $(G, U, \Delta) = (0.3i, 0, 0)$, (b)  $(1, 1/3, 0)$, and (c)  $(1, 1/3, -2)$, with $\kappa = 1$.
    In all cases, the Wigner function is non-negative. The color scales are omitted for visibility. These three examples exhibit the rich landscape of possible states that can be studied in Example D.
    }
    \label{fig:ExampleD_wigner}
\end{figure}

If $U=0$, the model is said to be Gaussian (to be studied in Sec.~\ref{sec:Gaussian}) and can be solved analytically. 
The steady-state in this case is a squeezed thermal state with
\begin{equation}\label{ExampleD_aves_ss}
    \langle a^\dagger a \rangle = \frac{2 |G|^2}{\kappa^2 + 4 \Delta^2 - 4 |G|^2},
    \qquad 
    \langle a a \rangle = - \frac{ G(2\Delta + i \kappa)}{\kappa^2 + 4 \Delta^2 - 4|G|^2}.
\end{equation}
The steady-state is only stable when  $4|G|^2 < \kappa^2 + 4 \Delta^2$, as is clear from the denominators. 
Conversely, if $U\neq 0$, the steady-state is always stable. 
In this case--- known as the Parameterically Pumped Kerr (PPK) model--- a rich variety of properties emerge.
We illustrate this in Fig.~\ref{fig:ExampleD_wigner}, where we 
plot the Wigner function of the steady-state for different choices of the parameters. 
Figure \ref{fig:ExampleD_wigner}(a) is the squeezed state~\eqref{ExampleD_aves_ss}, while (b) and (c) are different states that emerge when $U\neq 0$. 
For more details on the phase diagram of this model, see Refs.~\cite{kewming_diverging_2022,Mirrahimi2014}.
\rev{We also mention that the steady-state density matrix of this model can be found analytically, either using the generalized P-function~\cite{Drummond_1980,bartolo_exact_2016}, or  the coherent quantum absorber method~\cite{Stannigel2012,roberts_driven-dissipative_2019}. However, this does not give access to current fluctuations. Whether these solution methods can be generalized to also encompass that is, to the best of our knowledge, an open and interesting question.}

\subsection{Quantum jumps}
\label{sec:output_currents}

\begin{figure}
    \centering
    \includegraphics[width=0.4\textwidth]{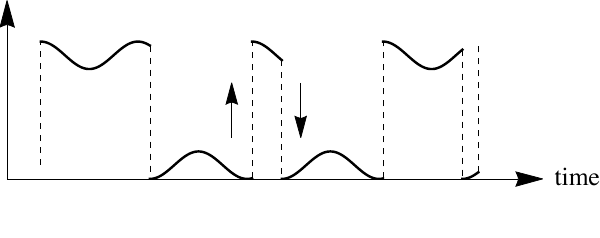}
    \caption{Depiction of an unraveled master equation into a stochastic jump equation Eq.~(\ref{quantum_jumps_SME}). The dynamics of an observable will  exhibit smooth evolution punctuated by sudden jumps.}
    \label{fig:jump_drawing}
\end{figure}

We return now to the general master equation~\eqref{M}, and introduce the notion of quantum jumps and output currents.
Inside the dissipators $\mathcal{D}[L_k]$, 
\rev{the terms of the form $L_k \rho L_k^\dagger$ play a very different role than the terms $L_k^\dagger L_k \rho$ and $\rho L_k^\dagger L_k$. 
For instance, in the case of a qubit, if $L = \sigma_-$ and $\rho = \upket \upbra$ then
$\sigma_- \upket \upbra \sigma_+ = \dwket\dwbra$,  causes the qubit to transition to the down state. 
Conversely $\sigma_+\sigma_- \upket \upbra  = \upket\upbra$ does not.
As we will see in what follows, the terms $L_k \rho L_k^\dagger$ will be associated to a quantum jump happening in channel $k$. 

To make quantum jumps precise, one must recall the concepts of quantum operations and generalized measurements~\cite{nielsen_2012}.
Any map taking density matrices to density matrices (known as completely positive and trace preserving, CPTP) can be written as the quantum operation 
\begin{equation}\label{Kraus_channel}
    \rho \to \sum_j M_j \rho M_j^\dagger, 
\end{equation}
where $\{M_j\}$ can be any set of operators satisfying the normalization condition $\sum_j M_j^\dagger M_j = 1$. 
A quantum operation, in turn, can always be interpreted as a generalized measurement, where the labels $j$ represent the different measurement outcomes (the number of which can be arbitrary). 
The probability that outcome $j$ is observed is
\begin{equation}\label{Kraus_outcome_probability}
    p_j = \tr(M_j \rho M_j^\dagger).
\end{equation}
And, if the outcome was $j$, the state of the system is updated as 
\begin{equation}\label{Kraus_conditional_outcome}
    \rho \to \frac{M_j\rho M_j^\dagger}{p_j},
\end{equation}
where the factor of $p_j$ is included to make sure the state is properly normalized. 
This represents the updated state of the system, \emph{conditioned} (or post-selected) on knowing that the outcome was $j$. 
This ``conditional state'' is therefore random, in the sense that $j$ is sampled randomly with probability $p_j$. 
Conversely, if we do not record the outcome, the updated state would be the statistical average 
\begin{equation*}
    \sum_j p_j \frac{M_j \rho M_j^\dagger}{p_j} = \sum_j M_j \rho M_j^\dagger,
\end{equation*}
which is precisely the original CPTP map~\eqref{Kraus_channel}. 
We refer to 
Eq.~\eqref{Kraus_channel} as the \emph{unconditional} state, in the sense that it is not conditioned on the actual outcome. 
The above decomposition in terms of generalized measurement outcomes is called an \emph{unravelling}~\cite{carmichael:book} of the quantum operation.
Unravellings are not unique~\footnote{If $N_\alpha = \sum_j U_{j\alpha} M_j$ with coefficients $U_{j\alpha}$ forming a unitary matrix, then the quantum operation will be exactly the same, but the outcomes $\alpha$ will be different. We will in fact exploit this when we talk about quantum diffusion, which correspond to a different unravelling than quantum jumps.}.
And, of course, implementing a specific one in the laboratory might not be simple at all. 
But the above discussion establishes that, at least in principle, every quantum operation can be unravelled into a set of of possible measurement outcomes and corresponding conditional states. 
}

\rev{We now apply the above ideas to the evolution~\eqref{M_sol} of the master equation. 
Focusing on an infinitesimal time $dt$, we can expand the exponential and write  
}
\begin{align}\label{exponential_evolution_infinitesimal}
    e^{\mathcal{L} dt} \rho 
    = \rho - i dt\big(\Hnh \rho - \rho \Hnh^\dagger\big) + dt \sum_{k=1}^\nops L_k \rho L_k^\dagger + \mathcal{O}(dt^2),
\end{align}
where 
\begin{equation}\label{H_non_hermitian}
    \Hnh = H - \frac{i}{2}\sum_k L_k^\dagger L_k,
\end{equation}
is a \rev{kind of} non-Hermitian Hamiltonian.
\rev{Equation~\eqref{exponential_evolution_infinitesimal} can be written  as \rev{a quantum operation~\eqref{Kraus_channel}}
\begin{equation}\label{kraus_channel}
    e^{\mathcal{L} dt} \rho =
     M_0 \rho M_0^\dagger +
     \sum_{k=1}^\nops M_k \rho M_k^\dagger,
\end{equation}
with Kraus operators 
\begin{equation}
\begin{aligned}\label{kraus}
    M_0 &= 1 - i \Hnh dt, \\[0.2cm]
    M_k &= \sqrt{dt} L_k, \qquad
    k = 1,\ldots,\nops.
\end{aligned}
\end{equation}
}One may verify that $M_0^\dagger M_0 + \sum_{k=1}^\nops M_k^\dagger M_k = 1 + \mathcal{O}(dt^2)$, so that the Kraus operators are indeed  normalized.
\rev{We have therefore unravelled one infinitesimal time step of the master equation evolution. But, of course, we can now  keep doing this for multiple steps. 
This is what is called the quantum jump unravelling of the master equation. 
At each time step there will either be a jump in channel $k$, or no jump (associated with $M_0$). 
The probability that a jump happens, according to Eq.~\eqref{Kraus_outcome_probability}, is }
\begin{equation}\label{pk}
    p_{k} = \tr\big\{ M_k \rho M_k^\dagger\big\} 
    = dt~\tr\big\{ L_k \rho L_k^\dagger\}.
\end{equation}
\rev{Since $dt$ is infinitesimal, it is always much more likely that no jump will occur (probability $p_0 = 1- \sum_{k=1}^\nops p_k$). 
In this unravelling, therefore, the dynamics is described by a series of random jumps occurring at random times (and in random channels), followed by a smooth no-jump evolution (as illustrated in Fig.~\ref{fig:jump_drawing}). Interestingly,}
the no-jump evolution (under $M_0$) is not unitary, but is described by the non-Hermitian Hamiltonian $\Hnh$. 
\rev{This happens because this evolution is conditional; that is, it is the evolution \emph{given} that no jump happened. And not observing a jump is still information, so that we still update our knowledge about the system.}

Quantum jumps are a consequence of the open nature of the problem; i.e., the interaction between system and environment. 
\rev{But in many experiments, they can  be connected with outcomes (clicks) in specific detectors placed within the environment.}
For instance, in a two-level atom, $\sigma_- \rho \sigma_+$ represents the emission of a photon which could be detected \rev{by placing a photo-detector close to it. A similar idea occurs in electron transport across quantum dots (example C), where the term $c \rho c^\dagger$ in Eq.~\eqref{ExampleB_M} extracts an electron from the system, while $c^\dagger \rho c$ injects an electron. In principle, both of these processes could be detected by observing the charge of the reservoirs. Of course, whether or not this is experimentally feasible is problem-specific. Furthermore, it might involve a non-ideal efficiency (that is, some jumps might be missed). 
Notwithstanding, what matters is that, at least in principle, we can associate quantum jumps with specific detectors located in the environment, with one detector for each channel $k$. 
}
This is \rev{fundamental the paradigm that we will use here} to bridge the gap between quantum dynamics and classical outcomes. 

The Kraus decomposition~\eqref{kraus_channel} allows us to describe the stochastic effects that monitoring these clicks would have on the system evolution. We denote the density matrix \textit{conditioned} on a particular sequence of clicks as $\rho_c(t)$.
At each time step, we choose one of the channels $k = 0,1,\ldots, \nops$ with probabilities $p_k^c=dt~\tr\big\{L_k^\dagger L_k \rho_c \}$ (and $p_0^c = 1- \sum_{k=1}^\nops p_k^c$).
\rev{If the outcome is $k = 1,\ldots,\nops$ we change the state of the system as 
\begin{equation}
    \rho_c(t+dt) = \frac{L_k \rho_c(t) L_k^\dagger}{\tr(L_k \rho(t) L_k^\dagger)}.
\end{equation}
Or if the outcome is $0$, we change it as 
\begin{equation}
    \rho_c(t+dt) = \rho_c(t) -i dt\big(\Hnh \rho_c(t) - \rho_c(t) \Hnh^\dagger\big).
\end{equation}
This yields a quantum trajectory: a stochastic evolution of the system density matrix conditioned on specific sets of clicks. 
In practice, this is not the most efficient way of simulating quantum trajectories. A more efficient method is the Monte Carlo wavefunction approach~\cite{Molmer1996} which we discuss in Appendix~\ref{app:MCWF}.
}

Let us introduce classical random variables $dN_k(t)=0,1$, which take the value $1$ when a jump occurs in channel $k$ and $0$ otherwise. 
That is 
\begin{equation}\label{conditional_prob_dN}
    P\big(dN_k(t)=1|\rho_c(t)\big) \equiv p_k^c(t) = dt~\tr\big\{L_k^\dagger L_k \rho_c(t) \} .
\end{equation}
Here we write $P\big(dN_k(t)=1|\rho_c(t)\big)$ with a conditional in $\rho_c(t)$ to emphasize that it depends on the entire stochastic trajectory $dN_k(t')$ for $t'<t$.
These random variables $dN_k(t)$, as we will see, play a fundamental role as they describe the connection between the quantum dynamics and classical observable outcomes. 
In terms of the $dN_k(t)$ we can also write the conditional updating as a stochastic master equation
\begin{align}
    \label{stochjump1}
      \rho_c(t+dt) &= \frac{(1- \sum_{k=1}^\nops dN_k(t))}{1-\sum_{k=1}^\nops p_k^c(t)} M_0 \rho_c(t) M_0^\dagger \\[0.2cm]
\nonumber & + \sum_{k=1}^\nops dN_k(t) \frac{M_k\rho_c(t)M^\dagger_k}{p_k^c(t)},
\end{align}
Since $p_k^c \propto dt$, we never observe more than a single jump during a time-step of $dt$, which implies the relations
\begin{equation}
    \label{eq:increment}
    dN_k dN_l = dN_k\delta_{kl},\hspace{1cm}dtdN_k = 0,
\end{equation}
with $\delta_{kl}$ denoting the Kronecker delta. The last equation describes the defining properties of a stochastic point (or jump) process \cite{jacobs_2010}, \rev{and will be explained in more detail below, in Eq.~\eqref{Ito_calculus_dN}. 
}

With the help of Eq.~\eqref{eq:increment}, we can expand the stochastic master equation~\eqref{stochjump1} into leading order in $dt$, allowing us to write it as~\cite{dalibard_1992,carmichael:book,Wiseman_2009}
\begin{equation}
    \label{quantum_jumps_SME}
    d\rho_c = dt\mathcal{L}\rho_c+\sum_{k=1}^\nops\left(dN_k-dt\langle L_k^\dagger L_k\rangle_c\right)\left(\frac{L_k\rho_cL_k^\dagger}{\langle L_k^\dagger L_k\rangle_c}-\rho_c\right),
\end{equation}
where $d\rho_c=\rho_c(t+dt)-\rho_c(t)$ and $\langle O\rangle_c = \tr\{O\rho_c\}$. This is a non-linear equation due to the appearance of the averages $\langle L_k^\dagger L_k\rangle_c$. 
It is also worth mentioning that \rev{if a state is initially pure, it will remain so under the evolution~\eqref{stochjump1}.} 
If this is the case, then instead of~\eqref{stochjump1} we can write 
\begin{equation}\label{stochjump1_pure_state}
    |\psi_c(t+dt)\rangle = \sum_{k=0}^\nops dN_k(t) \frac{M_k |\psi_c(t)\rangle}{\sqrt{p_k^c}}.
\end{equation}
Expanding in the same way that we did to get to~\eqref{quantum_jumps_SME}, this becomes~\cite{dalibard_1992,carmichael:book}
\begin{equation}\label{quantum_jumps_SME_pure_state}
\begin{aligned}
    d|\psi_c\rangle &= dt\Bigg(-i \Hnh + \frac{1}{2}\sum_{k=1}^\nops \langle L_k^\dagger L_k \rangle_c 
    \Bigg) |\psi_c\rangle \\[0.2cm]
    &\qquad + \sum_{k=1}^\nops dN_k(t) \Bigg( \frac{L_k}{\sqrt{\langle L_k^\dagger L_k\rangle}_c} -1\Bigg)|\psi_c\rangle.
\end{aligned}
\end{equation}
For most time steps the second term will vanish, and the system will evolve smoothly according to the first term, which essentially involves the non-Hermitian Hamiltonian $\Hnh$ in Eq~\eqref{H_non_hermitian}. 
Conversely, when a jump occurs one of the $dN_k$ will be 1 and the first term in Eq.~\eqref{quantum_jumps_SME_pure_state} can be neglected, since it is of order $dt$. 


The unconditional density matrix, $\rho(t) = e^{\mathcal{L} t}\rho(0)$, is recovered  by averaging over the ensemble of all possible sequences of clicks; i.e., by repeating the dynamics multiple times, with the same initial condition and different random sequences of clicks. 
We will denote this expectation value by
\begin{equation}
\label{quantum_jumps_ensemble_average}
    \rho(t) = E[\rho_c(t)].
\end{equation}
A similar average must therefore hold at the level of $p_k$ and $p_k^c$, which allows us to write
\begin{equation}\label{prob_dN}
    p_k = E[p_k^c]  = dt \tr\big\{ L_k^\dagger L_k \rho\big\} \equiv P\big(dN_k = 1\big).
\end{equation}
If we average Eq.~\eqref{stochjump1} over the $dN_k(t)$ we should therefore recover the unconditional map~\eqref{kraus_channel}. 
To actually carry out this computation however, care must be taken with the fact that $dN_k(t)$ and $\rho_c(t)$ are not statistically independent since, as highlighted in~\eqref{conditional_prob_dN}, we use $\rho_c(t)$ to generate $dN_k(t)$. 
Instead, the following identity holds~\footnote{This can be proven using the law of total expectation, which reads $E(X) = E_Y\big( E(X|Y) \big)$.
In simplified notations, let $dN$ be a random variable with $E(dN) = \lambda(X)$. Then
$E\big[ dN g(X)\big] = E_X\big[ E(dN g(X)|X) \big] = E_X\big[ g(X) E(dN|X) \big]  = E_X\big[g(X) \lambda(X)\big]$.
This is exactly the structure in the present problem, with $X$ actually representing $\rho_c(t)$ or, what is equivalent, the set of all past outcomes $dN_k(t')$ with $t'<t$. 
}:
\begin{equation}\label{Ito_calculus_dN}
    E\Big[ dN_k(t) g\big(\rho_c(t)\big)\Big] = E\Big[ p_k^c(t) g\big(\rho_c(t)\big) \Big],
\end{equation}
for any function $g(\rho_c)$. 
Using this relation in Eq.~\eqref{stochjump1} yields 
\begin{equation}
    E\Bigg[ dN_k(t) \frac{M_k\rho_c(t) M_k^\dagger}{p_k^c(t)} 
    \Bigg] = 
    E\Big[ M_k \rho_c(t) M_k^\dagger\Big] = M_k \rho(t) M_k^\dagger, 
\end{equation}
so that the unconditional dynamics~\eqref{kraus_channel} is indeed recovered. 
\rev{Eq.~\eqref{Ito_calculus_dN} also shows that, if we take $g(\rho_c) = dt$ (a constant) then $E[dN_k(t) dt] = E[p_k^c(t) dt] \propto dt^2$. The quantity $dN_k dt$ is therefore always of order $dt^2$, which explains why we can set $dN_k dt =0$, as in Eq.~\eqref{eq:increment}.}

\subsection{Output currents}

The increments $dN_k$ can be understood as the change in a \emph{counting variable} $N_k(t)$, which counts the net number of jumps that took place in channel $k$ between $[0,t]$ (or the number of clicks in a detector that measures those jumps). 
Often, though, we are interested in quantities that involve multiple channels.
For instance, we might wish to count the \emph{particle current}, which is the number of excitations exchanged between system and bath in a dissipator such as  Eq.~\eqref{ExampleA_M}. In this case, $\mathcal{D}[\sigma_+]$ should count as $+1$, and $\mathcal{D}[\sigma_-]$
as $-1$. 
To encompass this kind of situation, we define the \emph{total charge}
\begin{equation}\label{total_charge}
    N(t) = \sum_k \nu_k N_k(t), 
\end{equation}
where $\nu_k$ are weights associated to whatever physical process one is dealing with ($\nu_\pm = \pm1$ in the previous example). 
The stochastic current, in turn, is defined as the rate of change of the charge:
\begin{equation}\label{current_charge}
    I(t) = \frac{dN}{dt}, 
    \qquad 
    N(t) = \int\limits_0^t dt' I(t').
\end{equation}

The choice of $\nu_k$ is determined by what one wishes to describe. Here are some examples:
\begin{itemize}
    \item {\bf Particle current:} In Eq.~\eqref{ExampleA_M}, we choose $\nu_- = -1$ for $\mathcal{D}[\sigma_-]$ (emission) and $\nu_+=+1$ for $\mathcal{D}[\sigma_+]$ (absorption). 
    
    \item {\bf Energy current:} if each excitation carries a well-defined energy $\epsilon$, 
    we might similarly define $\nu_\pm = \pm \epsilon$. 
    
    \item {\bf Photon current:} in  Example D, Eq.~\eqref{ExampleD_M},  photons can only be lost to the environment, so $\nu=1$ gives the net photon current. Unlike the particle current, in this case we always have $N(t) \geqslant 0$ (the photon count can only go up). If there are 2-photon losses, we can also choose $\nu=1$ for $\mathcal{D}[a]$ and $\nu = 2$ for $\mathcal{D}[a^2]$, so that the jumps $a^2 \rho (a^{\dagger })^2$ count as two photons.
    
    \item {\bf Dynamical activity:} 
    This represents the net number of jumps, irrespective of their channels. It is obtained by setting $\nu_k=1$ for all $L_k$ in the QME. The dynamical activity has recently become popular in connection with the so-called  Kinetic Uncertainty Relations~\cite{Di_Terlizzi_2018,Prech_2023}.
\end{itemize}
One might also be interested in describing multiple current specimens at the same time, e.g.~particle currents to the left bath and to the right bath. 
These can be constructed as $N_\alpha(t) = \sum_k \nu_{\alpha k} N_k(t)$, with coefficients $\nu_{\alpha k}$, where $\alpha$ label the current type.
For simplicity, we will focus for now on a single current specimen but in Sec.~\ref{sec:multiple_currents} we show how the results can be generalized. 

The current $I(t)$ [or the charge $N(t)$] is a stochastic quantity. The average current is 
\begin{equation}\label{average_current_def}
    J(t) := E\big[ I(t) \big ] = \frac{1}{dt} E\big[dN(t)\big] = \frac{1}{dt}\sum_k \nu_k E\big[dN_k(t)\big ].
\end{equation}
Using Eq.~\eqref{prob_dN} we find 
\begin{equation}
\label{average_current}
    J(t) = \sum_k \nu_k \tr\big\{ L_k^\dagger L_k \rho(t)\big\}. 
\end{equation}
This is only an average, however, and therefore conveys only a limited amount of information about the stochastic current. 
Our goal in this tutorial is to develop the tools to go beyond the average, and also look at fluctuations.
In the remainder of this section, we discuss examples, and also introduce the paradigm of quantum diffusion, which is complementary to quantum jumps. 
The reader interested in skipping ahead, may go straight to Sec.~\ref{sec:fluctuations}. 

\subsubsection{Imperfect detection}

In actual experiments we often cannot detect the clicks with perfect efficiency. 
Introducing partial detection efficiency in the model is therefore crucial for describing various realistic situations. 
In our framework, this can be done seamlessly using the following trick.
Suppose the $\nops$ jump operators $L_k$ are each measured with an efficiency $\eta_k \in [0,1]$, such that $\eta_k=1$ means perfect efficiency. 
We can then double the number of jump operators by defining $L_k^0 = \sqrt{1-\eta_k} L_k$ and $L_k^1 = \sqrt{\eta_k} L_k$, so that Eq.~\eqref{M} is rewritten as
\begin{equation}
    \mathcal{L}\rho = -i[H,\rho] + \sum_{k=1}^\nops \mathcal{D}[L_k^0]\rho + \mathcal{D}[L_k^1]\rho.
\end{equation}
We now have $2\nops$ jump operators, 
and we can interpret $L_k^1$ as the set which is accessible for detection, and $L_k^0$ as a set that is inaccessible. 
In practice, this simply means that we need to construct the charge~\eqref{total_charge} using $\nu_k^0 = 0$, for the part that is inaccessible.

\subsubsection{Output current for Example A}
\label{sec:Rabi oscillations in a single qubit}

There are two dissipators in Eq.~\eqref{ExampleA_M}, so there will be two counting variables $dN_-$ and $dN_+$, associated to the jump operators $\sigma_-$ and $\sigma_+$. 
To construct the particle current, for example, we use weights $\nu_\pm = \pm 1$, leading to $I(t) =  \frac{dN_{+}}{dt} - \frac{dN_{-}}{dt}$.
The average current is then
\begin{equation}
\begin{aligned}
    J(t) &=  \gamma \bar{N}\langle \sigma_{-}\sigma_{+}\rangle -\gamma(\bar{N}+1)\langle \sigma_{+}\sigma_{-}\rangle\\
    &= \gamma\Big( \bar{N}- (2\bar{N}+1) \langle \sigma_+\sigma_-\rangle\Big).
    \label{ExampleA_J}
\end{aligned}
\end{equation}
In the steady-state this reduces to 
\begin{equation}
    J = -\frac{\gamma  \Omega ^2}{ \left(\Delta ^2+2 \Omega ^2\right)+\gamma^2(\bar{N}+\tfrac{1}{2})^2}.
    \label{ExampleA_Jss}
\end{equation}
The average current is finite due to the Rabi drive, which is constantly re-exciting the qubit. 
It is also negative for the same reason: since excitations are constantly being created in the qubit, it is more likely for it to emit an excitation than to absorb it.

A much richer behavior is obtained by looking at $I(t)$ for individual trajectories; i.e., simulating $\rho_{c}$ using \rev{e.g. the algorithm in Appendix~\ref{app:MCWF}}.
The top panel in 
Fig.~\ref{fig:ExampleA_trajs} exemplifies the quantum jumps mixed in with the smooth, no-jump dynamics. 
The corresponding trajectory $N(t) = \int_0^t I(t') dt'$ is shown in the bottom panel. 
This represents the net  charge $N(t)$ that would be observed by an experimentalist able to count both absorption and emission events.
In this simulation we have chosen the parameters such that $J/\gamma\approx -0.4$.

\begin{figure}
    \centering
    \includegraphics[width=\columnwidth]{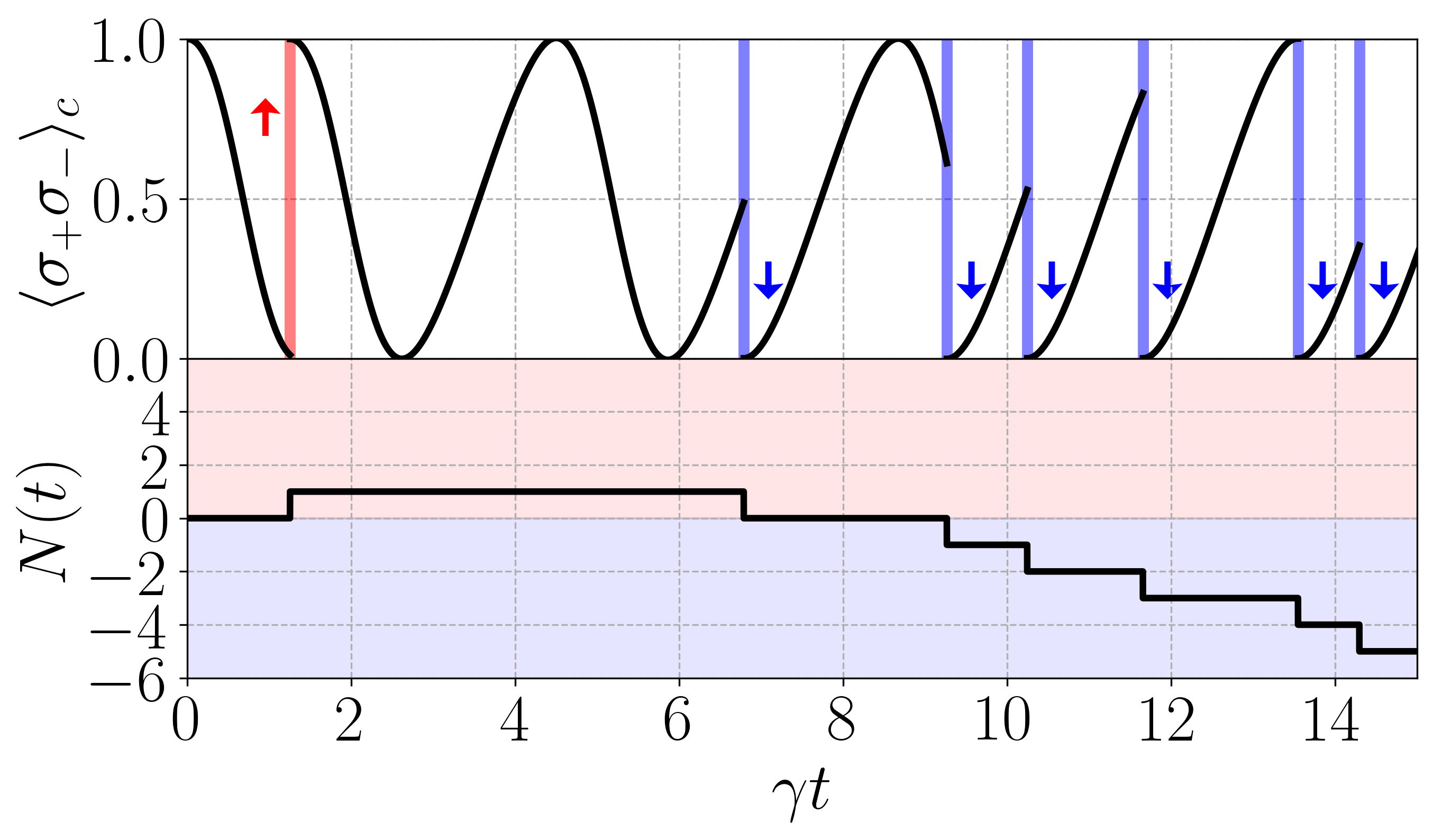}
    \caption{Quantum jump stochastic dynamics for Example A \rev{(Sec.~\ref{ssec:exampleA})}, with $\Delta=0$, $\Omega=\gamma$ and $\bar{N}=0.2$.
    The top figure shows the quantum trajectory associated with the conditional evolution $\langle \sigma_+\sigma_-\rangle_{c}$ as a function of time, where individual jump events are denoted by the arrows (red=absorption, blue=emission).
    The lower figure shows the net particle current $N(t)$ accumulated after a time $t$. 
    Due to the Rabi drive, which is constantly creating excitations, the current is always negative on average [Eq.~\eqref{ExampleA_Jss}]. 
    In this particular example  $J/\gamma \approx -0.4$.
    }
    \label{fig:ExampleA_trajs}
\end{figure}

We could also study the dynamical activity (jumps per unit time), obtained by using $\nu_+ = \nu_- = 1$. 
Instead of Eq.~\eqref{ExampleA_J}, in this case we have $K= \gamma(\bar{N}+1) \langle \sigma_+\sigma_-\rangle + \gamma \bar{N} \langle \sigma_- \sigma_+\rangle$.
An interesting particular case is when the Rabi drive vanishes, $\Omega = 0$.
In Eq.~\eqref{ExampleA_Jss}, this leads to $J=0$ since the system equilibrates with the bath; that is, because there is no Rabi drive to create excitations, the number of jumps into or out of the bath must balance out.
The dynamical activity, on the other hand, reads (for $\Omega=0$)
\begin{equation}
    K = \frac{2\gamma \bar{N} (\bar{N}+1)}{2\bar{N} +1},
\end{equation}
which is only zero if $\bar{N}=0$ (zero temperature). 
This happens because even in the absence of any drive, the system will continue to jump up and down due to the thermal fluctuations, so that even if the average current is zero (net in = net out), the activity overall is not zero.

\subsubsection{Output current for Example B}
\label{sec:Q quantum dot refrigerator}

\begin{figure}
    \centering
    \includegraphics[width=\columnwidth]{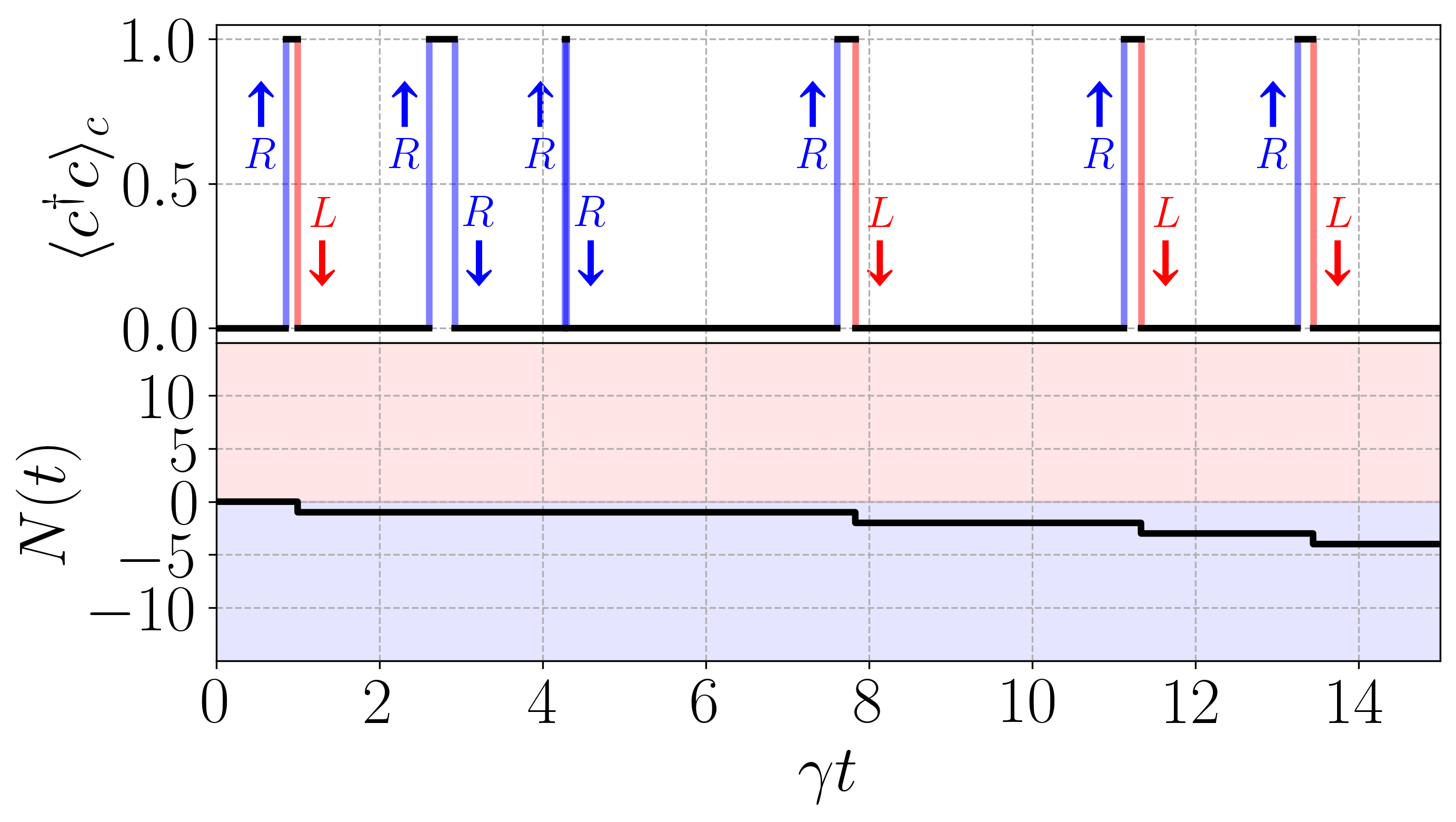}
    \caption{Stochastic trajectories for Example B \rev{(Sec.~\ref{ssec:exampleB})}, with $\gamma = \gamma_{L} = \gamma_{R}$, $T_{L} = 2\gamma, T_{R} = \gamma$, $\omega=21\gamma$, $\mu_{R}=20\gamma$, $\mu_{L}=10\gamma$. We have chosen the parameters such that there is a large chemical potential bias $\mu_{R} > \mu_{L}$, which causes the current to be  pumped against the temperature gradient. The top figure depicts the occupation of the dot $\langle c^{\dagger}c\rangle$ where we have labeled the color coded jumps from the right reservoir in blue, and red for the left. The bottom plot depicts the net charge $N(t)$ monitored from the left reservoir which decreases showing that particles are moving from the right reservoir (cold) to the hot reservoir (left). We count only particle events from the left reservoir $\nu_-^{L} = -1$, $\nu_+^{L} = +1$ and $\nu_\pm^{R} = 0$. Together with $\omega>\mu_R$, this implies the behavior of a refrigerator \cite{Potts_2019}.
    }
    \label{fig:ExampleB_trajs}
\end{figure}
Quantum trajectories for Example B are shown in  Fig.~\ref{fig:ExampleB_trajs}.
There are four dissipators
$\gamma_-^{R} \mathcal{D}[c]$,
$\gamma_-^{R} \mathcal{D}[c]$,
$\gamma_+^{R} \mathcal{D}[c^{\dagger}]$ and
$\gamma_+^{R} \mathcal{D}[c^{\dagger}]$, with the rates parametrized as in Eq.~\eqref{ExampleB_parametrization}. 
To compute the particle current from the left bath, for example, we may set $\nu_-^{R} = -1$, $\nu_+^{R} = +1$ and $\nu_\pm^{R} = 0$.
In the steady-state, we get 
\begin{equation}\label{ExampleB_Jss}
    J = \gamma_L(\langle c^\dagger c \rangle-f_L)= \frac{\gamma_L\gamma_R(f_R - f_L)}{\gamma_L+\gamma_R}.
\end{equation}
In some parameter ranges, this model can behave as a refrigerator, where a chemical potential bias $\mu_{R} > \mu_{L}$  pumps current against a temperature bias $T_{L} < T_{R}$~\cite{Esposito_2009_2, Esposito_2010,Potts_2019}.
This can be seen from the fact that $J$ depends on the  difference in Fermi distributions, not the difference in temperature. A well adjusted chemical potential could therefore change the sign of $J$. 
In Fig.~\ref{fig:ExampleB_trajs} we look at this from the perspective of individual quantum trajectories.
The dynamics in this case is clearly incoherent. 
Every time the dot receives an electron ($\langle c^\dagger c\rangle_c = 1$), it will remain there for a certain residence time, after which it will tunnel out to one of the two baths. 
At the level of the system's state, it is immaterial to which bath the electron tunnels to. But physically it matters, as this is what determines if heat flows to the hot or the cold bath. 
Because of the different Fermi distributions, emissions to one bath and absorptions from the other will be favored, resulting in the net average current~\eqref{ExampleB_Jss}.

\subsubsection{Output current for Example D}
\label{sec:exampleD_photo}

\begin{figure}
    \centering
    \includegraphics[width=\columnwidth]{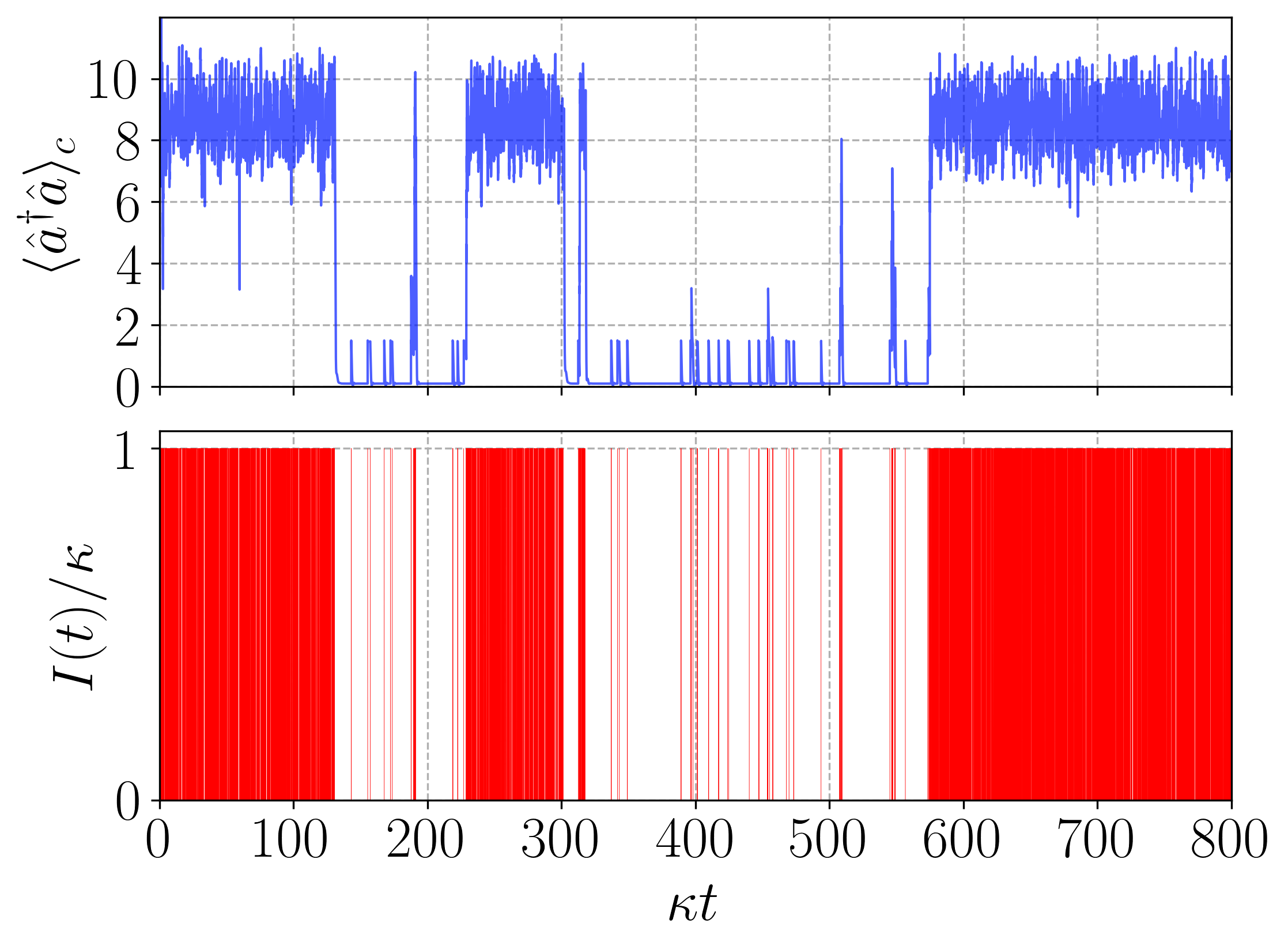}
    \caption{Quantum trajectories for direct photodetection of Example D \rev{(Sec.~\ref{ssec:exampleD})}.
    The parameters are fixed at $G/\kappa=1$, $U/\kappa=1/3$,  $\Delta/\kappa =2$, corresponding to the onset of the discontinuous transition (Fig.~\ref{fig:ExampleD_wigner}(c)). 
    (Top) Conditional mean photon number $\langle \hat{a}^{\dagger}\hat{a}\rangle_{\rm c}$  as a function of time during a single trajectory.
    (Bottom) Observed photocurrent $I(t)$, corresponding to a series of Dirac delta functions at each detection event.
    \rev{The rate of photodection events can tell us if the system is in a meta-stable state (outer lobes in Fig.~\ref{fig:ExampleD_wigner}(c)) but it cannot distinguish between the two meta-stable states}. Figure reproduced from Ref.~\cite{kewming_diverging_2022}.}
    \label{fig:ExampleD_photodetection}
\end{figure}

In Fig.~\ref{fig:ExampleD_photodetection} we plot an example trajectory of the PPK model with the same parameters as in Fig.~\ref{fig:ExampleD_wigner}(c). 
In the top panel we plot the conditional mean photon number $\langle a^{\dagger}a\rangle_{c}$ and below we plot the photocurrent $I(t)/\kappa$.
The dynamics here is seen to alternate between a bright and a dark phase.
It will be bright (many photon emissions) when the state is close to the outer lobes of the Wigner function, and it will be dark  when it is in the central lobe.
The system keeps transitioning  back and forth between these lobes, causing the current to stochastically alternate between bright and dark.

\subsection{Quantum diffusion and diffusive currents}
\label{sec:quantum_diffusion}
The \emph{quantum jump} approach connects clicks in a detector to the terms $L_k \rho L_k^\dagger$.
The master equation~(\ref{M}), however, is invariant under the transformation 
\begin{equation}\label{gauge}
    L_k \to L_k + \alpha_k, 
    \qquad 
    H \to H - \frac{i}{2} \sum_k\big(\alpha_k^* L_k - \alpha_k L_k^\dagger\big),
\end{equation}
where $\alpha_k$, henceforth referred to as \textit{reference currents}, are arbitrary constants (they can even be time-dependent).
We could therefore also consider quantum jumps associated to $(L_k + \alpha_k)\rho (L_k + \alpha_k)^\dagger$, which correspond to a different unravelling of the quantum master equation with 
$M_k = \sqrt{dt} (L_k + \alpha_k)$ ($M_0$ is also modified).
These modified jumps can also represent  clicks in a physical detector, as discussed in more detail below.

While the $\alpha_k$ are arbitrary, 
the big advantage, as we will see, is to consider the case where their magnitudes are very large. This causes the jumps to become very frequent.
To understand why this regime is interesting, let us look at how the average current~\eqref{average_current} changes: Writing $\alpha_k = |\alpha_k| e^{i \phi_k}$, we get 
\begin{align}\nonumber
    J(t) &= \sum_k \nu_k \tr\big\{ (L_k + \alpha_k)^\dagger (L_k + \alpha_k) \rho(t)\big\}
    \\[0.2cm]
    &= \sum_k \nu_k \Big(|\alpha_k|^2 + |\alpha_k| \langle x_k \rangle + \langle L_k^\dagger L_k\rangle\Big),
\label{average_current_gauge}
\end{align}
where we defined the \emph{quadrature} associated to each jump operator 
\begin{equation}
\label{quadrature}
    x_k = L_k e^{-i \phi_k} + L_k^\dagger e^{i \phi_k}.
\end{equation}
The first term in Eq.~\eqref{average_current_gauge} is just a constant shift $\nu_k |\alpha_k|^2$. 
The interesting part is the interplay between the second and third terms: If $|\alpha_k|$ is very large, the current will essentially be measuring $\langle x_k \rangle$ instead of $\langle L_k^\dagger L_k\rangle$. 
By adding a large reference current, we can thus access an entirely different observable. 

This motivates us to define a \emph{diffusive} stochastic current as 
\begin{equation}\label{Idiff}
    I_{\rm diff}(t) = \sum_{k=1}^\nops\frac{\nu_k}{|\alpha_k|} \Bigg( \frac{dN_k}{dt} - |\alpha_k|^2\Bigg),
\end{equation}
that is, we subtract from $dN_k/dt$ the constant offset $|\alpha_k|^2$, and then normalize the results by $|\alpha_k|$. 
The advantage of this is that the average current becomes, when $|\alpha_k|$ is large, 
\begin{equation}\label{diffusive_average_current}
   J_{\rm diff}(t)= E[I_{\rm diff}(t)]=
\sum_k \nu_k \langle x_k \rangle ,
\end{equation}
so that, on average, we are sampling a linear combination of the $x_k$.

The stochastic master equation~\eqref{quantum_jumps_SME} continues to hold after the transformation~\eqref{gauge}.
However, a more appropriate equation can be derived in the limiting case of large $|\alpha_k|$~\cite{belavkin:1989,gardiner_1992, carmichael:book, Wiseman_2009}.
The result, derived in Appendix~\ref{app:stochasticwiener}, reads
\begin{equation}
    \label{Diffusion_SME}
    d\rho_c = dt\mathcal{L}\rho_c+\sum_{k=1}^\nops dW_k \left[\mathcal{H}_k\rho_c-\langle x_k\rangle_c\rho_c\right],
\end{equation}
where 
\begin{equation}\label{H_k_superop}
    \mathcal{H}_k\rho_c = L_ke^{-i\phi_k}\rho_c+\rho_cL_k^\dagger e^{i\phi_k},
\end{equation}
are such that $\tr\big\{\mathcal{H}_k\rho_c\big\} = \langle x_k \rangle_c$. 
In Eq.~\eqref{Diffusion_SME}, $dW_k$ are independent Wiener increments, i.e., Gaussian random variables with $E[dW_k] = 0$, $E[dW_k^2] = dt$ and  \cite{jacobs_2010}
\begin{equation}
    \label{eq:itorule}
    dW_kdW_l=dt\delta_{kl}.
\end{equation}
We obtain Eq.~\eqref{Diffusion_SME} by time-averaging the stochastic master equation over the effect of many jumps: In the limit of large reference currents, the jumps are dominated by the term proportional to $|\alpha_k|^2$ and even a large number of jumps only results in an infinitesimal change in the density matrix, which can be described by a Gaussian stochastic process \cite{Wiseman_1993}. Analogously, the rate of change of the counting variables $dN_k/dt$  behaves as (see~Appendix~\ref{app:stochasticwiener})
\begin{equation}\label{current_relation_jump_diffusion}
    \frac{dN_k}{dt} \simeq |\alpha_k|^2 + |\alpha_k| \Big( \langle x_k \rangle + \frac{dW_k}{dt}\Big).
\end{equation}
Hence, the stochastic diffusive current~\eqref{Idiff} becomes 
\begin{equation}
\label{Idiffwiener}
    I_{\rm diff}(t) = \sum_k \nu_k \Bigg( \langle x_k\rangle_c + \frac{dW_k}{dt}\Bigg).
\end{equation}
Since $E[dW_k(t)]=0$ and $dW_k(t)$ is statistically independent of $\rho_c(t)$, averaging Eq.~\eqref{Diffusion_SME} over all trajectories recovers the original master equation~\eqref{M}, i.e., $E[\rho_c(t)] = \rho(t)$. 
Similarly, averaging Eq.~\eqref{Idiffwiener} yields the average diffusive current~\eqref{diffusive_average_current}.

Equations \eqref{Diffusion_SME} and \eqref{Idiffwiener} can also be obtained from an entirely different physical approach: 
namely, the continuous application of weak Gaussian measurements~\cite{Jacobs_2006}. 
Let $Y$ denote a Hermitian observable, and consider the measurement operators
\begin{equation}
\label{gaussian_POVM_kraus}
{M}_ z =\left(\frac{2\lambda dt}{\pi}\right)^\frac{1}{4}e^{-\lambda dt( z  -Y)^2},
\end{equation}
where $ z \in \mathbb{R}$ denotes the possible outcomes and $\lambda>0$ is the measurement strength.
To gain some physical intuition as to  what these measurements do, decompose $Y = \sum_y y|y\rangle\langle y|$. 
Given any state $\rho$, the probability of obtaining outcome $ z $ is then 
\begin{equation}\label{gaussian_POVM_probability}
    \tr \big\{ M_ z  \rho M_ z ^\dagger\big\} = \left(\frac{2\lambda dt}{\pi}\right)^{1/2} \sum_y e^{-2\lambda dt( z -y)^2}\langle y | \rho | y\rangle,
\end{equation}
which is a sum of very broad Gaussians, each with standard deviation $1/\sqrt{2\lambda dt}$,  centered on the eigenvalues $y$, and of height proportional to $\langle y |\rho |y\rangle$. 
Moreover, the unconditional action of the map, after averaging over all possibilities, is
\begin{equation}
    \int\limits_{-\infty}^\infty d z ~ M_ z  \rho M_ z ^\dagger = \sum_{y,y'} e^{-\lambda dt (y-y')^2/2}|y\rangle\langle y|\rho|y'\rangle\langle y'|.
\end{equation}
The map therefore does not affect the diagonals of $\rho$ in the eigenbasis of $Y$, but dampens the coherences (off-diagonals) by a small amount $e^{-\lambda dt (y-y')^2/2}$.

Consider now a scenario where we continuously apply the measurement operators~\eqref{gaussian_POVM_kraus}. 
We can imagine a stroboscopic dynamics, where a new measurement is applied after each step $dt$, leading to a series of random outcomes $ z _1, z _2,\ldots$. 
The continuous measurement is thus described by infinitely many, infinitely weak measurements. 
As we show in Appendix~\ref{app:povm}, this will lead precisely to the quantum diffusion stochastic master equation~\eqref{Diffusion_SME}, with jump operator $L = \sqrt{\lambda} Y$. 
Moreover, choosing $\nu = 1/(2\sqrt{\lambda})$ makes the sequence of outcomes $ z $ exactly equal to $I_{\rm diff}(t)$ in Eq.~\eqref{Idiffwiener}.
Notice that since $L^\dagger = L$, this will give rise to dephasing; moreover, the quadrature~\eqref{quadrature} is $x = 2L$. And with our choice of $\nu$, we also have that $\nu x=Y$ is the original operator we are measuring. 
More compactly, the resulting stochastic master equation can be written as 
\begin{equation}\label{Belavkin}
    d\rho_c = dt~\Bigg(-i [H,\rho_c] + \lambda \mathcal{D}[Y]\rho_c\Bigg)+ \sqrt{\lambda} dW \{Y -\langle Y \rangle, \rho_c\},
\end{equation}
and the output of each measurement is 
\begin{equation}
    z = \langle Y \rangle + \frac{1}{2\sqrt{\lambda}} \frac{dW}{dt}.
\end{equation}
This is sometimes referred to as the Belavkin equation~\cite{belavkin:1989}.
We finish by mentioning that quantum diffusion can also be formulated using a path integral approach. 
For more information on this, see~\cite{Pilgram2003,Jordan2004,Chantasri2013,Chantasri2015}.

\subsubsection{Homodyne, heterodyne and charge detection}
\label{sec:diffusion_homo_hetero_charge}

\begin{table*}
    \centering
    \caption{Common paradigms in quantum diffusion. The corresponding stochastic currents are given in Eq.~\eqref{Idiffwiener}.
    }
    \begin{tabular}{l|l|l|l|l}
         Name 
         &  
         Jump operators 
         & 
         Reference 
         &
         Weights
         &
         $J_{\rm diff}$
         \\[0.15cm]
         \hline 
         Homodyne
         & $L = \sqrt{\kappa} a$
         & $\alpha = |\alpha| e^{i\phi}$
         & $\nu = 1/\sqrt{2\kappa}$
         & $J_{\rm diff} = \langle x_\phi \rangle, \quad x_\phi = \tfrac{1}{\sqrt{2}} (a e^{-i \phi} + a^\dagger e^{i \phi})$
         \\[0.2cm]
         Heterodyne
         & $L_1 = L_2 = \sqrt{\tfrac{\kappa}{2}} a$
         & $\alpha_1 = |\alpha|,\quad \alpha_2 = i |\alpha|$
         & $\nu_1 = \nu_2 = 1/\sqrt{\kappa}$
         & $J_{\rm diff}^1 = \langle x_0 \rangle, \quad J_{\rm diff}^2 = \langle x_{\pi/2} \rangle, \qquad $
         \\[0.2cm]
         Heretodyne - time dep.
         & $L = \sqrt{\kappa} a$
         & $\alpha = |\alpha| e^{i \Delta t}$. 
         & $\nu = 1/\sqrt{2\kappa}$
         & $J_{\rm diff} = \cos(\Delta t)\langle x_0 \rangle + \sin(\Delta t) \langle x_{\pi/2} \rangle.$
         \\[0.2cm]
         Charge detection/
         & $L = \sqrt{\Gamma} \sigma_z$
         & $\alpha = |\alpha|$
         & $\nu=1/(2\sqrt{\Gamma})$
         & $J_{\rm diff} =  \langle \sigma_z \rangle$
         \\ 
         Gaussian measurements
         & & & &
    \end{tabular}
    \label{tab:diffusion_experiments}
\end{table*}

We now detail some common types of experiments involving diffusive currents.
For a summary, see Table~\ref{tab:diffusion_experiments}.

\paragraph{Homodyne detection:} We consider an optical system with a single jump operator $L =\sqrt{\kappa} a$, describing photons leaving the system. This outgoing light is combined at a beamsplitter with a
reference laser drive, called the 
local oscillator. This yields the reference current $\alpha=|\alpha|e^{i\phi}$, proportional to the laser amplitude, so that $(L+\alpha)\rho(L^\dagger+\alpha^*)$ corresponds to clicks in a photo-detector placed after the beamsplitter \cite{gardiner_book}. While most clicks result from the photons that originate from the local oscillator, information on the quadratures of the field will still be contained in the diffusive current. Choosing $\nu=1/\sqrt{2\kappa}$, we find $J_{\rm diff}(t)=\langle x_\phi\rangle$ with the quadrature operator
    \begin{equation}
        \label{quadrature_a}
        x_\phi = \frac{1}{\sqrt{2}}\left(a e^{-i\phi}+a^\dagger e^{i\phi}\right).
    \end{equation}
   We note that for $\alpha$ to be time-independent (as we assumed here), the master equation needs to be written in a frame rotating at the frequency of the local oscillator. The quadratures given in Eq.~\eqref{quadrature_a} are thus in this rotating frame. Usually we are interested in the quadratures in a frame rotating with the frequency of an external drive, $\omega_d$. The frequency of the local oscillator should then be chosen to match $\omega_d$.
    
\paragraph{Heterodyne detection:}  Consider splitting the light leaving the system in two parts using a beamsplitter, and applying homodyne detection to each outgoing beam. This can be described by splitting the jump operator in two, $L_1 = L_2 =\sqrt{\kappa/2}a$. For the local oscillators, we may choose $\alpha_1=|\alpha|$ and $\alpha_2=i|\alpha|$. Choosing $\nu_1=\nu_2=1/\sqrt{\kappa}$, this detection scheme results in two diffusive currents with averages $J_{\rm diff}^1(t) = \langle x_0\rangle$ and $J_{\rm diff}^2(t)=\langle x_{\pi/2}\rangle$, corresponding to orthogonal quadratures. 
Note that there is a price to pay for measuring two non-commuting observables simultaneously (even though not projectively): when splitting the outgoing light into two beams, vacuum fluctuations enter the unused port, enhancing the noise in the measurement. This can be seen from the fact that $\nu$ is larger by a factor of $\sqrt{2}$ compared to homodyne detection of a single quadrature. As a weighting factor, an enhanced $\nu$ will also enhance the fluctuations, as will be discussed in Sec.~\ref{sec:noise}.
    
Another method for implementing heterodyne detection is by mixing the light leaving the cavity with a single local oscillator, with a strong detuning $\Delta$ (from the frame of interest). In this case, we have $L=\sqrt{\kappa}a$ and a time-dependent reference current $\alpha=|\alpha|e^{i\Delta t}$. Choosing $\nu=1/\sqrt{2\kappa}$, the diffusive current then reads
\begin{equation}
    \label{eq:diffhet}
    J_{\rm diff}(t) = \cos(\Delta t)\langle x_0\rangle + \sin(\Delta t)\langle x_{\pi/2}\rangle.
\end{equation}
Assuming that $\Delta$ is sufficiently big, such that we may neglect changes in the system over the time-scale $1/\Delta$, we can recover the two diffusive currents from before as 
\begin{equation}
    \label{eq:diffhet2}
    \begin{aligned}
    &J_{\rm diff}^1(t) = \frac{\Delta}{\pi}\int_t^{t+2\pi/\Delta} dt'\cos(\Delta t') J_{\rm diff}(t)=\langle x_0\rangle,\\[0.2cm]
    &J_{\rm diff}^2(t) = \frac{\Delta}{\pi}\int_t^{t+2\pi/\Delta} dt'\sin(\Delta t') J_{\rm diff}(t)=\langle x_{\pi/2}\rangle.
    \end{aligned}
\end{equation}
Of course this scheme also suffers from the additional noise in comparison to homodyne detection. In this case, the added noise results from the averaging in Eq.~\eqref{eq:diffhet2} \cite{Wiseman_2009}.
    
\paragraph{Charge detection:} 
We consider for concreteness a two-level system. 
We could have, for instance, the eigenstates of $\sigma_z$ corresponding to two different charge configurations in a quantum dot system. 
As will be shown in Sec.~\ref{sec:QPC}, the charge can be detected with a QPC, a current carrying element placed close to the dot.
The current in the QPC is sensitive to any nearby charges and therefore can be used to detect if there is an electron in the dot or not~\cite{gustavsson_2006,flindt_2009}. In turn, the presence of the QPC  provides a fluctuating environment for the system, resulting in dephasing described by the jump operator $L=\sqrt{\Gamma}\sigma_z$ (Example C, Eq.~\eqref{ExampleC_M}). 
The reference current $\alpha$ is the current through the QPC when the system is in one of the eigenstates of $\sigma_z$~\cite{goan_2001}.
Choosing $\nu = 1/(2\sqrt{\Gamma})$, the diffusive current provides a measurement of the charge configuration, $J_{\rm diff} = \langle\sigma_z\rangle$. In contrast to optical homodyne detection, the electrical current through the detector therefore constitutes both the reference current, $\alpha$, as well as the source of dephasing, determining $\Gamma$ (there may, of course, also be other sources of dephasing).

We labeled this kind of process as ``charge detection`` but, from this discussion, it is clear that this is in fact more general. 
Indeed, in the paradigm of Gaussian measurements [Eq.~\eqref{gaussian_POVM_kraus}], it is clear that what we referred to as ``charge detection`` really applies to the detection of any Hermitian operator. 

\subsubsection{Output current for Example C}
\label{sec:Homodye detection of a qubit}

We study the diffusive dynamics of Eq.~\eqref{Diffusion_SME} for the master equation~\eqref{ExampleC_M}, where we monitor the population $\sigma_z$ of the qubit. 
This can be done experimentally with quantum dots, or superconducting qubits~\cite{Murch2013,Campagne_Ibarcq2016,Didier_2015,Blais_2021,He_2023}. Equation~\eqref{Diffusion_SME} reduces to 
\begin{equation}
\label{ExampleC_M_diff}
    d\rho_{c} = - i [H,\rho_{c}] + \Gamma \mathcal{D}[\sigma_{z}]\rho_{c} + \sqrt{\Gamma}(\sigma_{z}\rho_{c} + \rho_{c} \sigma_{z} - 2\langle \sigma_{z}\rangle_{c}\rho_{c})dW\,,
\end{equation}
where $\Gamma$ is the dephasing rate and $dW$ is a Wiener increment.
The diffusive stochastic current~\eqref{Idiffwiener} is thus
\begin{equation}\label{ExampleC_I_stoch}
    I_{\rm diff}(t) =  \langle \sigma_{z}\rangle_{c} + \frac{1}{2\sqrt{\Gamma}}\frac{dW}{dt}\,.
\end{equation}
In the steady-state,  $\rho_{\rm ss} = \mathbb{I}/2$ is the maximally mixed state, yielding a trivial average current $J_{\rm diff}(t) = \langle \sigma_{z} \rangle=0$.

\begin{figure}
    \centering
    \includegraphics[width=\columnwidth]{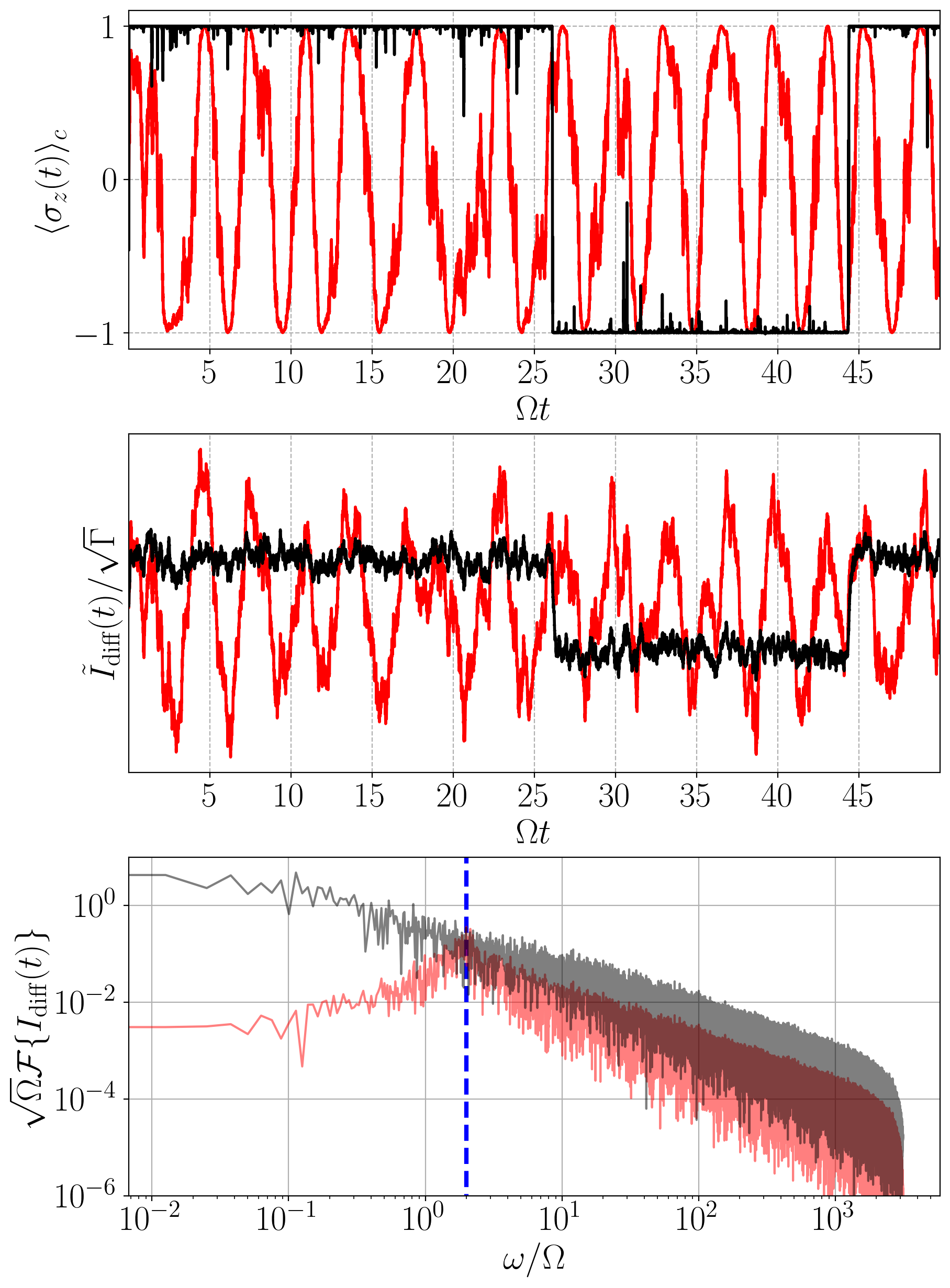}
    \caption{\rev{Output current for} Example C \rev{(Sec.~\ref{ssec:exampleC})} with $\Delta=0$, $\Gamma = (0.2 \Omega, 20\Omega)$, which correspond to the red and black curves respectively. In the top panel we plot both the oscillatory (red) and jump phases (black) of the quantum trajectory $\langle \sigma_{z}(t)\rangle_{c}$. In the middle panel we plot both filtered homodyne currents $\tilde{I}_{\rm diff}(t)$ corresponding to either regime, where the filtering is done with a Butterworth lowpass and bandpass filters respectively. To see the effect of the filtering on $\tilde{I}_{\rm diff}(t)$, we plot the Fourier transform of the filtered signals showing the suppression of unwanted frequencies. Here the blue dashed line corresponds to the Rabi frequency $2 \Omega$. For further details on this filtering see Appendix.~\ref{sec:Filtering basics} for further details. }
    \label{fig:quantum_diffusion}
\end{figure}

This model can be  used to demonstrate the quantum Zeno effect under continuous measurement \cite{Gagen_1993}, which arises when the rate of measurement $\Gamma$ becomes much larger than the Rabi frequency $\Omega$ \cite{He_2023}.
This causes the system to be pinned into one of the eigenstates $\upket, \dwket$ of the measured operator $\sigma_{z}$.
In Fig.~\ref{fig:quantum_diffusion} (top) we plot the quantum trajectories corresponding to both the weak and strong measurement regimes.
We see that $\Gamma\ll \Omega$ leads to Rabi oscillations,
which are suppressed when $\Gamma \gg \Omega$, being replaced by abrupt jumps. 

The raw signal $I_{\rm diff}(t)$ in Eq.~\eqref{ExampleC_I_stoch} is primarily dominated by the white noise $dW/dt$. 
In order to extract useful information, such as an estimate of the trajectories,  $I_{\rm diff}(t)$  must be filtered in post-processing yielding a filtered current $\tilde{I}_{\rm diff}(t)$.
An example of the resulting filtered trajectory is shown in the middle panel of Fig.~\ref{fig:quantum_diffusion}.
While not the subject of this tutorial, we note that the optimal  filtering strategy is highly dependent on the system dynamics and experimental parameters, and must therefore be chosen carefully.
In this example we used a first order Butterworth low-pass filter \cite{Williams2013}.
For $\Gamma \gg \Omega$, we used a cut-off at the Rabi frequency $2\Omega$ to attenuate high frequency signals while leaving low-frequency signals unaffected.
For $\Gamma \ll \Omega$, on the other hand, we applied the filter centered at $2\Omega$, in order to suppress all other frequencies. 
We can also see the effect of these filters by taking the Fourier transform $\mathcal{F}\{\tilde{I}_{\rm diff}(t)\}$ of the filtered current (Fig.~\ref{fig:quantum_diffusion}, lower panel), which shows the suppression of the unwanted frequencies.
For a more in depth discussion of Butterworth filters, please refer to Appendix~\ref{sec:Filtering basics}.

In the limit where $\Omega = 0$, this setup realises a (continuous) quantum non-demolition (QND) measurement~\cite{Braginsky1980} of the qubit population. A QND measurement is one that yields the same outcome when the measurement is repeated, which can only occur if the dynamics preserves the value of the measured observable. In this case, after a time $t\gg \Gamma^{-1}$ has elapsed, the conditional dynamics pins the qubit to one of its energy eigenstates, so that subsequent measurements provide no additional information on $\sigma_z$. Indeed, it is easy to check that the eigenstates of $\sigma_z$ are steady-state solutions of the \textit{conditional} master equation~\eqref{ExampleC_M_diff} when $\Omega=0$, because the final noise term vanishes when $\langle \sigma_z\rangle = \pm 1$.


\subsubsection{Homodyne detection for Example D}
\label{sec:exampleD_homo}

\begin{figure}
    \centering
    \includegraphics[width=\columnwidth]{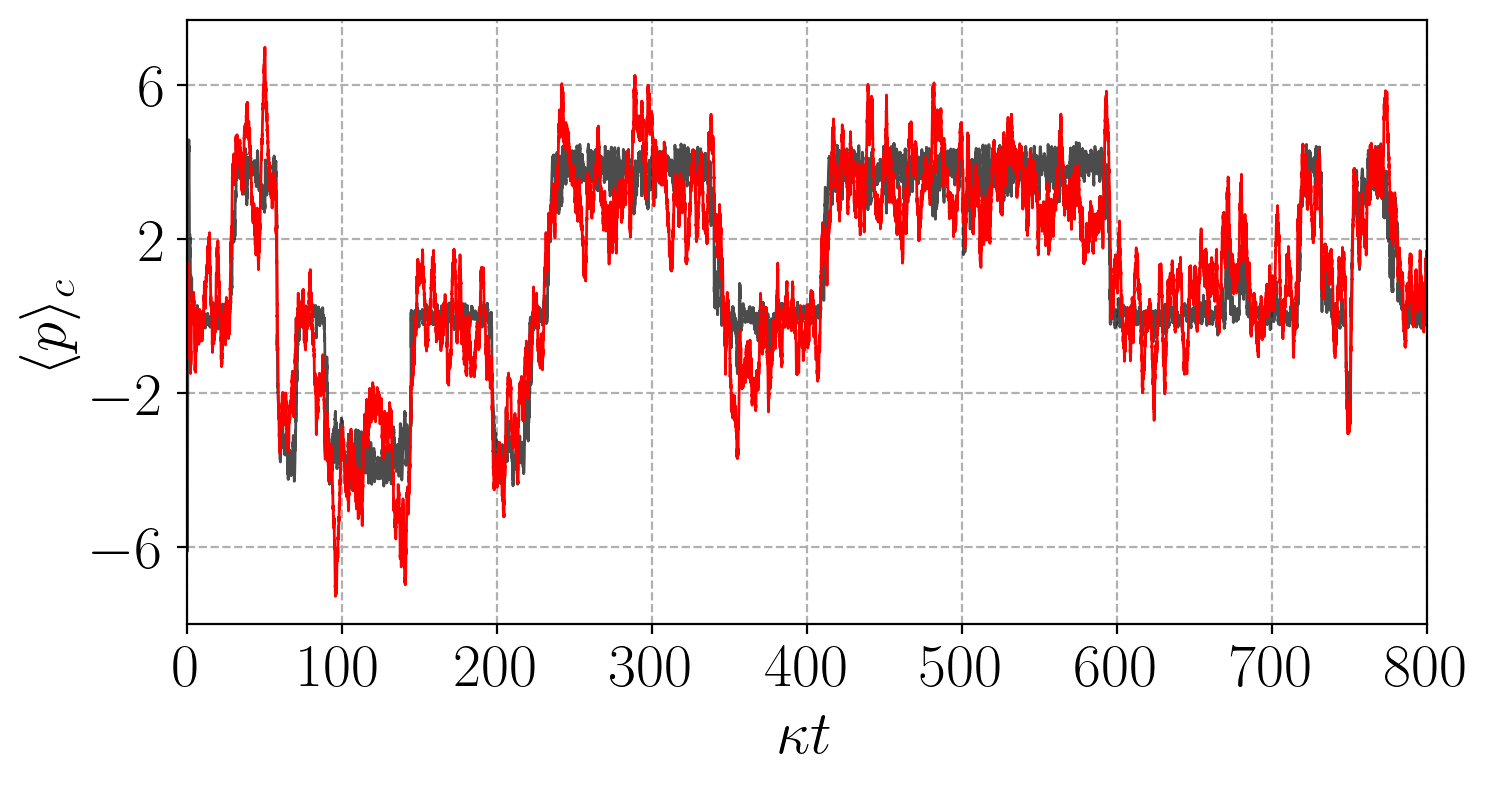}
    \caption{Quantum trajectories for homodyne detection along $p$ in Example D \rev{(Sec.~\ref{ssec:exampleD})}.
    The parameters are fixed at $G/\kappa=1$, $U/\kappa=1/3$,  $\Delta/\kappa =2$, corresponding to the configuration in Fig.~\ref{fig:ExampleD_wigner}(c). 
    The curves showcase the tunneling dynamics of the system, between three metastable states. 
    The filtered homodyne current $I_{\rm diff}(t)$ (red) and 
    underlying conditional moment $\langle p\rangle_{c}$ (black); 
    the former was processed by a lowpass filter to remove high frequency noise. Figure has been replicated from Ref.~\cite{kewming_diverging_2022}.}
    \label{fig:ExampleD_hom}
\end{figure}

We again consider Example D, but instead of counting photons like in Sec.~\ref{sec:exampleD_photo}, we perform homodyne detection along $p  = i(a^{\dagger}-a)/\sqrt{2}$ (see Table~\ref{tab:diffusion_experiments}), again for the same configuration as Fig.~\ref{fig:ExampleD_wigner}(c).
The results are shown in Fig.~\ref{fig:ExampleD_hom}.
The measurement along  $p$ can now also resolve  the two outer lobes, so that we see tunneling between all three of them.
We can again pass the raw homodyne current $I_{\rm diff}(t)$ through a lowpass Butterworth filter yielding a filtered current $\tilde{I}_{\rm diff}(t)$ which approximates the underlying trajectory $\langle p \rangle_{c}$.

\section{Fluctuations in the output current}
\label{sec:fluctuations}

The previous section introduced the idea of fluctuating (stochastic) currents as the output of an open quantum system's dynamics [Eqs.~\eqref{current_charge} and ~\eqref{Idiffwiener}, for jump and diffusive processes respectively].  
We also computed their corresponding average currents, Eqs.~\eqref{average_current} and~\eqref{diffusive_average_current}.
These averages, however, only convey a limited amount of information about the process. 
A much broader picture is obtained by looking at fluctuations. 
Crucially, since the currents $I(t)$ form a time series, one can look at the correlations between fluctuations at different times, i.e., \emph{two-point correlations. }
This is, in fact, where the true richness of the stochastic approach lies: the clicks in the detectors are not independent because they stem from the same quantum system. Their correlations therefore teach us about the system dynamics.

\subsection{Two-point correlation function}
\label{sec:two_point_function}
The correlations between $I(t)$ and $I(t+\tau)$ are captured by the two-point correlation function
\begin{align}\label{F_def}
    F(t,t+\tau) &= E\Big( \delta I(t) \delta I(t+\tau)\Big) \\[0.2cm]\nonumber&= E\big( I(t) I(t+\tau)\big)- J(t) J(t+\tau),
\end{align}
where $\delta I(t) = I(t) - J(t)$ is a shorthand for the current fluctuations (recall that $J(t) = E(I(t))$.
Notice that $I(t)$ is a classical random variable, so $I(t)$ and $I(t+\tau)$ commute; hence, $F(t,t+\tau) = F(t+\tau,t)$ and it suffices to take $\tau>0$. 

For quantum jumps, we show in Appendix~\ref{app:F} that the two-point function can be written as
\begin{equation}\label{F_jump}
    F(t,t+\tau) = \delta(\tau) K(t) + \tr\Big\{ \mathcal{J} e^{\mathcal{L}\tau} \mathcal{J} \rho(t)\Big\} - J(t) J(t+\tau),
\end{equation}
where 
\begin{equation}\label{K}
    K(t) = \sum_k \nu_k^2 \tr \big\{ L_k^\dagger L_k \rho(t)\big\} \geqslant 0, 
\end{equation}
and 
\begin{equation}\label{J_op}
    \mathcal{J}\rho = \sum_k \nu_k L_k \rho L_k^\dagger.
\end{equation}
This superoperator $\mathcal{J}$ will play a crucial role. 
For example, notice how the average current~\eqref{average_current} can be written as
\begin{equation}\label{average_current_J_op}
J = \tr \big\{\mathcal{J} \rho\big\}.    
\end{equation}

The first term in Eq.~\eqref{F_jump} is proportional to a Dirac delta function, which in the Fourier domain corresponds to \textit{white noise}: it is present at all frequencies with equal strength. This singular correlation appears due to the approximation that the jumps occur instantaneously, which is implicit in the master equation. Physically, of course, nothing diverges.
The prefactor $K(t)$ is a measure of how frequently jumps occur overall, and is closely related to the dynamical activity~\cite{Garrahan2007,Maes2008}, coinciding with it whenever $\nu_k = \pm 1$. 

The unequal-time correlations are captured by the  other two terms in Eq.~\eqref{F_jump}. 
The second term, in fact, is related to the probability of a jump occurring at time $t+\tau$ given that a jump was observed at time $t$ (see Appendix~\ref{app:F}) 
\begin{equation}
\label{conditional_jump_prob}
    P\Big(dN_q(t+\tau) = 1 | dN_k(t) = 1\Big) = \frac{dt^2}{p_k(t)}  \tr\big\{ \mathcal{L}_q e^{\mathcal{L} \tau} \mathcal{L}_k \rho(t)\big\},
\end{equation}
where $p_k(t) = dt \tr\big\{ \mathcal{L}_k \rho(t)\big\}$. 
In a quantum-optics context, this kind of correlation is related to Glauber's second-order coherence function $\gtwo(\tau)$ \cite{Glauber_1963}, as will be discussed in Sec.~\ref{sec:Coherence}. 
In the extreme case of a pure Poisson process, where each jump is statistically independent of all the others, the second and third terms in Eq.~\eqref{F_jump} cancel out, leaving only the white noise (see Appendix~\ref{app:Poisson}). 
Hence, white noise is associated with the occurrence of temporally uncorrelated fluctuations.

Next we turn to quantum diffusion [Eq.~\eqref{Idiffwiener}]. As shown in Appendix~\ref{app:hom}, the two-point function reads instead
\begin{equation}\label{diffusion_F}
    F_{\rm diff}(t,t+\tau) = \delta(\tau) K_{\rm diff} + \tr\big\{ \mathcal{H} e^{\mathcal{L}\tau} \mathcal{H} \rho(t)\big\} - J_{\rm diff}(t) J_{\rm diff}(t+\tau).
\end{equation}
This is  similar in structure to Eq.~\eqref{F_jump}, but with 
\begin{equation}
\label{eq:Kdiff}
    K_{\rm diff} = \sum_k \nu_k^2,
\end{equation}
and 
\begin{equation}\label{H_hom_op}
    \mathcal{H}\rho = \sum_k \nu_k \Big( e^{-i \phi_k} L_k \rho + e^{i \phi_k} \rho L_k^\dagger\Big)
    = \sum_k \nu_k \mathcal{H}_k \rho,
\end{equation}
with $\mathcal{H}_k$ given in Eq.~\eqref{H_k_superop}.
The superoperator $\mathcal{H}$ plays a very similar role to $\mathcal{J}$ in Eq.~\eqref{J_op}. For instance, just like $J = \tr\big\{ \mathcal{J}\rho\big\}$, we  see from Eq.~\eqref{diffusive_average_current} that $J_{\rm diff} = \tr \big\{ \mathcal{H}\rho\big\}$. In a quantum optics context, the second term in $F_{\rm diff}$ is related to Glauber's first order coherence function $g^{(1)}(\tau)$ (Sec.~\ref{sec:Coherence}). 

As before, the first term in Eq.~\eqref{diffusion_F} represents white noise. However, unlike in Eq.~\eqref{F_jump}, the white-noise intensity $K_{\rm diff}$ is independent of the present state of the system. 
This is because it derives from fluctuations of the reference current used to perform the measurement. 
In quantum optics, this is sometimes referred to as shot noise. 
Here, we avoid this term to prevent confusion with its other meanings in mesoscopic physics.

Henceforth, we will present most of the formulas in terms of the quantum-jump notation in Eq.~\eqref{F_jump}. 
However, unless stated otherwise, all expressions also hold for diffusion, provided we replace $K \to K_{\rm diff}$ and $\mathcal{J}\to \mathcal{H}$. 

\subsection{Power spectrum}
\label{sec:power_spectrum}

Equation~\eqref{F_jump} holds for any state $\rho(t)$. 
In the steady-state, though,  $F(t,t+\tau)$ becomes a function only of the time-difference $\tau$
\begin{align}
\label{F_steady_state}
    F(\tau) &= \delta(\tau) K + \tr\big\{ \mathcal{J} e^{\mathcal{L} |\tau|} \mathcal{J} \rhoss\big\} - J^2
\end{align}
It is then natural to look at its Fourier transform, which yields the \emph{power spectrum}
\begin{align}\label{power_spectrum2}
    S(\omega) 
    &= \int_{-\infty}^{\infty}e^{-i\omega \tau} F(\tau)d\tau
    \\[0.2cm]
    &= K + \int_{-\infty}^{\infty}e^{-i\omega \tau}
    \Big(
    \tr\big\{ \mathcal{J} e^{\mathcal{L} |\tau|} \mathcal{J} \rhoss\big\} - J^2
    \Big)
    d\tau.
\label{power_spectrum3}    
\end{align}
The power spectrum is real and even, $S(\omega) = S(\omega)^* = S(-\omega)$, which follows from the corresponding properties of $F(\tau)$, and the fact that the currents here are classical objects.
We see from Eq.~\eqref{power_spectrum3} that the white noise yields a frequency independent background. 
On top of this, temporal correlations give rise to peaks and dips in $S(\omega)$.
The position, width and height of these features convey information about the system. 
A detailed  guide is presented in Sec.~\ref{sec:s_weak} for the case of weak dissipation.
There we show that the positions of peaks are associated to the energy differences (transition frequencies) of the system and the widths are related to dissipation. 
Moreover, whether a dip or a peak appears \rev{indicates whether jumps are correlated or anti-correlated}. 
We also remark that the power spectrum is different from the emission spectrum, e.g., of an optical cavity, which describes the energies of the emitted quanta rather than the distribution of emissions in time. Emission and absorption spectra are discussed in Sec.~\ref{sec:emission}.

The power spectrum can also be related to the Fourier transform of the stochastic current $I(t)$ itself. 
Consider a large enough integration time $T$ and let 
\begin{equation}
    \delta \tilde{I}(\omega) = \rev{\frac{1}{\sqrt{T}}}\int\limits_0^T dt e^{i \omega t} \delta I(t),
\end{equation}
where, recall, $\delta I(t) = I(t) - J(t)$. 
Then it follows that 
$E\Big[ \big| \delta\tilde{I}(\omega)\big|^2\Big] = \rev{1/T}\int_0^T dt \int_0^T dt' e^{i \omega (t-t')} F(t-t')$. 
Changing variables to $\tau = t-t'$ and $s = (t+t')/2$ and carrying out the integral over $s$, we find that 
\begin{equation}\label{S_alternative}
    E\Big[ \big| \delta\tilde{I}(\omega)\big|^2\Big] = \int\limits_0^T d\tau \rev{\frac{T-\tau}{T}} F(\tau) e^{i \omega \tau} + \int\limits_{-T}^0 d\tau \rev{\frac{T+\tau}{T}} F(\tau) e^{i \omega \tau} .
\end{equation}
If $T$ is much larger than the time over which $F(\tau)$ decays to zero, then we can approximate $T\pm \tau \simeq T$, from which it follows that
\begin{equation}\label{S_alternative2}
    S(\omega) = \lim\limits_{T\to \infty} E\Big[ \big| \delta\tilde{I}(\omega)\big|^2\Big].
\end{equation}
This expression is \rev{known as the Wiener-Khintchine theorem \cite{Wiener_1930,Khintchine_1934,Clerk_2010} and  is} useful for computing $S(\omega)$ from finite-length time series (c.f.~Fig.~\ref{fig:ExampleA_FS}).

\subsubsection{Power spectrum for Example A}
\label{sec: spectrum of the resonantly driven qubit}

\begin{figure}
    \centering
    \includegraphics[width=\columnwidth]{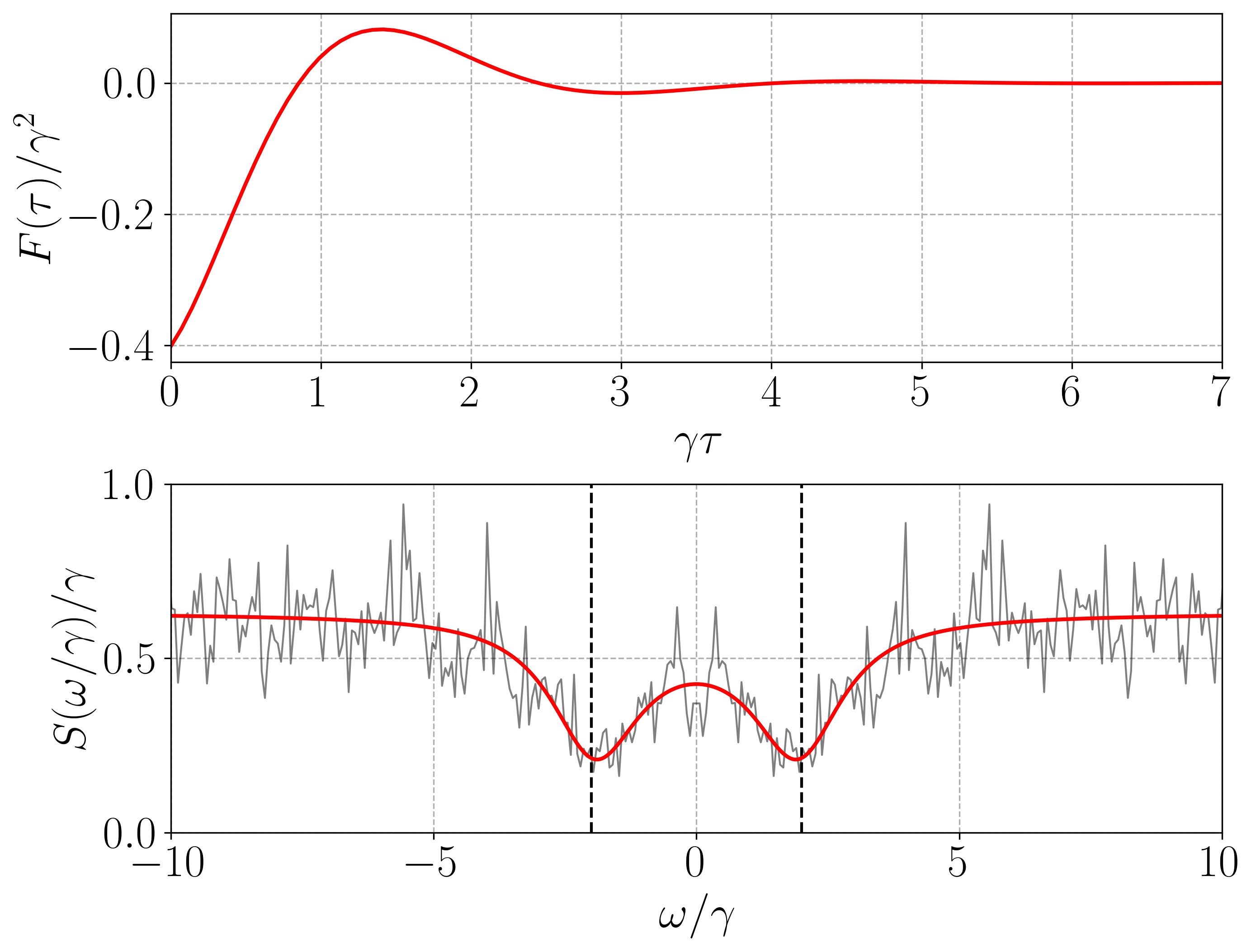}
    \caption{(top) Two-point function $F(\tau)$  and (bottom) power spectrum $S(\omega)$ \rev{[Eq.~\eqref{example_A_power_spectrum}]} for a quantum jump stochastic dynamics for Example A \rev{(Sec.~\ref{ssec:exampleA})} with $\Delta=0$, $\Omega=\gamma$ and $\bar{N}=0.2$. The two-point correlation function is initially negative indicating anti-correlated behavior but becomes positive at $\gamma\tau\sim1.5$, before quickly decaying to zero. The power spectrum has two dips which signify the presence of Rabi oscillations. The black dashed lines correspond to the Rabi frequency. The red curve corresponds to the power spectrum whereas the grey curve corresponds to the power spectrum as measured from the homodyne current of a single trajectory, with step $\gamma \delta t=0.001$ and total time $\gamma t=2000$, divided into $N_{\rm samp} = 20$ samples of length $N_{\rm steps} = 10^5$.}
    \label{fig:ExampleA_FS}
\end{figure}

We return to the resonantly driven qubit considered in Sec.~\ref{sec:Rabi oscillations in a single qubit} and Fig.~\ref{fig:ExampleA_trajs}. 
The two-point function $F(\tau)$ and power spectrum $S(\omega)$ are shown in  Fig.~\ref{fig:ExampleA_FS} using the same set of parameters as Fig.~\ref{fig:ExampleA_trajs} (we omit the Dirac delta contribution from $F(\tau)$).
The methods for computing these quantities  will be outlined in Sec.~\ref{sec:methods}.
\rev{The two-time correlation function is initially negative (anti-correlation), and takes its smallest values as $\tau\to 0$. This means that we are unlikely to see another detection event immediately following the first one, and we say the emissions are anti-bunched (see Sec.~\ref{sec:Coherence})}. This happens because whenever an emission occurs, it takes some time  for the system to be rotated back to its excited state by the drive.
The half-period of the bare Rabi oscillations is $\pi/2\Omega$, but the peak in $F(\tau)$ does not occur exactly at $\tau=\pi/2\Omega$ due to the perturbing effect of thermal excitations. 
In the long-time limit, $F(\tau)\rightarrow0$, indicating that detection events are uncorrelated with the initial detection event.

\rev{Anti-correlated emissions} cause dips in $S(\omega)$, which we plot in the lower panel of Fig.~\ref{fig:ExampleA_FS}.
The dips are positioned at $\omega = \pm 2\Omega$, corresponding to the energy gap of the Hamiltonian~\eqref{ExampleA_H}. The width of the peaks are proportional to the damping rate $\gamma$. 
In the limit $\bar{N}=0$, the power spectrum can be written analytically as 
\begin{equation}\label{example_A_power_spectrum}
    S(\omega) = J \left\{ 1 
-\frac{24 \gamma ^2 \Omega ^2}{\gamma ^4+\gamma ^2 \left(5 \omega ^2+16 \Omega
   ^2\right)+4 \left(\omega ^2-4 \Omega ^2\right)^2}
    \right\},
\end{equation}
In Fig.~\ref{fig:ExampleA_FS} we also illustrate how one might reconstruct the power spectrum from the stochastic current $I(t)$ \rev{in a single shot}, mimicking a real experimental situation. 
To do this, we must work with the discretized version of $I(t)$ which we will denote as $I_{t_{n}}$ where $t_{n}$ corresponds to the discrete index in time that is incremented by $t_{n} = t_{0} + n \delta t$, where $\delta t = N_{\rm steps}/T$ is the time increment, $N_{\rm steps}$ is the total number of time steps and $T$ is the total integration time.
We begin by collecting $N_{\rm samp}$ unique series measurements of the current $I_{t_{n}}^{i}$, labeled $i=1,\ldots,N_{\rm samp}$, each of length $N_{\rm step}$.
To compute the power spectrum, one first computes the signal $\delta I_{t_{n}}^{i}{ = I_{t_{n}}^{i} - {\rm E}_{t}[I_{t_{n}}^{i}]}$, where the average is taken on the time axis $t_{n}$.
One then computes the absolute square of numerical Fourier transform of $\delta I_{t}$ to get the power spectrum [c.f.~Eq.~\eqref{S_alternative2}]
\begin{equation}
    S_{\omega}^{i} = \frac{1}{T}\left| \sum_{n=0}^{N_{\rm steps}-1} \delta I_{t_{n}}^i e^{-i \omega n \delta t}\right|^{2}\,,
\end{equation}
and then concludes by averaging over all $N_{\rm samp}$ samples ${\rm E}_{i}[S_{\omega}^{i}]$.
The power spectrum is thus reconstructed from multiple samples of the signal, which can also be obtained by partitioning a single long observation into $N_{\rm samp}$ samples
\footnote{
This involves a ergodic hypothesis on the data, which is justified whenever $F(\tau)$ decays exponentially for large $\tau$. 
As we show in Sec.~\ref{sec:numerics_fluctuations_vectorized_notation}, this is always true for stable Lindblad dynamics with a single steady-state. 
}.

\subsubsection{Power spectrum for Example C}
\label{sec:powerspecC}
\begin{figure}
    \centering
    \includegraphics[width=\columnwidth]{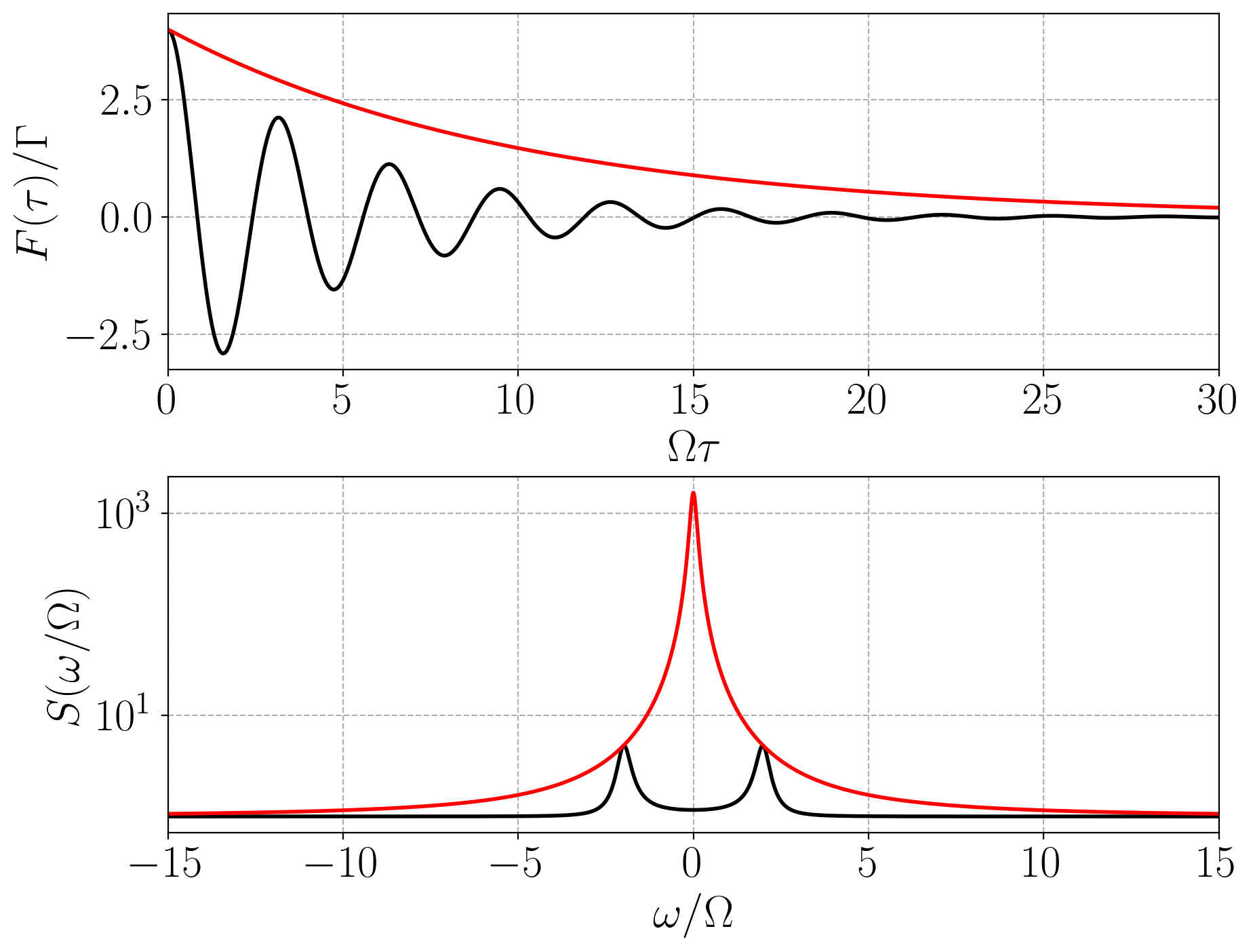}
    \caption{Two time correlation function $F(\tau)$  \rev{[Eq.~\eqref{Example_C_F}]} for Example C \rev{(Sec.~\ref{ssec:exampleC})} with homodyne detection along $\langle \sigma_{z}(t)\rangle_{c}$ with $\Omega=1$ and  $\Gamma = (0.2, 20)$ (black and red curves respectively). In the weak measurement phase we see an oscillating $F(\tau)$ with a decaying envelope. However, in the strong-measurement (i.e.~jump) phase all oscillations are suppressed and we simply see a positive, monotonically decaying $F(\tau)$. This positive correlation means that if the system is observed in either state $\upket$ or $\dwket$, then it is more likely to be observed in the same state at later time. The bottom plot shows the power spectrum $S(\omega)$ \rev{[Eq.~\eqref{ExampleC_S}]}. The two peaks for the oscillatory regime correspond to the eigenergies of the system. Note the Log scale of the y axis. For the jump regime, the slowly decaying $F(\tau)$ yields a sharply peak delta like power spectrum at the origin.}
    \label{fig:quantum_diffusion_power}
\end{figure}

Next we revisit the dephased qubit example of Sec.~\ref{sec:Homodye detection of a qubit} and Fig.~\ref{fig:quantum_diffusion}.
Since this is a diffusive measurement,  we still use formulas~\eqref{F_steady_state} and~\eqref{power_spectrum2}, but with the replacements $K \to K_{\rm diff}$ and $\mathcal{J} \to \mathcal{H}$.
Recall that this model exhibited two drastically different regimes depending on the strength of the measurement $\Gamma$: the oscillatory phase (weak measurement, $\Gamma \ll \Omega$) and the jump phase (strong measurement, $\Gamma \gg \Omega$).
We focus on the steady-state, where $\rho_{\rm ss}=\mathbb{I}/2$ and $J_{\rm diff} = 0$.
The two-point function in this case can be computed analytically and reads
\begin{align}\label{Example_C_F}
    F(\tau) = \delta(\tau) + 4 \Gamma  e^{-\kappa \tau} \left(\frac{\Gamma  \sinh \left(\Omega' \tau\right)}{\Omega'}+\cosh \left(\Omega' \tau \right)\right),
\end{align}
where $\Omega' = \sqrt{\Gamma ^2-4 \Omega ^2}$.
Inserting this in Eq.~\eqref{power_spectrum2} yields
\begin{align}
\label{ExampleC_S}
    S(\omega) = 1 + \frac{64 \Gamma ^2 \Omega ^2}{4 \Gamma ^2 \omega ^2+\left(\omega ^2-4 \Omega ^2\right)^2}\,.
\end{align}
These results are plotted in Fig.~\ref{fig:quantum_diffusion_power} for the same two measurement strengths used in Fig.~\ref{fig:quantum_diffusion}.
In the oscillatory phase the sign of $F(\tau)$ also oscillates, hence alternating between positively- and negatively-correlated.
This indicates the presence of coherent Rabi oscillations, which decay in time due to the dephasing induced by the measurement. 
Likewise, $S(\omega)$ exhibits two peaks at $\omega = \pm 2\Omega$, as these frequencies dominate the homodyne signal.
In the jump phase  the oscillatory dynamics is replaced with  a monotonic decay in $F(\tau)$. Indeed, if  $\Gamma \gg \Omega$ we get $F(\tau) \simeq 4 \Gamma e^{-2 t \Omega^2/\Gamma}$:
The two-point function is therefore always positive and very large, signaling a very large correlation. Moreover, it exhibits a very slow decay. 
The measurement therefore pins the system to one of the eigenstates and no change is ever possible, which is the limit of the quantum Zeno effect.
This is further seen in the power spectrum $S(\omega)$ which becomes a Lorentzian in the jump phase, and approaches a delta function in the limit of $\Gamma\rightarrow\infty$ (note the log scale in the bottom panel of Fig.~\ref{fig:quantum_diffusion}).

\subsection{The noise}
\label{sec:noise}

The fluctuations in the total charge $N(t)$  are captured by the variance ${\rm Var}(N(t)) = E\big(N(t)^2\big)-E\big(N(t)\big)^2$. 
It turns out to be  more convenient to study its rate of change
\begin{equation}\label{D}
    D(t) := \frac{d}{dt} {\rm Var}(N(t)).
\end{equation}
This quantity is widely studied in FCS (Sec.~\ref{sec:FCS}).
Confusingly, in the literature it goes by many names: scaled variance, diffusion coefficient, noise (some authors refer to it simply as the variance, although it is clear from Eq.~\eqref{D} that this is not the case).
We will call it \emph{noise}.
Using Eqs.~\eqref{current_charge} and~\eqref{F_def}, we can also write 
\begin{equation}
    {\rm Var}(N(t)) = 
     \int\limits_0^t dt' \int\limits_0^t dt'' F(t',t'').
\end{equation}
Plugging this in Eq.~\eqref{D} and recalling that $F(t,t') = F(t',t)$, yields
\begin{align}\label{D_integral_time}
    D(t) &= 2 \int\limits_0^t  F(t,t-\tau)d\tau.
\end{align}
This result is general, in that it holds even in transient regimes. 
In the steady state we can replace $F(t,t-\tau) = F(-\tau) = F(\tau)$, and \rev{(for sufficiently long times $t$)} extend the upper integration limit to $+\infty$.
As a result, we find that the noise is directly related to the zero-frequency component of the power spectrum~\eqref{power_spectrum2}
\begin{align}\nonumber
    D = S(0) &= 2\int\limits_0^\infty F(\tau) d\tau 
    \\[0.2cm]
    &=K + 2\int_0^{\infty}
    \Big(
    \tr\big\{ \mathcal{J} e^{\mathcal{L} \tau} \mathcal{J} \rhoss\big\} - J^2
    \Big)
    d\tau.
    \label{D_integral}
\end{align}
Since in \rev{this case} $D$ becomes independent of time, the variance will grow linearly in time
\begin{equation}
\label{varN_D}
    {\rm Var}\big(N(t)\big) = D t.
\end{equation}
This is why $D$ is also sometimes referred to as the ``scaled variance'': In the steady state, it essentially represents ${\rm Var}\big(N(t)\big)$, except for the fixed scaling with $t$.
Away from the steady state, $D(t)$ has no definite sign but Eq.~\eqref{varN_D} shows that in the steady state, $D\geqslant 0$.
However, the 2nd term in Eq.~\eqref{D_integral} is not necessarily positive, so that current correlations can both increase or decrease the noise. 
In fact, this is related to whether $S(\omega)$ has a peak or a dip at $\omega = 0$. 

The noise can be attributed the following interesting interpretation:
\emph{It yields the time it takes to reduce the uncertainty in estimating the current to one unit.} 
That is, suppose we wish to measure $J$. 
Physically, this is tantamount to measuring the expectation value of the system observable 
$\sum_k \nu_k  L_k^\dagger L_k$ for quantum jumps, or $ \sum_k \nu_k( e^{-i\phi_k}L_k + e^{i \phi_k} L_k)$ for quantum diffusion.
We can do this in a single shot, by collecting the current $I(t)$ for a total time $T$ and computing the time average
\begin{equation}\label{single_shot_accumulated_current}
\bar{I}(T) = \frac{1}{T}\int_0^T dt\, I(t).
\end{equation}
Clearly, $E[\bar{I}(T)] = J$ for any $T$, so this estimator is unbiased. 
The mean-squared error, on the other hand, is $\Delta \bar{I}^2(T) := E\big[(\bar{I}(T)-J)^2\big]$, which can be written as \footnote{
To derive this result, we first write it as $\Delta \bar{I}^2(T) = \frac{1}{T^2} \int_0^T dt_1 \int_0^T dt_2~F(t_1-t_2)$. We then change variables to $\tau = t_1-t_2$ and $t_s = (t_1+t_2)/$, and carry out the trivial integral over $t_s$. 
Moreover, the estimation process converges when  $\Delta \bar{I}^2(T) \to 0$ which, from Eq.~\eqref{Noise_Variance_estimator}, will occur whenever $F(\tau)$ decays faster than $1/\tau$. As will be shown in Sec.~\ref{sec:vectorization}, this is always true for stable Lindblad dynamics with a single steady-state. 
} 
\begin{equation}\label{Noise_Variance_estimator}
    \Delta \bar{I}^2(T) = \frac{2}{T^2} \int\limits_0^T d\tau (T-\tau) F(\tau).
\end{equation}
For sufficiently large $T$, we can approximate $T-\tau \sim T$ and extend the upper limit of integration to infinity, yielding 
\begin{equation}\label{Noise_variance_asymptotic}
    \Delta \bar{I}^2(T) \simeq \frac{D}{T}.
\end{equation}
Thus, $D$ determines how the error in the estimation diminishes with the integration time. 



\subsubsection{Fluctuations in Example C}

\begin{figure}
    \centering
    \includegraphics[width=\columnwidth]{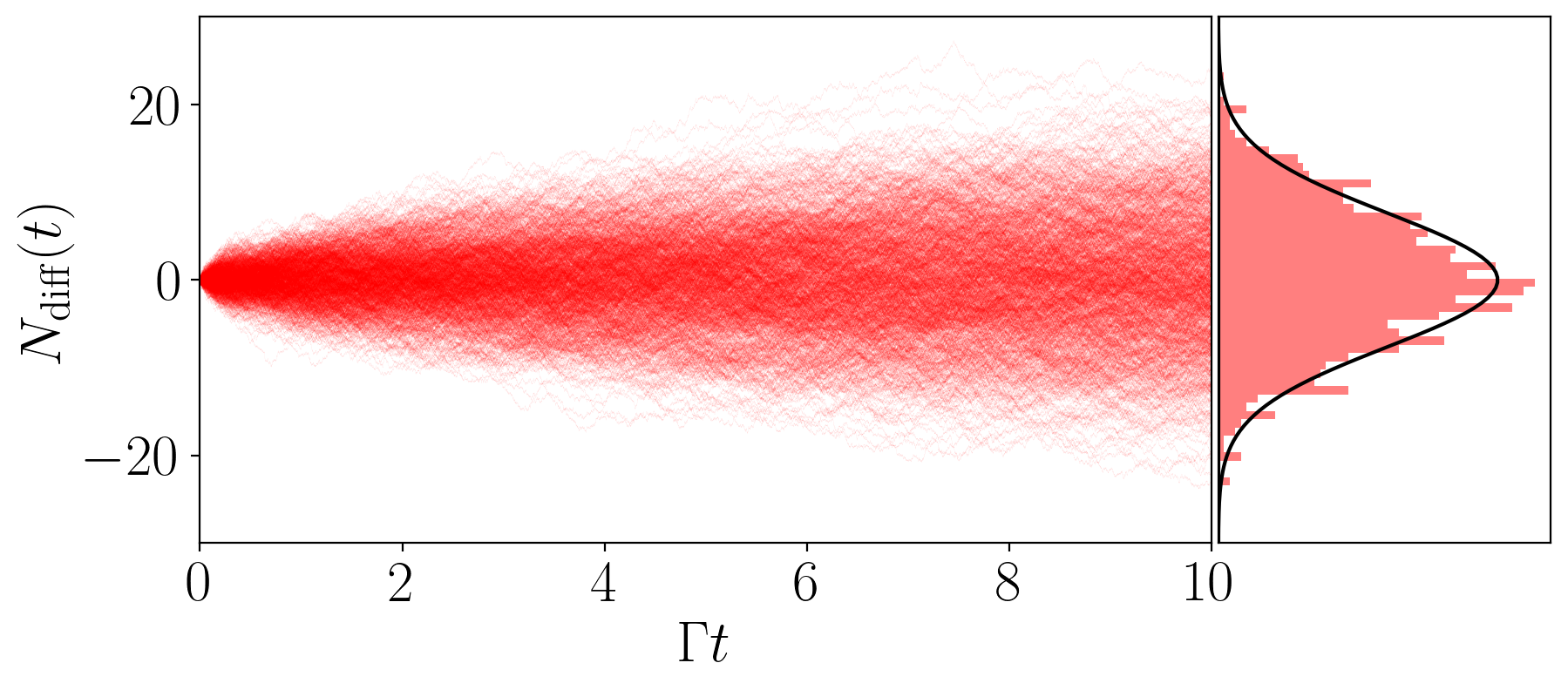}
    \caption{
    Integrated diffusive current $N_{\rm diff}(t)$ for Example C \rev{(Sec.~\ref{ssec:exampleC})}, for over 5000 simulations with $\Gamma=0.2\Omega$.
    The variance of the distribution, at \rev{long times}, is $Dt$ [Eq.~\eqref{varN_D}], where $D$ is is given by Eq.~(\ref{ExampleC_D}).}
    \label{fig:ExampleC_D}
\end{figure}

The trajectories for Example C were studied in Sec.~(\ref{sec:Homodye detection of a qubit}) and  $F(\tau)$ and $S(\omega)$ in Sec.~(\ref{sec:powerspecC}).
In fact, $S(\omega) $ was given in Eq.~(\ref{ExampleC_S}). 
Taking $\omega=0$ leads to 
\begin{equation}
\label{ExampleC_D}
    D = 1 + \frac{4 \Gamma^{2}}{\Omega^{2}}\,.
\end{equation}
In Fig.~\ref{fig:ExampleC_D} we plot the histogram of $N(t)$ for multiple trajectories as a function of $t$.
As is clear, the distribution becomes broader and broader with time. 
The variance, at \rev{long times} is, according to Eq.~\eqref{varN_D}, precisely $Dt$. 
Thus, in the oscillatory phase, $D$ is practically unity, while in the jump phase it grows as $\Gamma^2$.
We also note that, at long times, the distribution looks like a Gaussian due to the central limit theorem. Typically, cumulants higher than second order also scale linearly in time but their effects can only be seen in the tails of the distribution.
These higher-order cumulants can be treated using the tools of FCS (Sec.~\ref{sec:FCS}).

\subsection{Quantum regression theorem and coherence functions}
\label{sec:Coherence}

The two-point function $F$ in Eq.~\eqref{F_def} describes  correlations between the classical stochastic outputs $I(t)$ at different times. 
We can translate this into temporal correlations of the quantum jump operators $L_k$.
To do this, we must invoke the quantum regression theorem (QRT)~\cite{Lax1963,gardiner_book}, which provides a way of calculating two-time correlation functions from the QME.
The QRT states that, for any three system operators $A,B,C$,
\begin{equation}\label{QRT}
    \langle A(t) B(t+\tau) C(t)\rangle
    =
    \tr \Big\{ B e^{\mathcal{L}\tau} \Big(C \rho(t) A \Big) \Big\},
\end{equation}
(for fermionic operators, this formula requires a slight modification, as discussed in Ref.~\cite{Schwarz2016}, appendix B). It is often convenient to express Eq.~\eqref{QRT} in terms of the adjoint Liouvillian, $\mathcal{L}^\dagger$, which is defined implicitly by $\tr \left[ A \mathcal{L}(B)\right] = \tr \left[  \mathcal{L}^\dagger (A) B\right]$, and takes the explicit form
\begin{equation}
    \label{adjoint_liouvillian}
    \mathcal{L}^\dagger (\bullet) = i[H,\bullet] + \sum_j \mathcal{D}^\dagger[L_j](\bullet),
\end{equation}
where $\mathcal{D}^\dagger[L] (\bullet) = L^\dagger(\bullet)L - \tfrac{1}{2}\{L^\dagger L,\bullet\}$ is the adjoint dissipator. Then correlation functions can also be written as
\begin{equation}
    \label{two_time_function}
    \langle A(t) B(t+\tau )  C(t)\rangle = \tr \left[  A e^{\mathcal{L}^\dagger\tau}(B) C \rho(t)  \right].
\end{equation}
If the state $\rho(t)$ is already known, this only requires finding the operator $B(\tau) = e^{\mathcal{L}^\dagger\tau}(B)$ or, equivalently, solving the equation $dB/d\tau = \mathcal{L}^\dagger B$. This can be understood as a kind of Heisenberg-picture evolution for open quantum systems, although some caveats apply, e.g., the evolution operator $e^{\mathcal{L}^\dagger\tau}$ is not distributive over the operator product (see Appendix~\ref{app:adjoint_diss} for a discussion).

The QRT allows us to rewrite the typical terms in the quantum jump two-point function Eq.~\eqref{F_jump} as
\begin{align}
    \tr\Big\{ \mathcal{J} e^{\mathcal{L}\tau} \mathcal{J} \rho(t)\Big\} 
    &= \sum_{k,q} \nu_k \nu_q \Big\langle L_q^\dagger (t) L_k^\dagger (t+\tau) L_k(t+\tau) L_q(t)\Big\rangle.
\label{jump_F_QRT}        
\end{align}
Similarly, for quantum diffusion [Eq.~\eqref{diffusion_F}], we get 
\begin{align}\label{diffusion_F_QRT}
    \tr\Big\{ \mathcal{H} e^{\mathcal{L}\tau} \mathcal{H} \rho(t)\Big\} 
    = \sum_{k,q} \nu_k \nu_q {\rm Re}\Big\{ &e^{i (\phi_k - \phi_q)} \langle L_k^\dagger(t+\tau) L_q(t)\rangle 
    \\[0.2cm]\nonumber
    &+ e^{-i (\phi_k +\phi_q)} \langle L_k(t+\tau) L_q(t)\rangle \Big\}.
\end{align}
We now go into more detail on the interpretation of these terms. 
For clarity, we will assume a single jump operator $L$ with weight factor $\nu = 1$; the extension to multiple jump operators is immediate. We also focus on the steady state.

\subsubsection{Quantum jumps and second-order coherence}
\label{sec:Coherence jumps}

The average quantum jump current is $J = \langle L^\dagger L \rangle$.
The two-point function for quantum jumps can be written as 
\begin{align}
\nonumber
    F(\tau) &= \delta(\tau) J + \big\langle L^\dagger(t) L^\dagger(t+\tau) L(t+\tau) L(t)\big\rangle - J^2
    \\[0.2cm]
    &= \delta(\tau) J + J^2 \big[\gtwo(\tau)-1\big],
    \label{jump_F_g2}
\end{align}
where 
\begin{equation}\label{g2}
    \gtwo(\tau) = \frac{\langle L^\dagger(t) L^\dagger(t+\tau) L(t+\tau) L(t)\rangle}{\langle L^\dagger L\rangle^2}.
\end{equation}
In the context of quantum optics and photo-detection, $L = \sqrt{\kappa}a$ and this becomes Glauber's second-order coherence function~\cite{Glauber_1963}.
This function played a central role in establishing the particle nature of light, beginning  with the  experiment of Hanbury-Brown and Twiss \cite{BROWN_1956} and then later  used to characterize photon emission via resonance fluorescence \cite{Kimble_1977} and Hong-Ou-Mandel interference \cite{Hong_1987}; the interested reader will find the history of this work summarized in Refs.~\cite{Scully_1997, Lounis_2005, Walls_2008}.
Despite its history in quantum optics, the above formulation makes it clear that Eq.~\eqref{g2} is applicable to arbitrary counting processes, e.g., electron counting in quantum dots. 

From Eq.~\eqref{jump_F_g2}, we see that the Poisson process (where the jumps are completely independent of each other) happens when $\gtwo(\tau) = 1$ for all $\tau$. 
Moreover, as we show in Sec.~\ref{sec:vectorization}, $\gtwo(\tau\to\infty)\to 1$, so that $F(\tau)$  vanishes at infinity. 
The interesting question to ask, therefore, is whether $\gtwo(\tau)$ is smaller or larger than 1. 
From~\eqref{jump_F_g2} we see that if at any given time $\gtwo(\tau) > 1$, then $I(t)$ and $I(t+\tau)$ will be positively correlated, while $\gtwo(\tau) < 1$ means they are negatively correlated. 
Recall that positive correlations mean that if one variable is above average, there is a tendency for the other to be above average, and vice-versa. Conversely, negative correlation means that if one is above average, then the other has a tendency to be below average. 
In our case  $dN(t) = 0,1$ can take only two values, and the mean must lie somewhere in between. 
Hence, ``above average'' means it is more likely to observe a click, and ``below average'' means it is less likely. 
\rev{We can see this quite clearly by noting that $\gtwo$ is actually proportional  to the joint probability of observing clicks at $t$ and $t+\tau$, i.e.
\begin{equation}
    \label{gtwo_joint_prob}
    \gtwo(\tau) = \frac{P\left(dN(t+\tau)=1,dN(t)=1\right)}{ (Jdt)^2},
\end{equation}
which follows from Eq.~\eqref{conditional_jump_prob}. Since $(Jdt)^2$ would be the joint probability for two completely uncorrelated clicks, we can conclude that $\gtwo$ yields information on delayed coincidences:
\begin{itemize}
    \item {\bf Correlation} $\gtwo(\tau)>1$ (or $F(\tau) > 0$): a click at time $t$ implies it is \emph{more likely} to  observe another a click at time $t+\tau$.
    \item {\bf Anti-correlation} $\gtwo(\tau)<1$ (or $F(\tau) < 0$): a click at time $t$ means it is \emph{less likely} to  observe another a click at time $t+\tau$.
\end{itemize}
These regimes are related to the effects of bunching/anti-bunching and super/sub-Poissonian statistics, which are widely studied in the quantum optics literature. There are a few subtleties, however, and these have been the source of some confusion in the literature. 
We will explain this connection in greater detail below.
}

One often focuses on $\gtwo(0)$. 
The reason is that $\gtwo(\tau)$ is a smooth function, so $\gtwo(0)$ \rev{has} the same interpretation as above, but between two clicks that are infinitesimally close to each other.
The big advantage of $\gtwo(0)$ is that it is much simpler to compute, since it reduces to an expectation value in the steady state: 
\begin{equation}\label{g2_0}
    \gtwo(0) = \frac{\langle L^\dagger L^\dagger L L\rangle}{\langle L^\dagger L\rangle^2}.
\end{equation}
\rev{When $\gtwo(0) > 1$, the observation of one click increases the probability of seeing a second click immediately afterwards. Intuitively, this captures the phenomenon of bunching, where pairs of quanta are more likely to arrive together than apart. Anti-bunching is the converse phenomenon, where clicks are more likely to occur apart than together. }

\rev{To illustrate this, consider} a qubit system with $L = \sqrt{\gamma} 
\sigma_-$. 
Since $\sigma_-^2 = 0$ it immediately follows that $\gtwo(0)=0$. 
Qubit systems are therefore examples of extreme anti-bunching: if a click is observed, there is zero probability of observing another click immediately afterwards. In a fermionic picture, \rev{the fact that $\gtwo(0)=0$} can be viewed as a manifestation of the Pauli exclusion principle. 

As another example, consider the optical cavity dissipator $L = \sqrt{\kappa} a$. Using $[a,a^\dagger]=1$ we can write 
\begin{equation}
    \gtwo(0) = \frac{\langle a^\dagger a^\dagger a a \rangle}{\langle a^\dagger a \rangle^2}
    = 1 + \frac{{\rm Var}(a^{\dagger}a) - \langle a^{\dagger}a\rangle}{\langle a^{\dagger}a\rangle^{2}}\,,
\end{equation}
where ${\rm Var}(a^{\dagger}a)  {=} \langle(a^{\dagger}a)^{2} \rangle - \langle a^{\dagger}a \rangle^{2}$.
\rev{The probability of two clicks coinciding is therefore enhanced when 
${\rm Var}(a^{\dagger}a) > \langle a^{\dagger}a\rangle$, and suppressed otherwise}. 
For example, for thermal radiation $\rho = (1- e^{-\beta \omega})e^{-\beta \omega a^\dagger a}$, we get $\gtwo(0) = 2$ for any $\beta$. \rev{Thermal photons are therefore more likely to be detected together. This is the essence of the Hanbury-Brown and Twiss effect~\cite{BROWN_1956}}.

To get a deeper understanding on bunching and anti-bunching, consider the stochastic quantum jump dynamics for the cavity [e.g.~Eq.~\eqref{stochjump1}]. 
If a photon is emitted, the state of the system is modified from $\rho \to \rho' = \frac{a \rho a^\dagger}{\tr(a^\dagger a \rho)}$.
The average number of photons after the jump, $\langle a^\dagger a \rangle' := \tr(a^\dagger a \rho')$, will then be
\begin{equation}\label{g2_light_after_jump}
    \langle a^\dagger a \rangle' = \frac{\tr(a^\dagger a^\dagger aa \rho)}{\tr(a^\dagger a \rho)} = \gtwo(0)~ \langle a^\dagger a \rangle.
\end{equation}
Therefore, $\gtwo(0)$ specifies how our knowledge of the average photon number is updated after a jump. 
\rev{If $\gtwo(0)>1$,} the updated state will have more photons than before.
This may seem counter-intuitive, as one might think that if a photon emission is observed, the photon number should go down by exactly one unit. 
However, the photon number in a state $\rho$ generally fluctuates, and observing the jump updates our knowledge about the system. 
To make sense of this, consider the incoherent mixture $\rho = q|0\rangle\langle 0 | + (1-q) |n\rangle \langle n|$, so that $q$ represents the probability that there are no photons inside the cavity. 
In this state, $\langle a^\dagger a \rangle = (1-q)n$,
and after the jump $\rho' = |n-1\rangle \langle n-1|$ so $\langle a^\dagger a\rangle' = (n-1)$. 
Light, in this state, can be \rev{either} bunched or anti-bunched. For instance, if $n=2$ then Eq.~\eqref{g2_light_after_jump} yields $\gtwo(0) = \tfrac{1}{2(1-q)}$, which will be bunched for $q>1/2$, and anti-bunched otherwise. 
This happens because before the jump there was a probability that there were no photons at all in the cavity. 
But once we observe a jump that possibility is eliminated, and so the average number of photons increases.

\rev{Strictly speaking, however, the value of $\gtwo(0)$ alone is not sufficient to infer whether clicks occur preferentially in bunches or spaced apart. 
The issue is related to the (lack of) monotonicity of $\gtwo(\tau)$. 
For example, the value of $\gtwo(0) = 2$ for thermal light implies bunching in the above sense only if $\gtwo(\tau)$ decays monotonically to its long-time value of $\gtwo(\infty)=1$ (this was assumed implicitly in the discussion above).  If, instead, $\gtwo(\tau)$ \textit{increases} to some maximum value $\gtwo(\tau_{\rm max}) >2$ (in principle possible in a driven, nonlinear system), then photodetections are more likely to be separated by a time delay of $\tau_{\rm max}$ than to occur together. 

To avoid such paradoxes, some authors~\cite{Zou1990,emary_2012} insist on a stricter definition of bunching and anti-bunching that takes into account the entire behaviour of $\gtwo(\tau)$: 
\begin{itemize}
    \item {\bf Bunching}: $\gtwo(0)>\gtwo(\tau)$ for all $\tau$: the \textit{most likely} time to observe a second click is immediately after the first one ($\tau=0$).
    \item  {\bf Anti-bunching}: $\gtwo(0)<\gtwo(\tau)$ for all $\tau$: the \textit{least likely} time to observe a second click is immediately after the first one ($\tau=0$).
\end{itemize}
Historically, this definition arose in quantum optics to distinguish quantum from classical states of light. Classically, $\gtwo(\tau) = \langle I_{\rm cl}(t+\tau) I_{\rm cl}(t)\rangle/\langle I_{\rm cl}(t)\rangle \langle  I_{\rm cl}(t+\tau)\rangle$ is the normalised autocorrelation function of a fluctuating classical intensity $I_{\rm cl}(t)$. Assuming stationary statistics, the Cauchy-Schwarz inequality $\langle I_1 I_2\rangle^2 \leq \langle I_1^2\rangle \langle I_2^2\rangle$ then directly implies that $\gtwo(0)\geq \gtwo(\tau)$~\cite{Reid1986}. The experimental observation of anti-bunched photon emission by two-level atoms was therefore crucial for establishing the quantum nature of light, since such intensity correlations cannot arise from any stochastic classical field~\cite{Kimble_1977}. 

Nevertheless, $\gtwo(0)$ itself still precisely quantifies whether click coincidences are enhanced or suppressed relative to random (Poisson) noise, and it is common to see bunching (anti-bunching) defined by the condition $\gtwo(0)>1$ ($\gtwo(0)<1$) in the literature~\cite{Walls1979, Paul1982}. Since $\gtwo(\infty)=1$ for a continuously measured quantum system, this is equivalent to the stricter definition above whenever $\gtwo(\tau)$ is a monotonic function.

On a more coarse-grained level, click correlations can be quantified by the Fano factor, defined by~\cite{Fano_1947}
\begin{equation}
    \label{eq:fano}
    f(t) = \frac{{\rm Var}(N(t))}{E(N(t))}.
\end{equation}
[The denominator is usually chosen so that $f\geqslant 0$. And if the $\nu_k$ have units as, e.g., in electric charge transport, $f$ is usually divided by a convenient factor to remain dimensionless.] 
In the long-time limit, the Fano factor reduces to the ratio
\begin{align}
    \label{eq:fano_lt}
    f = \frac{D}{J},
\end{align}
and, using Eq.~\eqref{jump_F_g2}, we have
\begin{align}
    \label{fano_g2}
    f = 1+2J\int_0^\infty d\tau \left[g^{(2)}(\tau)-1\right].
\end{align}
Therefore, when the clicks are correlated (anti-correlated) on average, the integral above is positive (negative) and we have $f>1$ ($f<1$). The marginal case $f=1$ corresponds to ${\rm Var}(N) = E(N)$ (c.f.~Eq.~\eqref{eq:fano}), which is a characteristic of pure Poisson statistics (this is also shown explicitly in Appendix~\ref{app:Poisson}). This motivates the following common terminology:
\begin{itemize}
    \item {\bf Super-Poisson noise} $f>1$ (or $D>J$): clicks are (positively) correlated on average.
    \item {\bf Sub-Poisson noise} $f<1$ (or $D<J$): clicks are anti-correlated on average.
\end{itemize}
The Fano factor is particularly prominent in electronic charge transport \cite{blanter_2001,burkard_2000,emary_2012}, where time-integrated currents can be readily measured but the absence of single-electron detectors make a direct measurement of $g^{(2)}(\tau)$ extremely challenging. In the quantum optics literature, one sometimes comes across the equivalent Mandel parameter $\mathfrak{Q}(t) = f(t) - 1$, defined such that positive (negative) values pertain to super-Poisson (sub-Poisson) noise~\cite{Mandel1979}.
}

\rev{Intuitively, a super-Poissonian Fano factor is associated with bunching, while sub-Poisson noise is associated with anti-bunching. However, this is not always the case, as discussed in Refs.~\cite{Singh1983,Zou1990,emary_2012}. In fact, it is possible to construct examples where $\gtwo(\tau)$ is non-monotonic in time, such that $\gtwo(0)<\gtwo(\tau)$ yet nonetheless $f>1$; likewise, bunched detections can lead to sub-Poisson noise. Moreover, it is important to emphasise that while pure Poisson noise implies that $f=1$, the converse is not true: correlated clicks may still result in a Fano factor $f=1$ upon time-averaging, whenever $\gtwo(\tau)$ is non-monotonic.

A key takeaway message of this section is that care must be taken with the concepts of bunching, anti-bunching, and sub/super-Poisson noise. While they all quantify correlations between clicks, they have different meanings and sometimes inequivalent definitions in the literature.}

\subsubsection{$g^{(2)}$ function for Example D}

We consider the second-order coherence function for Example D. In the particular case where $U = 0$ the system can be solved analytically using Gaussian tools, as discussed below in Sec.~\ref{sec:parametric_oscillator}. 
The $g^{(2)}$ function, computed using those results, reads 
\begin{equation}\label{ExampleD_g2}
 \gtwo(\tau) = 1 + \frac{
        (\kappa-2G)^2 e^{-\tau(\kappa + 2 G)} + (\kappa+2G)^2 e^{-\tau(\kappa- 2G)} 
    }{8G^2}.
\end{equation}
In particular, we see that $\gtwo(0) = 2 + \kappa^2/4G^2$. 
This describes the statistical properties of radiation squeezing.

The case $G\to 0$ is somewhat delicate, as the above formula seems to imply $\gtwo$ diverges. 
What actually happens is that, since this is the vacuum state, both $\langle a^\dagger a \rangle \to 0$ and $\langle a^\dagger a^\dagger a a \rangle \to 0$, so that according to Eq.~\eqref{g2_0} we get an indeterminacy. 
To solve this, it is illuminating to also include the effects of temperature by changing the dissipator to  $\kappa (\bar{N} +1) \mathcal{D}[a] + \kappa \bar{N} \mathcal{D}[a^\dagger]$, where $\bar{N}$ is the Bose-Einstein distribution reflecting thermal fluctuations [similar to Eq.~\eqref{ExampleA_M}].
This will change $\gtwo(0)$ to 
\begin{equation}
    \gtwo(0) = 2+ \frac{
        G^2\kappa^2 (2\bar{N}+1)^2
    }{
        (2 G^2 + \bar{N} \kappa^2)^2
    }.
\end{equation}
If we take $\bar{N} \equiv 0$ we recover the previous result. 
But if we take $G \to 0$ with $\bar{N}$ finite, we get $\gtwo(0) = 2$, which is the result for thermal light. 
The order of limits therefore matters.

\subsubsection{Quantum diffusion and first-order coherence}
\label{sec:coherence_diffusion}

We now turn to quantum diffusion and consider the measurement of a single quadrature $x_\phi = L e^{-i \phi} + L^\dagger e^{i \phi}$, of a jump operator $L$.
We leave $\nu$ unspecified. 
For light ($L= \sqrt{\kappa}a$) this corresponds to homodyne detection (Table~\ref{tab:diffusion_experiments}) and $\nu = 1/\sqrt{2\kappa}$.
Using the QRT result in Eq.~\eqref{diffusion_F_QRT}, the diffusion two-point function $F_{\rm diff}$ in Eq.~\eqref{diffusion_F} becomes 
\begin{equation}
    \label{homodyne_two_point}
    F_\phi(\tau)/\nu^2 =  \delta(\tau) +  {\rm Re}\left[ \langle L^\dagger (\tau) L\rangle + e^{2i \phi}  \langle L (\tau) L\rangle \right] - \langle x_\phi\rangle^2.
\end{equation}
Each term in the square brackets is important in its own right. 
In quantum optics, the first term  is proportional to Glauber's first-order coherence function
\begin{equation}
    \label{first-order-coherence}
    \gone(\tau) = \frac{\langle L^\dagger (\tau) L\rangle}{|\langle L\rangle|^2}.
\end{equation}
As discussed in Sec.~\ref{sec:emission}, the Fourier transform of $\gone(\tau)$ describes the frequency spectrum of the emitted light~\cite{Gardiner_1985}. Hence, $\gone$ is related to the monochromaticity of a light beam: a long coherence time (slowly decaying $\gone(\tau)$) corresponds to a sharply peaked frequency distribution. 
Again in the optical context, the term $\langle L(\tau) L \rangle$ is related to \textit{squeezing} of radiation. We can isolate it by comparing the  fluctuations along two orthogonal quadratures~\cite{Carmichael1987}, e.g.
\begin{equation}
\label{quad_compare_squeezing}
\big[F_{0}(\tau) - F_{\pi/2}(\tau)\big]/\nu^2 =  2 {\rm Re}\left[\langle L (\tau) L\rangle \right] + \langle x_{\pi/2}\rangle^2 - \langle x_{0}\rangle^2.
\end{equation}
While the above results are used in the quantum optics context ($L = \sqrt{\kappa} a$ and $\nu = 1/\sqrt{2\kappa}$), they are also important in other scenarios, as we show in the following example.

\subsubsection{$g^{(1)}$ function for Example A}

\begin{figure}
    \centering\includegraphics{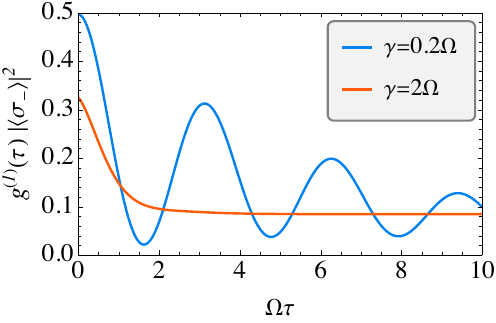}
    \caption{First-order coherence function \rev{[Eq.~\eqref{first-order-coherence}]} for Example A \rev{(Sec.~\ref{ssec:exampleA})}, with $\Delta = 0$ and $\bar{N}=0.1$. For weak dissipation  ($\gamma = 0.2\Omega$) we see damped oscillations, whereas for strong dissipation ($\gamma = 2\Omega$) we see pure exponential decay. \label{fig:g1}}
    
\end{figure}

We consider the first-order coherence function for a driven qubit, Example A. This function is directly connected to the coherence properties of photons emitted by the qubit into its surroundings, as we discuss in Sec.~\ref{sec:emission}. Fig.~\ref{fig:g1} shows $g^{(1)}(\tau)$ for two different driving strengths. We observe either underdamped or overdamped behaviour in time, depending on whether the Rabi frequency $\Omega$ is larger or smaller than the damping rate $\gamma$. These two different regimes give rise to very different frequency spectra of the emitted photons (see Sec.~\ref{sec:emission}).

\section{Full Counting Statistics}\label{sec:FCS}

So far we have focused on the first and second moments of the current/charge. 
We can also take a step further and consider the full statistics of the net charge $N(t)$ [Eq.~\eqref{total_charge}], i.e., the probability distribution $P(n,t)\equiv P\big(N(t)=n\big)$.
From this one can compute, for example,  \rev{$dE\big(N(t)\big)/dt = J(t)$ and $d{\rm Var}\big(N(t)\big)/dt =  D(t)$}, as studied in the previous section. 
In addition, the tools we now develop will allow us to generalize our previous results, and put quantum jumps and quantum diffusion within a unified framework.

\subsection{Quantum jumps}
\label{sec:FCS quantum jumps}

Let $\mathcal{N}$ denote the set of allowed values that $N(t)$ might take (note that even though the $N_k$ are integers, the $\nu_k$ are arbitrary, so the set $\mathcal{N}$ is not necessarily restricted to the integers).
To obtain $P(n,t)$, we introduce the so-called \emph{$n$-resolved density matrix} $\rho_n(t)$ \rev{\cite{cook_1981,gurvitz_1998}}, i.e., the unnormalized density matrix given that after a time $t$ the net charge is $n$. From Eq.~\eqref{kraus_channel}, we can infer that this must evolve as
\begin{equation}
    \label{eq:nresolvedkraus}
    \rho_n(t+dt) = M_0 \rho_n(t) M_0^\dagger + \sum_{k=1}^\nops M_k \rho_{n-\nu_k}(t) M_k^\dagger.
\end{equation}
This equation simply states that a jump in channel $k$ changes the total charge by $\nu_k$. The probability distribution for the total charge is then given by the normalization, $P(n,t) = \tr\{\rho_n(t)\}$, and the original density matrix can be recovered as 
$\rho = \sum_{n \in \mathcal{N}} \rho_n(t)$. 
Equation~\eqref{eq:nresolvedkraus} is considerably simplified by Fourier transforming the density matrix
\begin{equation}
    \label{FCS_rhochi}
    \rho_\chi(t) = \sum_{n \in \mathcal{N}} e^{in\chi}\rho_n(t),
\end{equation}
where $\chi$ is known as the \textit{counting field}. Fourier transforming Eq.~\eqref{eq:nresolvedkraus} and using Eq.~\eqref{kraus} for the Kraus operators, we find that the counting-field dependent density matrix $\rho_\chi$ obeys a generalized QME
\begin{equation}\label{gQME}
    \frac{d\rho_\chi(t)}{dt} = \mathcal{L}_\chi \rho_\chi(t), 
    \qquad \rho_\chi(0) = \rho_0,
\end{equation}
with the so-called \textit{tilted} Liouvillian
\begin{align}\label{tilted_liouvillian}
    \mathcal{L}_\chi\rho &= -i[H,\rho] + \sum_k \bigg( 
    e^{i \chi \nu_k} L_k \rho L_k^\dagger - \frac{1}{2} \{L_k^\dagger L_k, \rho\}\bigg).
\end{align}
Notice that for $\chi=0$ the generalized QME~\eqref{gQME} reduces to the usual QME in Eq.~\eqref{M}, since $\mathcal{L}_{\chi=0} = \mathcal{L}$.
The tilted Liouvillian is a crucial object that greatly simplifies the analysis. 
Note how the counting field appears only in the jump part, and pinpoints precisely the terms we associate to each jump channel. It also weights them with the appropriate $\nu_k$.
In fact, $\mathcal{L}_\chi$ encodes all relevant superoperators in its derivatives at $\chi=0$. 
For example, 
\begin{equation}\label{FCS_L_prime_jump}
    \mathcal{L}'\rho := \partial_\chi \mathcal{L}_\chi \rho \big|_{\chi=0} =
    i \sum_k \nu_k~L_k \rho L_k^\dagger
    = i \mathcal{J}\rho
\end{equation}
is the $\mathcal{J}$ superoperator  defined in Eq.~\eqref{J_op}. 
Similarly, the second derivative reads
\begin{equation}\label{FCS_L_double_prime_jump}
    \mathcal{L}''\rho = - \sum_k \nu_k^2 L_k \rho L_k^\dagger,
\end{equation}
which, \rev{upon taking the trace}, is minus the term $K$ in Eq.~\eqref{K}.

The solution of Eq.~\eqref{gQME} is $\rho_\chi(t) = e^{\mathcal{L}_\chi t} \rho_0$, which does not have unit trace because $\mathcal{L}_\chi$ is not trace preserving for $\chi \neq 0$.
From this solution, the desired probability distribution can be obtained as the inverse Fourier transform
\begin{equation}\label{FCS_P}
    P(n,t) = \int\limits_{-\infty}^\infty \frac{d\chi}{2\pi} e^{-i n \chi}~\tr \{\rho_\chi(t)\big\}.
\end{equation}
To be precise, this will not exactly yield $\tr\big\{\rho_n(t)\big\}$, but rather a sum of Dirac-$\delta$ functions at the allowed values $\mathcal{N}$, each with weight $\tr\big\{\rho_n(t)\big\}$.
If the set $\mathcal{N}$ comprises integers only, e.g., for particle current and dynamical activity, then $P(n,t)$ can be obtained without the Dirac $\delta$ functions by changing the integration limits to
\begin{equation}\label{FCS_P2}
    P(n,t) = \int\limits_{-\pi}^\pi \frac{d\chi}{2\pi} e^{-i n \chi}~\tr \{\rho_\chi(t)\big\}.
\end{equation}

\subsubsection{$P(n,t)$ for Example A}

\begin{figure}
    \centering
    \includegraphics[width=0.45\textwidth]{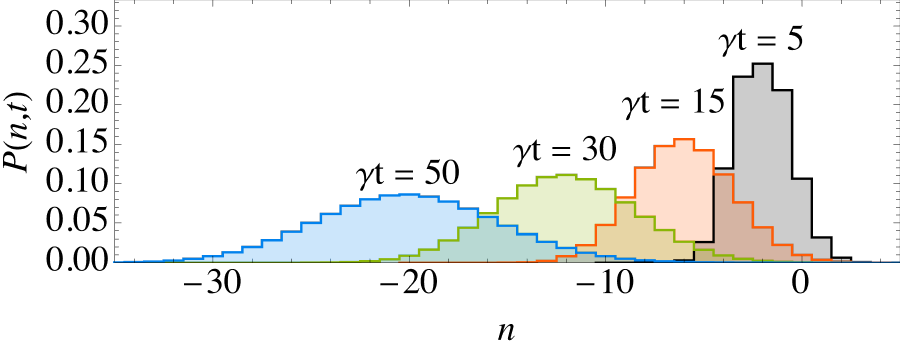}
    \caption{FCS probability $P(n,t)$, Eq.~\eqref{FCS_P2} for Example A \rev{(Sec.~\ref{ssec:exampleA})}. The parameters chosen are the same as those of Fig.~\ref{fig:ExampleA_trajs}. The negative count corresponds to the particle current flowing from cold to hot.
    }
    \label{fig:fcs_prob}
\end{figure}

The stochastic trajectories for Example A were studied in Sec.~\ref{sec:Rabi oscillations in a single qubit} and  Fig.~\ref{fig:ExampleA_trajs}. 
For that configuration, the charge $N(t)$ was more likely to be negative, although it could also be positive for some stochastic trajectories. 
We can compute $P(n,t)$ numerically
using Eq.~\eqref{FCS_P2}, as follows.
We first set a grid of $\chi$ values  from $-\pi$ to $\pi$. For each $\chi$, and for each time $t$, we compute $\rho_\chi(t) = e^{\mathcal{L}_\chi t}\rho_0 $ and its trace. Finally, we use this to integrate Eq.~\eqref{FCS_P2} numerically.  
Figure~\ref{fig:fcs_prob} illustrates the result for a grid of 1000 $\chi$ values.
For short times, the distribution is centered around $0$, and has a non-negligible portion with $n>0$. 
For later times, it then moves to the left and  $n>0$  becomes less and less likely. 
In this example, excitations continue to be emitted indefinitely, so that $P(n,t)$ will continue to move left. 
It also becomes more and more spread with time.

\subsubsection{$P(n,t)$ for Example D}

\begin{figure}
    \centering
    \includegraphics[width=\columnwidth]{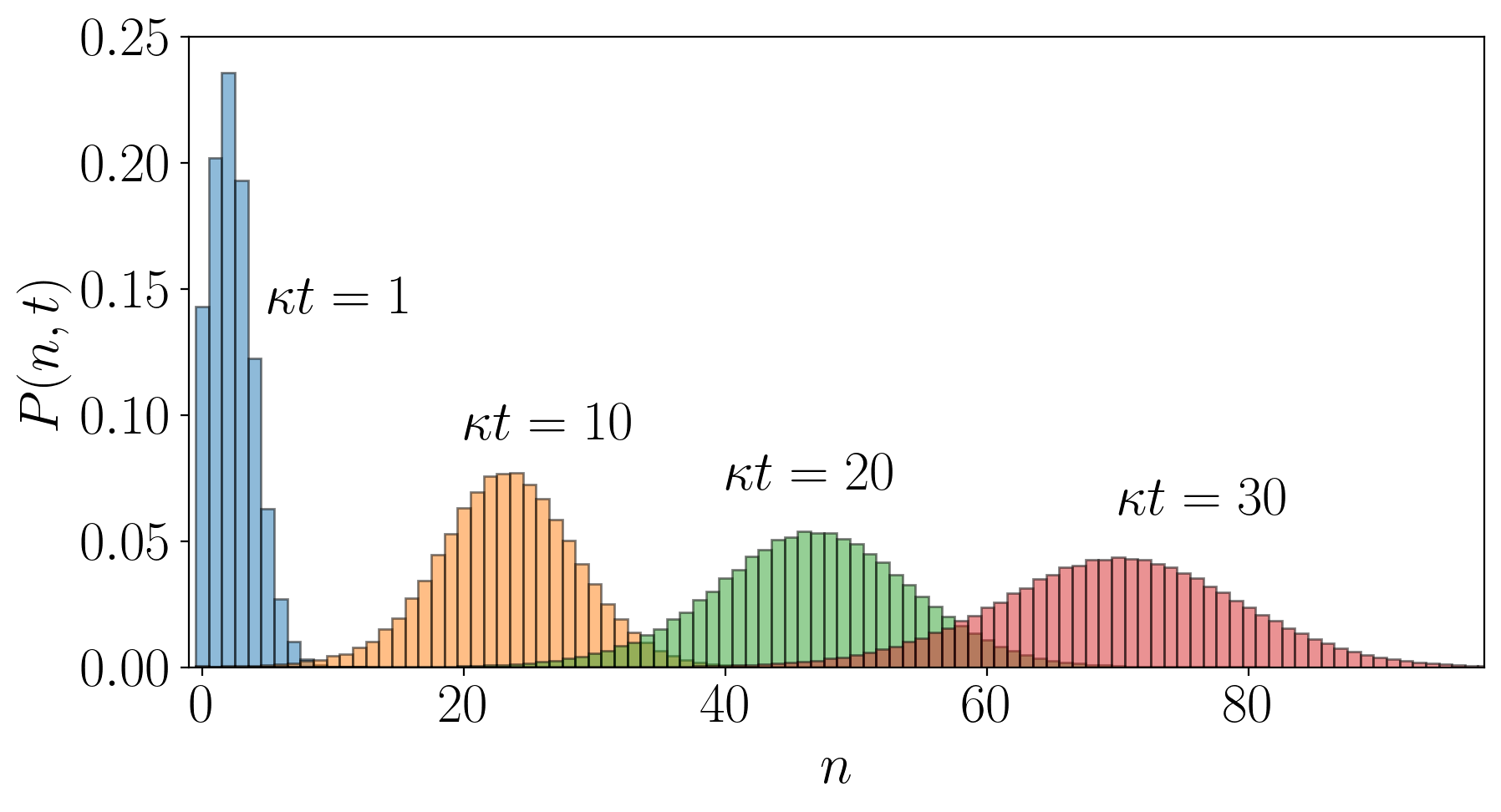}
    \caption{FCS probability $P(n,t)$ \rev{[Eq.~\eqref{FCS_P2}]} for Example D \rev{(Sec.~\ref{ssec:exampleD})} with with $G=1$, $U=1/3$, $\Delta=0$, and $\kappa=1$ for varying times.  }
    \label{fig:Pnt_exampleD}
\end{figure}

We follow the same procedure as the previous section to compute the FCS distribution for photon counting in Example D. The results are plotted in Fig.~\ref{fig:Pnt_exampleD}, which shows the average number of counts $n$ increasing with time. 
Here we use parameters that are different to those that generate the steady state in Fig.~\ref{fig:ExampleD_wigner}(c) with $G=1$, $U=1/3$, $\Delta=0$, and $\kappa=1$.

\subsubsection{Classical master equations}
\label{sec:classical_ME}

FCS was originally developed in the context of classical master equations. 
These are  included in the quantum result as a particular case. 
Consider a QME with $H = \sum_i E_i |i\rangle\langle i|$ and jump operators $L_{ij} = \sqrt{W_{ij}} |i\rangle\langle j|$, describing jumps between energy eigenstates, with transition rate $W_{ij}$ (for $i \neq j$).
From the QME, the diagonal elements $p_n = \langle n | \rho | n \rangle$ will evolve according to the classical (Pauli) master equation
\begin{equation}\label{Pauli_master_equation}
    \frac{dp_n}{dt} = \sum_j W_{nj} p_j - \Gamma_n p_n, 
\end{equation}
where $\Gamma_n = \sum_{j} W_{jn}$ is called the escape rate from level $n$. 
If the system starts without any coherences in this basis, it will remain so throughout. Hence, to fully describe the system it suffices to look at the populations $p_n$. 

The first term in Eq.~\eqref{Pauli_master_equation} describes jumps (since it describes a process which goes from $j\to n$) while the second describes the "no-jump events".
The tilted Liouvillian is still given by Eq.~\eqref{tilted_liouvillian}. 
But since we only want the dynamics of the populations, it is simpler to define a matrix $\mathbb{W}_\chi$ with entries  
\begin{equation}
    (\mathbb{W}_{\chi})_{nj} = \begin{cases} 
        W_{nj} e^{i \chi \nu_{nj}} & j\neq n,
        \\[0.2cm]
        -\Gamma_n & j = n,
    \end{cases}
\end{equation}
where $\nu_{nj}$ is the weight factor associated to the jump operator $L_{nj}$.
Letting $p_{\chi,n}(t) = \langle n | \rho_\chi(t) |n\rangle$, the vector $\bm{p}_\chi(t)$ will evolve according to
$\bm{p}_\chi(t) = e^{\mathbb{W}_\chi t} \bm{p}(0)$. 
Hence 
\begin{equation}
    \tr\{\rho_\chi(t)\} = \sum_{nj} (e^{\mathbb{W}_\chi t})_{nj} p_j(0),
\end{equation}
from which $P(n,t)$ can be computed using Eq.~\eqref{FCS_P}.

In a classical master equation all transitions are associated to jumps and can thus be monitored, at least in principle. 
The crucial difference of the quantum case is the presence of the unitary dynamics $-i[H,\rho]$, which generates transitions that, fundamentally, cannot be monitored and hence will never correspond to a click in a detector.

\subsection{Quantum diffusion}
\label{sec:FCS_quantum_diffusion}

The integrated charge in quantum diffusion, 
$N_{\rm diff}(t) = \int_0^t dt'~I_{\rm diff}(t')$ varies continuously.
Nevertheless, the resulting probability distribution $P(n,t)$ can still be written as in Eq.~\eqref{FCS_P}, where $\rho_\chi$ still satisfies~\eqref{gQME}. 
What changes is the shape of the tilted Liouvillian $\mathcal{L}_\chi$. 
Instead of Eq.~\eqref{tilted_liouvillian}, we now get 
\begin{equation}
\label{L_chi_diffusion}
    \mathcal{L}_\chi \rho = \mathcal{L} \rho + i \chi \mathcal{H}\rho - \frac{\chi^2}{2} K_{\rm diff}\rho,
\end{equation}
where $K_{\rm diff}$ and $\mathcal{H}$ are defined in Eqs.~\eqref{eq:Kdiff} and~\eqref{H_hom_op}.
This result is derived in Appendix~\ref{app:fcs_diffusive} in two ways: as a limit of the quantum jump model when $|\alpha_k|\to \infty$, and for the Gaussian POVM model discussed at the end of Sec.~\ref{sec:quantum_diffusion}. 
In particular, recall that the latter corresponds to a Hermitian jump operator $L = \sqrt{\lambda}Y$ (where $Y$ is the observable being measured), and a weight $\nu = 1/(2\sqrt{\lambda})$ for the current. 
The corresponding tilted Liouvillian therefore reads
\begin{equation}
    \mathcal{L}_\chi \rho = -i[H,\rho] +\lambda \mathcal{D}[Y]\rho+i\frac{\chi}{2}\{Y,\rho\}-\frac{\chi^2}{8\lambda}\rho.
\end{equation}
This equation nicely illustrates the trade-off between the measurement backaction --- associated to the dissipator $ \lambda\mathcal{D}[Y]$ --- and the white  noise $K_{\rm diff} \propto 1/\lambda$:
A large measurement strength (large $\lambda$) causes significant backaction, but reduces the white noise, while for small $\lambda$, it is the other way around.

\subsubsection{$P(n,t)$ for Example C}

\begin{figure}
    \centering
    \includegraphics[width=0.5\textwidth]{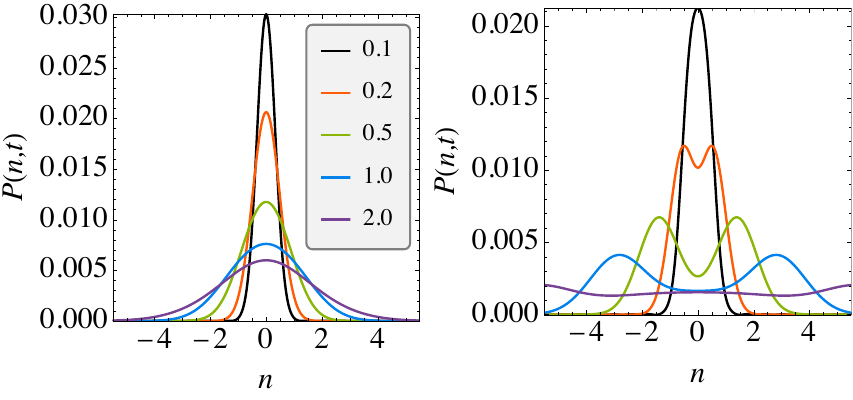}
    \caption{FCS probability distributions $P(n,t)$ for Example C \rev{(Sec.~\ref{ssec:exampleC})}, in the same regimes as studied in Fig.~\ref{fig:quantum_diffusion}. 
    Left: $\Gamma/\Omega = 0.2$.
    Right: $\Gamma/\Omega = 2.0$.
    Each curve is for a different time $\Omega t$, as denoted in the legend. 
    }
    \label{fig:fcs_prob_diffusion_example}
\end{figure}

Stochastic trajectories for Example C were studied in Sec.~\ref{sec:Homodye detection of a qubit} and  Fig.~\ref{fig:quantum_diffusion}. 
There, we saw that depending on the relative damping strength $\Gamma/\Omega$, the quantum trajectories were fundamentally different, presenting an oscillatory behavior for small $\Gamma$ and a jump behavior for large $\Gamma$. 
Figure \ref{fig:fcs_prob_diffusion_example} shows the corresponding FCS probability distribution $P(n,t)$ for the integrated charge. 
In this case the operator being measured is $\sigma_z$. Since the system oscillates symmetrically between the eigenstates of $\sigma_z$, we find a distribution that is symmetrical around $n=0$ so that on average $J_{\rm diff} = 0$. 
However, we see that $P(n,t)$ behaves in fundamentally different ways in the two regimes:
For small $\Gamma$ (left plot) the oscillatory nature of the trajectories results in a distribution of the integrated current made up of a single centered peak. Conversely, for large $\Gamma$ (right plot), the jump behavior results in a bimodal distribution. This happens because in this case, $\langle \sigma_z\rangle_c$ spends most of its time close to $\pm 1$, as in the red curve in Fig.~\ref{fig:quantum_diffusion}.

\subsubsection{Statistics of an undriven cavity}
\label{sec:cavity_statistics}

Consider the cavity model in Example D. 
However, let us assume that there is no external drive and no Kerr non-linearity. 
All we have is an optical cavity with dissipation. 
In the interaction picture with respect to $H = \omega_c a^\dagger a$ the Liouvillian reads simply 
$\mathcal{L}\rho = \kappa \mathcal{D}[a]\rho$.
Suppose that the cavity is prepared in some non-trivial state $\rho_0$ and, at $t=0$, we begin measuring the photons that leak out.
Each time we repeat this experiment we will therefore count a certain number $n$ of photons, which will be described by the FCS probability~$P(n,\infty)$ given by Eq.~\eqref{FCS_P2}.
In Appendix~\ref{app:cavity_probabilities} we show that in this particular model $P(n,\infty)$  yields the same statistics as a projective measurement of $\rho_0$ on the Fock basis,
\begin{equation}\label{empty_cavity_photo_detection}
    P(n,\infty)= \langle n | \rho_0 | n\rangle.
\end{equation}
While this may seem intuitive, it is not at all obvious. After all, a projective measurement is instantaneous, while the continuous measurement is extended in time.

Next, consider the case of homodyne measurements (Table~\ref{tab:diffusion_experiments}). 
We show, again in Appendix~\ref{app:cavity_probabilities}, that Eq.~\eqref{FCS_P} yields 
\begin{equation}\label{empty_cavity_homodyne}
    P(x,\infty) = \langle x | \rho_0 |x\rangle,
\end{equation}
which is a projective measurement of the initial state in the basis of the quadrature operator $x = (a e^{-i\phi} +a^\dagger e^{i\phi})/\sqrt{2}$ (these are the marginals of the Wigner function).
In this case, there is a caveat, though: 
The initial field present in the cavity leaks out at the amplitude damping rate $\kappa/2$. As time goes on, the fraction of signal coming from the initial state thus diminishes and at long times, only photons from the local oscillator are measured.
To amend for this, we must use a time-dependent weight factor $\nu(t) = \sqrt{\kappa/2} e^{-\kappa t/2}$~\cite{Wiseman_2009}, that matches the amplitude damping rate and suppresses the counts as time progresses. 
Only for this specific choice of $\nu(t)$ will Eq.~\eqref{empty_cavity_homodyne} follow.

Finally, we can consider heterodyne  detection. 
This will also require a similar time-dependent weight factor $\nu(t) = \sqrt{\kappa} e^{-\kappa t/2}$. 
The joint measurement of the two orthogonal quadratures $x_0$ and $x_{\pi/2}$ will then yield a joint distribution which coincides with the Husimi function 
\begin{equation}\label{empty_cavity_heterodyne}
    P( x_0  , x_{\pi/2} ,\infty) = \frac{1}{\pi} \langle \alpha | \rho_0| \alpha\rangle,
\end{equation}
where $|\alpha\rangle$ is a coherent state at $\alpha = ( x_0  + i  x_{\pi/2} )/\sqrt{2}$.
Since coherent states are not orthogonal, this measurement  is not projective, but instead corresponds to a generalized measurement with operators $M_\alpha = \frac{1}{\sqrt{\pi}} |\alpha\rangle \langle \alpha|$.

We note that alternative derivations of Eqs.~(\ref{empty_cavity_photo_detection}-\ref{empty_cavity_heterodyne}) starting from the quantum Langevin equation can be found in Sec.~4.7.6 of Ref.~\cite{Wiseman_2009}. These complement our derivations starting from the Lindblad master equation given in Appendix~\ref{app:cavity_probabilities}.

\subsection{Fluctuations for arbitrary $\mathcal{L}_\chi$}

The dependence of $\mathcal{L}_\chi$ with $\chi$ in Eq.~\eqref{L_chi_diffusion} is fundamentally different from the jump case in Eq.~\eqref{tilted_liouvillian}.
Nevertheless, as in Eqs.~\eqref{FCS_L_prime_jump} and~\eqref{FCS_L_double_prime_jump}, we can still obtain the relevant superoperators by taking derivatives with respect to $\chi$; namely,
\begin{align}
    \mathcal{L}'\rho &= i \mathcal{H}\rho,
    \qquad 
    \mathcal{L}''\rho = - K_{\rm diff} \rho,
\end{align}
which is in one-to-one correspondence to
Eqs.~\eqref{FCS_L_prime_jump} and~\eqref{FCS_L_double_prime_jump}. 
This suggests that all quantities studied in Sec.~\ref{sec:fluctuations} -- namely $J$, $F$, $S$ and $D$ -- can actually be written in terms of $\mathcal{L}'$ and $\mathcal{L}''$. 
Indeed, we show in Appendix~\ref{app:FCS_moments} that all equations of Sec.~\ref{sec:fluctuations} still hold for a \emph{general} tilted Liouvillian, provided we identify 
\begin{align}
\label{J_K_general_tilted}
    \genJ \rho := -i \mathcal{L}'\rho,
    \qquad 
    \genK := - \tr\big\{\mathcal{L}''\rho\big\}.
\end{align}
For quantum jumps $\genJ$ and $\genK$ reduce to Eqs.~\eqref{J_op} and~\eqref{K}. 
For diffusion $\mathcal{J}$ becomes $\mathcal{H}$ in Eq.~\eqref{H_hom_op} and $K$ becomes $K_{\rm diff}$ in Eq.~\eqref{eq:Kdiff}.
Equation~\eqref{J_K_general_tilted} provides a substantial generalization to our previous results: it not only unifies the formulas for quantum jumps and quantum diffusion, but also holds for \emph{any} tilted Liouvillian (and therefore can encompass other scenarios that are not covered in this tutorial).

\subsection{Cumulant generating function}
\label{sec: Cumulant generating function}

The Fourier transform of a probability distribution is called its characteristic function.
Returning to Eq.~\eqref{FCS_P}, we see that this is precisely what we have:  the quantity inside the integral is the characteristic function of $P(n,t)$:
\begin{equation}\label{FCS_M}
    M(\chi,t) := E\big[ e^{i N(t) \chi} \big] = \tr \{ \rho_\chi(t)\} = \tr\big\{ e^{\mathcal{L}_\chi t} \rho(0)\big\}.
\end{equation}
From $M$ we can extract all moments as 
\begin{equation}\label{FCS_moments}
    E\big[N(t)^j\big] = (-i \partial_\chi)^j M(\chi,t)\Bigg|_{\chi=0}.
\end{equation}
When interested in higher moments, it is often convenient to consider the cumulants, here denoted by $\llangle N(t)^j\rrangle$, which can be obtained from the cumulant generating function $C(\chi,t)$, defined as
\begin{equation}
    \label{eq:cgf}
    C(\chi,t) = \ln M(\chi,t),\hspace{.5cm}\llangle N(t)^j\rrangle=(-i \partial_\chi)^j C(\chi,t)\Big|_{\chi=0}.
\end{equation}
The first three cumulants are equal to the mean, the variance, and the skewness respectively.
In the long-time limit, the cumulant generating function takes on a particularly simple form, 
\begin{equation}\label{CGF_long_times}
    C(\chi,t) \simeq \lambda_0(\chi) t,
\end{equation}
where $\lambda_0(\chi)$ is the eigenvalue of $\mathcal{L}_\chi$ with the largest real part (the spectral properties of Liouvillians will be discussed in  Sec.~\ref{sec:methods}, and the proof of this result is given in Appendix~\ref{app:scgf}).
This eigenvalue is such that $\lambda_0(0) = 0$. 
\rev{In the past, the eigenvalue $\lambda_0(\chi)$ has been the subject of considerable attention, as it turns out to contain substantial information about the system, including properties such as e.g. phase transitions~\cite{Flindt2013}.}

Equation~\eqref{CGF_long_times}, together with~\eqref{eq:cgf}, shows that for large times, all cumulants will be proportional to $t$. 
This is consistent with  what we found in Eq.~\eqref{varN_D}, and is consistent with the results of Fig.~\ref{fig:fcs_prob}.
For this reason, we define the scaled cumulant generating function (SCGF) as 
\begin{equation}
\label{SCGF_2}
    C(\chi)=\lim\limits_{t\to \infty} \partial_t C(\chi,t)  = \lambda_0(\chi).
\end{equation}
From $C(\chi)$ we can determine all scaled cumulants by taking derivatives as $(-i \partial_\chi)^j C(\chi)\big|_{\chi=0}$. The first scaled cumulant is $J$, and the second is $D$. 
We also note that even when the eigenvalues of $\mathcal{L}_\chi$ cannot be computed analytically, the cumulants can sometimes be obtained by expanding the characteristic polynomial of $\mathcal{L}_\chi$ \cite{Bruderer_2014}.

For large times the distribution $P(n,t)$ will often look Gaussian. However, the fact that all cumulants scale with $t$ means that there will be deviations from Gaussianity, which can be observed by looking at the tails of the distribution (usually with a logscale plot). 
These so-called \emph{large deviations} represent a field of study in itself; see for instance Ref.~\cite{Touchette2009}.

The asymptotic distribution $P(n,t)$ can be obtained by inverse Fourier transforming Eq.~\eqref{CGF_long_times}, as in Eq.~\eqref{FCS_P2}:
\begin{equation}\label{SCGF_reconstructing_P}
    P(n,t) = \int \frac{d\chi}{2\pi}e^{-in\chi}e^{C(\chi)t}. 
\end{equation}
We note, however, that this may result in negative probabilities if $t$ is chosen such that the long-time limit is not justified \rev{(an example will be given below in Fig.~\ref{fig:saddle})}.

\subsubsection{SCGF for Example B}
\label{sec:scgfexb}
Consider Example B, a quantum dot coupled to two fermionic reservoirs [cf.~Sec.~\ref{ssec:exampleB}].
This example is sufficiently simple that the SCGF can be obtained analytically using the vectorization methods that will be discussed in Sec.~\ref{sec:methods}.  
The result is 
\begin{equation}
\label{eq:scgfb}
\begin{aligned}
    &C(\chi) = -\frac{\gamma_L+\gamma_R}{2} \\&+ \sqrt{\left(\frac{\gamma_L+\gamma_R}{2}\right)^2+\gamma_L\gamma_R\left[\big(e^{i\chi}-1\big) f_R\bar{f}_L +\big(e^{-i\chi}-1\big) f_L\bar{f}_R\right]},
    \end{aligned}
\end{equation}
where $\bar{f}_i = 1-f_i$.
Computing derivatives as in Eq.~\eqref{eq:cgf} yields all scaled cumulants.
The first is the average current $J$ in Eq.~\eqref{ExampleB_Jss}.
The second is the noise, which in this case reads 
\begin{equation}
\begin{aligned}\label{ExampleB_D}
    D = \frac{\gamma_L\gamma_R}{(\gamma_L+\gamma_R)^3}\bigg[(\gamma_L+\gamma_R)^2&(f_L\bar{f}_L+f_R\bar{f}_R)\\&+(\gamma_L^2+\gamma_R^2)(f_R-f_L)^2\bigg],
    \end{aligned}
\end{equation}
and so on for higher order cumulants. 
In studies of electronic transport, the first term of $D$ is usually referred to as thermal noise, as it remains present in equilibrium ($f_L=f_R$) and only vanishes at zero temperature. The second line, which vanishes in equilibrium, is referred to as shot noise \cite{blanter_2001}. In contrast, in quantum optics the white noise $K_{\rm diff}$ is referred to as shot noise (c.f.~Sec.~\ref{sec:two_point_function}). In both senses of the term, shot noise arises from the discrete nature of the particles that carry the observed current, i.e., electrons and photons respectively.

If one tunnel rate is much smaller than the other, the SCGF considerably simplifies and reduces to (choosing $\gamma_R\ll\gamma_L$)
\begin{equation}
\label{scgfbip}
C(\chi) = \gamma_R\left[\big(e^{i\chi}-1\big) f_R\bar{f}_L +\big(e^{-i\chi}-1\big) f_L\bar{f}_R\right].
\end{equation}
Note that the largest of the tunnel rates drops out of this expression. This is a general feature of long-time transport statistics: The smallest tunnel rate constitutes a bottleneck for transport and therefore determines the statistics. Before discussing the probability distribution associated to Eq.~\eqref{scgfbip}, we consider the regime where $f_R=1$ and $f_L = 0$ which is sometimes called the large bias regime, as it can be obtained by a large voltage bias across the quantum dot. In this regime the second term in Eq.~\eqref{scgfbip} vanishes and Eq.~\eqref{SCGF_reconstructing_P} reduces to a Poisson distribution
\begin{equation}
    \label{poisson}
    \begin{aligned}
    P(n,t) &= \int_{-\pi}^{\pi}\frac{d\chi}{2\pi}e^{-in\chi}e^{C(\chi)t} 
    \\& = e^{-\gamma_Rt}\int_{-\pi}^{\pi}\frac{d\chi}{2\pi}e^{-in\chi}e^{\gamma_Rte^{i\chi}} \\&= e^{-\gamma_Rt}\sum_{k=0}^\infty \frac{(\gamma_Rt)^k}{k!}\int_{-\pi}^{\pi}\frac{d\chi}{2\pi}e^{i(k-n)\chi}
    \\ &= \frac{(\gamma_Rt)^n}{n!}e^{-\gamma_Rt}.
    \end{aligned}
\end{equation}
Thus, in this regime the transport can be understood as being mediated by independent tunneling events into the right reservoir, see also Appendix~\ref{app:Poisson}. The Poisson distribution has the peculiar feature that all (scaled) cumulants are equal, here $J = \gamma_R$. 

Outside the regime $f_R=1, f_L=0$, the probability distribution corresponding to the SCGF in Eq.~\eqref{scgfbip} still has a closed form expression, which is given by 
\begin{equation}
    \label{pnbipoisson}
    P(n,t) = e^{-\gamma_Rt(f_R\bar{f}_L+f_L\bar{f}_R)}\left(\frac{f_R\bar{f}_L}{f_L\bar{f}_R}\right)^\frac{n}{2}I_n\left(2\gamma_Rt\sqrt{f_L\bar{f}_Lf_R\bar{f}_R}\right),
\end{equation}
where $I_n$ denotes the modified Bessel function of the first kind. This is a bi-directional Poisson distribution \cite{Esposito_2009}, which is determined by two rates: the rate for electrons to go from right to left $\gamma_Rf_R\bar{f}_L$ and from left to right $\gamma_Rf_L\bar{f}_R$. Similar to the Poisson distribution, its cumulants have a peculiar feature: all odd cumulants are equal to the difference of the rates, $J=\gamma_R(f_R-f_L)$, and all even cumulants are equal to the sum of the rates, $D=\gamma_R(f_L\bar{f}_R+f_R\bar{f}_L)$.

\subsubsection{SCGF for quantum diffusion}

For $\mathcal{L}_\chi$ given in Eq.~\eqref{L_chi_diffusion}, we can define a new state 
\begin{equation}
    \label{eq:fcsdiff}
    \tilde{\rho}_\chi(t) = e^{\frac{\chi^2}{2}K_{\rm diff}t}\rho_\chi(t).
\end{equation}
The corresponding generalized QME~\eqref{gQME} will be modified to depend on a new tilted Liouvillian 
\begin{equation}\label{SCGF_L_tilde_diffusion}
    \tilde{\mathcal{L}}_\chi = \mathcal{L} + i \chi \mathcal{H}.
\end{equation}
The SCGF $C(\chi)$ of the original process can then be related to the SCGF $\tilde{C}$ associated to $\tilde{\mathcal{L}}_\chi$ according to 
\begin{equation}
    \label{eq:scgfdiff}
    C(\chi) = - \frac{\chi^2}{2} K_{\rm diff} + \tilde{C}(\chi).
\end{equation}
The first term is the CGF of a Gaussian variable and will thus simply increase $D$ by $K_{\rm diff}$ [Eq.~\eqref{eq:Kdiff}]. All other cumulants are determined  by $\tilde{C}(\chi)$.
Note, though, that despite its simplicity, this term  $-\frac{\chi^2}{2}K_{\rm diff}$ is still crucial, as neglecting it can cause the probability distribution to acquire negative values.

Next, consider Eq.~\eqref{SCGF_L_tilde_diffusion} 
and assume we have a single jump operator $L$, which is also Hermitian. 
We will then have 
\begin{equation}
    \tilde{\mathcal{L}}_\chi \rho = -i [H,\rho] + \mathcal{D}[L]\rho + i \chi \nu \{L, \rho\}.
\end{equation}
We can now imagine a scenario where the dissipator $\mathcal{D}[L]$ is very small but, because of the weight $\nu$, the term proportional to $\chi$ is not. 
This is the case for the Gaussian POVM model ($L = \sqrt{\lambda} Y$ and $\nu = 1/(2\sqrt{\lambda})$) when $\lambda$ is very small.
In this limit we can neglect $\mathcal{D}[L]$, which allows us to  write down the solution for Eq.~\eqref{gQME}, leading to 
\begin{equation}
    \label{eq:fcscoh}
    \tilde{M}(\chi,t) = \tr\left\{e^{-iHt+i\frac{\chi}{2} Yt}\rho_0e^{iHt+i\frac{\chi}{2} Yt}\right\},
\end{equation}
where $Y = 2\nu L$ is the observable being measured.
This is the characteristic function that features in the theory of FCS of phase coherent systems \cite{levitov_1996,nazarov_2003}. 
It is well known that it can result in negative probabilities \cite{belzig_2001,Clerk_2011,hofer_2017}, which are a consequence of interference effects between different eigenstates of the observable $Y$ \cite{hofer_2016}.
These negativities actually happen because we are omitting the noise term; the characteristic function $M(\chi,t) = e^{-\chi^2 \nu^2 t/2} \tilde{M}(\chi,t)$ that describes an actual measurement leads to proper physical probabilities.
The general tilted Liouvillian in Eq.~\eqref{L_chi_diffusion} shows how Eq.~\eqref{eq:fcscoh} can be extended to include measurement noise and imprecision, as well as non-Hermitian jump operators.

As a particular case, if $[H,L]=0$ we can write $\tilde{M}(\chi,t) = \sum_y e^{i \chi y} p_y$, where $p_y = \langle y | \rho_0 |y\rangle$ and $Y =\sum_y y|y\rangle\langle y|$. 
Plugging $M(\chi,t) = e^{- \chi^2 \nu^2 t/2} \tilde{M}(\chi,t)$ in Eq.~\eqref{FCS_P} and carrying out the Gaussian integrals yields 
\begin{equation}
    P(n,t) = \sum_{y} p_y \frac{e^{-\frac{(n-yt)^2}{2 \nu^2 t}}}{\sqrt{2\pi \nu^2 t}}.
\end{equation}
The charge distribution is therefore a mixture of Gaussians, weighted by $p_y$, each with mean $yt$ and variance $\nu^2 t = t/4\lambda$.

\section{Solution methods}
\label{sec:methods}



\subsection{Computing $D$ using an auxiliary equation}
\label{sec:noise_auxiliary_MEQ}

There is a simple method to compute the noise $D(t)$ [Eq.~\eqref{D_integral_time}], which also works away from the steady state and even in driven systems~\cite{Schaller2014}. 
Let us first write it more explicitly using Eq.~\eqref{F_jump}
\begin{equation}\label{sigma_method_D_original}
    D(t) = K(t) + 2 \int\limits_0^t d\tau \Bigg( \tr\Big\{ \mathcal{J} e^{\mathcal{L}\tau} \mathcal{J} \rho(t-\tau)\Big\} - J(t) J(t-\tau)\Bigg).
\end{equation}
Now define the (traceless) superoperator $\mathcal{G}(\rho) := \mathcal{J} \rho - \rho \tr(\mathcal{J} \rho)$ and  an auxiliary variable $\sigma(t)$, which is the solution of
\begin{equation}\label{sigma_method_auxiliary_MEQ}
    \frac{d\sigma}{dt} = \mathcal{L}\sigma(t) + \mathcal{G}(\rho(t)),
    \qquad \sigma(0) = 0.
\end{equation}
This is similar to the original QME~\eqref{M}, but with an inhomogeneous RHS depending on $\rho(t)$ (the instantaneous  true state at time $t$). 
Even though $\mathcal{G}$ is a non-linear superoperator in $\rho$, this is still a linear equation in $\sigma$. The solution is therefore 
\begin{equation}\label{sigma_method_auxiliary_equation}
    \sigma(t) = \int\limits_0^t d\tau e^{\mathcal{L}\tau} \mathcal{G}(\rho(t-\tau)).
\end{equation}
One can now readily verify that Eq.~\eqref{sigma_method_D_original} can be written as 
\begin{equation}\label{sigma_method_D}
    D(t) = K(t) + 2 \tr\big\{ \mathcal{J} \sigma(t)\big\}. 
\end{equation}
Thus, we can obtain $D(t)$ by solving two master equations: the original one, yielding $\rho(t)$, and the auxiliary one given by Eq.~\eqref{sigma_method_auxiliary_equation}. 
\rev{Notice that while $\rho(t)$ affects $\sigma(t)$, the converse is not true. 
One may therefore view $\sigma(t)$ as a monitoring operator, which register information about $\rho(t)$.}

In the steady-state Eq.~\eqref{sigma_method_auxiliary_MEQ}  becomes the algebraic equation
\begin{equation}
    \mathcal{L}\sigma = - \mathcal{G}\rhoss = - \Big[ \mathcal{J} \rhoss - \rhoss~\tr(\mathcal{J}\rhoss)\Big].
\end{equation}
Thus, in addition to finding $\rhoss$, this method only requires solving an algebraic equation.
The method also works for driven systems, i.e., when $\mathcal{L}_t$ is time-dependent. 
One must simply account for this time-dependence when solving both the original QME~\eqref{M}, and the auxiliary equation~\eqref{sigma_method_auxiliary_equation}. 

\subsection{Vectorization}
\label{sec:vectorization}

One of the difficulties in working with the results developed in the previous section is that they always involve superoperators, such as $\mathcal{L}$ and $\mathcal{J}$, which multiply the state both on the left and right. 
We have already been able to simplify this somewhat by using the convention that $\mathcal{A} \mathcal{B} \rho \equiv \mathcal{A}\big( \mathcal{B}(\rho)\big)$, i.e., superoperators always act on whatever is on the right. This suggests that (linear) superoperators behave like matrices, while $\rho$ behaves like a vector. 
This connection is made precise by a procedure called \emph{vectorization} and, as we will see, it leads to a significant simplification of all the formulas. 
It also makes their numerical computation very efficient.

Vectorization is the operation which maps density matrices to vectors, and superoperators to matrices. 
There are many ways of doing this. The simplest is to \emph{stack columns}
\begin{equation}\label{vec_example}
{\rm vec} \begin{pmatrix}
a & b \\ c & d \end{pmatrix} 
=
\begin{pmatrix} a \\ c \\ b \\ d \end{pmatrix}.
\end{equation}
This allows us to write the density matrix as a vector.
To make the calculations cleaner, we will  use the following two notations interchangeably: 
\begin{equation}
    |\rho\rrangle = {\rm vec}(\rho).
\end{equation}

Given any 3 matrices $A, B, C$, the operation~\eqref{vec_example} satisfies the following useful property
\begin{equation}\label{vec_property}
    {\rm vec}(ABC) = (C\trans \otimes A) {\rm vec}(B)\,.
\end{equation} 
Notice that the RHS contains $C\trans$ and not $C^\dagger$. This happens because vectorization is a basis-dependent operation. 
Using Eq.~\eqref{vec_property} we can convert, for instance
\begin{equation}
    {\rm vec}(L \rho L^\dagger) = (L^* \otimes L) |\rho\rrangle.
\end{equation}
The result is therefore a bigger matrix $L^*\otimes L$, acting on the vector $|\rho\rrangle$.
We can also do the same thing for terms such as $[H,\rho]$: 
we simply write $H\rho = H \rho \id$, where $\id$ is the identity. 
We then get 
\begin{equation}
    {\rm vec}\big([H,\rho]\big) = \big(\id \otimes H - H\trans \otimes \id\big)|\rho\rrangle.
\end{equation}
Our original Liouvillian~\eqref{M} is therefore mapped into 
\begin{IEEEeqnarray}{rCl}
\label{vec_vectorized_L}
\mathcal{L} &=& -i (\id\otimes H - H\trans \otimes \id) 
\\[0.2cm]&&
+ \sum\limits_{k=1}^\nops  \Bigg[L_k^* \otimes L_k - \frac{1}{2} \id \otimes L_k^\dagger L_k - \frac{1}{2} (L_k^\dagger L_k)\trans \otimes \id\Bigg]\,,
\nonumber
\end{IEEEeqnarray}
which is a matrix of size $d^2$ (where $d$ is the dimension of the Hilbert space).
Sometimes it might be convenient to write the vectorized superoperator with a hat, as $\hat{\mathcal{L}}$. However, as we will see, it is usually clear from the context if the superoperator is vectorized or not, so for simplicity we will use the same symbol for both.

Vectorization takes $|\psi\rangle\langle \phi| \to |\phi\rangle^* \otimes |\psi\rangle$. 
Moreover, its inner product coincides with the Hilbert-Schmidt product between operators, $\llangle A | B \rrangle = \tr(A^\dagger B)$.
In particular, the normalization condition becomes 
\begin{equation}
    \tr(\rho) = \llangle \id | \rho \rrangle,
\end{equation}
where $|\id\rrangle = \sum_i |i\rangle^* \otimes |i\rangle$ is the vectorized identity. 

The steady-state Eq.~\eqref{steady_state} is now written as 
\begin{equation}\label{steady_state_vec}
    \mathcal{L} \rhossV = 0.
\end{equation}
In addition,  $\mathcal{L}$ is traceless, which means that  $\tr\big\{ \mathcal{L}(\rho)\big\} = 0$ for any state $\rho$. 
In vectorized notation this can be written as
\begin{equation}\label{traceless_vec}
    \idV \mathcal{L} = 0.
\end{equation}
Eqs.~\eqref{steady_state_vec} and~\eqref{traceless_vec} are in fact eigenvalue-eigenvector equations, with eigenvalue 0.
Since $\mathcal{L}$ is not Hermitian, it will have different left and right eigenvectors for each eigenvalue. Thus, $\rhossV$ is the \emph{right} eigenvector of $\mathcal{L}$ with eigenvalue 0, while  
$\idV$ is the \emph{left} eigenvector with the same eigenvalue. 
More generally, the eigenvalue-eigenvector equations will have the form
\begin{align}\label{L_eigenthings}
    \mathcal{L} |x_j\rrangle = \lambda_j |x_j\rrangle,
    \qquad 
    \llangle y_j | \mathcal{L} = \lambda_j \llangle y_j |. 
\end{align}
Since $\mathcal{L}$ is not Hermitian,  $\llangle y_j | \neq (|x_j\rrangle)^\dagger$, but they still satisfy 
\begin{equation}\label{L_orthogonality_vecs}
    \llangle y_j | x_k \rrangle = \delta_{jk}.
\end{equation}
We will adopt the convention that $\lambda_{0}=0$, so $|x_0\rrangle = \rhossV$ and $\llangle y_0| = \idV$. 
\rev{We also mention that if the eigenvalues $\lambda_j$ are complex, they will always come in complex-conjugate pairs~\cite{Albert2014}.}

The spectrum of the Liouvillian ultimately determines all dynamical properties. 
Because $\mathcal{L}$ is traceless, it always has \emph{at least} one eigenvalue 0, which therefore means there is also at least one steady state. 
Some systems might have multiple steady states (see~\cite{Buca_2012,Lieu2020} for examples).
For simplicity, we will always assume the steady-state is unique. 
As a consequence, all eigenvalues except $\lambda_{0}$, will be non-zero. 
Another point we implicitly assumed above was that $\mathcal{L}$ was in fact diagonalizable, meaning it has a set of $d^2$ linearly independent eigenvectors. Some non-Hermitian matrices might not be \rev{\cite{Minganti2019}}.
But again, this is  rare and we will assume for simplicity that $\mathcal{L}$ is diagonalizable. 

With these assumptions, the vectorized Liouvillian can be decomposed as 
\begin{equation}\label{L_eigen_decomp}
    \mathcal{L} = \sum_{j\neq0} \lambda_j |x_j\rrangle\llangle y_j|,
\end{equation}
where we exclude $j=0$ since $\lambda_{0}=0$. 
The full set of eigenvectors also satisfy the completeness relation
\begin{equation}\label{identity_matrix_decomposition}
    \rhossV \idV + \sum_{j\neq0}  |x_j\rrangle\llangle y_j| =\id_{d^2}.
\end{equation}
And finally, the matrix exponential appearing in Eq.~\eqref{M_sol} can be written as 
\begin{equation}\label{exp_L}
    e^{\mathcal{L}t } = \rhossV \idV + \sum_{j\neq0} 
    e^{\lambda_j t}|x_j\rrangle\llangle y_j|.
\end{equation}
This is a crucial result, as it completely specifies the dynamics of the system. 
The imaginary part of the $\lambda_j$ gives rise to oscillations. 
And if the dynamics is stable, we must have ${\rm Re}(\lambda_j) < 0$,
which will cause the second term in Eq.~\eqref{exp_L} to decay exponentially in time, until eventually all that survives is the first term.
In fact, since $\idV \rho_0\rrangle = 1$ by normalization, we have that 
\begin{align}    
|\rho(t) \rrangle &= e^{\mathcal{L}t} |\rho_0\rrangle
    = \rhossV + 
    \sum_{j\neq0} 
    e^{\lambda_j t}|x_j\rrangle\llangle y_j|\rho_0\rrangle,
\end{align}
which tends to $\rhossV$ when $t\to \infty$.

Since $\lambda_0 = 0$, the Liouvillian is not invertible. 
Instead, very often the results can be expressed in terms of the Drazin pseudo-inverse, defined as 
\begin{equation}\label{eigen_Drazin}
    \mathcal{L}^+ := \sum\limits_{j\neq0} \frac{1}{\lambda_j} |x_j\rrangle \llangle y_j|.
\end{equation}
i.e., we ``invert what we can.''
The properties of the Drazin inverse are discussed in detail in Appendix~\ref{app:pseudoinverse}, where we show that it can also be equivalently written as 
\begin{equation}
\label{Drazin_integral_def}
    \mathcal{L}^+  = -\int\limits_0^\infty d\tau e^{\mathcal{L}\tau}\Big(1 - \rhossV \idV \Big).
\end{equation}

\subsection{Computation of fluctuations in vectorized notation}
\label{sec:numerics_fluctuations_vectorized_notation}

We are now in a position to rewrite all of our main results in terms of the vectorized Liouvillian and its eigenvalues and eigenvectors. 
The average current~\eqref{average_current_J_op} reads 
\begin{equation}\label{vec_Jave}
    J = \idV  \mathcal{J} \rhossV.
\end{equation}
The steady-state two-point function ~\eqref{F_jump} becomes, in terms of Eq.~\eqref{exp_L}, 
\begin{align}\label{vec_F}
    F(\tau) &= 
    \delta(\tau) K + \idV \mathcal{J} e^{\mathcal{L} |\tau|} \mathcal{J} \rhossV- J^2
    \\[0.2cm]
    &= \delta(\tau) K + \sum_{j\neq0}
    e^{\lambda_j |\tau|} \idV \mathcal{J} |x_j\rrangle\llangle y_j|\mathcal{J} \rhossV.
    \label{F_eigen}
\end{align}
Notice how the first term in Eq.~\eqref{exp_L} cancels  the term $-J^2$. 
Since ${\rm Re}(\lambda_j) < 0$, we can  carry out the Fourier transform and obtain the power spectrum~\eqref{power_spectrum2}: 
\begin{align}\label{eigen_S}
    S(\omega) &= K - 2 \sum\limits_{j\neq0} \frac{\lambda_j}{\lambda_j^2 + \omega^2} \idV \mathcal{J} |x_j\rrangle\llangle y_j|\mathcal{J} \rhossV
    \\[0.2cm]
    &= K - \sum\limits_{j\neq0} \left(
    \frac{1}{\lambda_j-i\omega}+\frac{1}{\lambda_j+i\omega}
    \right)
    \idV \mathcal{J} |x_j\rrangle\llangle y_j|\mathcal{J} \rhossV.
    \nonumber
\end{align}
This result gives a very clean interpretation for the observed behavior of $S(\omega)$ in e.g.~Fig.~\ref{fig:ExampleA_FS}.
Namely, there will be peaks or dips near $\omega = {\rm Im}(\lambda_j)$, with the width of the peaks being proportional to ${\rm Re}(\lambda_j)$. 

We can also write Eq.~\eqref{eigen_S} more compactly as follows. For any $\omega\neq0$, the matrix $\mathcal{L}^2 + \omega^2$ is invertible and, 
using Eq.~\eqref{L_eigen_decomp}, can be written as
\begin{equation}
    \frac{1}{\mathcal{L}^2 + \omega^2} = \frac{1}{\omega^2} \rhossV \idV + \sum\limits_{j\neq 0}
    \frac{1}{\lambda_j^2 + \omega^2} |x_j\rrangle\llangle y_j|.
\end{equation}
Multiplying this by $\mathcal{L}$ and using Eqs.~\eqref{steady_state_vec} and~\eqref{L_eigen_decomp}, we get 
\begin{equation}
    \frac{\mathcal{L}}{\mathcal{L}^2 + \omega^2} = \sum\limits_{j\neq 0}
    \frac{\lambda_j}{\lambda_j^2 + \omega^2} |x_j\rrangle\llangle y_j|.
\end{equation}
Thus, we see that Eq.~\eqref{eigen_S} can also be written as 
\begin{align}\label{inverse_S}
    S(\omega) &= K - 2  \idV \mathcal{J}  \left( \frac{\mathcal{L}}{\mathcal{L}^2 + \omega^2} \right)\mathcal{J} \rhossV
    \\[0.2cm]
    &= K - 
    \idV \mathcal{J} \left(
    \frac{1}{\mathcal{L}-i \omega}+\frac{1}{\mathcal{L}+i \omega}\right)\mathcal{J} \rhossV.
    \nonumber
\end{align}
Finally, we turn to the noise $D$ in Eq.~\eqref{D_integral} by taking the limit $\omega \to0$. In terms of the Drazin inverse~\eqref{eigen_Drazin}, it becomes 
\begin{align}\label{eigen_D}
    D & = K - 2 \idV \mathcal{J} \mathcal{L}^+ \mathcal{J} \rhossV
    \\[0.2cm]
    &=K - 2 \sum\limits_{j\neq0} \frac{1}{\lambda_j} \idV \mathcal{J} |x_j\rrangle\llangle y_j|\mathcal{J} \rhossV.
\end{align}
The above results can be used to efficiently compute $S(\omega)$ and $D$. 
Details on concrete numerical implementations  are provided in  Appendix~\ref{app:numerics}.



\subsubsection{Drazin inverse for Example A}

To illustrate the Drazin inverse, consider Example A. For simplicity we assume $\bar{N}=\Delta = 0$.
The vectorized Liouvillian reads 
\begin{equation}
   \mathcal{L} =  \left(
\begin{array}{cccc}
 -\gamma  & -i \Omega  & i \Omega  & 0 \\
 -i \Omega  & -\frac{\gamma }{2}+i \Delta  & 0 & i \Omega  \\
 i \Omega  & 0 & -\frac{\gamma }{2}-i \Delta  & -i \Omega  \\
 \gamma  & i \Omega  & -i \Omega  & 0 \\
\end{array}
\right).
\end{equation}
Taking the Drazin inverse, using either~\eqref{Drazin_integral_def} or~\eqref{eigen_Drazin}, we get 
\begin{equation}\label{ExampleA_Drazin_inverse}
    \mathcal{L}^+ = \left(
\begin{array}{cccc}
 -\frac{\gamma  \left(\gamma ^2-4 \Omega ^2\right)}{\left(\gamma ^2+8 \Omega ^2\right)^2} &
   \frac{2 i \Omega }{\gamma ^2+8 \Omega ^2} & -\frac{2 i \Omega }{\gamma ^2+8 \Omega ^2} &
   \frac{12 \gamma  \Omega ^2}{\left(\gamma ^2+8 \Omega ^2\right)^2} \\
 \frac{8 i \Omega  \left(\gamma ^2+2 \Omega ^2\right)}{\left(\gamma ^2+8 \Omega ^2\right)^2}
   & -\frac{\gamma }{\gamma ^2+8 \Omega ^2}-\frac{1}{\gamma } & -\frac{8 \Omega ^2}{\gamma
   ^3+8 \gamma  \Omega ^2} & \frac{4 i \Omega  \left(\gamma ^2-4 \Omega
   ^2\right)}{\left(\gamma ^2+8 \Omega ^2\right)^2} \\
 -\frac{8 i \Omega  \left(\gamma ^2+2 \Omega ^2\right)}{\left(\gamma ^2+8 \Omega ^2\right)^2}
   & -\frac{8 \Omega ^2}{\gamma ^3+8 \gamma  \Omega ^2} & -\frac{\gamma }{\gamma ^2+8 \Omega
   ^2}-\frac{1}{\gamma } & -\frac{4 i \Omega  \left(\gamma ^2-4 \Omega
   ^2\right)}{\left(\gamma ^2+8 \Omega ^2\right)^2} \\
 \frac{\gamma  \left(\gamma ^2-4 \Omega ^2\right)}{\left(\gamma ^2+8 \Omega ^2\right)^2} &
   -\frac{2 i \Omega }{\gamma ^2+8 \Omega ^2} & \frac{2 i \Omega }{\gamma ^2+8 \Omega ^2} &
   -\frac{12 \gamma  \Omega ^2}{\left(\gamma ^2+8 \Omega ^2\right)^2} \\
\end{array}
\right).
\end{equation}

\subsubsection{Power spectrum for quantum jumps with weak dissipation}
\label{sec:s_weak}

To gain more intuition on the behavior of $S(\omega)$ for quantum jumps, we consider the case where the dissipative terms in Eq.~\eqref{M} are very small.
Let $H |n\rangle = E_n |n\rangle$, and define $\omega_{nm} = E_n - E_m$ as the set of all transition (Bohr) frequencies allowed in the system.
\rev{As shown in Appendix~\ref{app:weak_dissipation}, the peaks or dips at $\omega\neq 0$ obey the following rules:}
\begin{itemize}
    \item Position: $\omega \simeq \omega_{nm}$.
    \item Width: $\gamma_{nm} \simeq - \langle m| \mathcal{L}_D\Big(|m\rangle\langle n|\Big)|n\rangle $.
    \item Height: $g_{nm} \simeq \sum_{k,q} \nu_k \nu_q \langle m| L_k^\dagger L_k |n \rangle\langle n| L_q \rho L_q^\dagger |m\rangle$,
\end{itemize}
where $\mathcal{L}_D = \sum_k \mathcal{D}[L_k]$ is the total dissipator of the master equation.
If $g_{nm}$ is real, there will either be a peak when $g_{nm} >0$ or a dip when $g_{nm}<0$.
If $g_{nm}$ is complex, there will be a peak to the left of $\omega_{nm}$, immediately followed by a dip to the right, or vice-versa.
Peaks in $S(\omega)$ therefore reflect correlated jumps on that specific transition, while dips reflect anti-correlated jumps. 
This provides more detail than looking only at $\gtwo(0)$ [Sec.~\ref{sec:Coherence}], as it allows us to address individual transitions.
\rev{It is important to note that very often many of the $g_{nm}$ are zero (for $|n\rangle \neq |m\rangle$), so that not all non-zero $\omega_{nm}$ will indeed appear as peaks.
In fact, we see that $g_{nm} \propto \langle m | L_k^\dagger L_k |n\rangle$, and in many models $L_k^\dagger L_k$ is diagonal in the eigenbasis of $H$, so that this term vanishes. 
The existence of peaks/dips at $\omega \neq 0$ is therefore conditioned on having jump operators for which $[H,L_k^\dagger L_k] \neq 0$. 
}

\rev{On the other hand, the height and width of the central peak/dip, at $\omega = 0$, is more difficult to asses, even in this weak dissipation limit. The reason is that this peak will be related to the subset of system transition frequencies having $\omega_{nm} = 0$ (which include $\omega_{nn}$, but might also include $m\neq n$ when the spectrum is degenerate). 
Because of the degeneracy of this set, it is not possible in general to evaluate the height and width of the resulting peak.
A concrete illustration of all these ideas for the case of Example A is given in Appendix~\ref{app:weak_dissipation}.
}




\subsection{Computing cumulants using recursive methods}
\label{sec:cumulants_recursive}

Different methods exist for computing higher order cumulants~\cite{Hagele_2018,Flindt_2010} of the FCS [Sec.~\ref{sec: Cumulant generating function}].
Here we discuss the recursive method of Ref.~\cite{Flindt_2010}, that applies to both Markovian and non-Markovian counting statistics for arbitrary measurement processes. 
We will not cover the derivation in detail, but will rather outline the main results for the Markovian case.

Consider the tilted Liouvillian $\mathcal{L}_\chi$ of Eq.~(\ref{tilted_liouvillian}).
We can recast it in terms of the unperturbed Liouvillian $\mathcal{L}$ plus some perturbation $\delta \mathcal{L}_{\chi}$
\begin{equation}
    \mathcal{L}_{\chi} = \mathcal{L} + \delta \mathcal{L}_{\chi}\,,
\end{equation}
where $\delta\mathcal{L}_\chi$ vanishes smoothly as $\chi\to 0$. Following a similar spirit as the previous section on vectorization, we will define a new set of perturbed eigenvalues and left and right eigenvectors of the tilted Liouvillian using the vectorization introduced in Sec.~\ref{sec:vectorization} 
\begin{align}
    \mathcal{L}_{\chi}\vert x_{i}(\chi) \rrangle &= \lambda_{i}(\chi)\vert x_{i}(\chi) \rrangle,\\
    \llangle y_{i}(\chi) \vert \mathcal{L}_{\chi} &=\llangle y_{i}(\chi) \vert\lambda_{i}(\chi)\,,
\end{align}
which in the limit of $\chi=0$, recover the eigenvectors and eigenvalues of the unperturbed Liouvillian $\mathcal{L}$.
We are primarily interested in the largest eigenvalue $\lambda_{0}(\chi)$ 
\begin{align}
    \mathcal{L}_{\chi} \vert \rho_{\rm ss}(\chi) \rrangle &= \lambda_{0}(\chi)\vert \rho_{\rm ss}(\chi) \rrangle\,,
\end{align}
which is such that $\lambda_0(0) = 0$.
The key insight of Ref.~\cite{Flindt_2010} is to perform a Taylor expansion around $\chi=0$ of the eigenvalue $\lambda_{0}(\chi)$, the right eigenvector $\vert \rho_{\rm ss}(\chi) \rrangle$, and the perturbation $\delta \mathcal{L}_{\chi}$
\rev{
\begin{equation}
    \begin{aligned}
        \lambda_0(\chi) &= \sum_{n=1}^\infty \frac{(i\chi)^n}{n!}\llangle I^n\rrangle,\\
        \vert \rho_{\rm ss}(\chi) \rrangle & =  \sum_{n=0}^\infty\frac{(i\chi)^n}{n!} \vert \rho_{\rm ss}^{(n)}\rrangle,\\
        \delta \mathcal{L}_{\chi} &=\sum_{n=1}^\infty \frac{(i\chi)^n}{n!}\mathcal{L}^{(n)},
    \end{aligned}
\end{equation}
where $\llangle I^n\rrangle$ are the steady-state scaled cumulants of the current and $\vert \rho_{\rm ss}^{(0)}\rrangle=\vert \rho_{\rm ss}\rrangle$ is the ($\chi$-independent) steady state.}
This permits one to define a recursion relation for cumulants
\begin{equation}
\label{eq:recursion_cumulant}
    \llangle I^{n} \rrangle = \sum_{m=1}^{n}\binom{n}{m} \idV \mathcal{L}^{(m)}\vert \rho_{\rm ss}^{(n-m)} \rrangle\,,
\end{equation}
where 
\begin{align}
    \vert \rho_{\rm ss}^{(n)} \rrangle &= \mathcal{L}^{+}\sum_{m=1}^{n}\binom{n}{m}\left(\llangle I^{m} \rrangle - \mathcal{L}^{(m)}\right)\vert \rho_{\rm ss}^{(n-m)}\rrangle.
\end{align}
To showcase how this can be implemented, we will now compute the first and second cumulants in general as
\begin{align}
\label{eq:rec_cum_1}
    \llangle I \rrangle &= \idV \mathcal{L}^{(1)} \vert \rho_{\rm ss} \rrangle\,,\\
\label{eq:rec_cum_2}
    \llangle I^{2} \rrangle &= \idV \mathcal{L}^{(2)} \vert \rho_{\rm ss} \rrangle - 2 \idV \mathcal{L}^{(1)}  \mathcal{L}^{+}  \mathcal{L}^{(1)} \vert \rho_{\rm ss} \rrangle\,,
\end{align}
where we have used the fact that $\llangle 0 \vert \mathcal{L}^{+} = \mathcal{L}^{+}\vert \rho_{\rm ss} 
\rrangle = 0$ (see Appendix~\ref{app:pseudoinverse} for more details).

Given these results, it is straightforward to check whether these two definitions agree with our description of continuous quantum-jump measurements.
To do this, we can recast Eq.~(\ref{tilted_liouvillian}) into unperturbed and perturbed parts using vectorized notation, where the perturbation is given by 
\begin{equation}
    \delta \mathcal{L}_{\chi} = \sum_{k}(e^{i \chi \nu_{k}}-1) L_{k}^{*}\otimes  L_{k} \,.
\end{equation}
One can easily show that 
\begin{equation}
    \mathcal{L}^{(n)} = \sum_{k} \nu_{k}^{n} L_{k}^{*}\otimes L_{k}\,,
\end{equation}
and thus the first two terms are directly related to those in Eq.~(\ref{FCS_L_prime_jump}) and Eq.~(\ref{FCS_L_double_prime_jump})
\begin{equation}
    \mathcal{L}^{(1)} = -i \mathcal{L}'\,,\quad \mathcal{L}^{(2)} = - \mathcal{L}''\,.
\end{equation}
Plugging these two expressions into the Eq.~(\ref{eq:rec_cum_1}) and Eq.~(\ref{eq:rec_cum_2}) we obtain
\begin{align}
    \llangle I \rrangle &= \idV  \mathcal{J} \rhossV = J\,,\\
    \llangle I^{2} \rrangle
    & = K- 2 \idV \mathcal{J} \mathcal{L}^+ \mathcal{J} \rhossV = D\,. 
\end{align}
which agree with Eq.~(\ref{vec_Jave}) and Eq.~(\ref{eigen_D}) respectively.

We can also use this method to compute the first two moments of the tilted Liouvillian~\eqref{L_chi_diffusion} for quantum diffusion, as was considered above in Sec~(\ref{sec:FCS_quantum_diffusion}). In this example, the perturbed part of the Liouvillian corresponds to \rev{
\begin{equation}
    \delta \mathcal{L}_{\chi} = i \chi\mathcal{H} - \frac{\chi^{2}}{2}K_{\rm diff}\,,
\end{equation}
then, $\mathcal{L}^{(1)} = \mathcal{H} $ and $\mathcal{L}^{(2)}  =  K_{\rm diff}$}, which plugging back into our expressions leads to the first two cumulants
\begin{align}
    \llangle I_{\rm diff} \rrangle &= \idV  \mathcal{H} \rhossV = J_{\rm diff}\,,\\
    \llangle I_{\rm diff}^{2} \rrangle
    & = K_{\rm diff}- 2 \idV \mathcal{H} \mathcal{L}^+ \mathcal{H} \rhossV = D_{\rm diff}\,. 
\end{align}

\subsection{Computing probability distributions using the saddle-point approximation}
In most scenarios, no analytical expression for the distribution $P(n,t)$ can be found (for exceptions, see Sec.~\ref{sec:scgfexb}). The saddle-point approximation \cite{daniels_1954, butler_2007} provides a simple-to-evaluate approximation to $P(n,t)$ if the cumulant generating function is known. To derive the probability distribution in the saddle-point approximation, we write
\begin{equation}
\label{eq:saddle1}
P(n,t) = \int_{-R}^R\frac{d\chi}{2\pi} e^{-in\chi+C(\chi,t)} = \int_{-iR}^{iR}\frac{dz}{2\pi i} e^{-nz+C(-iz,t)},
\end{equation}
where $R\rightarrow \infty$ for continuous distributions and $R=\pi$ for discrete ones.

To obtain the saddle-point approximation, we deform the path of integration such that it passes a saddle-point of the function
\begin{equation}
    \label{eq:funcg}
    g(z) = C(-iz,t)-nz,
\end{equation}
along the path of steepest descent, i.e., where ${\rm Re}[g(z)]$ decreases fastest. After this deformation, the integral in Eq.~\eqref{eq:saddle1} is dominated by the neighborhood around the saddle point. This is justified in the long-time limit, where both $C(\chi,t)$ and the relevant $n$ scale linearly in time. The integrand in Eq.~\eqref{eq:saddle1} is then exponentially suppressed away from the saddle point. To employ the saddle-point approximation, we thus need to find the saddle point, as well as the path of steepest descent. 

To find the saddle point, we note that at a saddle point, the derivative of $g(z)$ vanishes. This results in the saddle-point equation
\begin{equation}
    \label{eq:speq}
    \partial_z C(-iz,t)|_{z=k} =-i\partial_\chi C(\chi,t)|_{\chi = -ik} = n,
\end{equation}
where we denote by $k$ the value for $z$ at the saddle point. In Ref.~\cite{daniels_1954}, it is proven that the saddle-point equation has a unique solution $k$ that is real (note that $C(-iz,t)$ is real for $z$ real).

To find the path of steepest descent, we approximate $g(z)$ around the saddle point by its second-order Taylor expansion
\begin{equation}
    \label{eq:gtaylor}
    g(z) \simeq g(k) - \frac{1}{2}\partial_\chi^2C(\chi,t)|_{\chi=-ik}(z-k)^2.
\end{equation}
From the definition of $C(\chi,t)$ in Eq.~\eqref{eq:cgf} we find
\begin{equation}
\label{eq:secneg}
    \partial_\chi^2C(\chi,t)|_{\chi=-ik} = -\frac{E\big[ N^2(t)e^{kN(t)}\big] E\big[ e^{kN(t)}\big]-E\big[ N(t) e^{kN(t)}\big]^2}{E\big[ e^{kN(t)}\big]^2}\leq 0,
\end{equation}
where the inequality holds for real $k$ and follows from the Cauchy-Schwarz inequality.
From Eqs.~\eqref{eq:gtaylor} and~\eqref{eq:secneg}, we can infer that the path of steepest descent corresponds to $z-k$ being purely imaginary. Close to the saddle point, the path of steepest descent may therefore be parametrized as $z = k+is$ with $s$ real. 

In the saddle-point approximation, the integral in Eq.~\eqref{eq:saddle1} is thus approximated as
\begin{align}
\label{eq:saddle0}
    P(n,t)& \simeq  e^{-nk+C(-ik,t)} \int_{-\infty}^\infty ds\, e^{-\frac{1}{2}|C''(-ik,t)|s^2}\notag 
    \\
        & =\frac{1}{\sqrt{2\pi|C''(-ik,t)|}}e^{-nk+C(-ik,t)},
\end{align}
where $k$ is obtained by solving Eq.~\eqref{eq:speq}. Extending the range of the integral in the last expression is justified when the integral is indeed dominated by the neighbourhood around the saddle point, i.e., values of $s$ close to zero. This is the case in the long-time limit, where $C(\chi,t)\propto t$. We note that the saddle-point approximation does not guarantee normalized probability distributions. If desired, the distribution can be normalized after performing the approximation.

We further note that whenever the saddle-point approximation is justified, it implies a large-deviation principle as
\begin{equation}
    \label{eq:largedev}
    -\lim_{t\rightarrow\infty}\frac{1}{t}\ln P(n,t) = k\frac{n}{t}-C(-ik),
\end{equation}
where $C(\chi)$ denotes the SCGF, c.f. Eq.~\eqref{SCGF_2}.

\subsubsection{Saddle-point approximation for Example B}
\label{sec:saddlexB}
As an illustration of the saddle-point approximation, we consider a quantum dot coupled to two fermionic reservoirs [c.f.~Sec.~\ref{ssec:exampleB}] in the long-time limit. The SCGF for this system is given in Eq.~\eqref{eq:scgfb}. For simplicity, we here consider the case $\gamma_L=\gamma_R=\gamma$ and the large bias limit, i.e., $f_R=1$ and $f_L=0$. In this case, the SCGF reduces to
\begin{equation}
 \label{eq:scgfbsimple}
    C(\chi) = \gamma \left(e^{i\chi/2}-1\right), \hspace{.25cm}\text{for}\hspace{.25cm} -\pi \leq \chi<\pi.
\end{equation}
We note that $C(\chi)$ is periodic in $\chi$ with period $2\pi$. However, for $\pi\leq\chi<2\pi$, the right-hand side of Eq.~\eqref{eq:scgfbsimple} does not correspond to the root of Eq.~\eqref{eq:scgfb} with the largest real part. For this scenario, the integral in Eq.~\eqref{eq:saddle1}, with $R=\pi$, can be expressed in closed form as
\begin{equation}
    \label{eq:pnsymmex}
    P(n,t) = e^{-\gamma t}\bigg[\frac{(\gamma t)^{2n}}{(2n)!}-\frac{2\gamma t(-1)^n}{(2n-1)\pi} {}_1{F}_2\left(\frac{1}{2}-n;\frac{3}{2},\frac{3}{2}-n;-\frac{(\gamma t)^2}{4}\right)\bigg],
\end{equation}
where ${}_1{F}_2$ denotes a generalized hypergeometric function.

\begin{figure}
    \centering
    \includegraphics[width=\columnwidth]{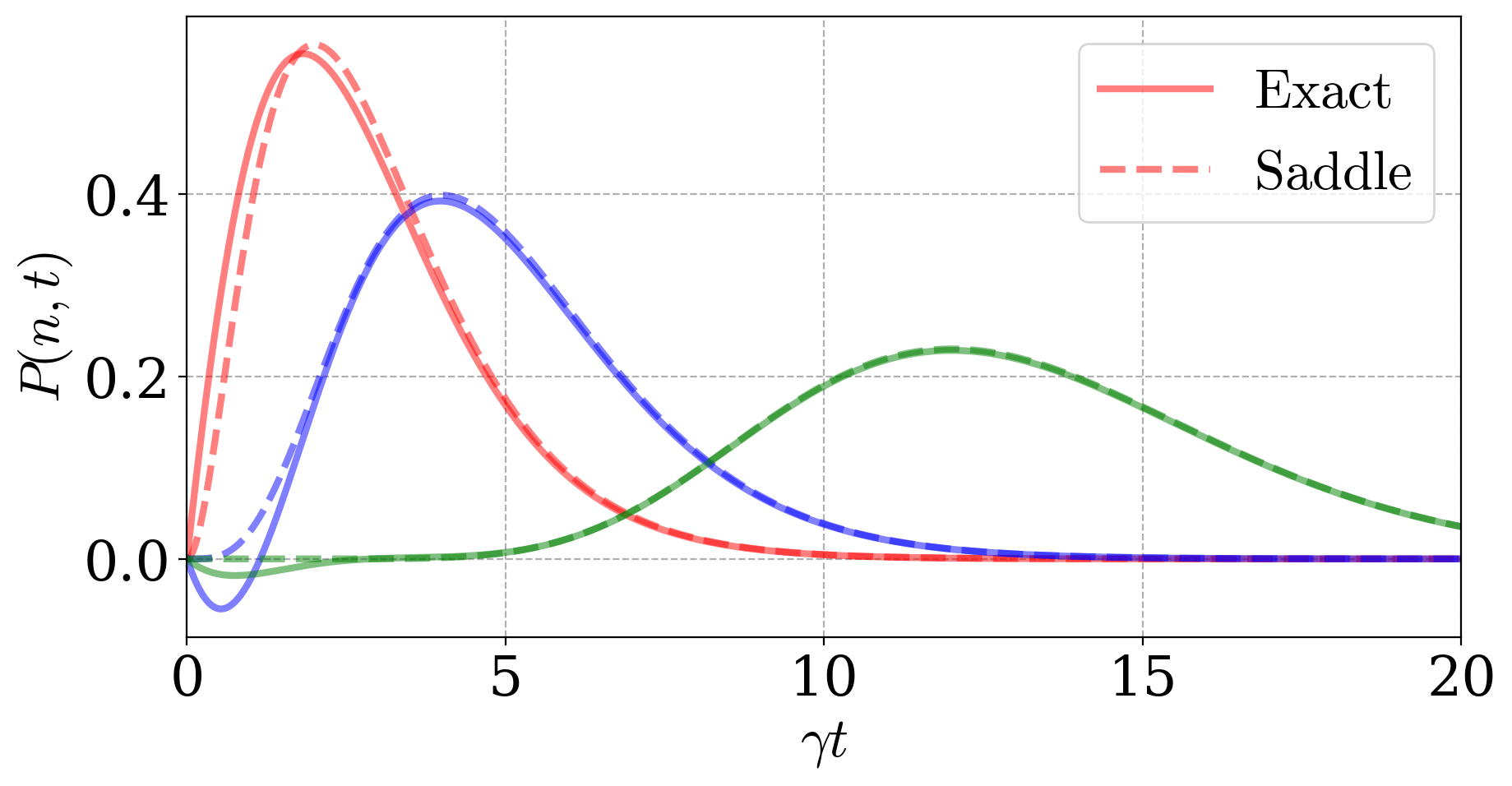}
    \caption{FCS Probability distribution for Example B \rev{(Sec.~\ref{ssec:exampleB})}. The solution (solid) given by Eq.~(\ref{eq:pnsymmex}) is compared to the saddle-point approximation (dashed) given by Eq.~(\ref{eq:saddle}) for $n=1, 2, 6$ (red, blue, green) respectively. For early times, the solution in Eq.~\eqref{eq:pnsymmex} has negative values which is an artifact of long-time limit taken to derive the SCGF in Eq.~\eqref{eq:scgfb}. For later times (i.e., $\gamma t\gg 1$) we find excellent agreement between the two methods. }
    \label{fig:saddle}
\end{figure}

A much nicer expression can be obtained using the saddle-point approximation. From the saddle-point equation in Eq.~\eqref{eq:speq}, we find the saddle point to be located at $k = 2\ln \frac{2n}{\gamma t}$. Equation~\eqref{eq:saddle0} then results in the probability distribution in the saddle-point approximation
\begin{equation}
\label{eq:saddle}
   P(n,t)\simeq \frac{e^{2n-\gamma t}}{\sqrt{n\pi}}\left(\frac{\gamma t}{2n}\right)^{2n}.
\end{equation}
A comparison between Eqs.~\eqref{eq:pnsymmex} and \eqref{eq:saddle} is shown in Fig.~\ref{fig:saddle}.

\subsection{Gaussian processes}\label{sec:Gaussian}

When considering systems of bosons or fermions, we can find compact closed formulas for $J, F(\tau), S(\omega)$ and $D$ for the particular case where the states and processes involved are Gaussian.
These formulas only depend on matrices which are of size at most $2N$, where $N$ is the number of modes. 
Hence, they can be used efficiently, even for complex configurations involving multiple modes. 
To provide a unified description of bosons and fermions, we consider a system described by $N$ annihilation operators $b_i$, which can be either bosonic ($[b_i,b_j^\dagger]=\delta_{ij}$) or fermionic ($\{b_i,b_j^\dagger\}=\delta_{ij}$).
Throughout, when we write $\pm$, the upper sign is for bosons and the lower for fermions. 

We study a general Gaussian dynamics given by the Liouvillian
\begin{equation}\label{Gaussian_Liouvillian}
    \mathcal{L}\rho = -i[H,\rho] + \sum_i \Big\{ \gamma_i^- \mathcal{D}[b_i]\rho + \gamma_i^+ \mathcal{D}[b_i^\dagger]\rho\Big\},
\end{equation}
where 
\begin{equation}\label{Gaussian_H}
    H = \sum_{i,j} \Big\{A_{ij} b_i^\dagger b_j + \frac{1}{2}\big( B_{ij} b_i^\dagger b_j^\dagger + B_{ij}^* b_j b_i\big) \Big\} + \sum_i \big(\epsilon_i b_i^\dagger + \epsilon_i^* b_i\big),
\end{equation}
with matrices $A= A^\dagger$ and $B = \pm B\trans$, and a vector $\bm{\epsilon}$.
For fermions one must set $\epsilon_i \equiv 0$ due to parity conservation~\cite{Wick1952,Bartlett2007}. 
The factors proportional to $B$ are called squeezing terms in the context of bosons, and pairing terms for fermions. 

The general jump operator $\mathcal{J}$ in Eq.~\eqref{J_op} is now parametrized as 
\begin{equation}\label{Gaussian_J_op}
    \mathcal{J}\rho = \sum_i \Big\{ \nu_i^- \gamma_i^- b_i \rho b_i^\dagger + \nu_i^+ \gamma_i^+ b_i^\dagger \rho b_i\Big\},
\end{equation}
with weights $\nu_i^-$ associated to the extraction of particles ($b_i \rho b_i^\dagger$), and $\nu_i^+$  associated to injection ($b_i^\dagger \rho b_i$). 
Similarly, the general diffusion superoperator~\eqref{H_hom_op} reads 
\begin{equation}\label{Gaussian_H_op}
\begin{aligned}
    \mathcal{H}\rho = \sum_i\Bigg[ &
    \nu_i^-\sqrt{\gamma_i^-} \big( e^{-i \phi_i^-} b_i \rho + e^{i \phi_i^-} \rho b_i^\dagger\big) \\[0.2cm]
    &+ \nu_i^+ \sqrt{\gamma_i^+}\big( e^{-i \phi_i^+} b_i^\dagger \rho + e^{i \phi_i^+} \rho b_i\big)\Bigg],
\end{aligned}
\end{equation}
with possibly different angles $\phi_i^-$ and $\phi_i^+$ for each channel. 
In practice, though, one will generally have $\phi_i^+ = - \phi_i^-$.

The formulas for quantum diffusion and quantum jumps will be provided below in Secs.~\ref{sec:Gaussian_diffusion} and~\ref{sec:Gaussian_jumps}. 
Before that, however, we must first introduce the notions of covariance matrix and Lyapunov equation.

Gaussian systems are fully characterized by the first and second moments of the operators $b_i$. 
When $B = 0$, the entire process can be modeled more simply in terms of a vector of means $\bm{\mu} = (\langle b_1 \rangle, \ldots, \langle b_N \rangle)$ and a $N\times N$ covariance matrix (sometimes also called ``correlation matrix'') 
\begin{equation}\label{Gaussian_C}
    C_{ij} = \langle b_j^\dagger b_i \rangle - \langle b_j^\dagger \rangle \langle b_i \rangle.
\end{equation}
Using the Liouvillian~\eqref{Gaussian_Liouvillian} one can show that the vector of means evolves as 
\begin{equation}\label{Gaussian_mu_eq}
    \frac{d\bm{\mu}}{dt} = - W \bm{\mu} -i \bm{\epsilon}, 
\end{equation}
where
\begin{equation}\label{Gaussian_W_N}
    W = i A + \tfrac{1}{2} \Gamma, \qquad 
    \Gamma = (\gamma_- \mp \gamma_+). 
\end{equation}
Here $\gamma_\pm$ are diagonal matrices with entries $\gamma_i^\pm$. 
The steady-state solution of Eq.~\eqref{Gaussian_mu_eq} is $\bm{\mu} = -i W^{-1} \bm{\epsilon}$, while for fermions $\bm{\mu} \equiv 0$ at all times.
The covariance matrix, on the other hand, evolves according to the Lyapunov equation 
\begin{equation}\label{Gaussian_lyapunov_C}
    \frac{dC}{dt} = - (W C + C W^\dagger) + \gamma_+. 
\end{equation}
At the steady-state this reduces to the algebraic equation
\begin{equation}\label{Gaussian_lyapunov_C_ss}
W C + C W^\dagger = \gamma_+    
\end{equation}
\rev{Using the vectorization ideas from Sec.~\ref{sec:vectorization} we can write this as 
$(\mathbb{I}\otimes W + W^* \otimes \mathbb{I}) |C\rrangle = |\gamma_+\rrangle$, which shows that a Lyapunov equation is just a linear equation (of the form $Ax=b$) for the vector $|C\rrangle$. In practice, there are more  efficient solvers for the Lyapunov equation, which can be found in most numerical libraries. 
}

When $B \neq 0$, we also require correlations of the form $\langle b_i b_j\rangle$. 
To handle this, it is convenient to introduce the Hermitian  operators 
\begin{equation}\label{Gaussian_quadrature_majorana}
    q_i = \frac{1}{\sqrt{2}} (b_i + b_i^\dagger), 
    \qquad 
    p_i = \frac{i}{\sqrt{2}} (b_i^\dagger - b_i).
\end{equation}
For bosons these are called quadratures, while for fermions they are called Majorana operators. 
It is important to bear in mind that different authors sometimes define them with other prefactors such as $1/2$ or $1$.
We use this definition so that they look as similar as possible to the original position and momentum operators. 

We define a vector of Hermitian operators $\bm{R} = (q_1,\ldots,q_N, p_1,\ldots,p_N)$. 
The mean vector is written as $\bm{r} = \langle \bm{R}\rangle$, and is now real by construction. 
It can be related to $\bm{\mu} = (\langle b_1 \rangle, \ldots, \langle b_N \rangle)$ as 
\begin{equation}
    \bm{r} = (\varphi \otimes \id_N) 
    \begin{pmatrix}
        \bm{\mu} \\ \bm{\mu}^*
    \end{pmatrix},
    \qquad 
    \varphi = \frac{1}{\sqrt{2}}
    \begin{pmatrix}
        1 & 1 \\ -i & i 
    \end{pmatrix}.
\end{equation}
Here and henceforth we will use often matrices such as $\varphi \otimes \id_N$, mixing a $2\times 2$ matrix with a $N\times N$ one. 
This is based on the ordering of the operators in $\bm{R}= (q_1,\ldots,q_N, p_1,\ldots,p_N)$. 
Sometimes it is more convenient to order them as
$(q_1,p_1,\ldots,q_N,p_N)$. For this case, one need simply change the matrices to $\id_N \otimes \varphi$.

The operators $\bm{R}$ satisfy a canonical algebra which can be written compactly as 
\begin{equation}
    [R_i, R_j]_\mp = R_i R_j \mp R_j R_i = i\Omega_{ij},
\end{equation}
where, for bosons $\Omega = (i\sigma_y)\otimes \id_N$, which is known as the symplectic form, whereas for fermions $\Omega = -i \id_{2N}$. 
The covariance matrix is now a $2N\times 2N$ matrix, and can be defined as~\footnote{From $\Theta$ we can extract the matrix $C$ as 
\begin{equation*}
    C = \pm \Big\{ 
    \tr_2 \big[ \Theta (\varphi \sigma_+ \sigma_- \varphi^\dagger \otimes \id_N)\big] - \id_N/2
    \Big\},
\end{equation*}
where $\tr_2$ is the partial trace over the $2\times 2$ side of a tensor structure such as $\id_2 \otimes \id_N$. 
Similarly, the matrix $C'_{ij} = \langle b_i b_j\rangle$ can be extracted as 
\begin{equation*}
    C' = \tr_2 \big[
    \Theta (\varphi \sigma_+  \varphi^\dagger \otimes \id_N)
    \big].
\end{equation*}} 
\begin{equation}
    \Theta_{ij} = \frac{1}{2} \langle [R_i, R_j]_\pm \rangle - \langle R_i \rangle\langle R_j \rangle 
    = \frac{i}{2} \Omega_{ij} \pm \langle R_j R_i \rangle - \langle R_i \rangle\langle R_j \rangle .
\end{equation}
Moreover, the Hamiltonian~\eqref{Gaussian_H} can be written as 
\begin{equation}
    H = \frac{1}{2} \bm{R}\trans \mathbb{H} \bm{R} + \bm{R}\trans \bm{f},
\end{equation}
where $\bm{f} = (\varphi \otimes \id_N) (\bm{\epsilon}, \bm{\epsilon}^*)$ and 
\begin{equation}
    \mathbb{H} = 
    \frac{1}{2} \Big\{ \id_2 \otimes (A \pm A\trans) + \sigma_z \otimes (B \pm B^*) 
    \pm \sigma_y\otimes(A\trans \mp A) -i \sigma_x \otimes(B\mp B^*)\Big\}.
\end{equation}

The vector of means $\bm{r}$ evolves according to 
\begin{equation}\label{Gaussian_mean_evolution_2N}
    \frac{d\bm{r}}{dt} = - \mathcal{W} \bm{r} + \Omega \bm{f},
\end{equation}
where 
\begin{equation}\label{Gaussian_W_2N}
    \mathcal{W} = - \Omega \mathbb{H} + \frac{1}{2} \id_2 \otimes \Gamma. 
\end{equation}
It is also worth noting that even though $\mathbb{H}$ differs for bosons and fermions, it turns out that $\Omega\mathbb{H}$ looks exactly the same:
\begin{equation}
    \Omega \mathbb{H} = -\frac{i}{2}\Bigg[ I_2 \otimes (A-A^{\rm T}) -  \sigma_y \otimes (A + A^{\rm T})- i \sigma_x \otimes (B+B^*) + \sigma_z \otimes (B-B^*)\Bigg].
\end{equation}
The covariance matrix again evolves according to a Lyapunov equation, which now reads 
\begin{equation}\label{Gaussian_lyapunov_2N}
    \frac{d\Theta}{dt} = - (\mathcal{W} \Theta + \Theta \mathcal{W}^\dagger) + \Upsilon, 
\end{equation}
where 
$\Upsilon = \frac{1}{2}\id_2 \otimes (\gamma_+ +\gamma_-)$ 
for bosons and 
$\Upsilon = - \frac{1}{2} \sigma_y \otimes (\gamma_- - \gamma_+)$
for fermions. 

In the next sections we provide the formulas for $J, F(\tau), S(\omega)$ and $D$. 
We will give them in terms of  $2N\times 2N$  matrices like $\Theta$. 
This is not only more general but, as it turns out, the result is also simpler and more transparent. 
We start with quantum diffusion, as this involves only 2-operator  correlation functions, which are easier to handle.
Then we move on to quantum jumps in Sec.~\ref{sec:Gaussian_jumps}.

We also mention here the connection with stochastic quantum trajectories.
For bosons, quantum jumps are not Gaussian preserving because $b_i \rho b_i^\dagger$ breaks Gaussianity. 
This does not mean we cannot compute $J, F(\tau), S(\omega)$ and $D$, of course. All it means is that we cannot use Gaussian techniques to model the stochastic trajectories. 
Bosonic diffusion, on the other hand, is Gaussian, and a beautiful framework exists for describing the stochastic trajectories solely in terms of covariance matrices. For the interested reader, we recommend Ref.~\cite{Genoni2016}. 
The case of Fermions is almost the opposite. 
For quantum diffusion with fermions, it is impossible to measure $\langle b_i \rangle$ due to parity conservation. One could, of course, study the diffusive unraveling with jump operators such as $b_i^\dagger b_i$, but the resulting dynamics is no longer Gaussian. 
However, and somewhat surprisingly, quantum jump trajectories with fermions \emph{are} Gaussian preserving. This is because all fermionic Fock states are actually Gaussian~\cite{Bravyi_2005}. 

\subsubsection{Quantum diffusion}
\label{sec:Gaussian_diffusion}

As argued above, Gaussian quantum diffusion only  make sense for bosons. Define a diagonal matrix $V$ with entries
\begin{equation}\label{Gaussian_V_diff}
    \begin{aligned}
        V_{ii} &= \frac{1}{\sqrt{2}} \Big(e^{-i \phi_i^-} \nu_i^- \sqrt{\gamma_i^-} + e^{-i \phi_i^+} \nu_i^+ \sqrt{\gamma_i^+}
        \Big), 
        \\[0.2cm]
        V_{i+N,i+N} &= \frac{i}{\sqrt{2}} (e^{-i \phi_i^-} \nu_i^- \sqrt{\gamma_i^-} - e^{-i \phi_i^+} \nu_i^+ \sqrt{\gamma_i^+}),
    \end{aligned}
\end{equation}
for $i = 1,\ldots,N$.
Moreover, define a vector $\bm{o} = (1,\ldots,1)$, of length $2N$.
In terms of the quadrature/majorana operators~\eqref{Gaussian_quadrature_majorana}, and their vector $\bm{R}$, we can  rewrite the diffusion superoperator~\eqref{Gaussian_H_op} as 
\begin{equation}
    \mathcal{H}\rho = \sum_{i=1}^{2N} \Big( V_{ii} R_i \rho + \rho R_i V_{ii}^*). 
\end{equation}
From this it readily follows that the average diffusive current is 
\begin{equation}
    \label{Gaussian_J_diff}
    J_{\rm diff} = \tr\big\{ \mathcal{H} \rho\big\} 
    = \bm{o}\trans (V + V^*) \bm{r}.
\end{equation}

Next, we compute the two-point correlation function $F_{\rm diff}(\tau)$ in Eq.~\eqref{diffusion_F}. 
Using the language of the quantum regression theorem, developed in Sec.~\ref{sec:Coherence}, we can write the relevant term in $F_{\rm diff}(\tau)$ as 
\begin{equation}
    \tr\big\{ \mathcal{H} e^{\mathcal{L} \tau} \mathcal{H} \rho\big\} = \sum_{ij} (V_{ii} + V_{ii}^*) \Big[ 
    V_{jj} \langle R_i(\tau) R_j \rangle + V_{jj}^* \langle R_j R_i(\tau)\rangle
    \Big].
\end{equation}
We therefore require the 2-operator correlation functions for the $\bm{R}$ operators. 
We develop these expressions in Appendix~\ref{app:Gaussian}. 
Focusing on steady states, the results are
\begin{equation}\label{Gaussian_two_point_correlations}
\begin{aligned}
    \langle R_j(t) R_i(t+\tau)\rangle &= \Big[ \pm G(\tau) \tilde{\Theta} + \bm{r} \bm{r}\trans\Big]_{ij},
    \\[0.2cm]
    \langle R_j(t+\tau) R_i(t)\rangle &= \Big[ \pm \tilde{\Theta}G^\dagger(\tau)+\bm{r} \bm{r}\trans\Big]_{ij},
\end{aligned}
\end{equation}
where $\tilde{\Theta} = \Theta - i \Omega/2$ and $G(\tau) = e^{-\mathcal{W} \tau}$ is the propagator for the model in question.
The plus or minus signs here refer again to bosons and fermions (we write the results for fermions because we will also need them in the quantum-jump case).
Using these results we arrive, after some simplifications, at 
\begin{equation}\label{Gaussian_F_diff}
    F_{\rm diff}(\tau) = \delta(\tau) K_{\rm diff} + 2\text{Re} \Bigg(\bm{o}\trans  (V+V^*) G(\tau) \tilde{\Theta} V^*\bm{o} \Bigg).
\end{equation}
where $K_{\rm diff} = \sum_k (\nu_k^-)^2 + (\nu_k^+)^2$.
The time-dependence of the two-point function is therefore dictated entirely by the matrix $G(\tau)$, which is weighted by both the system's covariance matrix $\Theta$, as well as the matrix $V$ [Eq.~\eqref{Gaussian_V_diff}] describing the $\nu_i^\pm$.

From $F_{\rm diff}$ we compute the power spectrum $S_{\rm diff}(\omega) = \int_0^\infty d\tau (e^{i \omega \tau} + e^{-i \omega \tau})F_{\rm diff}(\tau)$ [Eq.~\eqref{power_spectrum2}].
Since the only $\tau$ dependence is in $G(\tau)$, all we need is to employ the transformation 
\begin{equation}\label{Gaussian_Fourier_transform_W}
    \int_0^\infty d\tau (e^{i \omega \tau} + e^{-i \omega \tau}) e^{-\mathcal{W} \tau} = \frac{2\mathcal{W}}{\mathcal{W}^2 + \omega^2}.
\end{equation}
We then arrive at 
\begin{equation}\label{Gaussian_power_spectrum_diff}
\begin{aligned}
    S_{\rm diff}(\omega) = K_{\rm diff} + &4 \text{Re}\Bigg(\bm{o}\trans (V+V^*)  \frac{\mathcal{W}}{\mathcal{W}^2 + \omega^2} \tilde{\Theta} V^* \bm{o} \Bigg).
\end{aligned}
\end{equation}
Setting $\omega=0$ yields the noise in Eq.~\eqref{D_integral}
\begin{equation}
    D_{\rm diff}= K_{\rm diff} + 4\text{Re}\Bigg(\bm{o}\trans  (V+V^*)  \mathcal{W}^{-1} \tilde{\Theta} V^* \bm{o} \Bigg).
\end{equation}

\subsubsection{Quantum jumps}
\label{sec:Gaussian_jumps}

For quantum jumps, we start by rewriting the jump operator~\eqref{Gaussian_J_op} as 
\begin{equation}
    \mathcal{J} \rho = \sum_{i,j} V_{ji} R_i \rho R_j,
\end{equation}
where we defined a \emph{new} $V$-matrix 
\begin{equation}\label{Gaussian_V_matrix_jumps}
    V = \frac{1}{2} \id_2 \otimes(\nu_+ \gamma_+ + \nu_- \gamma_-) + \frac{1}{2}\sigma_y \otimes (\nu_+\gamma_+ - \nu_-\gamma_-). 
\end{equation}
[We use the same notation $V$ as in Eq.~\eqref{Gaussian_V_diff} to emphasize how these two matrices play similar roles, despite being different.]
The average current again readily follows:
\begin{equation}\label{Gaussian_J_jumps}
    J = \tr(\mathcal{J}\rho) = \sum_{i,j} V_{ji} \langle R_j R_i \rangle = \pm \tr(V\tilde{\Theta}) + \bm{r}\trans V \bm{r},
\end{equation}
The relevant part of the two-point function~\eqref{F_jump} will now take the form 
\begin{equation}
    \tr\big\{ \mathcal{J} e^{\mathcal{L} \tau} \mathcal{J} \rho\big\} = \sum_{i,j,k,\ell} V_{ji} V_{\ell k} \langle R_\ell (t) R_j(t+\tau) R_i(t+\tau) R_k(t)\rangle. 
\end{equation}
We thus now need a 4-operator correlation function.
For Gaussian states, we can do this using Wick's theorem~\cite{Wick1950}.
Let $R_i$ denote operators that are arbitrary linear combinations of the $b$ and $b^\dagger$ operators.
Wick's theorem states that for any Gaussian state,
\begin{equation}
\begin{aligned}
    \langle R_1 R_2 R_3 R_4 \rangle  &= 
    \langle R_1 R_2 \rangle \langle R_3 R_4 \rangle \pm \langle R_1 R_3 \rangle \langle R_2 R_4 \rangle 
    \\[0.2cm]
    &
    + \langle R_1 R_4 \rangle\langle R_2 R_3 \rangle - 2 \langle R_1 \rangle \langle R_2 \rangle \langle R_3 \rangle \langle R_4 \rangle .
\end{aligned}    
\end{equation}
With this result at hand, we obtain after many simplifications 
\begin{equation}\label{Gaussian_F_2N}
\begin{aligned}
    F(\tau) &= \delta(\tau) K + {\rm tr}\Big\{ G^\dagger(\tau) (V^{\rm T} \pm V) G(\tau)  \tilde{\Theta} V \tilde{\Theta}\Big\} 
    \\[0.2cm]
   %
   & + 2\text{Re}\Bigg(\bm{r}^{\rm T} (V^{\rm T}+V) G(\tau)  \tilde{\Theta} V+ \bm{r}\Bigg),
   \end{aligned}
\end{equation}
The first term $K$ is computed like Eq.~\eqref{Gaussian_J_jumps}, but using $(\nu_\pm)^2$ instead of $\nu_\pm$ in the definition of the matrix $V$. 
Notice how the last term in $F(\tau)$ is structurally very similar to Eq.~\eqref{Gaussian_F_diff}, except that the vector involved here is the mean 
$\bm{r}$, instead of the vector of constant entries $\bm{o} = (1,\ldots,1)$. 
This term therefore always vanishes when the mean is zero (and for fermions). 
In addition, though, $F(\tau)$ now has a new term (the one containing the trace), which is independent of the mean and has a $\tau$-dependence which is quadratic in $G(\tau)$. 
For fermions, this is the only relevant term. 

To compute $S(\omega)$ is now slightly more complicated. 
The 2nd line in $F(\tau)$ is again linear in $G$ and hence we can  use the transformation~\eqref{Gaussian_Fourier_transform_W}. 
The first term, though, is quadratic. 
To compute it efficiently, define a new matrix 
\begin{equation}\label{Gaussian_mathbbQ_def}
    \mathbb{Q}(\omega) = \int\limits_0^\infty d\tau~e^{-i \omega \tau} G(\tau) \tilde{\Theta} V \tilde{\Theta} G^\dagger(\tau).
\end{equation}
Then we can write 
\begin{equation}\label{Gaussian_power_spectrum_jumps}
    \begin{aligned}
        S&(\omega) = K + \tr \Big\{ (V\trans \pm V) \big[ \mathbb{Q}(\omega) + \mathbb{Q}(-\omega)\big] \Big\}
        \\[0.2cm]
        &+ 4 \text{Re}\Bigg(\bm{r}\trans  (V+V\trans) \frac{\mathcal{W}}{\mathcal{W}^2 + \omega^2} \tilde{\Theta} V \bm{r}\Bigg).
    \end{aligned}
\end{equation}
The reason why it is convenient to define $\mathbb{Q}(\omega)$ is because we do not need to perform any integrals to compute it, as in Eq.~\eqref{Gaussian_mathbbQ_def}. 
Instead, we can notice that $\mathbb{Q}(\omega)$ is actually the solution of a generalized Lyapunov equation~\footnote{A normal Lyapunov equation has the form $Ax + x A^\dagger = M$. Eq.~\eqref{Gaussian_generalized_lyapunov} is a generalized Lyapunov equation because it has instead the form $A x + x B^\dagger = M$. }
\begin{equation}\label{Gaussian_generalized_lyapunov}
    (\mathcal{W}-i \omega/2) \mathbb{Q}(\omega) + \mathbb{Q}(\omega) (\mathcal{W}^\dagger-i \omega/2) = \tilde{\Theta} V \tilde{\Theta}.
\end{equation}
So once we have $\Theta$, we must solve this equation for each value of $\omega$. 
Finally, setting $\omega=0$ yields the noise. 
\begin{equation}
    \begin{aligned}
        D& = K + 2\tr \Big\{ (V\trans \pm V)  \mathbb{Q} \Big\}
        + 4 \text{Re}\Bigg(\bm{r}\trans  (V+V\trans) \mathcal{W}^{-1} \tilde{\Theta} V\bm{r}\Bigg),
    \end{aligned}
\end{equation}
where $\mathbb{Q} \equiv \mathbb{Q}(\omega=0)$ is now the solution of the standard Lyapunov equation 
\begin{equation}
    \mathcal{W} \mathbb{Q} + \mathbb{Q} \mathcal{W}^\dagger = \tilde{\Theta} V \tilde{\Theta}. 
\end{equation}
Obtaining the noise therefore requires solving two Lyapunov equations: one for actual covariance matrix $\Theta$, and one for the auxiliary matrix $\mathbb{Q}$. 
This is the Gaussian equivalent of the situation encountered in Sec.~\ref{sec:noise_auxiliary_MEQ}: namely, that to compute the noise one must solve two master equations. 

\subsubsection{Example: Parametric oscillator}
\label{sec:parametric_oscillator}

As an application of these Gaussian methods we study the parametric oscillator. It is given by the Hamiltonian of Example D, Eq.~\eqref{ExampleD_H} with $U = 0$. 
For concreteness, here we also take $\Delta = 0$ and $G \to i G$, as this simplifies the results. 
We study both the quantum jump and quantum diffusion  unravellings of Eq.~\eqref{ExampleD_M}. 
The former represents direct photo-detection of the photons leaking out of the cavity, while the latter represents homodyne detection, except for a constant overall factor (see Table~\ref{tab:diffusion_experiments}). 

The steady-state values of $\langle a^\dagger a \rangle$ and $\langle a a \rangle$ were presented in Eq.~\eqref{ExampleD_aves_ss}. 
These are computed using the Lyapunov equation~\eqref{Gaussian_lyapunov_2N}. 
Here we reproduce the idea, focusing on $\Delta = 0$ and $G\to iG$. 
The matrices $\mathcal{W}$ and $\Upsilon$ [Eq.~\eqref{Gaussian_W_2N}] for this model have the form 
\begin{equation}\label{Parametric_oscillator_W}
    \mathcal{W} = \begin{pmatrix} 
    \frac{\kappa}{2} - G & 0 
    \\ 
    0 & \frac{\kappa}{2} + G
    \end{pmatrix},
    \qquad 
    \Upsilon = \frac{\kappa}{2} \id_2. 
\end{equation}
At the steady-state, Eq.~\eqref{Gaussian_lyapunov_2N} yields
\begin{equation}\label{Parametric_oscillator_Theta}
    \Theta := \begin{pmatrix}
        \langle q^2 \rangle & \frac{1}{2} \langle \{q,p\}\rangle 
        \\
        \frac{1}{2} \langle \{q,p\}\rangle  & \langle p^2 \rangle 
    \end{pmatrix}
    =\frac{1}{2}
    \begin{pmatrix}
        \frac{\kappa}{\kappa - 2G} & 0 
        \\
        0 & \frac{\kappa}{\kappa+2G}
    \end{pmatrix}.
\end{equation}
The steady-state is therefore squeezed, where the momentum is compressed while the position expands.  
The choice $G\to i G$ was convenient because it makes $\Theta$ diagonal. 
From $\Theta$ we can extract $\langle a^\dagger a \rangle = \frac{2 G^2}{\kappa^2 - 4 G^2}$, which is a particular case of~\eqref{ExampleD_aves_ss}.
This, in turn, yields the average photon current 
$J = \frac{2 \kappa G^2}{\kappa^2 - 4 G^2}$. 
The diffusive current, on the other hand, vanishes since $\langle a \rangle = 0$ for this model.

\begin{figure}
    \centering
    \includegraphics[width=0.5\textwidth]{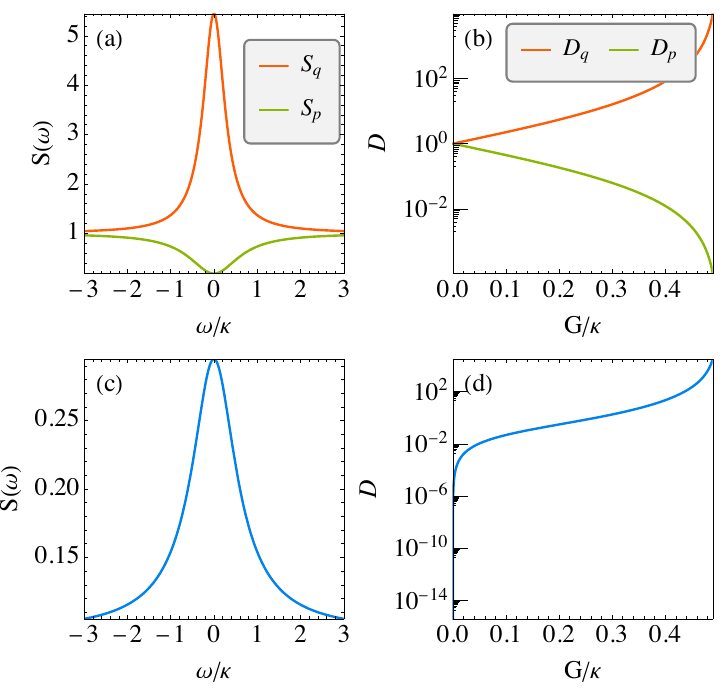}
    \caption{Power spectrum and noise in the parametric oscillator \rev{(Example D, Sec.~\ref{ssec:exampleD}, with $U=0$)}. 
    (a) Quantum diffusion power spectra $S_q$ and $S_p$ [Eq.~\eqref{Parametric_oscillator_Sq_Sp}]. 
    (b) Corresponding noises, Eq.~\eqref{Parametric_oscillator_Dq_Dp}. 
    (c) Power spectra for quantum jump unravelling, Eq.~\eqref{Parametric_oscillator_Sj}.
    (d) Corresponding noise, Eq.~\eqref{Parametric_oscillator_Dj}.
    }
    \label{fig:ExampleD_parametric_oscillator_Gaussian}
\end{figure}

Next we use the results of Sec.~\ref{sec:Gaussian_diffusion} to compute $F,S,D$ for quantum diffusion. 
The matrix $V$ in Eq.~\eqref{Gaussian_V_diff}, which is what defines the measurement in question, is given by 
$V = \sqrt{\frac{\kappa}{2}} \text{diag}\big( e^{-i \phi}, e^{i \phi}\big)$.
Plugging $\mathcal{W}$, $\Theta$ and $V$ in Eq.~\eqref{Gaussian_power_spectrum_diff} yields us the power spectrum.
For the particular choices of homodyning the position ($\phi = 0$) and momentum ($\phi = \pi/2$) we get~\footnote{The general result, for arbitrary $\phi$ is 
\begin{equation*}
    S(\omega) = 1 + \frac{8 G \kappa  \left(\cos (2 \phi ) \left(4 G^2+\kappa ^2+4 \omega ^2\right)+4 G \kappa
   \right)}{\left((\kappa -2 G)^2+4 \omega ^2\right) \left((2 G+\kappa )^2+4 \omega
   ^2\right)}.
\end{equation*}
}
\begin{equation}\label{Parametric_oscillator_Sq_Sp}
\begin{aligned}
    S_q(\omega) &= 1 + \frac{2 G \kappa}{\omega^2 + (G - \frac{\kappa}{2})^2},
    \\[0.2cm]
    S_p(\omega) &= 1 - \frac{2 G \kappa}{\omega^2 + (G + \frac{\kappa}{2})^2}.
\end{aligned}
\end{equation}
This is plotted in Fig.~\ref{fig:ExampleD_parametric_oscillator_Gaussian}(a). 
As can be seen, the momentum quadrature, which is squeezed, shows a dip below the white noise value of 1. 
The dip is strongest at  $G = \kappa/2$, called the threshold for parametric oscillation, where we get $S_p(\omega) = 1-\frac{\kappa^2}{\kappa^2+\omega^2}$. The spectrum therefore reaches identically zero at $\omega = 0$. 
That is, the noise $D = S(0)$ identically vanishes at the threshold for parametric oscillation. 
More generally, the noise reads
\begin{equation}\label{Parametric_oscillator_Dq_Dp}
    D_q = \frac{(\kappa+2G)^2}{(\kappa-2G)^2},
    \qquad 
    D_p = \frac{(\kappa-2G)^2}{(\kappa+2G)^2}.
\end{equation}
These are plotted in Fig.~\ref{fig:ExampleD_parametric_oscillator_Gaussian}(b) as a function of $G/\kappa$. 
As the threshold is approached, $D_q$ diverges while $D_p$ tends to zero. 

Next we turn to the quantum jump unravelling (direct photo-detection).
The relevant $V$-matrix is that defined in Eq.~\eqref{Gaussian_V_matrix_jumps}, which in this case reads 
\begin{equation}
    V = \frac{\kappa}{2} \begin{pmatrix}
    1 & i \\
    -i & 1
    \end{pmatrix}.
\end{equation}
Plugging this, together with Eqs.~\eqref{Parametric_oscillator_W} and~\eqref{Parametric_oscillator_Theta} into Eq.~\eqref{Gaussian_power_spectrum_jumps} we find 
\begin{equation}\label{Parametric_oscillator_Sj}
    S(\omega) = J + \frac{G^2\kappa^2}{(2 G+\kappa)^3 + (2G+\kappa)\omega^2}
    - 
    \frac{G^2\kappa^2}{(2 G-\kappa)^3 + (2G-\kappa)\omega^2},
\end{equation}
with $J = \frac{2 \kappa G^2}{\kappa^2 - 4 G^2}$.
This is plotted in Fig.~\ref{fig:ExampleD_parametric_oscillator_Gaussian}(c). As can be seen, the direct photo-detection spectrum always shows a peak around $\omega=0$. The corresponding noise is 
\begin{equation}\label{Parametric_oscillator_Dj}
    D = \frac{4 G^2 \kappa(8 G^4 + 2 G^2 \kappa^2 + \kappa^4)}{(\kappa^2-4 G^2)^3},
\end{equation}
and is shown in Fig.~\ref{fig:ExampleD_parametric_oscillator_Gaussian}(d) as a function of $G/\kappa$. 
Quite remarkably, we see that the noise associated to quantum jumps is extremely sensitive to $G$, changing by several orders of magnitude. 

\section{Connections with topical fields of research}\label{sec:topical}

\subsection{Waiting time distributions}
\label{sec:WTD}

Consider again the quantum jump scenario first introduced in Sec.~\ref{sec:output_currents}. 
The dynamics is described by a series of jumps, occurring at random times, followed by periods of no jumps, as illustrated in Fig.~\ref{fig:jump_drawing}. 
We are then naturally led to ask questions such as \emph{what is the time until the first jump} or \emph{what is the time between two jumps}?
Since these times are random, we will therefore have a waiting-time distribution (WTD).
WTDs have a long history in stochastic processes in general \cite{stratonovich_book} with applications including quantum optics \cite{vyas1988,carmichael_1989}, electronic transport \cite{brandes_2008,thomas2013,haack2014,Stegmann2021}, and even finance \cite{mainardi2000}. Recently, WTDs have been employed to estimate entropy production \cite{skinner2021} and to distinguish Majorana from Andreev bound states \cite{Schulz_2023}.  
Here we follow the development of Ref.~\cite{brandes_2008}, which focuses specifically on quantum master equations.

To compute the WTD we must decompose the dynamics into conditional evolutions involving jumps and no-jumps.
This can be accomplished using a Dyson series decomposition of $\rho(t) = e^{\mathcal{L} t} \rho(0)$. 
Consider the original QME~\eqref{M} and 
define 
$\mathcal{L}_k \rho = L_k \rho L_k^\dagger$.
These are the jump super-operators. 
Suppose that we only want to monitor a subset $\mathcal{M}$ of these jumps. 
For example, it might be that some channels are not accessible. 
We then define the no-jump superoperator as 
\begin{equation}\label{WTD_L0}
    \mathcal{L}_0 = \mathcal{L} - \sum_{k\in \mathcal{M}} \mathcal{L}_k. 
\end{equation}
In the case where all jump operators are  monitored, this reduces to
\begin{equation}\label{WTD_L0_Hnm}
    \mathcal{L}_0 \rho  = -i \big(\Hnh \rho - \rho \Hnh^\dagger).
\end{equation}
where $\Hnh$ is the non-Hermitian Hamiltonian defined in Eq.~\eqref{H_non_hermitian}.
Moreover, in this case we can connect $\mathcal{L}_0$ to  Kraus operator $M_0$ introduced in Eq.~\eqref{kraus}
\begin{equation}
    M_0 \rho M_0^\dagger = \rho + dt \mathcal{L}_0 \rho. 
\end{equation}
Notice how there is a subjective character to the definition of $\mathcal{L}_0$, in the sense that it depends on what we choose as our monitored channels $\mathcal{M}$. 
The decomposition is thus conditioned on what we assume can be monitored.
Different choices of $\mathcal{M}$ will lead to different $\mathcal{L}_0$, and hence different WTDs.

To  decompose the dynamics we use a Dyson series of $e^{\mathcal{L} t}$. 
This is exactly the same kind of expansion that is used in time-dependent perturbation theory. 
Labeling the initial state as $\rho_0$, it reads \rev{\cite{carmichael:book}}
\begin{align}\label{dyson}
    \rho(t) &= e^{\mathcal{L}_0 t} \rho_0 + \sum_{k \in \mathcal{M}} \int\limits_{0}^t dt_1 e^{\mathcal{L}_0 (t-t_1)} \mathcal{L}_k e^{\mathcal{L}_0 t_1} \rho_0 \\ \nonumber
    &+ \sum_{k,q \in \mathcal{M}} \int\limits_{0}^{t} dt_2 \int\limits_{0}^{t_2} dt_1 e^{\mathcal{L}_0 (t-t_2)} \mathcal{L}_k e^{\mathcal{L}_0 (t_2-t_1)} \mathcal{L}_q e^{\mathcal{L}_0 t_1}\rho_0 + \ldots.
\end{align}
Each term in the expansion contains a well defined number of jumps.
The first term has no jumps; the second has one jump occurring at time $t_1$, and then integrated over all possible $t_1$. 
The third term has two jumps, and so on and so forth. 
This can thus be interpreted as a decomposition of $\rho(t)$ into multiple conditional evolutions, each with a well defined number of jumps. 
This is similar in spirit to the decomposition used in Eq.~\eqref{eq:nresolvedkraus}, when we introduced FCS. 

The individual terms are not properly normalized, though. 
Their normalization gives precisely the probability that each conditional evolution occurs. 
For example,  
\begin{equation}\label{WTD_P_no}
    P_{\rm no}(t|\rho_0) := \tr\big\{ e^{\mathcal{L}_0 t} \rho_0 \big\},
\end{equation}
is the probability that, starting with the state $\rho_0$, no jump takes place up to time $t$. 
After proper normalization, the conditional state of the system, given that no jumps occurred, is therefore 
\begin{equation}\label{rho_no}
    \rho_{\rm no}(t) = \frac{e^{\mathcal{L}_0 t} \rho_0}{P_{\rm no}(t)}.
\end{equation}
We can also have similar definitions for the state conditioned on one jump, two jumps and so on. 

The above discussion makes it clear that during the no-jumps sections, the system evolves as $e^{\mathcal{L}_0t}$, while if a jump occurs we must apply $\mathcal{L}_k$.
We can therefore use this to define the WTD. 
More precisely, starting with an initial state $\rho_0$, the probability that the first jump occurs  at time $t$ and in channel $k$, is given by 
\begin{equation}\label{WTD_WTD_def_1}
    W(t,k|\rho_0) = \tr \Big\{ \mathcal{L}_k e^{\mathcal{L}_0 t} \rho_0\Big\}. 
\end{equation}
This quantity refers to a single jump, after a specific state preparation. It is distinct from the WTD between two jumps, which one might observe in steady-state conditions. The latter will be defined in Eq.~\eqref{WTD_WTD_two_jumps}.
\rev{As a sanity check, if $t=0$ we get $W(0,k|\rho_0) dt = dt \tr\big\{ L_k^\dagger L_k\rho_0\big\}$, which  is precisely the probabilities $p_k$ studied in Eq.~\eqref{pk}.
}

Marginalizing over the channels $k$ gives us the probability that it takes a time $t$ for the jump to occur, irrespective of where it occurs.
Using Eq.~\eqref{WTD_L0}, together with the fact that $\tr(\mathcal{L}\rho) = 0$ for any $\rho$, we then get 
\begin{equation}\label{WTD_W_survival}
    W(t|\rho_0) =  \sum_{k \mathcal{M}} W(t,k|\rho_0)= - \tr\big\{ \mathcal{L}_0 e^{\mathcal{L}_0 t} \rho_0\big\} = - \frac{d P_{\rm no}}{dt},
\end{equation}
where $P_{\rm no}$ is given in Eq.~\eqref{WTD_P_no}.
In the language of probability theory, $P_{\rm no}(t|\rho_0)$ is the survival probability, while $W(t|\rho_0)$ is the probability density function. 
Integrating $W(t|\rho_0)$ from $0$ to $\infty$, and using the fact that $P_{\rm no}(0) = 1$, we then get 
\begin{equation}
    \int\limits_0^\infty W(t|\rho_0)dt = 1 - P_{\rm no}(\infty|\rho_0). 
\end{equation}
The term $P_{\rm no}(\infty|\rho_0)$ reflects the possibility that no jump ever occurs. 
Whether this will vanish or not depends on the problem in question, and is related to the existence of dark states, i.e., initial states which are not affected by the jump operators in question. 
For simplicity, we are going to restrict the discussion to those cases where a jump always eventually occurs, so that $P_{\rm no}(\infty|\rho_0) = 0$. 
The WTD~\eqref{WTD_WTD_def_1} will then be normalized as 
\begin{equation}\label{WTD_normalization}
    \sum_{k \in \mathcal{M}} \int\limits_0^\infty dt~W(t,k|\rho_0) = 1.
\end{equation}
The assumption that $P_{\rm no}(\infty|\rho_0) = 0$ coincides with the assumption that $\mathcal{L}_0$ is invertible. 
Indeed, if this is true then normalization also follows from Eq.~\eqref{WTD_W_survival} and the identity $\int_0^\infty e^{\mathcal{L}_0 t} dt = - \mathcal{L}_0^{-1}$
\rev{(a relation between $\mathcal{L}_0^{-1}$ and the Drazin inverse $\mathcal{L}^+$ is explored in Appendix~\ref{app:drazin_L0_relation}).}

If we marginalize $W(t,k|\rho_0)$ in $t$, instead of $k$, we get the probability that the jump occurs in channel $k$, irrespective of when it happens: 
\begin{equation}\label{WTD_k_rho0}
    W(k|\rho_0) = \int\limits_0^\infty dt~W(t,k|\rho_0) = - \tr\big\{ \mathcal{L}_k \mathcal{L}_0^{-1} \rho_0\big\}.
\end{equation}
This is an interesting quantity in itself, because it can be used to assess which channel is more likely to click.

The moments of $W(t|\rho_0)$ [Eq.~\eqref{WTD_W_survival}] can be computed with the identity $\int_0^\infty dt ~ t^n e^{\mathcal{L}_0 t} = (-1)^{n+1} n! \mathcal{L}_0^{-(n+1)}$, and read
\begin{equation}\label{moments_WTD}
    E(T^n) = (-1)^n n! ~\tr\big\{ \mathcal{L}_0^{-n} \rho\big\}. 
\end{equation}
So, for example, $E(T)$ is the average time until the first jump (irrespective of the channel). 
\rev{Alternatively, we can also post-select on the channel and ask, for example, ``what is the average time until a jump, given that the jump occurred in channel $k$?''
The underlying distribution, from Bayes' rule, is 
$W(t|\rho_0,k) = \frac{W(t,k|\rho_0)}{W(k|\rho_0)}$.
Hence
\begin{equation}
    E(T^n|\rho_0,k) = \int\limits_0^\infty dt~W(t|\rho_0,k) t^n = 
    \frac{(-1)^n n!}{W(k|\rho_0)} \tr\big\{ \mathcal{L}_k \mathcal{L}_0^{-(n+1)}\rho_0 \big\}.
\end{equation}
The moments satisfy the law of total expectation, so that we can recover Eq.~\eqref{moments_WTD} as 
\begin{equation}
    E(T^n|\rho_0) = \sum_k W(k|\rho_0) E(T^n|\rho_0,k).
\end{equation}
However, the same is not true for the variance, for example, which instead satisfies the law of total variance:
\begin{equation}
\begin{aligned}
    {\rm var}(T) =&
    {\rm var}_k\Big( E(T|k)\Big) + E_k \Big({\rm var}(T|k)\Big)
    \\[0.2cm]
    =& \sum_k W(k|\rho_0) \Big[E(T|k)^2 - E(T)^2\Big] 
    \\&+ \sum_k W(k|\rho_0) \Big[ E(T^2|k) - E(T|k)^2\Big].
\end{aligned}    
\end{equation}
The dependence on $\rho_0$ was omitted from some of the terms, for readability. The first term describes how the average waiting times fluctuate between the different channels, while the second term describe the variance of the waiting time within each channel (and then averaged over all channels). 
}


Eq.~\eqref{WTD_WTD_def_1} is the WTD for a single jump after a specific state preparation $\rho_0$. 
One might also be interested in the waiting times between two consecutive jumps. 
This is common when we deal with the steady state, where many jumps are constantly happening and we just want to make a histogram of the time between two jumps. 
We can construct the corresponding WTD directly from Eq.~\eqref{WTD_WTD_def_1}. 
If the system is in $\rhoss$ and a jump occurs in a given channel $q$, then 
the state must be updated to
$\rhoss \to \frac{ \mathcal{L}_q \rhoss}{\tr(\mathcal{L}_q \rhoss)}$. 
Using this in Eq.~\eqref{WTD_WTD_def_1} then yields
\begin{equation}\label{WTD_WTD_two_jumps}
    W(t,k|q) = \frac{\tr\big\{ \mathcal{L}_k e^{\mathcal{L}_0 t} \mathcal{L}_q \rhoss\big\}}{\tr\big\{ \mathcal{L}_q \rhoss\big\}}.
\end{equation}
This WTD is normalized as in Eq.~\eqref{WTD_normalization}, for each $q$. 
\rev{One notices here a similarity to the two-point correlation function $F(\tau)$ in Eq.~\eqref{F_jump}, in the sense that both have the structure ``jump $\to$ evolve $\to$ jump.'' 
The main difference is that $F$ evolves with $e^{\mathcal{L} t}$, 
while $W$ has $e^{\mathcal{L}_0 t}$. 
This reflects the fact that $W$ describes correlations between two points in time, \emph{conditioned} on there being no jumps in between, while for $F$ there is no such restriction. }

\subsection{Path probabilities in the jump unravelling }

\rev{The WTD can be extended to describe a sequence of multiple jumps. This will produce a detection record of the form 
\begin{equation}\label{path_quantum_trajectory}
    (t_1,k_1), (t_2,k_2),\ldots,
\end{equation}
describing the set of channels where the jumps occurred, together with the time tags of when they occurred. 
Eq.~\eqref{path_quantum_trajectory} is, in fact, what an experimentalist with access to the quantum jump unravelling would observe. 
We use $t_n$ to denote absolute times; the relative times, between  jumps, is $\tau_n = t_n - t_{n-1}$.
Below we shall make use of both notations. 
The probability of observing a specific sequence of $N$ jumps can be read from the Dyson series~\eqref{dyson}:
\begin{equation}\label{WTD_multi_jump}
    W(t_1,k_1,\ldots,t_N,k_N|\rho_0) = \tr\Big\{ \mathcal{L}_{k_N} e^{\mathcal{L}_0 (t_N - t_{N-1})}
    \ldots \mathcal{L}_{k_1} e^{\mathcal{L}_0 t_1} \rho_0
    \Big\},
\end{equation}
where it is implicit that $t_{i+1} > t_i$. 
Moreover, if at any given time $t$, a total of $N$ jumps occurred, the (unnormalized) state of the system will be 
\begin{equation}\label{WTD_rho_multi_jump}
    \rho^c(t|t_1,k_1,\ldots t_N,k_N) = e^{\mathcal{L}_0 (t-t_N)} \mathcal{L}_{k_N} e^{\mathcal{L}_0 (t_N - t_{N-1})}
    \ldots \mathcal{L}_{k_1} e^{\mathcal{L}_0 t_1} \rho_0,
\end{equation}
where $t>t_N$ and the factor of $e^{\mathcal{L}_0(t-t_N)}$ accounts for the no-jump evolution of the system after the $N$-th jump. 
}

\rev{The two formulas above reflect a subtlety concerning two different ensembles one can work on: 
\begin{itemize}
    \item {\bf $t$-ensemble:} we fix a given final time $t$, but allow the number of jumps $N$ that take place to fluctuate;
    \item {\bf $N$-ensemble:} we fix a total number $N$ of jumps, but the final time $t_N$ at which the last jump takes place is allowed to fluctuate.
\end{itemize}
The distribution in Eq.~\eqref{WTD_multi_jump} is in the $N$-ensemble, because we are talking about specifically $N$ jumps and we place no restriction on what $t_N$ might be. 
In fact, Eq.~\eqref{WTD_multi_jump} is normalized as
\begin{equation}
    \sum_{k_1,\ldots,k_N \in \mathcal{M}} \int\limits_0^\infty dt_N \ldots \int\limits_0^{t_2} dt_1 W(t_1,k_1,\ldots,t_N,k_N) = 1,
\end{equation}
where the integrals are all nested in order to respect $t_{i+1}>t_i$.

The $t$-ensemble is, in many aspects, more natural since in physics we usually describe systems at specific times $t$. In fact, all previous results about the quantum jump unravelling in this tutorial have been implicitly in the $t$-ensemble.
Eq.~\eqref{WTD_rho_multi_jump} is also in the $t$-ensemble because $t$ is fixed, but $N$ is allowed to fluctuate. 
The normalization gives us the probability 
\begin{equation}\label{WTD_Wtilde_t_ensemble}
\tilde{W}_t(t_1,k_1,\ldots,t_N,k_N|\rho_0) = \tr\big\{ \rho^c(t|t_1,k_1,\ldots,t_N,k_N)\big\},    
\end{equation}
which differs from Eq.~\eqref{WTD_multi_jump} because $N$ is a random variable. 
This difference is better appreciated if we look at how $\tilde{W}_t$ is normalized:
\begin{equation}\label{WTD_t_ensemble_normalization}
    \sum_{N=0}^\infty \,\sum_{k_1,\ldots,k_N\in \mathcal{M}}  \int_0^t dt_N  \cdots \int_0^{t_2} dt_1 \tilde{W}_t(t_1,k_1,\ldots,t_N,k_N|\rho_0) = 1.
\end{equation}
This follows from the Dyson series~\eqref{dyson}, and the fact that sum over all terms yields the unconditional state  $\rho(t)$, which is normalized, $\tr(\rho(t)) =1$. }

\rev{Consider now a sequence of jumps, like in Eq.~\eqref{path_quantum_trajectory}, and let us analyze a stroboscopic dynamics, where we only look at the state of the system immediately after each jump. 
We will use relative times $\tau_i = t_i - t_{i-1}$.
If the $i$-th jump was associated to the pair $(\tau_i,k_i)$  then the state of the system will change as 
\begin{equation}\label{stroboscopic_map}
    \rho_{i}^c = \frac{\mathcal{L}_{k_i} e^{\mathcal{L}_0 \tau_i} \rho_{i-1}^c}{W(\tau_i,k_i|\rho_{i-1}^c)},     
\end{equation}
where $W(\tau,k|\rho)$ is precisely the WTD introduced in Eq.~\eqref{WTD_WTD_def_1} and $\rho_0^c = \rho_0$ is the initial state.
Using this result recursively we can write Eq.~\eqref{WTD_multi_jump} as 
\begin{equation}\label{WTD_multi_jump_hidden_markov}
    W(\tau_1,k_1,\ldots,\tau_N,k_N|\rho_0) = W(\tau_N,k_N|\rho_{N-1}^c) \ldots W(\tau_1,k_1|\rho_0).
\end{equation}
This formula seems to imply that the outcomes~\eqref{path_quantum_trajectory} form a Markov chain. 
This is not true, however, because each $\rho_i^c$ will depend on all previous outcomes  $(\tau_j,k_j)$ with $j\leqslant i$.
So what this shows is that the memory each outcome $(\tau_i,k_i)$ retains of the past is  encoded entirely in $\rho_{i-1}^c$.
This is therefore akin to a hidden Markov model, where $\rho_i^c$ is a hidden layer, while $(\tau_i,k_i)$ are the observations. 
}. 

\rev{We can also treat the $t$-ensemble similarly. Eq.~\eqref{WTD_rho_multi_jump} becomes 
\begin{equation}
    \rho^c(t|t_1,k_1,\ldots,t_N,k_N) = e^{\mathcal{L}_0 (t-t_N)} \rho_N^c,
\end{equation}
and Eq.~\eqref{WTD_Wtilde_t_ensemble} can be written (now using absolute times)
\begin{equation}
\begin{aligned}
    \tilde{W}_t(t_1,k_1,\ldots,t_N,k_N|\rho_0) =& P_{\rm no}(t-t_N|\rho_N^c) W(t_N-t_{N-1},k_N|\rho_{N-1}^c)\times
    \\[0.2cm]&\times \ldots W(t_2-t_1,k_2|\rho_1^c) W(t_1,k_1|\rho_0).
\end{aligned}    
\end{equation}
The first term $P_{\rm no}(t-t_N|\rho_N)$ reflects here the fact that there were no other jumps after $t_N$. 
}

\subsubsection{Stationarity}

\rev{Very often we are interested in the statistics of many-jumps in a steady-state scenario. For example, we discussed how Eq.~\eqref{WTD_WTD_two_jumps} represents what one would obtain by making a histogram of the waiting times in the steady-state. 
Intuitively, one expects that after many jumps Eq.~\eqref{WTD_multi_jump_hidden_markov} should become stationary; that is, it should become approximately (i)  independent of the initial state $\rho_0$ and (ii)  translationally invariant.
Stationarity does not happen for all models. 
For example, an optical cavity in the absence of any drive might produce a few jumps, depending on the initial state, but eventually these will stop. 
The conditions for a model to support a stationary jump distribution are related to the eigenvalues of the superoperator $-\mathcal{J} \mathcal{L}_0^{-1}$, as shown recently in~\cite{Landi2023}.
here we will ask instead a simpler question: is there a special state $\pi$ for which Eq.~\eqref{WTD_multi_jump} is stationary from the very beginning?
Consider, for instance, the two outcome distribution $W(\tau_1,k_1,\tau_2,k_2|\pi)$. If we marginalize over $\tau_2,k_2$ we recover $W(\tau_1,k_1|\pi)$ precisely as given in Eq.~\eqref{WTD_WTD_def_1}. 
Stationarity means that if we marginalize, instead, over $\tau_1,k_1$ we should obtain $W(\tau_2,k_2|\pi)$ with \emph{exactly} the same shape as $W(\tau_1,k_1|\pi)$. }

\rev{Naively, one might assume that this special state is the steady-state $\rhoss$. It turns out, however, that this is not the case. Indeed, using  Eq.~\eqref{WTD_multi_jump} and
marginalizing $W(\tau_1,k_1,\tau_2,k_2)$ over $\tau_1,k_1$ yields
\begin{equation}
    \begin{aligned}
        \sum_{k_1}\int d\tau_1 ~\tr\big\{ \mathcal{L}_{k_2} e^{\mathcal{L}_0 \tau_2} \mathcal{L}_{k_1} e^{\mathcal{L}_0 \tau_1} \pi\big\} = 
        \tr\big\{ \mathcal{L}_{k_2} e^{\mathcal{L}_0 \tau_2} \mathcal{J}(-\mathcal{L}_0^{-1}) \pi\big\},
    \end{aligned}
\end{equation}
where, recall, $\mathcal{J} = \sum_{k\in\mathcal{M}} \mathcal{L}_k$.
If this is to have the form~\eqref{WTD_WTD_def_1} then  $\pi$ must satisfy the fixed point equation 
\begin{equation}
    \pi = -\mathcal{J}\mathcal{L}_0^{-1} \pi. 
\end{equation}
One can show that the solution is the  so-called ``jump steady-state'' (JSS)~\cite{Landi2023}
\begin{equation}\label{JSS}
    \pi = \frac{\mathcal{J} \rhoss}{K}, 
\end{equation}
where 
\begin{equation}
    K = \sum_{k\in\mathcal{M}} \langle L_k^\dagger L_k\rangle_{\rm ss} = \sum_{k\in\mathcal{M}} \tr\big\{ \mathcal{L}_k \rhoss\big\} = - \tr\big\{ \mathcal{L}_0 \rhoss\big\} ,
\end{equation}
is called the dynamical activity, and describes the number of jumps per unit time in the steady-state. 
We use the same letter $K$ here to emphasize the connection with Eq.~\eqref{K}; in fact, the two coincide provided we take $\nu_k = 1$ for $k\in \mathcal{M}$ and zero otherwise. 
}

\rev{We can now establish several results concerning the steady-state statistics, which resonate directly with experiments. 
First, what is the relative frequency with which channel $k$ clicks? 
We start with $W(\tau,k|\pi)$ and marginalize over $\tau$, leading to 
\begin{equation}\label{WTD_frak_pk}
    \mathfrak{p}_k := \int\limits_0^\infty d\tau ~W(\tau,k|\pi) 
    =  - \frac{1}{K}\tr\big\{ \mathcal{L}_k \mathcal{L}_0^{-1} \mathcal{J}\rhoss\big\}
    = \frac{1}{K} \tr\big\{\mathcal{L}_k \rhoss\big\},
\end{equation}
where we used the fact that $\mathcal{J} = \mathcal{L}-\mathcal{L}_0$, and $\mathcal{L}\rhoss = 0$. 
These are closely related to the original jump probabilities $p_k$ in Eq.~\eqref{pk}. 
The difference is that $\mathfrak{p}_k$ is restricted only to the subset $\mathcal{M}$ of jumps, and therefore need a different normalization.}
If we marginalize Eq.~\eqref{WTD_WTD_two_jumps} over $t$ we obtain the transition probability that jump $q$ is followed by jump $k$
\begin{equation}
    W(k|q) = \frac{
        \tr \big\{ \mathcal{L}_k (-\mathcal{L}_0^{-1}) \mathcal{L}_q \rhoss \big\}
    }{
        \tr \big\{ \mathcal{L}_q \rhoss \big\}
    }.
\end{equation} 
One may verify that $\mathfrak{p}_k$ are the fixed points of this transition probability, $\sum_q W(k|q) \mathfrak{p}_q = \mathfrak{p}_k$.

Next, ``what is the time between jumps, irrespective of the chanels''?
Again, starting with $W(t,k|\pi)$ but now marginalizing over $k$ yields 
\begin{equation}
    W(t) = \tr\big\{ \mathcal{L}_0 e^{\mathcal{L}_0t} \pi\big\}
    =
    - \frac{
    \tr \big\{ \mathcal{L}_0 e^{\mathcal{L}_0 t} \mathcal{L}_0 \rhoss\big\}
    }{
    \tr(\mathcal{L}_0 \rhoss)
    },
\end{equation}
which is normalized as $\int_0^\infty W(t)dt = 1$.
This result could also have been obtained by averaging Eq.~\eqref{WTD_WTD_two_jumps} over $\mathfrak{p}_q$. 
From $W(t)$ it follows that the average time between jumps is
\begin{equation}\label{Mean_waiting_time}
    E(T) =  \frac{1}{K},
\end{equation}
which makes sense, since the dynamical activity is precisely the number of jumps per unit time in the steady-state.


\subsubsection{Renewal processes}\label{ssec:renewal}

A renewal process is any process for which, after a jump, the state of the system is completely reset; i.e., no memory of its previous past is retained. 
Mathematically, in a renewal process, for each jump operator $\mathcal{L}_k$ there is a corresponding (properly normalized) state $\sigma_k$  such that 
\begin{equation}\label{WTD_renewal_definition}
    \frac{\mathcal{L}_k \rho}{\tr(\mathcal{L}_k \rho)} = \sigma_k. 
\end{equation}
In vectorization language (Sec.~\ref{sec:vectorization}), this is tantamount to  $\mathcal{L}_k$ being rank-1 matrices of the form 
\begin{equation}
    \mathcal{L}_k = |\sigma_k \rrangle\llangle \xi_k|,
\end{equation}
for vectors $\llangle \xi_k|$. 

For example, in the case of a qubit, a jump $\sigma_-\rho \sigma_+$ is renewal since it completely resets any $\rho$. 
Conversely, for a bosonic system a jump  $a \rho a^\dagger$ is \emph{not} renewal since $a\rho a^\dagger$ still retains a memory of $\rho$.
Another example of a system that is not renewal is two coupled qubits (double quantum dot) with jump operators acting locally on each qubit. A jump will reset the local state of one qubit, but not the global state of the two.
Generally, therefore, renewal processes are quite rare. 

For renewal processes, the WTD between two jumps [Eq.~\eqref{WTD_WTD_two_jumps}] becomes independent of $\rho$ 
\begin{equation}
    W(t,k|q) = \tr\big\{ \mathcal{L}_k e^{\mathcal{L}_0 t} \sigma_q\big\} = \llangle \xi_k | e^{\mathcal{L}_0 t} |\sigma_k\rrangle.  
\end{equation}
Moreover, Eqs.~\eqref{JSS} and~\eqref{WTD_frak_pk} reduce to 
\begin{equation}
    \mathfrak{p}_k = \llangle \xi_k\rhossV/K,
    \qquad 
    \pi = \sum_k \mathfrak{p}_k \sigma_k.
\end{equation}
The JSS is therefore a simple a statistical mixture of the post-jump states $\sigma_k$. 
Recursively applying Eq.~\eqref{WTD_renewal_definition} allows us to decompose Eq.~\eqref{WTD_multi_jump} as 
\begin{equation}
    W(t_1,k_1,\ldots,t_N,k_N|\pi) = \sum_q W(t_N,k_N|k_{N-1}) \ldots W(t_1,k_1|q) \mathfrak{p}_q .
\end{equation}
Renewal processes therefore form a Markov chain: each outcome $t_i,k_i$ depends only on the previous jump channel $k_{i-1}$. 
If there is only one jump channel, the waiting times therefore become independent and identically distributed:
\begin{equation}
    W(t_1,\ldots,t_N) = W(t_N) \ldots W(t_1).
\end{equation}

For renewal processes with a single jump channel, the variance of the waiting-time distribution can be connected to the noise $D$ [Eq.~\eqref{D}]. 
Defining the mean waiting time $\mu = E(T)$ and the variance $\sigma^2 = E(T^2) - \mu^2$, the relation reads 
\begin{equation}\label{renewal_D_sigma}
    D = \frac{\sigma^2}{\mu^3},
\end{equation}
which provides a deep connection between fluctuations in the WTD and the variance of the integrated charge $N(t)$. 
This relation is proven in Appendix~\ref{app:WTD}, and is 
 well known in renewal theory~\cite{cox_renewal}.

\subsection{Multiple current specimens and current-current correlations}
\label{sec:multiple_currents}

\rev{\subsubsection{Correlations between different currents}}

Throughout this tutorial we have mainly considered the case where there is only a single current of interest, $I(t)$ or $N(t)$, defined by the set of weights $\nu_k$ that enter into Eq.~\eqref{total_charge}. 
All results readily generalize when we have two or more current specimens, or multiple detectors~\cite{Jordan2005}.
For instance, in systems with two reservoirs, this could be the particle current to each of the baths. Alternatively, it could be the energy and particle currents. For multiple currents, Eq.~\eqref{total_charge} generalizes to 
\begin{equation}
    \label{multiple_charge}
    N_\alpha(t) = \sum_k \nu_{\alpha k} N_k(t),
\end{equation}
where $\nu_{\alpha k}$ is the weight that current $\alpha$ attributes to the jump $L_k$. The corresponding current is then $I_\alpha(t) = dN_\alpha/dt$. For instance, in Example A, Eq.~\eqref{ExampleA_M}, we might use $\nu_{p-} = -1$, $\nu_{p+} = +1$ for the particle current and $\nu_{d-} = \nu_{d+} = 1$ for the dynamical activity. 

\rev{
With multiple currents, an interesting question concerns the correlations between them. For example, how are the emissions to one reservoir correlated with absorptions from the other? To address this question in general, we consider the correlations among the set of ``elementary currents'' $\{I_k\}_{k=1}^{\nops}$, where
\begin{equation}
    \label{elementary_current}
    I_k(t) = \frac{dN_k}{dt},
\end{equation}
is the current pertaining to one specific jump operator $L_k$ entering the master equation~\eqref{M}. The corresponding average current is simply
\begin{equation}
    \label{average_current_vector}
    J_k = E[I_k(t)] = \tr\left(\mathcal{L}_k \rhoss \right),
\end{equation}
where $\mathcal{L}_k\rho = L_k\rho L_k^\dagger$ and we focus on the steady state for simplicity. The steady-state two-point correlation function between any pair of elementary currents is (for $\tau\geq 0)$
\begin{align}
        \label{two_point_multiple}
    F_{kq}(\tau) & = E[I_k(t+\tau) I_q(t)] - J_kJ_q \notag \\
    & = J_k \delta_{kq} \delta(\tau) + \tr\left( \mathcal{L}_k e^{\mathcal{L}\tau} \mathcal{L}_q\rhoss \right) - J_k J_q,
\end{align}
which can be derived following the exact same steps used in Appendix~\ref{app:F}, but taking care with the indices $k,q$. For a set of elementary diffusive currents of the form 
\begin{align}
    \label{elementary_diffusive}
    I_k(t) = \left \langle e^{-i \phi_k}L_k + e^{i \phi_k}L_k^\dagger\right \rangle_c(t) + \frac{dW_k}{dt},
\end{align}
one obtains similar expressions to Eqs.~\eqref{average_current_vector} and~\eqref{two_point_multiple}, but with the replacement 
\begin{equation}
    \label{elementary_current_superop_diffusive}
\mathcal{L}_k(\rho)  \to  \mathcal{H}_k(\rho) = e^{-i\phi_k} L_k \rho + e^{i\phi_k}\rho L_k^\dagger,
\end{equation}
as can be shown following the same steps as in Appendix~\ref{app:hom}. For $\tau <0$, the autocorrelation function can be found using the symmetry relation $F_{kq}(-\tau) = F_{qk}(+\tau)$.

From Eq.~\eqref{two_point_multiple} we readily compute the power spectrum matrix as in Eq.~\eqref{power_spectrum},
\begin{equation}
    \label{power_spectrum_matrix}
    S_{kq}(\omega) = \int_{-\infty}^\infty d\tau \, e^{-i\omega \tau} F_{kq}(\tau),
\end{equation}
and the noise or diffusion matrix $D_{kq} = S_{kq}(0)$. The latter is interpreted similarly to Eq.~\eqref{D}: the diagonal entries are  $D_{kk} = \frac{d}{dt} {\rm Var}\big(N_k(t)\big)$. Conversely, the off-diagonals read 
\begin{equation}
    D_{kq} = \frac{d}{dt} {\rm Cov}\big(N_k(t), N_q(t)\big),
\end{equation}
where ${\rm Cov}(A,B) = E(AB)-E(A)E(B)$ is the covariance. The power spectrum matrix is Hermitian at each frequency, $S_{kq}(\omega) = S_{qk}^*(\omega) = S_{qk}(-\omega)$. The noise matrix is real and symmetric, $D_{kq} = D_{qk}$, and admits the explicit expression
\begin{equation}
    \label{noise_matrix_explicit}
    D_{kq} = J_k \delta_{kq} - \llangle \mathbbm{1}|\mathcal{L}_k \mathcal{L}^+\mathcal{L}_q|\rhoss \rrangle - \llangle \mathbbm{1}|\mathcal{L}_q \mathcal{L}^+\mathcal{L}_k|\rhoss \rrangle.
\end{equation}
Meanwhile, the real, symmetric matrix
\begin{equation}\label{coherence_power_spectrum}
    \mathbb{S}_{kq}(\omega) := \frac{|S_{k q}(\omega)|^2}{S_{kk}(\omega) S_{qq}(\omega)},
\end{equation}
is referred to as ``coherence'' in the language of signal processing, and provides a (frequency-resolved) measure of the correlations between any pair of elementary currents. With these definitions in hand, it is straightforward to find the corresponding quantities for any linear combination of the elementary currents: one simply sums over the corresponding coefficients as in Eq.~\eqref{multiple_charge}. For example, the long-time covariance matrix is given by 
\begin{equation}
    \label{diffusion_matrix}
    D_{\alpha\beta} = \frac{d}{dt} {\rm Cov}\left(N_\alpha(t),N_\beta(t)\right) = \sum_{k,q} \nu_{\alpha k} D_{kq}\nu_{\beta q}.
\end{equation}
}

FCS (Sec.~\ref{sec:FCS}) can be extended to study cross-correlations in a similar way, by introducing multiple counting fields $\chi_\alpha$: one for each current specimen, pertaining to a total charge transfer $n_\alpha$. Let us use numeric labels for the different currents, $\alpha = 1,2,\ldots$, allowing us to introduce a convenient vector notation $\bm{n} = (n_1, n_2, \ldots)$, $\bm{\chi} = (\chi_1, \chi_2,\ldots$), and $\bm{\nu}_k = (\nu_{1k}, \nu_{2k},\cdots)$. The multi-current analogue of Eq.~\eqref{FCS_P} is then 
\begin{equation}
    \label{multiple_FCS}
    P(\bm{n},t) = \left(\prod_\alpha \int \frac{d\chi_\alpha}{2\pi}\right) \, e^{-i\bm n\cdot \bm\chi} {\rm tr}[\rho_{\bm{\chi}}(t)],
\end{equation}
where $\rho_{\bm \chi}$ is the solution of $d\rho_{\bm \chi}/dt = \mathcal{L}_{\bm \chi}\rho_{\bm \chi}$, with
\begin{equation}
    \label{multiple_tilted_liouvillian}
\mathcal{L}_{\bm \chi}\rho =  -i [H,\rho] + \sum_{k=1}^{\sf r}\left( e^{i\bm{\nu}_k\cdot \bm{\chi}} L_k \rho L_k^\dagger - \tfrac{1}{2} \{ L_k^\dagger L_k,\rho\}\right).
 \end{equation}

An important example of current-current correlations arises in heterodyne detection of the light emitted by an optical cavity. As discussed in Sec.~\ref{sec:diffusion_homo_hetero_charge}, we split the jump operator entering the QME~\eqref{ExampleD_M} into two as $L_1 = L_2 = \sqrt{\kappa/2} a$, and take orthogonal phases for the reference currents $\phi_1 = 0$ and $\phi_2 = \pi/2$. The corresponding output currents are defined by choosing $\nu_{11} = \nu_{22} = 1/\sqrt{\kappa}$ and $\nu_{12} = \nu_{21} = 0$. From Eq.~\eqref{two_point_multiple}, we obtain the current-current correlations
\begin{align}
F_{11}(\tau) & = \kappa^{-1}\delta(\tau) + {\rm Re}\left[  \langle a^\dagger(\tau) a\rangle +  \langle a(\tau) a\rangle\right] - J_1^2,\notag \\
F_{22}(\tau) & = \kappa^{-1}\delta(\tau) + {\rm Re}\left[  \langle a^\dagger(\tau) a\rangle -  \langle a(\tau) a\rangle\right]  - J_2^2,\notag  \\
 F_{12}(\tau)& = {\rm Im}\left[ \langle a(\tau) a\rangle + \langle a^\dagger(\tau) a\rangle  \right], \notag \\
F_{21}(\tau) & = {\rm Im}\left[ \langle a(\tau) a\rangle - \langle a^\dagger(\tau) a\rangle \right].
\end{align}
 By combining these correlations appropriately, we recover the complex functions that characterize the first-order optical coherence, discussed in Sec.~\ref{sec:Coherence}.  This can be done succinctly by defining the complex current $I_{\rm Het}(t) = [I_1(t) + i I_2(t)]/\sqrt{2}$, whose average is $J_{\rm Het} = E[I_{\rm Het}(t)] = \langle a\rangle$, which is a complex number $J_{\rm Het} = |J_{\rm Het}| e^{i\varphi}$. It follows that
\rev{\begin{align}
    \label{heterodyne_fluctuations}
    & E[\delta I^*_{\rm Het}(t+\tau)\delta I_{\rm Het}(t)]  = \kappa^{-1}\delta(\tau) + |J_{\rm Het}|^2 \left(g^{(1)}(\tau) - 1\right),\notag \\
     &E[\delta I_{\rm Het}(t+\tau)\delta I_{\rm Het}(t)]  = |J_{\rm Het}|^2 \Big( \langle a(\tau) a\rangle  - \cos 2\varphi\Big).
\end{align}
where $\delta I_{\rm Het} = I_{\rm Het}-J_{\rm Het}$.} Therefore, both the emission spectrum, which depends on $g^{(1)}(\tau)$ [Eq.~\eqref{first-order-coherence}] and the spectrum of squeezing, related to $\langle a(\tau)a\rangle$, can be inferred by monitoring temporal fluctuations of the heterodyne current. 

\subsubsection{Higher-order current correlations}

\rev{So far we have considered only the first two cumulants of the currents: namely, their averages and second-order correlation functions. Recent years have seen growing interest in higher-order cumulants of output currents in continuously measured quantum systems. Going beyond second-order correlations gives access to more detailed information on the system, allowing for more stringent comparisons between experimental data and theoretical models~\cite{Sifft2021} or the detection of quantum measurement invasiveness~\cite{Bednorz2012}, for example. Exact formulas for arbitrary current correlation functions were derived using stochastic calculus methods in Refs.~\cite{Hagele_2018,Tilloy2018} for diffusive measurements, and then generalised to quantum-jump measurements in Ref.~\cite{Guilmin2023}. Analogous results for multi-time correlation functions have been derived using a quasi-probability approach in Ref.~\cite{Bednorz2012}, and directly from the quantum master equation in Ref.~\cite{Blasi2024}. Here, we simply describe the results using the formalism of this tutorial, and refer the interested reader to the original references for further details.}

\rev{To find a general expression for current correlation functions, consider the generating functional
\begin{equation}
    \label{generating_functional_def}
    \mathcal{Z}[\bm{\chi}(t)] = E\left[ \exp \left( \sum_{k=1}^\nops \int_{0}^{t} dt' \,\chi_k(t') I_k(t') \right)\right].
\end{equation}
Here, $I_k(t)$ is the elementary current corresponding to jump channel $k$, as in Eq.~\eqref{elementary_current}, and $\bm{\chi}(t) = [\chi_1(t),\cdots,\chi_\nops(t)]$ is a vector of arbitrary source functions which, as we will see, coincide with the counting fields in FCS  up to a factor $i$ (Sec.~\ref{sec:FCS}). An arbitrary $M$-point correlation function can then be generated by functional differentiation of $\mathcal{Z}[\bm{\chi}(t)]$, as~\footnote{\rev{Recall that the functional derivative of a functional $\Phi[u(t)]$ with respect to the function $u(t)$ is
$$\frac{\delta \Phi[u(t)]}{\delta u(t')} = \lim_{\varepsilon\to 0} \frac{\Phi[u(t)+\varepsilon \delta(t-t')] - \Phi[u(t)]}{\varepsilon}$$}}
\begin{align}
\label{correlation_functional_derivative}
    E[I_{k_1}(t_1) \cdots I_{k_M}(t_M)] = \left.\frac{\delta }{\delta \chi_{k_1}(t_1)}\cdots \frac{\delta }{\delta \chi_{k_M}(t_M)} \mathcal{Z}[\bm{\chi}(t)]\right\vert_{\bm{\chi}=0}.
\end{align}
We assume without loss of generality that the time arguments are ordered so that $0\leq t_1 \leq t_2 \leq \cdots \leq t_M \leq t.$ As shown in Ref.~\cite{Guilmin2023}, the generating functional can be expressed as
\begin{equation}
\label{generating_functional_solution}
    \mathcal{Z}[\bm{\chi}(t)] = \tr \left\{\mathcal{T}\exp \left[ \int_0^t dt'\, \mathcal{L}_{\bm{\chi}}(t') \right] \rho_0\right\}.
\end{equation}
where $\mathcal{L}_{\bm{\chi}}$ is a tilted Liouvillian. For quantum jumps it reads
\begin{equation}
    \label{tilted_Liouvillian_higher-order}
    \mathcal{L}_{\bm{\chi}}(t)  = \mathcal{L} + \sum_{k=1}^\nops \left(e^{\chi_k(t)} - 1\right) \mathcal{L}_k,
\end{equation}
which is the multi-current generalization of Eq.~\eqref{tilted_liouvillian}. 
Conversely, for diffusion [Eq.~\eqref{elementary_diffusive}] it becomes~\cite{Tilloy2018,Guilmin2023}
\begin{equation}
    \label{tilted_Liouvillian_higher-order_diffusive}
   \mathcal{L}_{\bm{\chi}}(t)  = \mathcal{L} + \sum_{k=1}^\nops \left(\chi_k(t) \mathcal{H}_k + \frac{\chi_k(t)^2}{2}\right),
\end{equation}
which is the generalization of Eq.~\eqref{L_chi_diffusion}.
All of these results can therefore be understood as an elegant generalization of FCS results of Sec.~\ref{sec:FCS_quantum_diffusion}, to describe also  the statistics of instantaneous currents.}

\rev{However, the generating functional~\eqref{generating_functional_solution} contains more information than the standard FCS approach, since we can obtain arbitrary temporal correlation functions. If we assume that all time arguments are \textit{distinct}, using Eq.~\eqref{correlation_functional_derivative} we obtain
\begin{align}
\label{multi_time_correlation_function}
   &  E\left[I_{k_1}(t_1)I_{k_2}(t_2)\cdots I_{k_M}(t_M) \right] \notag \\ & = \tr\left[\mathcal{L}_{k_M}\mathcal{S}(t_M,t_{M-1})\mathcal{L}_{k_{M-1}}\cdots \mathcal{L}_{k_1}\mathcal{S}(t_1,0) \rho_0\right],
\end{align}
where $\mathcal{S}(t,t') = e^{\mathcal{L}(t-t')}$ is the time-evolution operator, and the result holds for $0<t_1<\cdots <t_M<t$. Eq.~\eqref{multi_time_correlation_function} is the natural generalisation of Eqs.~\eqref{average_current_vector} and~\eqref{two_point_multiple} to arbitrary $M$-point functions, and it shares the same intuitive mathematical structure. For diffusive currents, the same expression holds but with the replacement $\mathcal{L}_k\to\mathcal{H}_k$~\cite{Tilloy2018}. This result can also be derived through more elementary methods, by generalising the argument of Appendix~\ref{app:hom} to multiple distinct time points~\cite{Hagele_2018}.}

\rev{When two or more time arguments coincide, however, the correlation function is singular, e.g.~the delta-function singularity of the two-point function, $F(t_1,t_2) \sim K\delta(t_1-t_2)$ [c.f.~Eq.~\eqref{F_jump}]. To regulate these divergences systematically, one can consider the smoothed currents
\begin{equation}
\label{filtered_current}
    I_k[f] = \int_{0}^t dt'\, f(t-t') I_k(t').
\end{equation}
Here, the filter function $f(t)$ could represent, for example, a lowpass filter that removes high-frequency components of the white noise [see Appendix~\ref{sec:Filtering basics}]. Such filtering emerges in any real measurement due to the finite response time of the measuring apparatus, and is therefore quite natural from an experimental as well as from a mathematical point of view. }

\rev{Smoothed correlation functions can be found from partial differentiation of a modified generating function, as~\cite{Tilloy2018}
\begin{align}
    E\left[I_{k_1}[f_1] \cdots I_{k_M}[f_M]\right] = \left.\frac{\partial}{\partial \epsilon_1} \cdots \frac{\partial}{\partial \epsilon_M} \mathcal{Z}_{\bm{\epsilon}}[\{f_m(t)\}]\right\vert_{\bm{\epsilon}=0},
\end{align}
where 
\begin{equation}
    \mathcal{Z}_{\bm{\epsilon}}[\{f_m(t)\}] = E\left[ \exp\left(\sum_{m=1}^M \epsilon_m I_{k_m} [f_m]\right)\right].
\end{equation}
This is nothing but a special case of the generating functional in Eq.~\eqref{generating_functional_def}, obtained by choosing the source functions 
\begin{equation}
    \label{special_source_functions}
    \chi_k(t) = \sum_{m=1}^M \epsilon_m f_m(t) \delta_{k,k_m}.
\end{equation}
Therefore, an explicit expression for $\mathcal{Z}_{\bm{\epsilon}}[\{f_m(t)\}]$ is obtained by substituting Eq.~\eqref{special_source_functions} into Eq.~\eqref{tilted_Liouvillian_higher-order} (or Eq.~\eqref{tilted_Liouvillian_higher-order_diffusive} in the diffusive case). } 

\rev{To clarify how this procedure works in practice, consider the following example. Suppose we are interested in the smoothed currents
\begin{equation}
    \tilde{I}_k(t) = \int_0^t dt'\, \Lambda e^{-\Lambda(t-t')} I_k(t'),
\end{equation}
where $\Lambda$ is the bandwidth of a lowpass filter, modeled here by a simple exponential response in the time domain. We wish to compute the correlation functions
\begin{align}
        \label{smoothed_corr}
    \tilde{F}_{k_{1}k_2}(t_1,t_2) = E[\tilde{I}_{k_1}(t_1)\tilde{I}_{k_2}(t_2)]-E[\tilde{I}_{k_1}(t_1)]E[\tilde{I}_{k_2}(t_2)].
\end{align}
This can be achieved by introducing the filter functions
\begin{equation}
\label{filter_function_example}
    f_m(t) = \Theta(t_m-t)  \Lambda  e^{-\Lambda(t_m-t)} ,
\end{equation}
where $\Theta(t)$ is the unit step function, so that $\tilde{I}_k(t_m) = I_k[f_m]$ in the notation of Eq.~\eqref{filtered_current}. The modified generator entering Eq.~\eqref{generating_functional_solution} becomes
\begin{equation}
\label{L_zeta_example}
    \mathcal{L}_{\bm{\chi}}(t) = \begin{cases}\displaystyle 
        \mathcal{L} + \sum_{m=1}^2 \left( e^{\epsilon_m f_m(t)}-1\right) \mathcal{L}_{k_m} & (k_1\neq k_2) \\
        \displaystyle \mathcal{L} + \left( e^{\epsilon_1 f_1(t) + \epsilon_2 f_2(t)}-1\right) \mathcal{L}_{k_1} & (k_1 = k_2),
    \end{cases}
\end{equation}
All other $\chi_k(t)$ have been set directly to zero, since we are only interested in correlation functions for the currents $\tilde{I}_{k_1}$ and $\tilde{I}_{k_2}$. Assuming for simplicity that $\rho_0 = \rhoss$ is the steady state, we find the average current~\footnote{\rev{To find derivatives of time-ordered exponentials such as $$\mathcal{S}_{\bm{\chi}}(t,0) = \mathcal{T}\exp \left\{\int_0^t dt' \mathcal{L}_{\bm{\chi}}(t')\right\},$$ it can be helpful to first discretize the exponential into a time-ordered product as
$$\mathcal{S}_{\bm{\chi}}(t,0) = \lim_{L\to \infty} e^{\mathcal{L}_L\Delta t}e^{\mathcal{L}_{L-1}\Delta t} \cdots e^{\mathcal{L}_1\Delta t},$$
where $\mathcal{L}_l = \mathcal{L}_{\bm{\chi}}(l\Delta t)$ and $\Delta t = t/L$. Differentiating with respect to some parameter $\epsilon$ then yields
\begin{align*}
\frac{\partial \mathcal{S}_{\bm{\chi}}}{\partial\epsilon} & = \lim_{L\to\infty}\sum_{l=1}^L \Delta t\, e^{\mathcal{L}_L\Delta t} \cdots  e^{\mathcal{L}_{l+1}\Delta t} \frac{\partial \mathcal{L}_l}{\partial\epsilon}e^{\mathcal{L}_l\Delta t} \cdots e^{\mathcal{L}_1\Delta t} \\
& = \int_0^t dt'\,\mathcal{S}_{\bm\chi}(t,t')\frac{\partial \mathcal{L}_{\bm{\chi}}(t')}{\partial\epsilon}\mathcal{S}_{\bm\chi}(t',0),
\end{align*}
so that we recover an expression such as the second line of Eq.~\eqref{average_smoothed_example}}}
\begin{align}
    \label{average_smoothed_example}
    E[\tilde{I}_{k_1}(t_1)] & = \left.\frac{\partial  \mathcal{Z}_{\bm{\epsilon}}}{\partial \epsilon_1}\right|_{\bm{\epsilon}=0} \notag \\
    & = \int_0^t dt'\,\tr \left[ \mathcal{S}(t,t') f_1(t')\mathcal{L}_{k_1} \mathcal{S}(t',0) \rhoss \right]\notag \\
    & = \int_0^{t} dt'\, f_1(t') \tr\left[\mathcal{L}_{k_1} \rhoss \right].
\end{align}
Clearly, this is consistent with Eq.~\eqref{average_current_vector}, and for $\Lambda t_1 \gg 1$ we obtain simply $ E[\tilde{I}_{k_1}(t_1)] \approx J_{k_1}$. Similarly, the steady-state two-point function is given by
\begin{align}
    \label{corr_smoothed_example}
    & E[\tilde{I}_{k_1}(t_1)\tilde{I}_{k_2}(t_2)] = \left.\frac{\partial^2  \mathcal{Z}_{\bm{\epsilon}}}{\partial \epsilon_1\partial \epsilon_2}\right|_{\bm{\epsilon}=0} \notag \\
    & = \delta_{k_1k_2} \Delta(t_1-t_2) J_{k_1}\notag \\
    & \quad + \int_0^t dt'\int_0^{t'} dt'' f_1(t') f_2(t'')\tr \left[ \mathcal{L}_{k_1} \mathcal{S}(t',t'') \mathcal{L}_{k_2} \rhoss \right]\notag \\
    & \quad + \int_0^t dt'\int_0^{t'} dt'' f_2(t') f_1(t'')\tr \left[ \mathcal{L}_{k_2} \mathcal{S}(t',t'') \mathcal{L}_{k_1} \rhoss \right],
\end{align}
where
\begin{equation}
    \Delta(t_1-t_2) = \int_0^t dt' f_1(t')f_2(t') \approx \frac{\Lambda}{2} e^{-\Lambda|t_1-t_2|},
\end{equation}
where the final approximation holds for $\Lambda(t_1+t_2)\gg 1$.

These results are, of course, fully consistent with Eq.~\eqref{two_point_multiple}, and they become exactly equivalent in the limit of infinite detector bandwidth, $\Lambda\to \infty$. In fact, we could have derived Eq.~\eqref{corr_smoothed_example} directly by convolution of Eq.~\eqref{two_point_multiple} with the filter functions~\eqref{filter_function_example}. However, the present approach allows systematic computation of higher-order cumulants as well, by appropriately modifying the source fields entering the generator $\mathcal{L}_{\bm{\chi}}(t)$. While a general expression for such cumulants can be formally written down~\cite{Tilloy2018}, it is typically easier to find the specific correlation function of interest by constructing the generating function and computing its derivatives explicitly. Finally, we note that the generating functional approach can be extended to account for dark counts (where jumps are erroneously registered by the detector), finite detection efficiencies, and mixed detection schemes with both quantum-jump and diffusive output currents; see Ref.~\cite{Guilmin2023} for details.
}

\subsection{Quantum sensing and metrology}
\label{sec:metrology}

Continuous quantum measurements offer the ideal platform for sensing and metrology. 
The output currents are associated to direct clicks in a detector and therefore form a classical time-series of random outcomes. 
Using methods such as maximum likelihood, one can use these outcomes to estimate parameters \rev{of the model; i.e., which enter in} the system Hamiltonian or the jump operators. 
\rev{In other words, we can use a continuously measured system} as a \emph{sensor}.
Quantum sensing and metrology is now a broad and fertile field of research~\cite{Paris2008,degen_2017}. 
In this section we focus on  aspects of it which pertain specifically to continuously measured systems. 
We will assume the reader has some knowledge of quantum metrology, in particular the concept of quantum Fisher information (QFI). 
For a concise introduction to the subject, see Ref.~\cite{Paris2008}.

In the standard paradigm of quantum metrology, a system is prepared in some state $\rho_\theta$, which depends on a certain unknown parameter $\theta$.
The standard procedure is then to measure $\rho_\theta$, obtaining a random outcome $X$. 
Afterwards, the system is reset, i.e., prepared again in $\rho_\theta$, and the protocol is repeated. 
The theory of quantum metrology aims to determine: (i) what kind of measurement is able to extract the most information about $\theta$, and (ii) how quantum resources such as entanglement and squeezing can provide a boost in sensitivity compared to the classical case.

The continuous measurement metrology paradigm is slightly different. 
The state of the system is never reset. Instead, it is continuously measured in the steady state, leading to a series of outcomes, which are now correlated with each other (as explored extensively in this tutorial).
These outcomes are, for instance, the random increments $dN_k$ obtained at each time step of a quantum jump trajectory (Sec.~\ref{sec:output_currents}). 
For simplicity, we will denote them more generally as $X_1,X_2,\ldots$. 
Notwithstanding this fundamental difference in paradigm, the two basic questions remain the same. For example, depending on the system and the parameter being estimated, different unravellings may lead to more or less precision.

Techniques for doing metrology with quantum continuous measurements have been studied in~\cite{Gammelmark_2013,Kiilerich_2014,Kiilerich_2016,Gammelmark_2014,Albarelli2018} \rev{and were recently applied to estimate temperature in a model akin to Example A with $\Omega=0$ \cite{Boeyens_2023}.}
These references have developed tools for computing e.g. the maximum likelihood estimators directly from quantum trajectories. 
In addition, they also discuss how to obtain the corresponding Fisher information (FI). 
Given a certain trajectory, with data points $X_1, X_2, \ldots$ and corresponding probability distribution $P(x_1,x_2,\ldots)$, the FI is defined as 
\begin{equation}
    \mathcal{F}(\theta) = \sum_{x_1,x_2,\ldots} P(x_1,x_2,\ldots) \Bigg(\frac{\partial}{\partial\theta} \ln P(x_1,x_2,\ldots)\Bigg)^2.
\end{equation}
The usefulness of the FI lies in the Cram\'er-Rao bound (CRB): let $Q(X_1,X_2,\ldots)$ denote any function of the data. 
It then follows that 
\begin{equation}\label{CRB}
    {\rm Var}\big(Q\big) \geqslant \frac{\big[ 
        \partial_\theta E(Q)
    \big]^2}{\mathcal{F}(\theta)}.
\end{equation}
The FI therefore bounds the fluctuations of any quantity. 
For example, suppose we choose $Q(X_1,X_2,\ldots)$ so 
that $E(Q) = \theta$, is the very parameter we are trying to estimate
(this is called an \emph{unbiased estimator}).
In this case the CRB~\eqref{CRB} reduces to ${\rm Var}(Q) \geqslant 1/\mathcal{F}(\theta)$; the FI therefore imposes a constraint on the error of the unbiased estimator. 
More generally, if the estimator is biased and we define the bias as $b(\theta) = E(Q) - \theta$, then Eq.~\eqref{CRB} yields ${\rm Var}(Q) \geqslant (1+\partial_\theta b)^2/\mathcal{F}(\theta)$, which is called the CRB for biased estimators. 

For any unravelling there will be an associated FI, since the shape of the data and its distribution will be different in each case. 
Computing the FI for specific unravellings can seldom be done analytically, and techniques for obtaining it numerically were developed in Refs.~\cite{Gammelmark_2013,Albarelli2018}.
The quantum Fisher information, on the other hand, represents the maximization of the FI over all possible unravellings. 
And, interestingly, this can be written down in a compact and elegant way, as first discovered in~\cite{Gammelmark_2014}.
The proof, which we defer to the original paper, uses methods very similar to those developed in Sec.~\ref{sec: Cumulant generating function} on FCS, as well as the perturbation theory methods used in Sec.~\ref{sec:cumulants_recursive}.  
Consider a general Lindblad equation of the form~\eqref{M}. 
Assume that both $H$ and the $L_k$ can depend on a certain parameter $\theta$. 
Then the QFI associated to measuring for a sufficiently large time $\tau$ is given asymptotically by~\cite{Gammelmark_2014} 
\begin{equation}\label{Metrology_QFI}
\begin{aligned} 
    \mathcal{F}(\theta) = 4 \tau \Bigg( \sum_k\langle (\partial_\theta L_k)^\dagger (\partial_\theta L_k)\rangle - &\idV  \mathcal{L}_L \mathcal{L}^+ \mathcal{L}_R \rhossV 
    \\
    &- \idV \mathcal{L}_R \mathcal{L}^+ \mathcal{L}_L \rhossV\Bigg).
\end{aligned}
\end{equation}
The first term is an expectation value in the steady state. 
In the other two, $\mathcal{L}^+$ is the Drazin inverse [Eq.~\eqref{Drazin_integral_def}], while
$\mathcal{L}_{L(R)}$ are superoperators defined as 
\begin{align}
\mathcal{L}_L \rho &= -i (\partial_\theta \Hnh) \rho + \sum\limits_k (\partial_\theta L_k) \rho L_k^\dagger,
\\[0.3cm]
\mathcal{L}_R \rho &= i \rho(\partial_\theta \Hnh^\dagger)+ \sum\limits_k L_k \rho (\partial_\theta L_k^\dagger),
\end{align}
where $\Hnh = H - \frac{i}{2} \sum_k L_k^\dagger L_k$ [Eq.~\eqref{H_non_hermitian}].
Thus, to compute the QFI we only need to have access to the Drazin inverse. 
This is similar in complexity to computing the noise [as per Eq.~\eqref{eigen_D}].

Notice how Eq.~\eqref{Metrology_QFI} depends linearly on the total measured time $\tau$. 
This means that the mean-squared error in the estimation is going to decrease as $1/\tau$. 
The longer we measure, the higher is the precision with which we can estimate the parameter. 
The origin of this linear dependence is related to the fact that, even though the outcomes of a continuous measurement are correlated, this correlation still decays with time.
That is, a click in a given time will be correlated with clicks at past times, but the correlation between clicks becomes negligible if they are very far apart. 
This was recently proven in  Ref.~\cite{Radaelli_2023}. 

Equation~\eqref{Metrology_QFI} often simplifies considerably, since the parameter $\theta$ is usually only encoded in either the $L_k$ or $H$. 
For example, if it is encoded in $H$ only, it reduces to 
\begin{equation}\label{Metrology_QFI_H}
    \mathcal{F}(\theta) = - 4 \tau~ \tr\Big[ (\partial_\theta H) \mathcal{L}^+ \Big(\big\{ \rhoss, \partial_\theta H \big\}\Big)\Big],
\end{equation}
which we now wrote in non-vectorized notation for clarity. 

In general, we interpret $\mathcal{F}(\theta)$ as the amount of information about $\theta$ contained in the continuous measurement data.
Equivalently, $\mathcal{F}(\theta)$ is the sensitivity of the statistics of the data to small changes in the parameter. 
The QFI corresponds to a maximization of this information over all possible unravelings. 
Thus, in general, it is not readily accessible experimentally, as we never have access to all possible unravellings. 
Notwithstanding, it provides a crucial upper bound: no unravelling can do any better.

\subsubsection{QFI for Example A}

To illustrate the idea, consider Example A. For simplicity we take $\bar{N} = \Delta = 0$, so we can use the Drazin inverse in Eq.~\eqref{ExampleA_Drazin_inverse}. 
The Hamiltonian is then simply $H = \Omega \sigma_x$. First,  suppose the parameter we wish to measure is the Rabi frequency $\Omega$ in Eq.~\eqref{ExampleA_H}. Then Eq.~\eqref{Metrology_QFI_H} applies and we get the remarkably simple result
\begin{equation}
    \mathcal{F}(\Omega) = \frac{16 \tau}{\gamma}.
\end{equation}
The QFI in this case is independent of $\Omega$, and inversely proportional to the bath coupling strength. 
This happens because large couplings tend to reduce the Rabi oscillations. 

Next suppose we wish to estimate $\gamma$. In this case the only dependence is on the jump operator $L = \sqrt{\gamma} \sigma_-$.
We then find 
\begin{equation}
    \mathcal{F}(\gamma) = \frac{4 \tau \Omega^2}{\gamma^3 + 8 \gamma \Omega^2}.
\end{equation}
As a sanity check,  this vanishes for zero Rabi drive ($\Omega = 0$):  no excitations are created in the system, and hence there is no current, even at the stochastic level. 

\rev{
\subsection{Fluctuation Theorems, the Fluctuation-Dissipation Theorem, and the Onsager Reciprocal Relations}
\label{sec:FTs}
}
\rev{\subsubsection{Fluctuation Theorems}}
\rev{
Fluctuation theorems are powerful relations that hold arbitrarily far from equilibrium and relate the probabilities for a system trajectory, $X$, to its time-reverse, $X_{\rm tr}$ \cite{Jarzynski_2011,seifert_2012,campisi_2011,Esposito_2009}. They take the general form
\begin{equation}
    \label{eq:fluctuation_theorem}
    \frac{P(X_{\rm tr})}{P(X)} = e^{-\sigma(X)},
\end{equation}
where $\sigma(X)$ denotes the entropy production along the trajectory $X$. The fluctuation theorem thus states that if a trajectory produces a sizable amount of entropy ($\sigma \gg 1$), it is exponentially unlikely for its time-reverse to occur. This explains why in macroscopic systems (where entropy production becomes large), we only observe processes with positive entropy production. 
}

\rev{
From Eq.~\eqref{eq:fluctuation_theorem} a so-called integral fluctuation theorem may be derived by multiplying with $P(X)$ and summing over all possible trajectories \cite{jarzynski_1997}
\begin{equation}
    \label{eq:integral_ft}
    \langle e^{-\sigma(X)} \rangle  = 1,
\end{equation}
where the average is over the distribution $P(X)$.
Using Jensen's inequality, it can be shown that Eq.~\eqref{eq:integral_ft} implies the second law of thermodynamics
\begin{equation}
    \label{eq:secondlaw}
    \langle \sigma(X) \rangle \geq 0.
\end{equation}
While negative entropy production is a possibility, the average value of $\sigma$ always has to be nonnegative. The fluctuation theorem may thus be seen as a generalization of the second law of thermodynamics to systems where fluctuations matter.
}

\rev{
In the setting of this tutorial, we may derive a fluctuation theorem from the FCS \cite{harbola_2007a,harbola_2007,saito_2008,Andrieux_2009}.
We consider a scenario where the system is connected to multiple reservoirs, which we label $\alpha$ (see, for instance, Example B and Eq.~\eqref{ExampleB_M}). 
This means that we can define a one-to-one function $\alpha(k)$, which associates each jump operator $L_{k}$ to a single reservoir $\alpha$. 
The trajectory $X$ in the fluctuation theorem will then be the net charges $\{N_\alpha\}$ flowing to each reservoir, which we can build using Eq.~\eqref{multiple_charge}, for multiple current specimens, but with weights $\nu_{\alpha k}$ which are such that a given $k$ only appears in one charge $N_\alpha$. 
}

\rev{Not all master equations will satisfy a fluctuation theorem; to do so, there is a fundamental condition which they must satisfy, called local detailed balance. 
It consists of the assumption that the jump operators always come in pairs: for each $L_k$, there always exists another jump operator $L_{k'}$ belonging to the same bath ($\alpha(k') = \alpha(k)$), such that
\begin{equation}
    \label{eq:local_db}
    L_{k} = L_{k'}^\dagger e^{\nu_{\alpha k}\sigma_{\alpha}/2}.
\end{equation}
In this and what follows, we will leave the dependence $\alpha = \alpha(k)$ implicit. 
The $\nu_{\alpha k}$ are the weight factors in $N_\alpha(t)$ and, for the formula to be consistent, they must satisfy $\nu_{\alpha k'} = - \nu_{\alpha k}$. 
The quantities $\sigma_\alpha$ appearing in Eq.~\eqref{eq:local_db} represent the  entropy production associated to the pair of jumps $k,k'$.
In the most common case, if $L_k$ exchanges an energy $\varepsilon_k$ with its bath, then $L_{k'}$ exchanges $\varepsilon_{k'} = -\varepsilon_k$.
The associated entropy production is $\sigma_\alpha = \varepsilon_k/T_\alpha$, so that the entropy produced by $L_k$ is $\nu_{\alpha k} \sigma_\alpha$, while that produced by $L_{k'}$ is $-\nu_{\alpha k} \sigma_\alpha$. 

}

\rev{
Here we focus on the case where both the Hamiltonian as well as the jump operators obey time-reversal symmetry
\begin{equation}
    \label{eq:timerev}
    \Theta  H\Theta^ {-1} = H,\hspace{1.5cm}\Theta  L_k\Theta^ {-1} =L_k,
\end{equation}
where $\Theta$ denotes the anti-unitary time-reversal operator \cite{sakurai:book,strasberg:book} which obeys $\Theta i = -i\Theta $ and $\Theta^2=\pm1$. For $\Theta^2=1$, which is the case for particles with integer spin (or no spin), Eq.~\eqref{eq:timerev} implies that we may construct a basis where $H$ and $L_k$ are real-valued matrices. In the following, we focus on this case for simplicity. The final results also hold for $\Theta^2=-1$, as shown in Appendix \ref{app:ftthetam1}.
}

\rev{
In vectorized notation, the tilted Liouvillian in Eq~\eqref{multiple_tilted_liouvillian} may be written as [c.f.~Eq.~\eqref{vec_vectorized_L}]
\begin{IEEEeqnarray}{rCl}
\label{vec_vectorized_Ltilt}
\mathcal{L}_{\bm \chi}  &=& -i (\id\otimes H - H\trans \otimes \id) 
\\[0.2cm]&&
+ \sum\limits_{k=1}^\nops  \Bigg[e^{i\nu_{\alpha k}\chi_{\alpha}}L_k^* \otimes L_k - \frac{1}{2} \id \otimes L_k^\dagger L_k - \frac{1}{2} (L_k^\dagger L_k)\trans \otimes \id\Bigg]\,,
\nonumber
\end{IEEEeqnarray}
where there is a single counting field $\chi_\alpha$ associated to each current $N_\alpha$, and we also used the fact that each $k$ appears only in a single  $\alpha$ to write $\bm{\nu}_k \cdot \bm{\chi} = \nu_{\alpha k} \chi_\alpha$ [c.f.~Eq~\eqref{multiple_tilted_liouvillian}].
For real-valued $H$ and $L_k$, the local detailed-balance condition in Eq.~\eqref{eq:local_db} implies
\begin{equation}
    \label{eq:transposejumps}
    e^{i\nu_{\alpha k}(-\chi_{\alpha}+i\sigma_\alpha)}L_k \otimes L_k = e^{i\nu_{\alpha k'}\chi_{\alpha}}L_{k'}\trans \otimes L_{k'}\trans,
\end{equation}
which in turn implies the symmetry (using $H=H\trans$) \cite{segal_2023}
\begin{equation}
    \mathcal{L}_{-\bm \chi+i\bm \sigma} = \mathcal{L}_{\bm \chi}\trans,
\end{equation}
where $\bm \sigma$ is a vector with entries $\sigma_\alpha$. Because transposition of a matrix does not affect its eigenvalues, the SCGF in Eq.~\eqref{eq:scgfb} (which is the eigenvalue of the tilted Liouvillian with the largest real part) exhibits a similar symmetry
\begin{equation}
    \label{eq:symmscgf}
    C(\bm \chi) = C(-\bm \chi+i\bm \sigma).
\end{equation}
For the long-time probability distribution, this implies via Eq.~\eqref{SCGF_reconstructing_P}
\begin{equation}
    \label{eq:ftfromliou}
    \frac{P(-\bm n, t)}{P(\bm n, t)}=e^{-\bm n \cdot \bm \sigma}.
\end{equation}
When $n_\alpha$ count excitations exchanged with the environment, Eq.~\eqref{eq:ftfromliou} is known as an exchange fluctuation theorem \cite{jarzynski_2004}.
From Eq.~\eqref{eq:ftfromliou}, an integral fluctuation theorem may be derived in analogy to Eqs.~\eqref{eq:integral_ft} and \eqref{eq:secondlaw}
\begin{equation}
    \label{eq:intftfromliou}
    \langle e^{-\bm n \cdot \bm \sigma}\rangle = 1\hspace{.5cm}\Rightarrow\hspace{.5cm} \langle \bm n\rangle \cdot \bm\sigma\geq 0,
\end{equation}
where the average is over the distribution $P(\bm n, t)$ and the inequality is obtained using Jensen's inequality.
We note that we only used Eqs.~\eqref{eq:local_db} and \eqref{eq:timerev} to derive the fluctuation theorem without assuming anything on $\sigma_\alpha$. Equation \eqref{eq:ftfromliou} may thus be applicable to systems describing non-thermal reservoirs.

We note that for broken time-reversal symmetry, similar fluctuation theorems may be derived \cite{manzano_2018,soret_2022}. In that case, the probability distribution in the denominator of Eq.~\eqref{eq:ftfromliou} has to be replaced with the distribution describing the time-reversed scenario.
}

\rev{
\subsubsection{Fluctuation Theorem for Example B}
\label{sec:ftforB}
Consider Example B, a quantum dot coupled to two electronic reservoirs [cf.~Sec.~\ref{ssec:exampleB}], where we count the electrons entering and leaving to each reservoir independently. With the master equation in Eq.~\eqref{ExampleB_M}, this corresponds to the weights $\nu_\pm^\alpha=\pm 1$. Since the rates in the master equation obey
\begin{equation}
    \label{eq:ratesdb}
    \frac{\gamma_+^\alpha}{\gamma_-^\alpha} = e^{-\beta_\alpha(\omega-\mu_\alpha)},
\end{equation}
the relation in Eq.~\eqref{eq:local_db} is fulfilled, with $\sigma_\alpha = -\beta_\alpha(\omega-\mu_\alpha)$. The quantity $\sigma_\alpha$ gives the entropy production associated to an electron entering the quantum dot from reservoir $\alpha$. The fluctuation theorem in Eq.~\eqref{eq:ftfromliou} then reduces to
\begin{equation}
    \label{eq:ftnlnr}
    P(-n_L,-n_R,t) = P(n_L,n_R,t)e^{\sum_{\alpha=L,R}n_\alpha\beta_\alpha(\omega-\mu_\alpha)}. 
\end{equation}
This relation may further be simplified by using the fact that in the long-time limit, $n_L\simeq -n_R$ due to current conservation (i.e., the current entering from the left reservoir matches the current leaving to the right reservoir). Formally, this can be shown by introducing the unitary superoperator
\begin{equation}
    \label{eq:unitsup}
    \mathcal{U}(\bullet) = e^{i\frac{\chi_R}{2}c^\dagger c}(\bullet) e^{i\frac{\chi_R}{2}c^\dagger c},\hspace{.5cm}\mathcal{U}^\dagger(\bullet) = e^{-i\frac{\chi_R}{2}c^\dagger c}(\bullet) e^{-i\frac{\chi_R}{2}c^\dagger c}.
\end{equation}
One may then show that
\begin{equation}
    \label{eq:liouunit}
    \mathcal{U}^\dagger \mathcal{L}_{\chi_L,\chi_R}\mathcal{U} = \mathcal{L}_{\chi_L-\chi_R,0}.
\end{equation}
Since a unitary transformation does not change the eigenvalues of a matrix, the SCGF has the symmetry
\begin{equation}
    \label{eq:scgfsymlt}
    C(\chi_L,\chi_R) = C(\chi_L-\chi_R,0).
\end{equation}
It is straightforward to show that, for the probability distribution, this implies
\begin{equation}
    \label{eq:probsymlt}
    P(n_L,n_R,t)= P(n_L)\delta_{n_L,-n_R}.
\end{equation}
With this relation, the fluctuation theorem in Eq.~\eqref{eq:ftnlnr} reduces to the more familiar form \cite{tobiska_2005,harbola_2007a,Andrieux_2006}
\begin{equation}
    \label{eq:ftexb}
    \frac{P(-n,t)}{P(n,t)} = e^{n[\beta_L(\omega-\mu_L)-\beta_R(\omega-\mu_R)]}\xrightarrow[\beta_L=\beta_R]{}e^{-n\beta eV},
\end{equation}
where $n$ denotes the number of electrons that passed from the left to the right reservoir (which may be measured at either reservoir) and we introduced the bias voltage $eV=\mu_L-\mu_R$. The fluctuation theorem in Eq.~\eqref{eq:ftexb} may also be shown to hold for the explicit probability distribution presented in Eq.~\eqref{pnbipoisson} (where $\gamma_R\ll\gamma_L$) by using $I_n(z)=I_{-n}(z)$ and the relation
\begin{equation}
    \label{eq:fermirel}
    \frac{f_\alpha}{1-f_\alpha} = e^{-\beta_\alpha(\omega-\mu_\alpha)}.
\end{equation}
}

\rev{
\subsubsection{Fluctuation-Dissipation Theorem}
The fluctuation-dissipation theorem relates fluctuations in thermal equilibrium to linear-response coefficients \cite{callen_1951,kubo_1966}. A prominent example is the thermal noise in a two-terminal electric conductor which reads \cite{blanter_2001}
\begin{equation}
    \label{eq:fdt_el}
    S(\omega) = \omega\coth\left(\frac{\omega}{2T}\right)G(\omega)\hspace{.2cm}\xrightarrow{\omega\rightarrow 0}\hspace{.2cm} S(0) = 2TG(0),
\end{equation}
where $G(\omega)$ denotes the frequency-dependent ac-conductance \footnote{Note that for the electrical noise, often a different convention is used, where $S(0)=2D$, resulting in the fluctuation-dissipation theorem $S(0)=4T G$, see for instance Sec.~1.4.2 in Ref.~\cite{nazarov_book}.}. The power spectrum in equilibrium thus contains information on the linear response of the system. More generally, if a continuity equation relates the measured currents to an observable in the system, a fluctuation-dissipation theorem may be exploited to access linear-response coefficients. For an example with energy currents, see Ref.~\cite{brange_2019}.
While the fluctuation-dissipation theorem traditionally describes linear response around thermal equlibrium, it has been extended to non-equilibrium steady states \cite{prost_2009,seifert_2010}. In the following, we will focus on the zero-frequency limit of the fluctuation-dissipation theorem (see Ref.~\cite{Marcos_2010} for a finite-frequency approach to FCS).
}

\rev{
From the fluctuation theorem, a fluctuation-dissipation theorem may be derived \cite{Esposito_2009}. To this end, we introduce the notion of equilibrium. In equilibrium, we expect any process to be equally likely to its time-reverse. This implies $P(\bm n,t) = P(-\bm n,t)$. Inspecting Eq.~\eqref{eq:ftfromliou}, this happens only if $\bm\sigma \cdot\bm n =0$, which is certainly not true for any values of $n_\alpha$. However, we note that Eq.~\eqref{eq:ftfromliou} only holds in the long-time limit, where conservation laws generally impose strong conditions on the probability distribution as exemplified in Eq.~\eqref{eq:probsymlt}. Here we assume that these conservation laws impose the condition
\begin{equation}
    \label{eq:problteq}
    P(\bm n,t) \propto \delta_{\bm \sigma_{\rm eq}\cdot\bm n,0},
\end{equation}
where $\bm \sigma_{\rm eq}$ denotes the value $\bm\sigma$ takes in equilibrium, which usually is of the form $\sigma_\alpha = \beta(\varepsilon_\alpha-\mu)$, determined by a single temperature and chemical potential for all reservoirs. In this case, the Kronecker delta in Eq.~\eqref{eq:problteq} ensures energy conservation, i.e., $\sum_\alpha n_\alpha \varepsilon_\alpha = 0 $. It also ensures that no entropy is produced in equilibrium, which allows us to effectively set $\bm \sigma_{\rm eq}\cdot \bm n=0$. 
Since this is zero, adding $\bm \sigma_{\rm eq}\cdot \bm n$ to the exponent of Eq.~\eqref{eq:ftfromliou} therefore allows us to recast the fluctuation theorem  as
\begin{equation}
    \label{eq:ftfromliouds}
    \frac{P(-\bm n, t)}{P(\bm n, t)}=e^{-\bm n \cdot  \bm\delta}\hspace{.5cm}\Rightarrow\hspace{.5cm}\langle e^{-\bm n \cdot \bm\delta}\rangle = 1.
\end{equation}
where $ \bm\delta = \bm \sigma-\bm \sigma_{\rm eq}$.
}

\rev{
Note that the quantity appearing in the integral fluctuation theorem in Eq.~\eqref{eq:ftfromliouds} has the form of a moment generating function [c.f.~Eq.~\eqref{FCS_M}]. Its logarithm may thus be expanded in the cumulants of $\bm n$ as \cite{jarzynski_1997}
\begin{equation}
    \label{eq:cumexp0}
    0 = \ln \langle e^{-\bm n\cdot \bm\delta}\rangle = \sum_{j=1}^\infty \frac{(-1)^j}{j!}\llangle (\bm\delta \cdot \bm n)^j\rrangle,
\end{equation}
where the first equality directly follows from the integral fluctuation theorem. 
Taking the derivative with respect to $\delta_\alpha$ and $\delta_\beta$ and subsequently setting $\bm\delta=0$ then results in
\begin{equation}
    \label{eq:fdisstfromliou}
    \partial_{\delta_\alpha}\llangle n_\beta\rrangle\big|_{\bm \delta =0}+ \partial_{\delta_\beta}\llangle n_\alpha\rrangle\big|_{\bm \delta =0}=\llangle n_\alpha n_\beta\rrangle\big|_{\bm \delta =0}.
\end{equation}
Taking a time-derivative as well as the long-time limit, this relation reduces to the fluctuation-dissipation relation (c.f.~Sec.~\ref{sec:multiple_currents})
\begin{equation}
    \label{eq:fdisstfromliou2}
    \begin{aligned}
   D_{\alpha\beta}\big|_{\bm \delta =0} &= \partial_{\delta_\alpha}J_\beta\big|_{\bm \delta =0}+ \partial_{\delta_\beta}J_\alpha\big|_{\bm \delta =0}\\&= 2\partial_{\delta_\alpha}J_\beta\big|_{\bm \delta =0},
   \end{aligned}
\end{equation}
where the last equation follows from the Onsager reciprocal relations in Eq.~\eqref{eq:onsagerrel} below. If we apply Eq.~\eqref{eq:fdisstfromliou} to Example B discussed in Sec.~\ref{sec:ftforB}, we recover the zero-frequency version of the fluctuation-dissipation theorem in Eq.~\eqref{eq:fdt_el}, with the conductance $G(0)=\partial_V J$ and voltage $eV=\mu_L-\mu_R$. For this result that involves the electrical current and noise, the weight factors should be chosen as $\nu_\pm = \pm e$, with $e$ denoting the charge of the electron.
}

\rev{
In analogy to Eq.~\eqref{eq:fdisstfromliou}, similar relations for non-linear response currents can be obtained by taking higher-order derivatives of Eq.~\eqref{eq:cumexp0} with respect to $\delta_\alpha$ \cite{tobiska_2005,Andrieux_2007,forster_2008}.
}

\rev{
\subsubsection{Onsager Reciprocal Relations}
We introduce the linear-response coefficients
\begin{equation}
    \label{eq:onsagerL}
    L_{\alpha\beta} = \partial_{\delta_\beta}J_\alpha\big|_{\bm \delta =0}.
\end{equation}
These coefficients define a positive semi-definite matrix. This can be shown from the positivity of the entropy production which follows from Eq.~\eqref{eq:ftfromliouds}. In linear response, we may write
\begin{equation}
    \label{eq:onsagerLpos}
    \langle \bm n\cdot \bm\delta\rangle = \sum_{\alpha,\beta}\delta_\alpha L_{\alpha\beta}\delta_\beta \geq 0,
\end{equation}
which, since $\delta_\alpha$ may be chosen arbitrarily, implies that the matrix with entries $L_{\alpha\beta}$ is positive semi-definite. The Onsager reciprocal relations state that, for time-reversal symmetric systems, this matrix is also symmetric \cite{onsager_1931,casimir_1945}
\begin{equation}
    \label{eq:onsagerrel}
    L_{\alpha\beta} = L_{\beta\alpha}.
\end{equation}
This follows from the fluctuation theorem \cite{Esposito_2009} in Eq.~\eqref{eq:ftfromliouds} by considering the SCGF, which obeys the symmetry 
\begin{equation}
    \label{eq:scgfsymdel}
    C(\bm\chi,\bm\delta)=C(-\bm\chi+i\bm\delta,\bm\delta)
\end{equation}
where we treat the SCGF as a function of the two variables $\bm\chi$ and $\bm \delta$. In terms of the SCGF, the Onsager coefficients may be written as
\begin{equation}
    \label{eq:onsagerfromscgf}
    L_{\alpha\beta} = -i\partial_{\delta_\beta}\partial_{\chi_\alpha}C(\bm\chi,\bm\delta)\big|_{\bm\chi=\bm\delta=0}.
\end{equation}
Differentiating Eq.~\eqref{eq:scgfsymdel}, one may show that
\begin{equation}
    \label{eq:onsagerfromscgf2}
    \begin{aligned}
    \partial_{\delta_\beta}\partial_{\chi_\alpha}C(\bm\chi,\bm\delta)\big|_{\bm\chi=\bm\delta=0}&= -\frac{i}{2}\partial_{\chi_\beta}\partial_{\chi_\alpha}C(\bm\chi,\bm\delta)\big|_{\bm\chi=\bm\delta=0}\\&=\partial_{\delta_\alpha}\partial_{\chi_\beta}C(\bm\chi,\bm\delta)\big|_{\bm\chi=\bm\delta=0}.
    \end{aligned}
\end{equation}
From this equation, together with Eq.~\eqref{eq:onsagerfromscgf}, the Onsager reciprocal relations in Eq.~\eqref{eq:onsagerrel} follow immediately. We note that from these equations, also the fluctuation-dissipation theorem in Eq.~\eqref{eq:fdisstfromliou2} follow as $2L_{\alpha\beta}=D_{\alpha\beta}$. Furthermore, by taking higher order derivatives of Eq.~\eqref{eq:scgfsymdel}, higher order reciprocal relations may be derived \cite{Andrieux_2007}.
}

\rev{
\subsubsection{Fluctuation-Dissipation Theorem and Onsager Reciprocal Relation for Example B}
As in Sec.~\ref{sec:ftforB}, we consider Example B, where $\delta_\alpha = -\beta_\alpha(\omega-\mu_\alpha)-\sigma_{\rm eq}$. Here, the equilibrium condition in Eq.~\eqref{eq:problteq} is met as long as $\sigma_L=\sigma_R=\sigma_{\rm eq}$, where $\sigma_{\rm eq}$ may be chosen arbitrarily. Interestingly, this does not only happen in thermal equilibrium, where $\beta_L=\beta_R$ and $\mu_L=\mu_R$, but it may occur when the voltage bias exactly counteracts the thermal bias \cite{humphrey_2002}. The average currents in the steady state read [c.f.~Eq.~\eqref{ExampleB_Jss}] 
\begin{equation}
    J_R = -J_L = \frac{\gamma_L\gamma_R(f_R-f_L)}{\gamma_L+\gamma_R}.
\end{equation}
writing $f_\alpha = 1/(e^{-\sigma_\alpha}+1)$, we find $\partial_{\delta_\alpha}f_\alpha = f_\alpha(1-f_\alpha)$. The Onsager coefficients then read
\begin{equation}
    \label{eq:ExampleB_Onsager}
    L_{LL}=L_{RR} = -L_{LR}=-L_{RL} = \frac{\gamma_L\gamma_R}{\gamma_L+\gamma_R}f_{\rm eq}(1-f_{\rm eq}).
\end{equation}
The Onsager reciprocal relations thus hold as expected. Furthermore, in equilibrium the noise of Example B is [c.f.~Eq.~\eqref{ExampleB_D}]
\begin{equation}
    \label{eq:noiseexampleB2}
    D_{LL} = D_{RR} = -D_{LR} = -D_{RL} = 2\frac{\gamma_L\gamma_R}{\gamma_L+\gamma_R}f_{\rm eq}(1-f_{\rm eq}),
\end{equation}
obeying the fluctuation-dissipation theorem.
}

\subsection{Thermodynamic and Kinetic Uncertainty relations (TURs \& KURs)}
\label{sec:TURs}

One area of research that has received considerable attention in recent years relates the behaviour of quantum trajectories to the framework of stochastic thermodynamics. 
Given the unraveled description of stochastic quantum trajectories,  thermodynamic quantities such as heat, work, and entropy production also become random variables.
This was first studied in Refs.~\cite{Horowitz_quantum_2012, Hekking_quantum_2013}.
Since this first insight there has been an extensive body of research on using quantum trajectories to study quantum thermodynamics. The most recent progress is reviewed in Ref.~\cite{Manzano_2022}.

One of the most significant developments in the field of stochastic thermodynamics has been the discovery of thermodynamic uncertainty relations (TURs)~\cite{Barato_2015,Gingrich_2016}. 
TURs establish a trade-off between the noise in a certain current and the entropy production rate $\dot{\sigma}$.
Entropy production \rev{was introduced in Sec.~\ref{sec:FTs}. In short, it }is the fundamental concept behind the second law of thermodynamics,
quantifying the degree of irreversibility, or dissipation, of a process. 
According to the second law, the \rev{average} entropy production rate should always be non-negative $\langle \dot{\sigma}\rangle \geqslant 0$. For a review on entropy production, see Ref.~\cite{Landi2021_RMP_entropy_production}.
The original TUR, which holds \emph{only} for classical Markovian stochastic processes, reads~\cite{Barato_2015} 
\begin{equation}\label{TUR_original}
    \frac{D}{J^2} \geqslant \frac{2}{\langle \dot{\sigma}\rangle }.
\end{equation}
This bound was derived for classical master equations of the form described in Sec.~\ref{sec:classical_ME}. 
In that notation, the entropy production acquires the form~\cite{schnakenberg:1976}
\begin{equation}\label{TUR_entropy_production}
    \langle \dot{\sigma}  \rangle = \sum_{n\neq j} W_{nj} p_j^{\rm ss} \ln \big[ 
        W_{nj} p_j^{\rm ss}/W_{jn} p_n^{\rm ss}
    \big].
\end{equation}
The fact that the right-hand side of the TUR  is inversely proportional to $\langle \dot{\sigma} \rangle$ means that in order to decrease the fluctuations, one must increase the dissipation.

In Ref.~\cite{Di_Terlizzi_2018} the authors derived a different bound, termed the Kinetic Uncertainty Relation (KUR). The bound reads 
\begin{equation}\label{KUR_original}
    \frac{D}{J^2} \geqslant \frac{1}{K},
\end{equation}
where $K$ is the dynamical activity; i.e., the number of jumps per unit time. 
In the notation of Sec.~\ref{sec:classical_ME}, $K = \sum_{n,j} W_{nj} p_j^{\rm ss}$. 
The TUR and KUR are independent bounds, and no strict relation exists between them. 
In fact, in some regimes the TUR may be rather tight and the KUR loose, or the other way around. 
Bounds unifying the two were recently developed in Refs.~\cite{Vo_2022,VanVu2023}. 

The TUR and KUR only hold for classical stochastic processes, and they can be violated in quantum systems.
For example, an extensive analysis in a double quantum dot system, including a discussion on how to compute the noise $D$ using FCS, can be found in~\cite{Prech_2023}. 
These violations are due to quantum-coherence effects and therefore mark a fundamental difference between quantum and classical stochastic processes. 
The fact that they can be violated in the quantum regime has led to intensive research in trying to derive new bounds that would represent the quantum version of the TUR~\cite{Carollow_unravelling_2019, Hasegawa_quantum_2020, Hasegawa_thermodynamic_2021, Miller_thermodynamic_2021, Vu_thermodynamic_2022}.
The  derivations of these bounds make use of many of the concepts discussed in this tutorial, in particular FCS (Sec.~\ref{sec:FCS}), vectorization (Sec.~\ref{sec:vectorization}), and the quantum Fisher information (Sec.~\ref{sec:metrology}). 

For concreteness, we will focus here on the results of Ref.~\cite{Hasegawa_quantum_2020}, as they match in a particularly nice way to the tools developed in this tutorial. 
Their first main result was the following bound for the steady-state noise:
\begin{equation}\label{TUR_hasegawa_2020}
    \frac{D}{J^2}  \geqslant \frac{h_0'}{\mathcal{F}},
\end{equation}
where $h_0'$ is a numerical factor which is 1 for jump currents and 1/2 for diffusive currents (see below). Moreover, 
\begin{equation}\label{TUR_hasegawa_f}
    \mathcal{F} = K-4 \idV  \mathcal{L}_L \mathcal{L}^+ \mathcal{L}_R \rhossV
    - 4\idV \mathcal{L}_R \mathcal{L}^+ \mathcal{L}_L \rhossV ,
\end{equation}
where $K = \sum_k \langle L_k^\dagger L_k \rangle$ is the dynamical activity, while 
\begin{equation}
 \mathcal{L}_L \rho = -i \Hnh \rho + \frac{1}{2} \sum_k L_k\rho L_k^\dagger,
 \quad 
 \mathcal{L}_R \rho = i \rho \Hnh^\dagger + \frac{1}{2} \sum_k L_k\rho L_k^\dagger,
\end{equation}
(here $\Hnh$ is defined in Eq.~\eqref{H_non_hermitian}).
The derivation of this bound uses the quantum Fisher information discussed in Eq.~\eqref{Metrology_QFI}, and is based on the clever idea of performing an infinitesimal deformation of both the Hamiltonian and the jump operators to
\begin{equation}\label{TUR_hasegawa_2020_deformation}
    H_\theta = (1+\theta)H\,,\quad L_{k,\theta} =\sqrt{1+\theta}L_{k}\,.
\end{equation}
One then applies the Cram\'er-Rao bound~\eqref{CRB}, taking the function $Q$ to be the integrated charge $N(t)$ [Eq.~\eqref{current_charge}]. 
The bound is to be evaluated at $\theta = 0$, so that 
${\rm Var}_\theta(N(t))\Big|_{\theta = 0} = {\rm Var}(N(t)) = Dt$
[Eq.~\eqref{varN_D}].
Conversely, the average current changes because we are deforming all jump operators by $\sqrt{1+\theta}$.
This causes
$E_\theta(N(t)) = \int_0^t dt' E_\theta (I(t')) = h_\theta J t$, where $h_\theta$ depends on which current we are studying: For jump currents $J(t)$ is quadratic in $L_k$ and so $h_\theta = 1+\theta$, while for diffusive currents $J(t)$ is linear in $L_k$ and so $h_\theta = \sqrt{1+\theta}$. 
In either case, because of this simple scaling we then have
$\partial_\theta E_\theta(N(t))\Big|_{\theta=0}= h_0'Jt$.
Finally, we evaluate $F(\theta)$ at $\theta = 0$ using Eq.~\eqref{Metrology_QFI}. 
Combining everything then yields Eq.~\eqref{TUR_hasegawa_2020}.

\rev{The first term in Eq.~\eqref{TUR_hasegawa_f} is the same dynamical activity as in the KUR~\eqref{KUR_original}. The other two terms, on the other hand, represent quantum corrections. The quantity $f$ is a kind of quantum total activity, in the sense it also captures the activity associated to the unitary dynamics. Indeed, }
if we specialize Eq.~\eqref{TUR_hasegawa_2020} to the classical master equation case discussed in Sec.~\ref{sec:classical_ME} -- i.e. $H = 0$ and  jump operators 
$L_{nj} = \sqrt{W_{nj}} |n\rangle\langle j|$  -- 
In this case it turns out that the 2nd and 3rd terms in Eq.~\eqref{TUR_hasegawa_f} vanish and we recover $f=K$; that is, the quantum bound~\eqref{TUR_hasegawa_2020} recovers the KUR~\eqref{KUR_original}. 

Reference \cite{Hasegawa_quantum_2020} also showed how, in the classical case, the TUR~\eqref{TUR_original} can be derived from the same method. 
Unfortunately, this holds only for classical MEs. 
Instead of the deformation~\eqref{TUR_hasegawa_2020_deformation}, one now changes the rates $W_{nj}$ in the jump operators as
\begin{equation}
    W_{nj}^\theta = W_{nj} \Bigg[ 
        1 + \theta \left(1- \sqrt{\frac{W_{jn} p_n^{\rm ss}}{W_{nj} p_j^{\rm ss}}}\right)
    \Bigg].
\end{equation}
Proceeding in exactly the same way yields the classical TUR in Eq.~\eqref{TUR_original}, with $\langle \dot{\sigma}\rangle$ defined as in Eq.~\eqref{TUR_entropy_production}.

\subsubsection{Hasegawa's bound for Example A}

\begin{figure}
    \centering
    \includegraphics[width=0.4\textwidth]{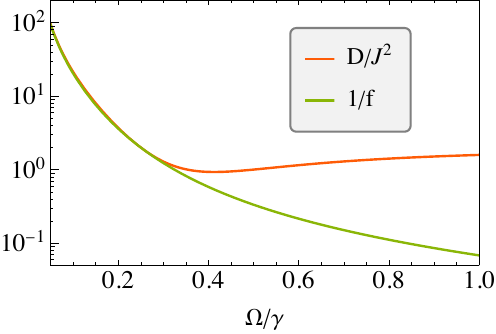}
    \caption{The ratio $D/J^2$ and the right-hand side $1/f$ of Hasegawa's bound~\eqref{TUR_hasegawa_2020}, for Example A \rev{(Sec.~\ref{ssec:exampleA})}, as a function of $\Omega/\gamma$ (with $\Delta = \bar{N} = 0$). 
    }
    \label{fig:ExampleA_hasegawa}
\end{figure}

Take Example A with $\Delta = \bar{N} = 0$. 
The average current is $J = 4\gamma\Omega^2/(\gamma^2 + 8 \Omega^2)$ and the noise is 
\begin{equation}
    D = \frac{4 \gamma  \Omega ^2 \left(\gamma ^4-8 \gamma ^2 \Omega ^2+64 \Omega
   ^4\right)}{\left(\gamma ^2+8 \Omega ^2\right)^3}.
\end{equation}
The right-hand side of Eq.~\eqref{TUR_hasegawa_2020}, on the other hand, has 
\begin{equation}
    f = \frac{4 \Omega ^2 \left(\gamma ^2+32 \Omega ^2\right)}{\gamma ^3+8 \gamma  \Omega ^2}.
\end{equation}
The left- and right-hand sides of Eq.~\eqref{TUR_hasegawa_2020} are shown in Fig.~\ref{fig:ExampleA_hasegawa} as a function of $\Omega/\gamma$. 
As can be seen, in certain regions the bound can be quite tight, while in others it can become rather loose (notice the log scale). 

\subsection{Quantum point contacts}
\label{sec:QPC}

A quantum point contact (QPC) is a tunneling device consisting of two fermionic reservoirs connected by a single tunnel junction \cite{blanter_2001}. 
When a QPC is placed next to a quantum dot, the electric field from the charges in the dot modulate the QPC's tunneling amplitude. 
The amount of current flowing through the QPC therefore gives a measure of the amount of nearby charges~ \cite{levinson_dephasing_1997, aleiner_dephasing_1997, stodolsky_measurement_1999, buttiker_charge_2000, gustavsson_2006}.
A similar charge sensor can be implemented by replacing the QPC with a single electron transistor (SET's) \cite{Schoelkopf_1998, Fujisawa_2004, Lu_2003, Bylander_2005}, or by employing an optical scheme based on resonance fluorescence \cite{kurzmann2019}.
As we have discussed, making a continuous charge measurement on the system leads to conditioning of the system dynamics on the measurement outcomes.
This effect has been studied extensively, first considering the ensemble average \cite{Gurvitz_measurements_1997, goan_2001}, then in terms of the unraveled quantum trajectories formalism \cite{goan_2001,Korotkov1999,Korotkov2001}, and subsequently for the repeated interactions framework \cite{Bettmann_2023}.
Since these early contributions, many have also become interested in using QPCs to access the FCS of the transported charge~\cite{Averin2005,Cuetara_2013, Cuetara_2015}. Experiments showcasing this include~\cite{gustavsson_2006, Gustavsson_2009, Fujisawa_2004, Fujisawa_2006, Ubbelohde_2012}.
Using the techniques introduced in the previous sections, we will bridge the gap between quantum trajectories and the FCS for QPCs.
We follow the approach developed in Refs. \cite{goan_2001}.
For concreteness, we will focus on the scenario of Example B; the extension to multiple quantum dots is straightforward.

\begin{figure}
    \centering
    \includegraphics[width=0.5\textwidth]{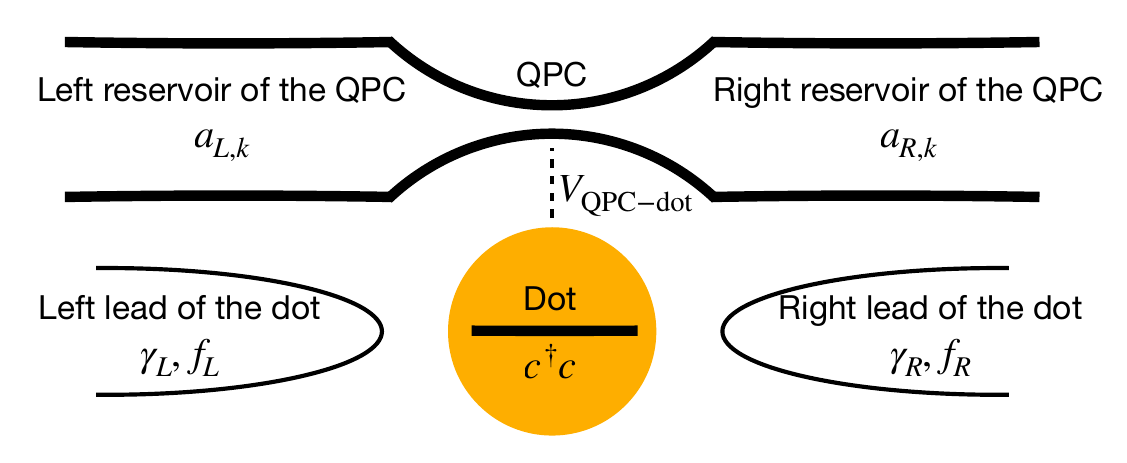}
    \caption{Schematics of a QPC placed next to a quantum dot in order to detect its charge. }
    \label{fig:QPC}
\end{figure}

The setup is illustrated in Fig.~\ref{fig:QPC}, which shows the
quantum dot, together with  its two leads (as described by the master equation~\eqref{ExampleB_M}).
The QPC, which is placed next to the dots, is  modeled by two \emph{additional} reservoirs,  described by annihilation operators $a_{\alpha, k}$, with energies $\omega_{\alpha,k}$, where $\alpha = L, R$.  
The two sides of the QPC are coupled according to  
\begin{align}
\label{eq: H_QPC_full}
\begin{aligned}
H_\mathrm{QPC} =& \sum_{k,\alpha \in \{\rm L,R\}} \omega_{\alpha,k}a_{k,\alpha}^\dagger a_{k,\alpha}\\& +  \sum_{k,k'}\left(\mathcal{T}_{k,k^\prime}~a_{\mathrm{R},k}^\dagger a_{\mathrm{L},k^\prime} 
+ 
\mathcal{T}_{k,k^\prime}^*~a_{\mathrm{L},k^\prime}^\dagger a_{\mathrm{R},k}\right)\,,
\end{aligned}
\end{align}
where $\mathcal{T}_{k,k^\prime}$ are the tunneling coefficients between states $k$ in the left reservoir  and $k'$ in the right one. 
The electrostatic potential of the quantum dot introduces an interaction term with the QPC that is of the form 
\begin{align}
V_{\rm QPC-dot} =& \sum_{k, k^\prime} c^\dagger c\Big(
\chi_{k,k^\prime}a_{\mathrm{R},k}^\dagger a_{\mathrm{L},k^\prime} + 
\chi_{k,k^\prime}^*a_{\mathrm{L},k^\prime}^\dagger a_{\mathrm{R},k}\Big),
\end{align}
for coefficients $\chi_{k,k'}$. 
Hence, the tunneling amplitudes of the QPC are modified depending on whether the dot is empty or occupied. 

Reference \cite{goan_2001} showed how to trace out the QPC in order to obtain a quantum master equation for the dot alone. 
The chemical potentials of the QPC reservoirs, $\mu_L = \mu + eV/2$ and $\mu_R = \mu - eV/2$, as well as their temperatures $T$, are assumed to satisfy 
$|eV|,k_\mathrm{B}T \ll \mu$.
They also assumed that around the chemical potential $\mu$, the tunneling rates  $\mathcal{T}_{k,k'}$, $\chi_{k,k'}$, as well as the densities of states $g_L$ and $g_R$ of the reservoirs, are roughly $k$-independent.
This means that the most relevant contribution will come from the tunneling rates $\mathcal{T}_{00}$ and $\chi_{00}$ that lie around the average chemical potential $\mu$. 
The resulting master equation then acquires the form 
\begin{equation}
    \label{QPC_M_unconditional}
    \frac{d \rho}{dt} = \mathcal{L} \rho + \mathcal{D}\left[ \mathcal{T}+\mathcal{X} c^\dagger c\right] \rho,
\end{equation}
where  
$\mathcal{T} = 2\pi e V g_L g_R \mathcal{T}_{00}$ 
and 
$\chi = 2\pi e V g_L g_R \chi_{00}$. 
Here $\mathcal{L}\rho$ is the Liouvillian~\eqref{ExampleB_M}, which encapsulates the effect of the two leads coupled to the dot (Fig.~\ref{fig:QPC}), with rates $\gamma_{L(R)}$ and Fermi occupations $f_{L(R)}$. 
Notice that 
\begin{equation}
\mathcal{D}[\mathcal{T} + \chi c^\dagger c]\rho = |\chi|^2 \mathcal{D}[c^\dagger c] \rho -i [\delta c^\dagger c, \rho] ,
\end{equation}
where $\delta = (\chi^* \mathcal{T} - \chi \mathcal{T}^*)/2i$.
The action of the QPC therefore produces a dephasing in the dot, as well as a  Lamb shift; i.e., a shift in the natural energy gap of the dot by a factor $\delta$.
The new term in Eq.~\eqref{QPC_M_unconditional} does not affect the dynamics, which has a steady-state  occupation 
\begin{equation}
    \langle c^\dagger c \rangle = \frac{\gamma_L f_L + \gamma_R f_R}{\gamma_L + \gamma_R}. 
\end{equation}
It also does not affect the electron current to the reservoirs, computed in Eq.~\eqref{ExampleB_Jss}.

The authors of Ref.~\cite{goan_2001} also considered the conditional master equations in both the jump and diffusion unravelling. 
The jump unravelling describes the case where the number of electrons tunneling through the QPC is small and is strongly affected by whether or not the dot is occupied. 
Conversely, the diffusion unravelling models the case where the tunnel current is very high and weakly affected by the occupation of the dot.
In practice, the latter regime is usually more experimentally relevant. 
In both cases, this provides a very neat and straightforward application of the formalism developed in this tutorial. 

In the jump unravelling regime, there is a current increment $dN(t)$ associated to the jump operator $L= \mathcal{T} + \chi c^\dagger c$ (with weight $\nu = 1$). 
An increment $dN(t) =1$ means an electron has tunneled through the QPC. 
The probability of observing this increment is 
$P(dN=1) = dt \langle L^\dagger L \rangle_c$.
If at a given time  the dot is empty, the tunneling probability will be $dt |\mathcal{T}|^2 := dt \mathcal{P}$;
and if the dot is occupied it will be  $dt |\mathcal{T} + \chi|^2:= dt \mathcal{P}'$. 
Due to electrostatic repulsion, the presence of an electron in the dot raises the tunneling barrier, 
implying that $\mathcal{P}' < \mathcal{P}$.

The resulting stochastic master equation is of the form~\eqref{quantum_jumps_SME}
\begin{equation}
\label{QPC_SME_jump}
    d\rho_c = dt \Bigg( \mathcal{L}\rho_c + 
    \mathcal{D}[L] \rho_c
    \Bigg)
    +
    \big(dN - dt \langle L^\dagger L \rangle_c) \left( 
        \frac{L \rho_c L^\dagger}{\langle L^\dagger L \rangle_c} - \rho_c
    \right).
\end{equation}
Please bear in mind that $\mathcal{L}$, as given in Eq.~\eqref{ExampleB_M}, has in itself another 4 jump operators, associated to currents to the left and right leads of the dot. 
When writing Eq.~\eqref{QPC_SME_jump}, it is assumed that these jump operators are not monitored; the only monitoring comes from the QPC. 
Of course, whether this is the case or not depends on the experiment. 
For example, in Sec.~\ref{sec:Q quantum dot refrigerator} we discussed quantum jump trajectories where the actual leads of the dots were monitored. 
In principle, we could also have all 5 jump operators to be monitored. 
Within our framework, these situations are all readily addressable by appropriately choosing the weight factors $\nu_k$. 

The average current through the QPC is given by Eq.~\eqref{average_current} with $L= \mathcal{T} + \chi c^\dagger c$; viz., 
\begin{equation}
    J_{\rm QPC} = \mathcal{P} + (\mathcal{P}'- \mathcal{P}) \langle c^\dagger c \rangle, 
\end{equation}
which is simply a weighted average of 
$\mathcal{P}$ when the dot is empty and $\mathcal{P}'$ when it is occupied. 
Notice, therefore, how this differs from the particle current $J$ flowing through the dot. 

The two-point correlation of the QPC current, Eq.~\eqref{F_jump} is given by 
\begin{equation}
\begin{aligned}
    F(\tau) = J_{\rm QPC}\delta(\tau) 
    + e^{- (\gamma_L+\gamma_R)\tau} (\mathcal{P}'-\mathcal{P})^2  \langle c^\dagger c \rangle (1- \langle c^\dagger c \rangle). 
\end{aligned}
\end{equation}
This is positive and therefore represent \rev{\emph{correlated}} emissions. 
The corresponding power spectrum is a simple Lorentzian around $\omega = 0$, with the noise given by 
\begin{equation}
    D = J_{QPC} + \frac{2}{\gamma_L + \gamma_R}(\mathcal{P}'-\mathcal{P})^2  \langle c^\dagger c \rangle (1- \langle c^\dagger c \rangle).
\end{equation}

If $|\mathcal{T}| \gg |\chi|$ 
the number of tunneling electrons through the QPC will be very large, and only weakly affected by the occupation of the dot. 
In this limit the conditional dynamics of the dot is more appropriately described by a diffusive master equation [Eq.~\eqref{Diffusion_SME}].
The fact that the jump operator in question is $L = \mathcal{T} + \chi c^\dagger c$ means this situation is  identical in spirit to shifting the jump operators through a local oscillator, as we first did when we introduced the diffusive unravelling in Sec.~\ref{sec:quantum_diffusion}.
For simplicity we assume $\mathcal{T}, \chi$ are real and hence $\chi < 0$ (since $\mathcal{P}'< \mathcal{P}$).
The resulting diffusive SME therefore reads 
\begin{equation}
    d\rho_c = dt \Bigg(\mathcal{L}\rho_c  + \chi^2 \mathcal{D}[c^\dagger c]\rho_c\Bigg) + \chi dW \Big[ c^\dagger c \rho_c + \rho_c c^\dagger c - 2 \langle c^\dagger c \rangle_c \rho_c\Big].
\end{equation}
The corresponding stochastic current
is given by Eq.~\eqref{current_relation_jump_diffusion} as 
\begin{equation}
    I_{\rm QPC} = \mathcal{T}^2 + 2\mathcal{T}\Bigg(\chi \langle c^\dagger c \rangle_c + \frac{dW}{dt}\Bigg).
\end{equation}
The diffusive current~\eqref{Idiffwiener}, which subtracts and normalizes the  offset $\mathcal{T}$, reads
\begin{equation}
    I_{\rm diff} = 2\chi \langle c^\dagger c \rangle_c + \frac{dW}{dt}.
\end{equation}

\subsection{Feedback: from Maxwell's demon to error correction}

Throughout this tutorial we have focused on the information that the output current gives us about the system. 
This information, in turn, can be fed back  to the system to manipulate its time-evolution \cite{Wiseman_2009,zhang_2017}. Such measurement and feedback scenarios enable a great deal of tasks including:
\begin{itemize}
    \item {\bf Maxwell's demon:} Maxwell envisioned a ``neat fingered being'' \cite{maxwell:book} that could detect individual particles in a gas. This being, now known as Maxwell's demon, could use this information to create a non-equilibrium situation out of equilibrium, e.g., by sorting the particles by their energy. As most prominently worked out by Szilard \cite{szilard:1929,szilard_eng}, this could be used to extract energy in the form of work from an equilibrium environment, seemingly violating the second law of thermodynamics. Recent advances in nanotechnology enabled the implementation of Maxwell's and Szilard's ideas in ground breaking experiments \cite{toyabe_2010,koski_2014,kumar_2018,cottet_2017}. By now, it is well established that the second law is in fact not violated if the demon is included in the thermodynamic analysis \cite{landauer_1961,bennett:1982}. Multiple ways of including both classical and quantum information into the thermodynamic bookkeeping have been put forward \cite{sagawa_2012,deffner_2013,horowitz_2014,potts_2018}. 
    
    \item  {\bf State preparation:} To perform experiments with quantum systems, it is often required to initiate the system in a low-entropy state. Furthermore, certain states have been shown to act as a resource for different tasks \cite{chitambar_2019}. The most prominent example is provided by entangled states which can be used for tasks in the fields of quantum sensing \cite{degen_2017} and quantum cryptography \cite{gisin_2002}. Initiating and/or stabilizing such quantum states can be a difficult task which in many cases may be enabled by feedback. Some examples include the cooling of a mechanical oscillator close to its ground state \cite{rossi_2018}, Fock state stabilization \cite{sayrin_2011}, and the stabilization of Rabi oscillations \cite{vijay_2012}.
    
    \item {\bf Quantum error correction:} A major obstacle in developing a large-scale quantum computer is provided by decoherence. The field of quantum error correction \cite{terhal_2015,roffe_2019} aims to counteract this problem by using redundant encodings of the quantum information, such that errors can be detected and corrected for without destroying the fragile superposition states that are required for quantum computation. Many strategies exist to use a measured current in a feedback loop for error correction \cite{ahn_2003,ahn_2002,sarovar_2004,borah_2022}. For an experiment on the subject, see~\cite{Livingston2022}.
\end{itemize}

The simplest example of feedback is provided by adding a term linear in the (stochastic) current $I(t) = \sum_k \nu_k \frac{dN_k}{dt}$, i.e., of the form $I(t)\mathcal{K}$, where $\mathcal{K}$ is an arbitrary superoperator describing the action taken from the feedback. 
This term enters as 
\begin{equation}
    \label{SME_feedback_basic}
    \rho_c+d\rho_c = e^{I(t)\mathcal{K}dt}\left[\rho_c+(d\rho_c)_0\right],
\end{equation}
where $(d\rho_c)_0$ denotes the increment of the conditional density matrix in the absence of feedback, 
e.g., Eq.~\eqref{quantum_jumps_SME} or~\eqref{Diffusion_SME} for jump or diffusion unravellings. 
Equation~\eqref{SME_feedback_basic} ensures that the time-evolution due to feedback is inserted at each time-step \textit{after} the measurement. 
\rev{And the fact that we are using only the instantaneous current, at that time step, ensures that the dynamics is still time-local (Markovian).}

For quantum jumps, expanding the exponential in Eq.~\eqref{SME_feedback_basic} and using the fact that $(dN_k)^l=dN_k$, we find the stochastic master equation for linear feedback \cite{wiseman_1994}
\begin{equation}
    \label{eq:masterfeedpoint}
    \begin{aligned}
    d\rho_c =& dt\mathcal{L}\rho_c+dt\sum_{k=1}^\nops\left(\langle L_k^\dagger L_k\rangle_c\rho_c-L_k\rho_cL_k^\dagger\right)\\&+\sum_{k=1}^\nops dN_k\left(e^{\nu_k\mathcal{K}}\frac{L_k\rho_cL_k^\dagger}{\langle L_k^\dagger L_k\rangle_c}-\rho_c\right).
    \end{aligned}
\end{equation}
Averaging over all measurement outcomes results in the unconditional master equation \cite{wiseman_1994}
\begin{equation}
    \label{eq:masterfeedpoint2}
    \frac{d\rho}{dt} = -i[H,\rho]+\sum_{k=1}^\nops\left(e^{\nu_k\mathcal{K}}L_k\rho L_k^\dagger-\frac{1}{2}\{L_k^\dagger L_k,\rho\}\right),
\end{equation}
which describes the ensemble-averaged effects of feedback. 
This equation has been employed to describe problems including charging a quantum battery \cite{yao_2021}, enhancing the precision of parameter estimation \cite{wang_2020}, manipulating EPR steering \cite{huang_2018}, entanglement generation \cite{carvalho_2007}, quantum state stabilization \cite{kieslich_2012}, and quantum error correction \cite{ahn_2003}.

Alternatively, we can also use linear feedback within the diffusive unravelling. 
To this end, we replace $I(t)\rightarrow I_{\rm diff}(t)$ in Eq.~\eqref{SME_feedback_basic} and we interpret $(d\rho_c)_0$ as the right-hand side of Eq.~\eqref{Diffusion_SME}. Expanding the exponential in Eq.~\eqref{SME_feedback_basic} to linear order in $dt$ (second order in $dW_k$), we find with the help of Eq.~\eqref{Idiffwiener} \cite{Wiseman_1993f}
\begin{equation}
    \label{eq:masterfeed2}
    \begin{aligned}
    d\rho_c =& dt\mathcal{L}\rho_c+dt\mathcal{K}\mathcal{H}\rho_c+dtK_{\rm diff}\mathcal{K}^2\rho_c\\&+\sum_{k=1}^\nops dW_k \left[\mathcal{H}_k\rho_c-\langle x_k\rangle_c\rho_c\right],
    \end{aligned}
\end{equation}
where $\mathcal{H}_k$ is given in Eq.~\eqref{H_k_superop} and $\mathcal{H}=\sum_k\nu_k\mathcal{H}_k$ in Eq.~\eqref{H_hom_op}.
Again, averaging over all trajectories results in the unconditional master equation \cite{Wiseman_1993f}
\begin{equation}
    \label{eq:masterfeed3}
    \frac{d\rho}{dt} = \mathcal{L}\rho+\mathcal{K}\mathcal{H}\rho+K_{\rm diff}\mathcal{K}^2\rho.
\end{equation}
This equation has been employed to tackle countless problems including quantum battery charging~\cite{mitchison_2021}, engineering many-body dynamics \cite{lammers_2016}, quantum error correction \cite{ahn_2002,sarovar_2004}, cooling of a trapped ion \cite{rabl_2005,steixner_2005} and a macroscopic oscillator \cite{mancini_1998}, entanglement generation \cite{wang_2005}, quantum state stabilization \cite{wang_2001,stockton_2004}, and to generate squeezing \cite{wiseman_1994s}.

Equation~\eqref{eq:masterfeed3} is extremely useful for feedback protocols where the measured signal is being fed back linearly to the system. However, we may be interested in protocols that are non-linear in the measurement outcome. For instance, consider a measurement of the location of an electron in a double quantum dot. Depending on the outcome, external voltages may suddenly be changed to alter the energy landscape of the double dot. Such feedback protocols can be employed to implement Maxwell's demon and Szilard's engine in an electronic system \cite{schaller_2011,koski_2014,chida_2017,annby_2020}. As illustrated in Fig.~\ref{fig:quantum_diffusion}, the diffusive current first needs to be filtered to remove the white noise that otherwise completely drowns the signal. 
Reference~\cite{annby_2022} has recently put forth fundamental new results for dealing with non-linear feedback, including filtered currents.  
We consider a low-pass filter with a bandwidth $\gamma$ described by
\begin{equation}
    \label{eq:filtd}
    \tilde{I}_{\rm diff}(t) = \gamma\int_0^\infty d\tau \, e^{-\gamma\tau} I_{\rm diff}(t-\tau).
\end{equation}
In this case, it can be shown that the filtered measurement outcome $\tilde{I}_{\rm diff}(t)$ follows an Ornstein-Uhlenbeck process, describing noisy relaxation towards the system-dependent value $\sum_k\nu_k\langle x_k\rangle_c$. With the help of the low-pass filter, a master equation can be derived for the quantity
\begin{equation}
    \label{eq:rhod}
    \rho(z) = E[\rho_c\delta(z-\tilde{I}_{\rm diff}(t))],
\end{equation}
which \rev{represents a current-resolved density matrix. It} provides a joint description of both the system state $\rho=\int dz~ \rho(z)$ and the observed measurement outcome $z$, which occurs with probability $p(z)=\tr\{\rho(z)\}$. The time-evolution of this object is governed by 
a hybrid equation, termed 
quantum Fokker-Planck master equation \cite{annby_2022}\footnote{Not to be confused with the quantum Fokker-Planck equation traditionally used in Quantum Optics to describe the phase-space representation of a master equation.}
\begin{equation}
    \label{Feedback_quantum_Fokker_Planck}
    \frac{d\rho(z)}{dt} = \mathcal{L}(z)\rho(z)-\gamma\partial_z\left(\mathcal{H}-z\right)\rho(z)+\frac{\gamma^2}{2}K_{\rm diff}\partial_z^2\rho(z).
\end{equation}
Here the feedback is included in a $z$-dependence of the Liouvillian which is not restricted to be linear in $z$. 
The second and third terms in the last equation correspond to drift and diffusion terms familiar from Fokker-Planck equations. However, the drift term is given by a superoperator, ensuring that the outcome $z$ drifts towards a system-dependent quantity. 

In the large bandwidth limit, where $\gamma$ is much larger than any energy scale of the system, a master equation for the reduced system state can be obtained, generalizing Eq.~\eqref{eq:masterfeed3} to feedback protocols that are nonlinear in the measurement outcome \cite{annby_2022}.
Beyond that, however, solving Eq.~\eqref{Feedback_quantum_Fokker_Planck} is generally challenging as one has both the operator character of $\rho$, as well as the continuous dependence on the variable $z$. 

\subsection{Emission and absorption spectra\label{sec:emission}}

So far we have primarily focused on the temporal fluctuations of the output currents, which are equivalently represented in the Fourier domain by the power spectrum $S(\omega)$. However, it is also of interest to consider the emission spectrum, which describes the frequency content of the emitted quanta themselves. One may also be interested in the absorption spectrum, which describes the rate at which the system absorbs energy from an external (coherent or thermal) field. While distinct from the power spectrum $S(\omega)$, the emission and absorption spectra can similarly be computed from the system dynamics under certain assumptions on the nature of the environment into which the quanta are emitted. The framework for achieving this is known as input-output theory, originally developed by Gardiner and Collett~\cite{Gardiner_1985} and described in detail in Ref.~\cite{gardiner_book}.

\subsubsection{Input-output theory}

We focus on the setting typical of quantum optics, where the environment comprises a set of bosonic modes that are coupled linearly to the system. 
For simplicity, let us first assume that there is a single dissipation channel with jump operator $L$. This situation can be modelled by a global system-environment Hamiltonian of the form 
\begin{equation}
\label{H_io}
    H_{\rm tot} = H + \int_{-\infty}^\infty d\omega\, \left[\omega b^\dagger_\omega b_\omega + \frac{i}{\sqrt{2\pi}}\left( b^\dagger_\omega L  - L^\dagger b_\omega \right)\right],
\end{equation}
where $H$ is the system Hamiltonian and $b^\dagger_\omega$ creates an excitation with frequency $\omega$ in the environment, satisfying the commutation relation $[b_\omega,b^\dagger_{\omega'}] = \delta(\omega-\omega')$. Note that here the frequencies are assumed to extend over all positive and negative values, and the system-environment coupling is taken to be frequency-independent 
\rev{(the resulting coupling constant is absorbed into $L$).}
Strictly speaking, these assumptions are unphysical because they imply the Hamiltonian is unbounded from below. However, they are justified in the quantum-optical context where the system-bath coupling is weak and the relevant transition frequencies of the system are relatively high. In that case, only environment modes near resonance with the system are significant and all others (including all those with negative frequency) make a negligible contribution. 

The benefit of this model is that it generates linear equations of motion for the environment operators in the Heisenberg picture:
\begin{equation}
    \label{io_b_eom}
    \frac{d b_\omega(t) }{dt} = -i \omega b_\omega(t) + \frac{1}{\sqrt{2\pi}}L(t).
\end{equation}
This equation can be formally solved in terms of an initial condition at a time $t_0 < t$, as
\begin{equation}
    \label{io_b_sol0}
    b_\omega(t) = e^{-i\omega(t-t_0)} b_\omega(t_0) + \frac{1}{\sqrt{2\pi}}\int_{t_0}^t dt' e^{-i\omega(t-t')} L(t'). 
\end{equation}
Now, the Heisenberg equation for an arbitrary system operator $A$ is given by
\begin{equation}
    \label{io_A_eqn}
    \frac{dA}{dt} = i[H,A] + \frac{1}{\sqrt{2\pi}}\int_{-\infty}^\infty d\omega \left([L^\dagger,A] b_\omega - b^\dagger_\omega [L,A] \right).
\end{equation}
Substituting Eq.~\eqref{io_b_sol0} into Eq.~\eqref{io_A_eqn}, interchanging the order of integration, and using the fact that
\begin{equation}
\label{io_delta_identity}
     \int_{t_0}^t dt'  \int_{-\infty}^\infty \frac{d\omega}{2\pi} e^{-i\omega (t-t')} L(t') = \int_{t_0}^t dt' \delta(t-t') L(t')  = \frac{1}{2}L(t),
\end{equation}
one obtains
\begin{equation}
\label{QLE}
     \frac{dA}{dt} = i[H,A] + \mathcal{D}^\dagger[L] A + [L^\dagger,A] b_{\rm in} - b^\dagger_{\rm in} [L,A].
\end{equation}
Here, we have identified the adjoint dissipator $\mathcal{D}^\dagger[L]$, defined below Eq.~\eqref{adjoint_liouvillian}, and defined the \textit{input field}
\begin{equation}
\label{b_in_def}
    b_{\rm in}(t) = \frac{1}{\sqrt{2\pi}}\int d\omega\, e^{-i\omega(t-t_0)} b_\omega(t_0).
\end{equation}

Eq.~\eqref{QLE} is known as a quantum Langevin equation (QLE), by analogy with the classical Langevin equation describing a system driven by a fluctuating force. The first term of Eq.~\eqref{QLE} represents the coherent Hamiltonian evolution and the second term describes damping. The final two terms describe the influence of environmental noise, which is represented by the input field $b_{\rm in}$. The noise statistics are determined by the correlation functions of $b_{\rm in}$, evaluated in the initial state at $t=t_0$, which is assumed to be a product state with respect to the system-environment partition. For example, a Markovian thermal reservoir is characterized by the input correlation function
\begin{equation}
    \label{thermal_noise_input}
    \langle b_{\rm in}^\dagger(t) b_{\rm in}(t')\rangle = \bar{N} \delta (t-t'),
\end{equation}
where $\bar{N}$ is the thermal occupation number. For this kind of input, it can be shown using the methods of quantum Ito calculus~\cite{gardiner_book} that the QLE~\eqref{QLE} is equivalent to the master equation
\begin{equation}
\label{QME_QLE}
     \frac{d}{dt} \rho = -i[H,\rho]+ (\bar{N}+1) \mathcal{D}[L] \rho  + \bar{N} \mathcal{D}[L^\dagger] \rho,
\end{equation}
which is the standard QME describing coupling to a bosonic thermal environment. A similar equivalence holds for a more general class of Gaussian noise inputs, including coherent displacement and squeezing~\cite{gardiner_book}.

The QLE~\eqref{QLE} describes the dynamics of the system in terms of the noise input. To find a connection to the output currents, we note that Eq.~\eqref{io_b_eom} can also be solved in terms of a final boundary condition at time $t_1 > t$, as 
\begin{equation}
    \label{io_b_sol1}
    b_\omega(t) = e^{i\omega(t_1-t)} b_\omega(t_1) - \frac{1}{\sqrt{2\pi}}\int_{t}^{t_1} dt' e^{i\omega(t'-t)} L(t'). 
\end{equation} 
Equating the right-hand sides of Eqs.~\eqref{io_b_sol0} and~\eqref{io_b_sol1}, integrating over all frequencies, and using Eq.~\eqref{io_delta_identity}, one obtains the \textit{input-output relation}
\begin{equation}
    \label{input_output}
    b_{\rm out}(t) = b_{\rm in}(t) + L(t),
\end{equation}
where we have defined the output field
\begin{equation}
\label{b_out}
    b_{\rm out}(t) = \frac{1}{\sqrt{2\pi}} \int d\omega\, e^{i\omega(t_1-t)} b_\omega(t_1).
\end{equation}
This describes properties of the environment at time $t_1$, which is usually taken to be in the far future. The input and output fields obey the commutation relations
\begin{equation}
    \label{b_in_commutation}
    [b_{\rm in}(t), b_{\rm in}^\dagger(t')] =  [b_{\rm out}(t), b_{\rm out}^\dagger(t')] = \delta(t-t'),
\end{equation}
reflecting the Markov assumption inherent to the model. 

The input-output formalism extends straightforwardly to a master equation with multiple jump operators and (possibly time-dependent) driving fields. For each jump operator $L_j$, one considers an independent bosonic reservoir described by a Hamiltonian of the form~\eqref{H_io}. Each reservoir gives rise to an input field $b_{\mathrm{in},j}$ and output field $b_{\mathrm{out},j}$, defined analogously to Eqs.~\eqref{b_in_def} and~\eqref{b_out}. Then, the system dynamics is described by a quantum Langevin equation
\begin{equation}
\label{QLE_multiple}
     \frac{dA}{dt} = i[H,A] + \sum_j \left(\mathcal{D}^\dagger[L_j] A + [L_j^\dagger,A] b_{\mathrm{in},j} - b^\dagger_{\mathrm{in},j} [L_j,A]\right),
\end{equation}
while the input-output relation generalises to
\begin{equation}
\label{input-Output-multiple}
     b_{\mathrm{out},j}(t) = b_{\mathrm{in},j}(t) + L_j(t).
\end{equation}

\subsubsection{Emission spectrum}

The input-output relation~\eqref{input-Output-multiple} allows one to compute spectral properties of steady-state currents using knowledge of the system dynamics. In general, a system producing a stationary current must have multiple input-output channels to sustain a flow of energy between them. For notational simplicity, however, we focus on one particular input-output channel and drop the subscript $j$ in the following.

The quantity $\langle b^\dagger_\omega(t_1) b_\omega(t_1)\rangle d\omega$ is the number of quanta (e.g.~photons) in the environment in a small frequency interval near $\omega$, as would be measured at time $t_1$. If we take $t_1$ to be large enough so that the steady state has been reached, the total number of emitted photons diverges. To regulate the divergence, we consider the Fourier transform over a finite observation time $T$,
\begin{equation}
\label{b_w_finite_T}
\tilde{b}_{\rm out}(\omega) = \frac{1}{\sqrt{2\pi}}\int_{-T/2}^{T/2}dt\, e^{i\omega(t-t_1)} b_{\rm out}(t).
\end{equation}
Clearly, this reduces to $b_\omega(t_1)$ as $T\to \infty$~\footnote{Note that $t_0 < -T/2$ and $t_1 > T/2$, so we must simultaneously take the limits $t_0\to -\infty$ and $t_1\to +\infty$.}. The asymptotic flux of output photons per unit frequency near $\omega$ is given by
\begin{widetext}
\begin{align}
\label{asymptotic_flux}
  &  \lim_{T\to\infty} \frac{1}{T}\left \langle \tilde{b}^\dagger_{\rm out}(\omega)\tilde{b}_{\rm out}(\omega)\right\rangle  =  \lim_{T\to\infty} \frac{1}{T}\left[\left \langle \tilde{b}^\dagger_{\rm in}(\omega)\tilde{b}_{\rm in}(\omega)  \right\rangle + \left \langle\tilde{b}^\dagger_{\rm in}(\omega)\tilde{L}(\omega) +\tilde{L}^\dagger(\omega)\tilde{b}_{\rm in}(\omega)\right\rangle + \left \langle\tilde{L}^\dagger(\omega)\tilde{L}(\omega)\right\rangle \right],
\end{align}
\end{widetext}
where we used the input-output relation and defined $\tilde{b}_{\rm in}(\omega)$ and $\tilde{L}(\omega)$ analogously to Eq.~\eqref{b_w_finite_T}. The first term on the RHS of Eq.~\eqref{asymptotic_flux} represents the input flux, the third term describes spontaneous emission by the system, while the second term describes both absorption and emission processes stimulated by the input field.

To isolate the emission spectrum, the corresponding input channel must be in the vacuum state (otherwise one cannot distinguish whether the detected photons were emitted by the system or were present in the environment already). For a vacuum input, the expectation value of any operator string with $b_{\rm in}$ on the right or $b_{\rm in}^\dagger$ on the left vanishes. Therefore, only the third term on the RHS of Eq.~\eqref{asymptotic_flux} survives, yielding the emission spectrum
\begin{align}
\label{emission_spectrum}
\mathcal{E}(\omega) & = \lim_{T\to\infty} \frac{1}{T} \left \langle\tilde{L}^\dagger(\omega)\tilde{L}(\omega)\right\rangle \notag \\
& = \lim_{T\to\infty} \frac{1}{2\pi T} \int_{-T/2}^{T/2}dt\int_{-T/2}^{T/2}dt' \, e^{-i\omega(t-t')} \langle L^\dagger(t) L(t')\rangle \notag \\
& = \frac{1}{2\pi}\int_{-\infty}^{\infty}d\tau \, e^{-i\omega\tau} \langle L^\dagger(\tau) L \rangle.
\end{align}
To obtain the third line of Eq.~\eqref{emission_spectrum}, we replace the correlation function by its steady-state value (valid since $T$ is large), which depends only on the time difference $\tau = t-t'$. The result is then obtained after changing integration variables to $\tau$ and $t_s = (t+t')/2$ and carrying out the trivial integral over $t_s$~\footnote{\rev{Strictly speaking, this derivation is valid only if $\langle L\rangle = 0$ in the steady state, so that $\langle L^\dagger(\tau) L \rangle \to |\langle L\rangle|^2 = 0$ as $\tau\to \infty$. In cases where $\langle L\rangle \neq 0$, the result still holds but the Fourier transform is singular at $\omega=0$ (e.g. see Eq.~\eqref{emission_spectrum_ExA}). To derive the result in this case, one adds and subtracts $|\langle L\rangle|^2$ from the integrand on the second line of Eq.~\eqref{emission_spectrum}, which yields
\begin{align*}
   \mathcal{E}(\omega) &  = \frac{1}{2\pi}\int_{-\infty}^{\infty}d\tau \, e^{-i\omega\tau} \left [\langle L^\dagger(\tau) L \rangle - |\langle L\rangle|^2\right] \\ 
   & \quad + \lim_{T\to\infty} \frac{1}{2\pi T} \int_{-T/2}^{T/2}dt\int_{-T/2}^{T/2}dt' |\langle L\rangle|^2 e^{-i\omega(t-t')},
\end{align*}
where the first line follows using the change of integration variables described below Eq.~\eqref{emission_spectrum}, which is valid because $\langle L^\dagger(\tau) L \rangle - |\langle L\rangle|^2$ decays to zero asymptotically. The second line above is proportional to 
$$ \frac{1}{2\pi T} \left\lvert\int_{-T/2}^{T/2}dt \,e^{i\omega t} \right\rvert^2 = \frac{T}{2\pi} {\rm sinc}^2(\omega T/2) \to \delta(\omega),$$
which tends to a delta function in the limit $T\to \infty$. Hence, we have 
\begin{equation*}
    \mathcal{E}(\omega) = |\langle L\rangle|^2 \delta(\omega) + \frac{1}{2\pi}\int_{-\infty}^{\infty}d\tau \, e^{-i\omega\tau} \left [\langle L^\dagger(\tau) L \rangle - |\langle L\rangle|^2\right],
\end{equation*}
which is formally equivalent to Eq.~\eqref{emission_spectrum}.}
}. The integrand of Eq.~\eqref{emission_spectrum} is proportional to the first-order coherence function, $g^{(1)}(\tau)$, which can be computed from the master equation using the quantum regression theorem as discussed in Sec.~\ref{sec:Coherence}. The total emitted flux is
\begin{equation}
    \label{total_emitted_flux}
    \int d\omega\, \mathcal{E}(\omega) = \langle L^\dagger L\rangle = J,
\end{equation}
which equals the total average output current (with $\nu=1$). Therefore, $\mathcal{E}(\omega)$ quantifies the contribution to $J$ from emitted quanta with frequency $\omega$.

\subsubsection{Absorption spectrum}

The rate at which photons at a given frequency are absorbed from an external field is quantified by the absorption spectrum. However, this term has at least two distinct meanings in the literature, depending on the nature of the input field driving the system.

\textit{a. Incoherent absorption:} In the case of a thermal noise input, the output flux can be shown to be~\cite{Gardiner_1985}
\begin{align}
    & \lim_{T\to\infty} \frac{1}{T}\left \langle \tilde{b}^\dagger_{\rm out}(\omega)\tilde{b}_{\rm out}(\omega)\right\rangle = \frac{\bar{N}}{2\pi}  + (\bar{N} +1) \mathcal{E}(\omega) - \bar{N}\mathcal{A}(\omega),
\end{align}
where the first term represents the input flux, the second term is the rate of stimulated and spontaneous emission, while the third term is the rate of absorption, with 
\begin{align}
    \label{incoherent_absorption}
    \mathcal{A}(\omega) & = \lim_{T\to\infty} \frac{1}{T} \left \langle\tilde{L}(\omega)\tilde{L}^\dagger(\omega)\right\rangle  = \frac{1}{2\pi}\int_{-\infty}^{\infty}d\tau \, e^{i\omega\tau} \langle L(\tau) L^\dagger \rangle.
\end{align}
This function quantifies the frequency content of quanta absorbed from the thermal input field. We refer to it here as the \textit{incoherent} absorption spectrum. Conceptually, $\mathcal{A}(\omega)$ is similar to the emission spectrum $\mathcal{E}(\omega)$: both describe the frequency distribution of quanta exchanged with a white-noise field. We see that 
\begin{equation}    \label{absorption_spectrum_current}
    \int d\omega\, \mathcal{A}(\omega) = \langle LL^\dagger\rangle,
\end{equation}
which is the average current associated with jump operator $L^\dagger$ appearing in the last term on the RHS of Eq.~\eqref{QME_QLE}.

\textit{b. Coherent absorption spectrum: }Alternatively, one may ask about the rate of photons absorbed from a coherent drive at a specific frequency. \rev{This  situation is described mathematically by considering a jump operator of the form $L = \sqrt{\gamma} c$, for  some dimensionless system operator $c$ and dissipation rate $\gamma$, and taking a coherent input field of the form
\begin{equation}
    \label{coherent_input}
    b_{\rm in}(t) = \frac{i \Omega e^{-i\omega_d t} }{\sqrt{\gamma}} + b'_{\rm in}(t).
\end{equation}
The first, $c$-number term represents the coherent drive amplitude with frequency $\omega_d$ and Rabi frequency $\Omega$ and the remainder $b'_{\rm in}(t)$ represents a vacuum input. Referring to the QLE~\eqref{QLE} (or Eq.~\eqref{QLE_multiple}), we see that the only effect of adding the coherent drive is to shift the Hamiltonian as $H\to H + \Omega (e^{i\omega_d t}c  + e^{-i\omega_d t} c^\dagger )$.}

\rev{To find the rate at which quanta are absorbed from the drive, we look at the difference between the total output and input flux}, which can be found by integrating Eq.~\eqref{asymptotic_flux} over frequency. \rev{In the steady state, this will be equal to the difference between the rates of emission and absorption. In particular, the total emission rate is just the output current $J$, given by the integral of the third term on the RHS of Eq.~\eqref{asymptotic_flux} (c.f. Eqs.~\eqref{emission_spectrum} and~\eqref{total_emitted_flux}). The absorption rate is therefore given by the second term on the RHS of Eq.~\eqref{asymptotic_flux}, integrated over frequency, i.e.}
\begin{align}
\label{absorption_spectrum}
   \mathcal{W}(\omega_d) & = -\lim_{T\to \infty} \frac{1}{T} \int_{-\infty}^\infty d\omega \left \langle\tilde{b}^\dagger_{\rm in}(\omega)\tilde{L}(\omega) +{\rm h.c.}\right\rangle \notag \\
    & = -\lim_{T\to \infty}\frac{1}{T} \int_{-T/2}^{T/2} dt \left \langle{b}^\dagger_{\rm in}(t){L}(t) +{\rm h.c.}\right\rangle \notag \\
    & = i\Omega \lim_{T\to \infty}\frac{1}{T} \int_{-T/2}^{T/2} dt \left\langle   e^{i\omega_d t}  c(t) - e^{-i\omega_d t}  c^\dagger(t)\right\rangle.
\end{align} We refer to this quantity as the \textit{coherent} absorption spectrum. Eq.~\eqref{absorption_spectrum} can also be interpreted as the average power absorbed from the drive~\cite{alicki_1979} in units of the energy quantum $\omega_d$, i.e.
\begin{equation}
\mathcal{W}(\omega_d) = \frac{1}{\omega_d}\overline{\left\langle \frac{\partial H_d(t)}{\partial t}\right \rangle},
\end{equation}
where $H_d(t) = \Omega (e^{i\omega_d t}c  + e^{-i\omega_d t} c^\dagger )$ is the driving Hamiltonian and the overline $\overline{\bullet}$ denotes a time average. Note that, because the evolution of $c(t)$ depends on the drive, reconstructing the coherent absorption spectrum generally requires solving a different master equation for each value of $\omega_d$. 

\subsubsection{Emission and absorption spectra for Example A: Mollow Triplet}

\begin{figure}
    \centering
    \includegraphics[width=\linewidth]{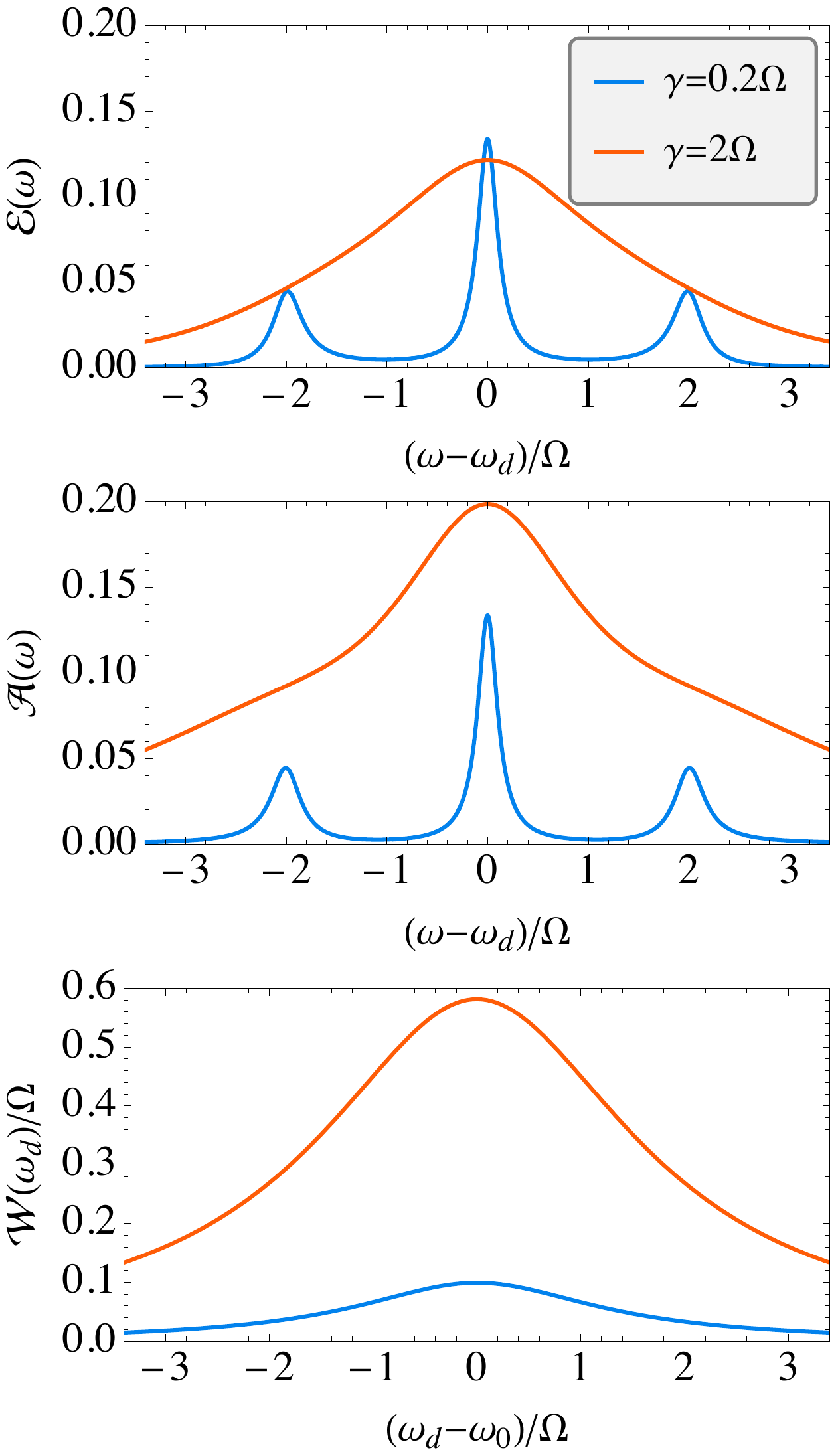}
    \caption{Emission and absorption spectra for Example A \rev{(Sec.~\ref{ssec:exampleA}) for resonant driving ($\omega_d = \omega_0$) and two different damping rates, $\gamma = 0.2\Omega$ (blue curves), and $\gamma = 2\Omega$ (red)}. Top panel: emission spectrum \rev{[Eq.~\eqref{emission_spectrum_ExA}]} Middle panel: incoherent absorption spectrum. Bottom panel: coherent absorption spectrum \rev{[Eq.~\eqref{absorption_spectrum_ExA}]}, as a function of the detuning.
    \label{fig:emission_spec}
    }   
\end{figure}

We now compute the emission and absorption spectra for Example A, a driven two-level system [Eq.~\eqref{ExampleA_M}]. This problem was theoretized by Mollow~\cite{Mollow_1969,Mollow_1972}, and first observed experimentally in Sodium vapor in Ref.~\cite{Schuda1974}. 
Within the input-output formalism, it is described by a qubit with Hamiltonian $H = \omega_0 \sigma_z/2$ and two input-output channels. The first input is a coherent driving field as in Eq.~\eqref{coherent_input}, with associated jump operator $L_1 = -i\sqrt{\epsilon} \sigma_- $ and $\epsilon\to 0$. The second input describes the environment into which photons are emitted, with associated jump operator $L_2 = \sqrt{\gamma}\sigma_-$. We assume this input is described by thermal white-noise statistics as in Eq.~\eqref{thermal_noise_input}, with a small thermal component of $\bar{N}=0.1$.

Moving to a rotating frame via the unitary transformation $e^{-i\omega_d t\sigma_z/2}$, the emission spectrum is given by\rev{
\begin{align}
    \label{emission_spectrum_ExA}
    \mathcal{E}(\omega) & = \frac{\gamma}{2\pi} \int_{-\infty}^\infty d\tau \, e^{i(\omega_d-\omega)\tau} \langle \sigma_+(\tau) \sigma_- \rangle, \\
    & = \gamma |\langle \sigma_-\rangle|^2 \delta(\omega-\omega_d) \notag \\ & \quad + \frac{\gamma}{2\pi} \int_{-\infty}^\infty d\tau \, e^{i(\omega_d-\omega)\tau} \left[\langle \sigma_+(\tau) \sigma_- \rangle - |\langle \sigma_-\rangle|^2 \right].\notag
\end{align}
Here, for completeness, we have explicitly separated the emission spectrum into its divergent and regular parts. The former arises due to the presence of a non-zero expectation value $\langle \sigma_-\rangle$, and can be understood in terms of elastic scattering of photons from the driving field into the output channel. A similar term appears in the emission spectrum for any system with $\langle L\rangle \neq 0$; see the footnote below Eq.~\eqref{emission_spectrum}. The second, regular part of Eq.~\eqref{emission_spectrum_ExA} represents the spectral content of photons scattered inelastically by the two-level system,} which is plotted in the top panel of Fig.~\ref{fig:emission_spec}. For weak driving, $\Omega \ll \gamma$, emission is peaked around the driving frequency with a linewidth proportional to $\gamma$. For strong driving, $\Omega \gg \gamma$, the emission spectrum splits into the characteristic Mollow triplet, with sidebands emerging at $\omega = \omega_d \pm \Omega$. This reflects a transition in the $g^{(1)}(\tau)$ coherence function from overdamped decay to underdamped oscillations as $\Omega$ increases, as shown in Fig.~\ref{fig:g1}. Similar features are seen in the incoherent absorption spectrum, \rev{whose regular part is} plotted in the middle panel of Fig.~\ref{fig:emission_spec}.

The coherent absorption spectrum for Example A can be computed analytically,
\begin{equation}
\label{absorption_spectrum_ExA}
    \mathcal{W}(\omega_d) = \Omega \langle \sigma_y\rangle = \frac{\gamma\Omega^2}{(\omega_0-\omega_d)^2 + 2\Omega^2 + (\gamma/2)^2},
\end{equation}
where the expectation value is taken with respect to the steady state of Eq.~\eqref{ExampleA_M} (which is time-independent in the rotating frame) with detuning $\Delta = \omega_0 - \omega_d$. This result is plotted in the bottom panel of Fig.~\ref{fig:emission_spec}. Unlike the emission and incoherent absorption spectra, $\mathcal{W}(\omega_d)$ always shows a single peak at the resonance frequency. For weak driving, $\Omega\ll \gamma$, the absorption linewidth is of order $\gamma$, while at stronger driving it is instead proportional to the Rabi frequency, $\Omega$.

\subsection{Quantum correlations and entanglement witnesses}

As we have discussed throughout this tutorial, the two-time correlation function provides a measure of temporal correlations between currents.
This makes it particularly useful for the study of quantum correlations and measures of entanglement. 
Here we will briefly discuss a few seminal examples which are relevant to the techniques we have described. 
For a tutorial on the subject, see~\cite{Lewalle2021}.

It is possible to use the $\gtwo(\tau)$ function  for multiple measurement ports, as described in Sec.~\ref{sec:multiple_currents}, as an entanglement criterion between observed currents.
This was first described in Ref.~\cite{B_rkje_2011} where the authors showed how entanglement between two driven optomechanical systems could be measured. 
In this proposal, there are two optomechanical cavities with a mechanical frequency $\omega_{M}$ which are coherently driven at a frequency $\Omega_{D}$. Here the position of a mechanical oscillator modulates the frequency of an optical cavity mode.
In the measured output current, the spectrum exhibits \emph{mechanical sidebands} which are displaced from the drive frequency $\Omega_{D}$ by the mechanical frequency $\omega_{M}$.
These sidebands are the result of Raman scattering processes where a photon either gains or loses energy by destroying or creating one phonon in the mechanical oscillator. The corresponding output modes are commonly known as the blue and red sidebands, respectively.
Moreover, the presence of a photon in either the red or blue sideband can be used to encode a logical qubit for the purpose of verifying entanglement. 
As shown Ref.~\cite{B_rkje_2011}, these sidebands can serve as the logical basis to entangle the two mechanical oscillators by mixing the output on a single 50:50 beam splitter, erasing any which-way information. 
In each arm of the experiment after the beam-splitter---denoted $A$ and $B$---there is a frequency filter, projecting out the red/blue sidebands and thus leading to 4 measurement outcomes $A_{r}, A_{b}, B_{r}, B_{b}$.
This creates entanglement in the following way: the detection of a red/blue sideband photon informs us that one phonon has been created/annihilated in one of the oscillators.
Given that path information has been erased, we cannot say which oscillator the phonon has been created in.
Thus, the final entangled state is a superposition of single-phonon states localised in each of the two mechanical oscillators.

As the authors show, the creation of a phonon in either of the mechanical oscillators by the coherent drive can be measured via the the multi-current $\gtwo$ function conditioned on a particular measurement outcome 
\begin{equation}
    \gtwo_{A_{b}|A_{r}}(\tau) = \frac{\langle b_{A_r}^{\dagger}(\tau)b_{A_b}^{\dagger}(\tau+t)b_{A_b}(\tau+t)b_{A_r}(\tau)\rangle}{\langle b_{A_r}^{\dagger}b_{A_r}\rangle \langle b_{A_b}^{\dagger}b_{A_b} \rangle}\,,
\end{equation}
which can be computed using the methods outlined in Sec.~\ref{sec:multiple_currents}.
This function provides one with a measure of the coherence between the measurement of a blue-sideband photon measured in detector $A_{b}$ at $t+\tau$ given that we measured a red-sideband photon at $\tau$ in $A_{r}$, providing a clear indication that a phonon excitation was coherently---as opposed to a thermal excitation---generated in one of the two oscillators. 
Using these multi-current $\gtwo$ functions, the author derive an upper bound on an entanglement witness denoted
\begin{equation}
    R_{\rm m}(\tau) = 4\left(\frac{ \gtwo_{A_{b}|A_{r}}(\tau) +  \gtwo_{B_{b}|A_{r}}(\tau) - 1}{ \left(\gtwo_{A_{b}|A_{r}}(\tau) -  \gtwo_{B_{b}|A_{r}}(\tau)\right)^{2}}\right)\,,
\end{equation}
in which a measurement of $R_{\rm m}(\tau)<1$ is evidence of entanglement between the two mechanical oscillators.
This entanglement has since been verified experimentally \cite{Riedinger_2018}.

Other research has focused on using the emission spectrum $\mathcal{E}(\omega)$ to study quantum correlations. 
For example, as was shown in \cite{Woolley_2014}, two mechanical oscillators with bosonic annihilation (creation) operators $a\,(a^{\dagger})$ and $b\,(b^{\dagger})$ can be entangled in a two-mode squeezed state---where the squeezing parameter is $r$---by independently coupling both oscillators to a single driven optical cavity mode $c\,(c^{\dagger})$.
The mechanical sidebands that are measured in the cavity emission spectrum are directly related to the occupations of mechanical two-mode Bogoliubov modes $ \langle \beta_{i}^{\dagger} \beta_{i}\rangle$, defined in terms of the squeezing parameter $r$ by
\begin{equation}
    \beta_{1} = a \cosh r + b^{\dagger} \sinh r\,,\qquad \beta_{2} = b \cosh r + a^{\dagger} \sinh r\,.
\end{equation}
As shown in \cite{Woolley_2014}, the entanglement witness for continuous modes \cite{Duan_inseparability_2000} is directly related to the occupation of these Bogoliubov modes $\langle \beta_{i}^{\dagger}\beta_{i}\rangle$.
Moreover, these occupations are directly related to the heights of the measured red (blue) sidebands in $\mathcal{E}(\omega)$.
Thus, from a direct measurement of $\mathcal{E}(\omega)$ one can witness the entanglement of mechanical squeezed states, which has in fact been accomplished experimentally \cite{Ockeloen_Korppi_2016}.

\section{Conclusions and outlook}

The first main goal of this tutorial is to put forth a toolbox for describing current fluctuations in continuously measured systems. 
We have done this by using a unified approach, where we start with a general quantum master equation and identify the jump operators with particular clicks in a detector. 
One can then build the corresponding output currents by weighting these clicks according to the physics one is interested in.
This allowed us to describe, within a single framework, various phenomena from quantum optics to mesoscopic transport. 
Our second goal was to provide a bridge between the stochastic master equations often used by quantum opticians, and the tools of FCS used in condensed-matter physics and statistical mechanics. 
Both of these goals, we believe, are unique and cannot be found elsewhere in the literature.

Overall, we hope that this tutorial will encourage researchers working with quantum master equations to go beyond the average current. 
There are countless studies in the literature describing fascinating physical systems but focusing only on the average current, as it is something that readily follows from the QME. 
Here we have shown that, with just a little more effort, one can go beyond the average to compute the two-point correlation function, the power spectrum, the noise, the FCS distribution, and the scaled cumulant generating function.

After studying this tutorial, the reader should have gained the following insights.
The reader has gained an appreciation for stochastic quantum trajectories which arise from continuous measurements in the basic setting introduced in Sec.~\ref{sec:currents}. 
Most importantly, they understand the intrinsic connection between the stochastic measurement back-action on the conditioned density matrix, and the measured output currents, for both jump and diffusive unravelings. 
The reader understands the connection between two-point correlation functions, the power spectrum, and the underlying noise in this measured current, as discussed in Sec.~\ref{sec:fluctuations}. 
Through Sec.~\ref{sec:FCS} the reader has become familiar with the theory of full counting statistics, epitomized by the tilted Liouvillian and the cumulant generating function.
An overview of the main formulas developed in Secs.~\ref{sec:currents}-\ref{sec:FCS} is shown in Fig.~\ref{fig:overview}. 
After completing Sec.~\ref{sec:methods}, the reader has learned techniques to compute all these new quantities. 
This involves a mixture of techniques which are not always conventional.
Finally, from Sec.~\ref{sec:topical}, the reader has gained a broad appreciation of the topical literature in which these techniques can be readily used.

Looking ahead, we believe the toolbox put forth in this tutorial may be used to advance various fields of research. One example is thermodynamics: how to account for the energetics of quantum continuous measurements~\cite{Naghiloo2018}, and the associated entropy production~\cite{Manzano_2022,Landi_2022}?
TURs and KURs~\cite{Hasegawa_quantum_2020} as well as the possibility of doing feedback control~\cite{annby_2022}, offer two interesting scenarios where continuous measurements meet thermodynamics. However, much remains unexplored. 
Another foundational field with exciting possibilities is that of Markovianity and memory in quantum systems~\cite{Rivas2014,Milz2021}. 
Throughout this tutorial the system itself evolves according to a Markovian master equation, while the outcome currents are non-Markovian --- this is the quantum analog of a hidden Markov model. 
Many features relating to the memory effects of the output current still remain open. 
Going further, very little is known of continuously measured systems whose underlying dynamics is non-Markovian. We hope that our readers will now feel empowered and inspired to tackle these fascinating open questions. 

\section*{Acknowledgments}

The authors acknowledge fruitful discussions with G. Schaller. 
GTL acknowledges the financial support of the S\~ao Paulo Funding Agency FAPESP (Grant No.~2019/14072-0.). MJK acknowledges the financial support from a Marie Sk\l odowska-Curie Fellowship (Grant No. 101065974). MTM is supported by a Royal Society-Science Foundation Ireland University Research Fellowship (URF\textbackslash R1\textbackslash 221571). PPP acknowledges funding from the Swiss National Science Foundation (Eccellenza Professorial Fellowship PCEFP2\_194268). This project is co-funded by the European Union (Quantum Flagship project ASPECTS, Grant Agreement No.~101080167). Views and opinions expressed are however those of the authors only and do not necessarily reflect those of the European Union, REA or UKRI. Neither
the European Union nor UKRI can be held responsible for them. 

\appendix

\section{Monte Carlo wavefunction method}
\label{app:MCWF}

\rev{In this appendix we describe a more efficient method of performing stochastic simulations of the quantum jump unravelling. 
First, a small parenthesis: let $X$ denote a random variable that can take values over a discrete alphabet $x = 1,\ldots, d$, with probabilities $p_x$. 
To sample $X$ efficiently one may use the standard inversion method~\cite{Devroye1986}:
\begin{enumerate}
    \item Construct the cumulative distribution 
    $C_x = \sum_{y=1}^x p_y$.
 \item Sample a uniformly distributed random number $r \in [0,1]$.
 \item Select the smallest  $x$ such that $C_x\geq r$. 
\end{enumerate}
This can, of course, also be adapted to continuous variables by replacing a sum with an integral. 
}

\rev{Quantum jumps can be simulated by applying the Kraus operators in Eq.~\eqref{kraus}: discretize time in small steps $dt$. At each time step:
\begin{enumerate}
    \item Sample one of $k = 0,1,\ldots,\nops$ with probabilities $p_k^c = dt \tr\big\{ L_k^\dagger L_k \rho_c\big\}$ and $p_0^c = 1- \sum_{k=1}^\nops p_k^c$;
    \item Apply the corresponding operator $\rho_c \to M_k \rho_c M_k^\dagger/p_k$ associated to the value of $k$ that was sampled.
\end{enumerate}
Despite its simplicity, this method is not at all efficient because most of the time the sampling will return $k=0$ (no jump).
We are thus using a large number of random numbers, but in most time steps very little happens. }

\rev{Here we describe the Monte Carlo wavefunction (MCWF) method~\cite{Molmer1996}, which is much more efficient.
The method is rooted on the waiting time distribution ideas discussed in Sec.~\ref{sec:WTD}. In particular, Eq.~\eqref{WTD_P_no} states that if the system is prepared in $\rho_0$, the probability that up to a time $t$ no jump has occurred is given by $P_{\rm no}(t) = \tr\big\{e^{\mathcal{L}_0 t} \rho_0\big\}$, where  
$\mathcal{L}_0$ is the no-jump superoperator defined in Eq.~\eqref{WTD_L0}.
The conditional state of the system, given that no jump occurred is $e^{\mathcal{L}_0 t} \rho_0/P_{\rm no}(t)$.
We will assume, as in Sec.~\ref{sec:WTD}, that only a subset $\mathcal{M}$ of the jump operators are monitored. 
The idea of MCWF is to first sample the random time $\tau$ where the jump will occur, and then sample the resulting channel $k$.
Both are sampled using the inversion method above.
}

\rev{First, the jump times: $P_{\rm no}(t)$ is a survival probability (prob. that it does not jump) so $1-P_{\rm no}(t)$ is the cumulative distribution. 
We therefore need to sample a uniform random number $r$ and solve $1-P_{\rm no}(t) = r$, which will give us the jump time $\tau$. 
Since $r$ is uniform over $[0,1]$, this is the same as solving for $P_{\rm no}(t) = r$. 
Unfortunately, this is often a complicated equation, that cannot be solved analytically. Instead, we evolve the system in small time steps as $\rho \to e^{\mathcal{L}_0 \Delta t}\rho$, and compute the trace at each step, which is precisely $P_{\rm no}$. We then keep going until $P_{\rm no} = r$, which gives us the jump time $\tau$. 
The state immediately before the jump will be $\rho^* = e^{\mathcal{L}_0 \tau} \rho_0/P_{\rm no}(\tau)$.
}

\rev{Next, we need to decide to which channel the jump occurs. We know from Eq.~\eqref{pk} that $p_k = dt~\tr\big\{L_k^\dagger L_k \rho^*\big\}$. 
But over the set $k\in\mathcal{M}$ these are not normalized.
We therefore define normalized probabilities  $\mathfrak{p}_k = p_k/K$, where $K = \sum_{k\in \mathcal{M}} p_k$. 
The jump channel is then chosen  by sampling $k \in \mathcal{M}$ from $\mathfrak{p}_k$. 
The state of the system is also updated to a post-jump state $L_k \rho^* L_k^\dagger/\tr\{L_k^\dagger L_k \rho^*\big\}$.
Then  the process restarts. 
}

\rev{To summarize, the $n^{\rm th}$ iteration of the MCWF algorithm reads (with $t_0 = 0$):
\begin{enumerate}
    \item Start from the state $\rho_{n-1}$ at time $t_{n-1}$.
    \item Sample two uniform random numbers $r,r' \in [0,1]$. 
    \item Evolve $\rho_{n-1}$ in time under $\mathcal{L}_0$, and compute its trace
    $P_{\rm no}(t) = \tr\big\{ e^{\mathcal{L}_0 t}\rho_{n-1}\big\}$
    at each timestep. 
    \item Stop when $P_{\rm no}(t) = r$. This defines the duration $\tau_n$ from the last jump (so the absolute time is $t_n = t_{n-1} + \tau_n$). The pre-jump state is $\rho_n^* = e^{\mathcal{L}_0 \tau_n} \rho_{n-1}/P_{\rm no}(\tau_n)$. 
    \item Use $r'$ with the inversion method to sample the channel $k_n\in \mathcal{M}$ from the distribution
    \begin{equation}
    \label{MCWF_jump_probs}
        \mathfrak{p}_k = \frac{\tr\big\{ L_k^\dagger L_k \rho_n^*\big\}}{\sum_{q\in \mathcal{M}} \tr\big\{ L_q^\dagger L_q \rho_n^*\big\}}.
    \end{equation}   
    \item Update the state as 
    \begin{equation}
    \label{rho_n_recursion_MCWF}
        \rho_{n}   = \frac{L_{k_n} \rho_n^* L_{k_n}^\dagger}{\tr\big\{ L_{k_n} \rho_n^* L_{k_n}^\dagger\big\}}      
        = \frac{\mathcal{L}_{k_n} e^{\mathcal{L}_0\tau_n} \rho_{n-1}}{ \tr\left(\mathcal{L}_{k_n} e^{\mathcal{L}_0\tau_n} \rho_{n-1}\right)}.
    \end{equation}
\end{enumerate}
Suppose the algorithm runs to a specified final time $t_f$, and it happened that there were a total of $N$ jumps. 
The unnormalized state of the system at time $t_f$ will then be $\rho(t_f) = e^{\mathcal{L}_0 (t_f-t_N)} \rho_N$. 
}

\rev{The dynamics will depend on the choice of channels $\mathcal{M}$ that are monitored, as this affects both $\mathcal{L}_0$ and all  $k$-sums.
If all jumps are monitored and the initial state $\rho_0$ is pure, then the conditional states $e^{\mathcal{L}_0\tau_n}\rho_{n-1}$ remain pure at all times. 
This follows from the fact that, in this case
$e^{\mathcal{L}_0 t}\rho = e^{-i \Hnh t} \rho e^{i \Hnh^\dagger t}$ [c.f. Eq.~\eqref{WTD_L0_Hnm}]
It is then possible to reformulate the entire procedure as a stochastic evolution of pure states $\rho_n = |\psi_n\rangle\langle \psi_n|$. In particular, the states are generated according to the Markov chain
\begin{equation}
    \label{MCWF_pure}
    |\psi_n\rangle = \frac{L_{k_n} e^{-i \Hnh \tau_n} |\psi_{n-1}\rangle}{\lVert L_{k_n} e^{-i \Hnh \tau_n} |\psi_{n-1}\rangle\rVert},
\end{equation}
with all other formulas being  adapted in a similar way.
This provides a major advantage, since only pure state vectors (wavefunctions) rather than density matrices need be stored, which is the origin of the name MCWF. In fact, this leads to a quadratic reduction (with respect to Hilbert-space dimension) in the memory cost of a simulation.}

\section{Derivation of the stochastic master equation for quantum diffusion}
\label{app:stochasticwiener}

To derive the stochastic master equation~\eqref{Diffusion_SME} for quantum diffusion, we employ the transformation given in Eq.~\eqref{gauge} to the stochastic master equation for quantum jumps, cf.~Eq.~\eqref{quantum_jumps_SME}. 
For simplicity, we focus here on a single channel, the generalization to $\nops$ channels is straightforward. 
Upon regrouping terms, we find
\begin{equation}
    \label{eq:stochwiener1}
    \begin{aligned}
    &d\rho_c =  dt \mathcal{L} \rho_c
    \\&+\left( \frac{dN}{|\alpha|}-dN\frac{\langle x\rangle}{|\alpha|^2}-|\alpha|dt\right) \left[Le^{-i\phi}\rho_c+\rho_cL^\dagger e^{i\phi}-\langle x\rangle_c\rho_c\right]\\&+\frac{1}{|\alpha|}\left( \frac{dN}{|\alpha|}-|\alpha|dt\right)\left(L\rho_c L^\dagger-\langle L^\dagger L\rangle_c\rho_c\right).
    \end{aligned}
\end{equation}
In addition, the transformation also affects the probability of observing a jump [cf.~Eq.~\eqref{average_current_gauge}]
\begin{equation}
    \label{eq:jumpprobgauge}
    P(dN = 1) = dt \left(|\alpha|^2+|\alpha|\langle x\rangle_c+\langle L^\dagger L \rangle_c\right).
\end{equation}
Eq.~\eqref{Diffusion_SME} can be obtained by taking the limit $|\alpha|\rightarrow\infty$. 
However, some subtleties arise concerning the time scales involved: 
Eq.~\eqref{eq:stochwiener1} is only accurate in the regime where, in each time step $dt$, at most one jump occurs. Hence, it is implicitly assumed that $dt$ remains sufficiently small so that 
$|\alpha|dt\ll1$. 
Conversely, to obtain Eq.~\eqref{Diffusion_SME} we want a scenario where in each time step many jumps occur.
Hence, we must integrate 
Eq.~\eqref{eq:stochwiener1} over a time $d\tau$ chosen such that $\rho_c$ only changes infinitesimally, while $d\tau|\alpha|\rightarrow\infty$. 

In the first line of Eq.~\eqref{eq:stochwiener1}, this simply results in the replacement $dt\rightarrow d\tau$. For the second line, we identify the Wiener increment as 
\begin{equation}
    \label{eq:wiener}
    dW = \int_0^{d\tau} \left[ \frac{dN(t)}{|\alpha|}-dN(t)\frac{\langle x\rangle_c}{|\alpha|^2}-|\alpha|dt\right].
\end{equation}
To see that this is indeed a Wiener increment, we note that the probability of observing $dN=1$ only depends on the conditional state $\rho_c$. 
If we can neglect the changes in $\rho_c$ during the time-interval $d\tau$, we may assume all $dN$ during this time-interval to be uncorrelated. Then it holds that
\begin{equation}
    \label{eq:probdn}
    P\left(\int_0^{d\tau} dN\right) = B\left[d\tau/dt,P(dN = 1)\right],
\end{equation}
where $B[n,p]$ denotes the binomial distribution with $n$ trials and success probability $p$. For a large number of trials, the binomial distribution is well approximated by a Gaussian distribution with mean $np$ and variance $np(1-p)$. From Eq.~\eqref{eq:wiener}, it then follows that $dW$ is also normally distributed with mean
\begin{equation}
    \label{eq:dwmean}
    E[dW] = \frac{d\tau}{|\alpha|}\left[\langle L^\dagger L \rangle_c\left(1-\frac{\langle x\rangle_c}{|\alpha|}\right)-\langle x\rangle_c^2\right]\rightarrow 0,
\end{equation}
where the arrow denotes the limit of large $|\alpha|$, and variance
\begin{equation}
    \label{eq:dwvar}
    E[dW^2] = d\tau\left(1+\frac{\langle x\rangle_c}{|\alpha|}+\frac{\langle L^\dagger L\rangle_c}{|\alpha|^2}\right)\left(1-\frac{\langle x\rangle_c}{|\alpha|}\right)^2+\mathcal{O}(dt)\rightarrow d\tau.
\end{equation}
For an illustration of why $dW^2=E[dW^2]$, we refer the reader to Ref.~\cite{jacobs_2010}.

Finally, we note that
\begin{equation}
    \label{eq:dnmin}
    \int_0^{d\tau} \left[\frac{dN}{|\alpha|}-|\alpha|dt\right] = \frac{dW+|\alpha|d\tau}{1-\frac{\langle x\rangle_c}{|\alpha|}}-|\alpha|d\tau \rightarrow dW+\langle x\rangle_c d\tau .
\end{equation}
The third line in Eq.~\eqref{eq:stochwiener1} thus vanishes as it is $\mathcal{O}(1/|\alpha|)$.
We hence arrive at Eq.~\eqref{Diffusion_SME}.
Equation~\eqref{eq:dnmin} also allows for deriving the smoothed version of the diffusive stochastic current given in Eq.~\eqref{Idiffwiener}
\begin{equation}
    \label{eq:diffstochapp}
    \frac{1}{d\tau} \int_0^{d\tau} I_{\rm diff}(t) dt = \nu \Bigg( \langle x\rangle_c + \frac{dW}{d\tau}\Bigg),
\end{equation}
where $I_{\rm diff}(t)$ is given in Eq.~\eqref{Idiff}. To recover the equations in the main text, we re-label $d\tau\rightarrow dt$ at the end of all calculations carried out in this appendix.

\section{Modeling a continuous measurement by a series of detectors}
\label{app:povm}

In this appendix we provide additional details on the Gaussian measurement approach to quantum diffusion, described in Eq.~\eqref{gaussian_POVM_kraus}. 
The Gaussian nature of the POVMs can be motivated by the central limit theorem \cite{Jacobs_2006}. Each Gaussian POVM models the effect of the continuous measurement over the time $d t$. We may equally well describe this by $N$ POVMs separated in time by the spacing $d t/N$, where $N$ can be taken arbitrarily large. By the central limit theorem, the sum of these $N$ measurement outcomes is approximately distributed by a Gaussian.

We first look at a single measurement.
From the probability distribution of outcomes in Eq.~\eqref{gaussian_POVM_probability} we get the average and variance of $z$ to be
\begin{equation}
    E(z) = \langle Y \rangle, 
    \qquad 
    {\rm Var}(z) = \frac{1}{4\lambda dt} + \langle Y^2 \rangle - \langle Y\rangle^2 \simeq \frac{1}{4\lambda dt},
\end{equation}
where the last equality holds since $dt$ is infinitesimal. 
From this, we see that the quantity 
\begin{equation}\label{gaussian_POV_wiener_definition}
    dW = 2\sqrt{\lambda} dt (z-\langle Y \rangle),
\end{equation}
will behave exactly like a Wiener increment; that is, $E(dW) = 0$ and $dW^2 = dt$. 
Let us now define $L = \sqrt{\lambda}Y$ and $\delta L = L - \langle L \rangle$. 
The map $M_z \rho M_z^\dagger$ can now be expanded, up to order $dt$ (and remembering that $dW^2 = dt$). 
As a result, one finds 
\begin{align}\label{gaussian_POVM_action_map}
    \frac{M_z \rho M_z^\dagger}{\tr\big\{ M_z\rho M_z^\dagger\}} &= 
     \rho + dt \Big[ \delta L \rho \delta L - \frac{1}{2} \{ \delta L^2, \rho\} \Big] + dW \{\delta L, \rho\}
    \\[0.2cm]
    &= \rho + dt \mathcal{D}[L]\rho + dW \big( \mathcal{H}\rho - \langle x \rangle \rho\big).
\end{align}
In the last line we simply recast the results in the language of the quantum diffusion SME in Eq.~\eqref{Diffusion_SME}: since $L^\dagger = L$, it follows that $\mathcal{D}[\delta L] = \mathcal{D}[L]$. 
Moreover $\mathcal{H}\rho = L \rho + \rho L = \{L, \rho\}$, as defined in Eq.~\eqref{H_k_superop} and $x = L + L^\dagger = 2L$ [Eq.~\eqref{quadrature}].

To finish, we now compose the action of the map~\eqref{gaussian_POVM_action_map}, with the remaining evolution of the system. 
Just for now, we define  $\mathcal{L}$ as the Liouvillian composing all other dynamical elements of the system, which includes the unitary part, as well as other potential dissipators. 
We then imagine a stroboscopic dynamics, where we alternate between the ``free'' evolution with $\mathcal{L}$, and the Gaussian map~\eqref{gaussian_POVM_action_map}. 
This will lead to a conditional density matrix evolving according to 
\begin{equation}
\label{eq:gaussian_POVM_updatec}
{\rho}_c(t+d t) = e^{\mathcal{L} dt}\frac{M_z \rho_c(t)M_z^\dagger}{\tr \big\{ M_z \rho_c(t)M_z^\dagger\big\}},
\end{equation}
which is conditioned on the outcome $z$, as well as all other past outcomes. 
Expanding $e^{\mathcal{L}dt}$ to first order in $dt$ and combining with the result of Eq.~\eqref{gaussian_POVM_action_map} then leads to 
\begin{equation}
\label{eq:belavkin}
\rho_c(t+dt) = \rho_c(t) +  
\Big(\mathcal{L} \rho + \mathcal{D}[L]\rho\Big)+ dW \big( \mathcal{H}\rho - \langle x \rangle \rho\big).
\end{equation}
This is precisely the Belavkin Equation~\eqref{Belavkin}. 
Note that now we can also absorb $\mathcal{D}[L]$ into the Liouvillian $\mathcal{L}$. 

Lastly, we compare the outcomes $z$ with the stochastic current $I_{\rm diff}(t)$ in Eq.~\eqref{Idiffwiener}.
In terms of $x = 2L = 2 \sqrt{\lambda} Y$, we can write Eq.~\eqref{gaussian_POV_wiener_definition} as 
$z = \frac{1}{2\sqrt{\lambda}} \big(\langle x \rangle + \frac{dW}{dt}\big)$. 
Comparing with~\eqref{Idiffwiener}, we therefore see that $z$ will reduce to $I_{\rm diff}$ provided we set the associated weight to $\nu = 1/(2\sqrt{\lambda})$. 


\section{Derivation of $F(t,t+\tau)$ in Eq.~(\ref{F_jump})}
\label{app:F}

To derive Eq.~\eqref{F_jump} we must compute 
\begin{equation}
\begin{aligned}
\label{autocorr_probability}
    E\big(I(t) I(t+\tau)\big) &= \frac{1}{dt^2} E\big(dN(t) dN(t+\tau)\big) \\
    &=\frac{1}{dt^2} \sum_{k,q} \nu_k \nu_q~ P\big(dN_k(t) = 1, dN_q(t+\tau) = 1\big), 
\end{aligned}
\end{equation}
where we used the fact that $dN$ takes on the value $\nu_k$ whenever $dN_k = 1$. 
To proceed, we therefore need the joint probability $P\big(dN_k(t) = 1, dN_q(t+\tau) = 1)$ that in time $t$ there is a jump in channel $k$ \emph{and} in time $t+\tau$ another jump in channel $q$ (irrespective of what happens in between). 
We assume $\tau > 0$.
To proceed, we first write it in terms of a conditional probability  
\begin{align}
\label{joint_click_prob}
    & P\big(dN_k(t) = 1, dN_q(t+\tau) = 1\big) \notag \\
    & \hspace{2.5cm} = P\big(dN_q(t+\tau) = 1 | dN_k(t) = 1\big) p_k(t),
\end{align}
\rev{where $p_k(t) = \tr\{L_k^\dagger L_k\rho(t)\} dt$ is the probability of a jump at time $t$.} The conditional probability is then computed as follows. 
If a jump occurred in channel $k$ at time $t$, then the state of the system must be updated to 
\begin{equation}
    \rho(t) \to \frac{ \mathcal{L}_k \rho(t)}{\tr\big\{ \mathcal{L}_k \rho(t)\big\}} =  \frac{dt}{p_k}\mathcal{L}_k \rho(t),
\end{equation}
where $\mathcal{L}_k\rho = L_k \rho L_k^\dagger$.
This is then used as the initial state, and the system evolves up to time $t+\tau$:
\begin{equation}
    \frac{dt}{p_k}\mathcal{L}_k \rho(t)
    \to 
    \frac{dt}{p_k} e^{\mathcal{L}\tau}\mathcal{L}_k \rho(t).
\end{equation}
The conditional probability is now given by the same formula~\eqref{pk} for $p_q(t+\tau)$, but using this as the initial state. That is
\begin{equation}
\label{conditional_click_prob}
    P\big(dN_q(t+\tau) = 1 | dN_k(t) = 1\big)  = \frac{dt^2}{p_k} \tr\big\{ \mathcal{L}_q e^{\mathcal{L}\tau}\mathcal{L}_k \rho(t)\big\}.
\end{equation}
Hence, all factors of $dt$ eventually cancel and we are left with 
\begin{align}
    E\big(I(t) I(t+\tau)\big) &=\sum_{k,q} \nu_k \nu_q  {\rm tr} \big\{ \mathcal{L}_q e^{\mathcal{L}\tau}\mathcal{L}_k \rho(t)\big\}
    \nonumber\\[0.2cm]
    &= \tr\big\{ \mathcal{J} e^{\mathcal{L}\tau} \mathcal{J} \rho(t)\big\}.
\end{align}
The case $\tau=0$ must be handled separately: 
\begin{equation}
    E\big(I(t)^2\big) = \frac{1}{dt^2} \sum_k \nu_k^2 p_k = \frac{1}{dt} \sum_k \nu_k^2 \tr\big\{ L_k^\dagger L_k \rho(t)\big\} := \frac{1}{dt} K(t),
\end{equation}
where $K(t)$ is  defined in Eq.~\eqref{K}. 
The result still depends on $1/dt$ and is therefore infinite in the limit $dt\to 0$.
We may thus formally replace it with a $\delta$ function, and therefore write 
\begin{equation}
    E\big(I(t) I(t+\tau)\big) = \delta(\tau) K(t) +  \tr\big\{ \mathcal{J} e^{\mathcal{L}\tau} \mathcal{J} \rho(t)\big\}.
\end{equation}
Subtracting $J(t) J(t+\tau)$ then yields Eq.~\eqref{F_jump}.

\section{Derivation of $F_{\rm diff}(t,t+\tau)$ in Eq.~(\ref{diffusion_F})}
\label{app:hom}

For simplicity, we will assume there is a single jump operator $L$ involved. The generalization to multiple operators is straightforward, but more cumbersome. 
The system will thus evolve according to the conditional dynamics 
\begin{equation}\label{hom_d_rho_c}
    d\rho_c = dt \mathcal{L} \rho_c + dW \Big[\mathcal{H}\rho_c - \langle x \rangle_c \rho_c\Big],
\end{equation}
where $\mathcal{H} \rho = L e^{-i \phi} \rho + \rho L^\dagger e^{i\phi}$ and $x = L e^{-i\phi} + L^\dagger e^{i\phi}$. Moreover, 
the associated stochastic current is simply 
\begin{equation}
    I_{\rm diff}(t) = \langle x\rangle_c(t) + \xi(t),
\end{equation}
where
$\langle x \rangle_c(t) = \tr\big\{x \rho_c(t)\big\}$ and 
$\xi(t) = dW/dt$. 

It is convenient to discretize time in steps of $dt$, and let $t_j = j dt$, $j =0,1,2,\ldots$. 
The randomness in the system stems from the Wiener increments $dW_j = dW(t_j)$. 
The expectation $\langle x \rangle_j = \langle x \rangle_c(t_j)$ depends, by definition, on $dW_0,\ldots,dW_{j-1}$. 
We will use the notation $dW_{0:j-1} = (dW_0,\ldots,dW_{j-1})$, to denote the collection of Wiener increments from $0$ to $j-1$. 

The evolution of $\rho_j = \rho_c(t_j)$ given by Eq.~\eqref{hom_d_rho_c} will be written more compactly as
\begin{equation}
    \rho_{j+1} = \mathcal{V}_{dW_j} \rho_j,
\end{equation}
where the map $\mathcal{V}_{dW_j} \rho$ is non-linear in $\rho$. The evolution from $t_j$ to $t_k > t_j$ can thus be written as 
\begin{equation}
    \rho_k = \mathcal{V}_{dW_{k-1}} \ldots 
    \mathcal{V}_{dW_{j+1}} \mathcal{V}_{dW_j} \rho_j.
\end{equation}

To obtain the two-point function we must compute (assuming $t_k > t_j$)
\begin{align}\label{hom_tmp_4_terms}
    E\Big(I_{\rm diff}(t_j) I_{\rm diff}(t_k)\Big) &=
    E\big(\langle x\rangle_j \langle x\rangle_k\big) + E\big( \langle x\rangle_j \xi_k \big) 
    \\[0.2cm]\nonumber
    &+E\big(\langle x \rangle_k \xi_j\big) + E\big(\xi_j\xi_k\big).
\end{align}
The 2nd term vanishes because $\langle x\rangle_j$ is independent of $\xi_k$ (since $t_k > t_j$), and $E(\xi_k) = 0$. 
The fourth term in~\eqref{hom_tmp_4_terms} 
yields $E(\xi_j \xi_k) = \delta_{jk}/dt \to \delta(t_j-t_k)$.

To compute the two remaining terms, we shall make use of the law of total expectation, which states that given a function $f(dW_{0:k-1})$, we can write 
\begin{equation}
    E\big( f(dW_{0:k-1})\big) 
    = E\Bigg( E\Big[ f(dW_{0:k-1}) \big| dW_{0:j-1}\Big] \Bigg),
\end{equation}
where 
$E(a|b) = \sum_a a p(a|b)$ is the conditional average, which is still a function of $b$. 
The above equation therefore establishes that to compute an average over $dW_{0:k-1}$, we can first compute an average over $dW_{j:k-1}$, with fixed $dW_{0:j-1}$, and afterwards average over $dW_{0:j-1}$.

For example, we can write 
\begin{equation}
    E\big(\langle x \rangle_k \langle x \rangle_j\big) = E\Bigg( \langle x \rangle_j E\Big[ \langle x \rangle_k \big| dW_{0:j-1}\Big]\Bigg).
\end{equation}
Now: 
\begin{align}
    E\Big[ \langle x \rangle_k \big| dW_{0:j-1}\Big]
    = \tr \Big\{ x E\Big[\mathcal{V}_{dW_{k-1}} \ldots \mathcal{V}_{dW_{j+1}} \mathcal{V}_{dW_j} \rho_j \big| dW_{0:j-1}\Big]\Big\},
\end{align}
where $\rho_j$ only depends on $dW_{0:j-1}$. 
Since $dW_{0:j-1}$ is fixed, we can compute the averages over $dW_{k-1},\ldots dW_{j+1}$ and $dW_{j}$. These averages are all independent and, for a given $\rho$, independent of $dW$,
\begin{equation}
    E(\mathcal{V}_{dW}\rho) = \rho + dt \mathcal{L} \rho \simeq e^{\mathcal{L} dt}\rho.
\end{equation}
We are thus left with 
\begin{equation}
    E\Big[ \langle x \rangle_k \big| dW_{0:j-1}\Big] = \tr \Big\{ x e^{\mathcal{L}(t_k-t_j)} \rho_j\Big\}.
\end{equation}
Hence 
\begin{equation}\label{hom_term1_final_result}
    E\big(\langle x \rangle_k \langle x \rangle_j\big)
    = 
    E\Big[ 
    \tr \big\{ x e^{\mathcal{L}(t_k - t_{j})} \rho_j \big\} 
    \tr \big\{ x \rho_j \big\} \Big],
\end{equation}
where this remaining average is over $dW_{0:j-1}$. 
Albeit complicated, we will see that this term will actually cancel out in the final expression.

Finally, we turn to the fourth term in Eq.~\eqref{hom_tmp_4_terms}. It reads 
\begin{align}
    E\big(\langle x \rangle_k \xi_j\big) &= 
    \tr\Big\{ x E\Big[ \xi_j \mathcal{V}_{dW_{k-1}} \ldots \mathcal{V}_{dW_j} \rho_j \Big]\Big\}
\end{align}
This is similar to before. But the term corresponding to $j$ is different since it is multiplied by $\xi_j = dW_j/dt$. 
In fact, from~\eqref{hom_d_rho_c} we see that 
\begin{equation}
    E(\xi_j \mathcal{V}_{dW_j}\rho) =  \mathcal{H} \rho - \rho \tr\big\{ x \rho\big\}.
\end{equation}
Hence 
\begin{align}
    E\big(\langle x \rangle_k \xi_j\big)
     &= 
    E\Big[ \tr \big\{ x e^{\mathcal{L} (t_k - t_{j})} \mathcal{H} \rho_j \big\}\Big] 
    \\[0.2cm]\nonumber
    &~~- E\Big[ 
    \tr \big\{ x e^{\mathcal{L}(t_k - t_{j})} \rho_j \big\} 
    \tr \big\{ x \rho_j \big\} \Big].
\end{align}
When we plug this in Eq.~\eqref{hom_tmp_4_terms}, the second line will cancel out the term obtained in Eq.~\eqref{hom_term1_final_result}. In fact, although the exponentials are slightly different in both, this difference becomes negligible when $dt\to 0$. 
Eq.~\eqref{hom_tmp_4_terms} therefore finally reduces to 
\begin{equation}
    E\Big(I_{\rm diff}(t_j) I_{\rm diff}(t_k)\Big) = 
    \delta(t_j-t_k) + E\Big[ \tr \big\{ x e^{\mathcal{L} (t_k - t_{j})} \mathcal{H} \rho_j \big\}\Big] .
\end{equation}
This finishes the proof. To obtain Eq.~(\ref{diffusion_F}) we must simply 
subtract $J_{\rm diff}(t_j) J_{\rm diff}(t_k)$, and take the limit $dt\to 0$.

\section{White noise}
\label{app:Poisson}

A stationary Poisson process is described by a stochastic increment $dN_k(t)$ whose average is a constant, $E[dN_k(t)] = p_k = dt\times {\rm const.}$, and which is independent of any previous jumps, so that 
\begin{equation}
    P(dN_k(t+\tau)=1|dN_q(t) = 1) = p_k,
\end{equation}
for any $\tau>0$. Consider now a current $I(t) = \sum_k \nu_k dN_k(t)/dt$ constructed from a linear combination of stationary Poisson increments. The average current is 
\begin{equation}
    \label{current_av_Poisson}
    J = \sum_k \frac{\nu_k p_k}{dt}.
\end{equation}
Following the arguments of Appendix~\ref{app:F}, for $\tau>0$ we have
\begin{align}
\label{Poisson_autocorr}
    E\big( I(t) I(t+\tau)\big) & = \frac{1}{dt^2}\sum_{k,q} \nu_k \nu_q P(dN_k(t+\tau)=1|dN_q(t) = 1) p_q \notag \\
    & = \frac{1}{dt^2}\sum_{k,q} \nu_k \nu_q p_k p_q \notag \\
    & = J^2.
\end{align}
Therefore, the connected autocorrelation function for arbitrary $\tau$ is simply 
\begin{align}
\label{Poisson_autocorr_connected}
    F(\tau) = E\big( I(t) I(t+\tau)\big)  - J^2 = K\delta(\tau),
\end{align}
where $K = \sum_k \nu_k^2 p_k/dt$. 
The two-point function of a Poisson white noise therefore only contains a Dirac-delta correlation. Deviations from white noise in the power spectrum of a stationary quantum-jump process correspond to a departure from pure Poisson statistics.

\rev{The Fano factor [c.f.~Eq.~\eqref{eq:fano}] for a linear combination of Poisson processes thus reads
\begin{equation}
    f=\frac{K}{J} = \frac{\sum_k \nu_k^2 p_k}{\sum_k \nu_k p_k}.
\end{equation}
For $\nu_k = 1$, the Fano factor reduces to one which explains why $f=1$ is associated with Poissonian statistics. Note that if all weights are equal, $\nu_k=q$, the Fano factor is usually defined as 
\begin{equation}\label{eq:fano2}
    f(t) = \frac{{\rm Var}(N(t))}{qE(N(t))},
\end{equation}which still results in $f=1$.}

The counterpart of white noise for a time-continuous process is simply the Wiener process, $W(t)$, whose increment obeys $E[dW(t)] = 0$, $dW(t)^2 = dt$, and $E[dW(t) dW(t+\tau)] = 0$ for $\tau\neq 0$. These properties ensure that fluctuations at different times are statistically independent while retaining a finite power spectrum. A current formed from a linear combination of independent Wiener increments can be written as $I(t) = J + \sum_k \nu_k dW_k(t)/dt$, for a constant $J$. It is then straightforward to see that, for $\tau>0$, 
\begin{align}
\label{Wiener_autocorr}
    E\big( I(t) I(t+\tau)\big) - J^2 = 0,
\end{align}
and that when $\tau=0$,
\begin{align}
\label{Wiener_fluct_tau_zero}
    E\big( I(t) I(t)\big) - J^2 & =  \frac{1}{dt^2}\sum_{k,q}\nu_k\nu_q E[dW_k(t) dW_q(t)] \notag \\
    & = \frac{1}{dt}\sum_{k} \nu_k^2,
\end{align}
using the property $E[dW_k(t) dW_q(t)] = \delta_{kq}dt$. In the limit $dt\to 0$, we can thus write
\begin{equation}
    F(\tau) = K_{\rm diff}\delta(\tau),
\end{equation}
with $K_{\rm diff}=\sum_k\nu_k^2.$

\section{Filtering basics}
\label{sec:Filtering basics}

\begin{figure*}[!t]
    \centering
    \includegraphics[width=\textwidth]{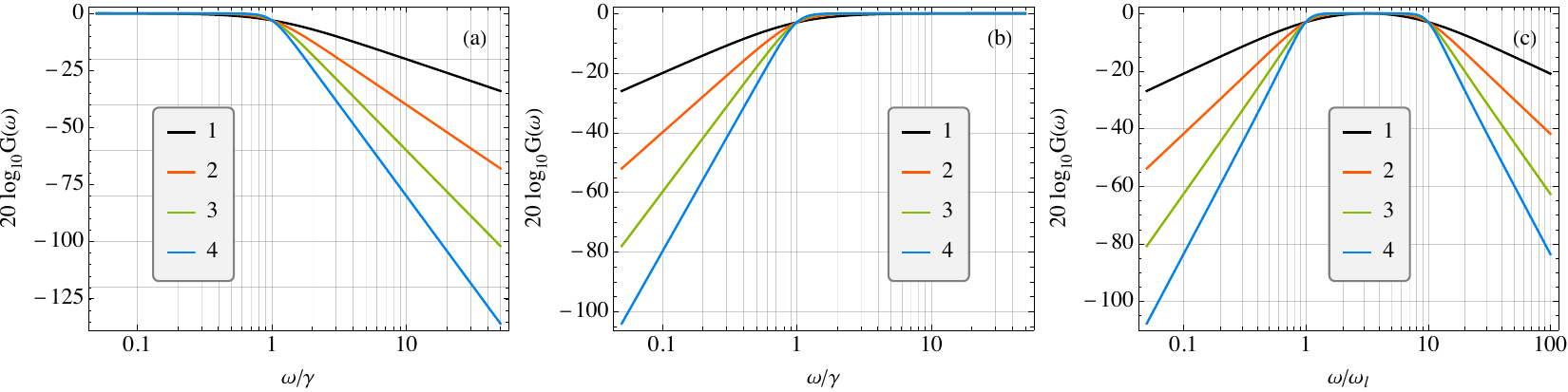}
    \caption{Gain function for Butterworth filters of different orders, $n = 1,2,3,4$.
    (a) Lowpass. (b) Highpass. (c) Bandpass with $\omega_h = 10\omega_l$.  
 }
    \label{fig:gain}
\end{figure*}

In this  appendix, we  touch on some basic concepts in signal processing that were used in Sec.~\ref{sec:Homodye detection of a qubit}.
A filter converts an input signal $X(t)$ into an output signal $Y(t)$ by means of a convolution
\begin{equation}\label{filter_time_space}
    Y(t) = \int\limits_{-\infty}^\infty h(t-t') X(t') dt',
\end{equation}
for a certain function $h(t)$. 
It is convenient to define $h(t)$ to be zero when $t<0$;
A commonly used filter in the quantum optics literature \cite{Warszawski_2002_1, Warszawski_2002_2, Sarovar_high_2005, Wheatley_adaptive_2010, Liu_deterministic_2010, Feng_generating_2011, annby_2022}---which was similarly defined in Eq.~(\ref{eq:filtd}) --- is that given by 
\begin{equation}\label{low_pass_basic}
    h(t) = \gamma e^{-\gamma t} \theta(t),
\end{equation}
where $\gamma$ is called the filter's bandwidth.
Here $\theta(t)$ is the Heaviside function, which makes this filter \emph{causal}; i.e., $Y(t)$ only depends on past values of $X(t)$.
Eq.~\eqref{low_pass_basic} is an example of a lowpass filter because, as we will see, it tends to remove high frequencies but allow for low frequencies to pass.  

Filters are more commonly defined in Fourier space. 
Using the convolution theorem we get 
\begin{equation}
    Y(\omega) = h(\omega) X(\omega),
\end{equation}
where $h(\omega)$, the Fourier transform of $h(t)$, is called the \emph{transfer function}.
For the filter~\eqref{low_pass_basic} we find 
\begin{equation}\label{low_pass_H_omega}
    h(\omega) = \frac{1}{1 - i\omega/\gamma}.
\end{equation}
This complex function is more conveniently analyzed in terms of its phase and absolute value. 
The latter, in particular, is called the ``gain'':
\begin{equation}
    G(\omega) = |h(\omega)|.
\end{equation}
Further, for visualization purposes, the gain is measured in decibels (Db) and is thus plotted on a log scale given by 
\begin{equation}
    G_{\rm Db}(\omega) = 20 \log_{10}\{G(\omega)\}\,.
\end{equation}
For the lowpass filter~\eqref{low_pass_H_omega}, 
\begin{equation}\label{low_pass_G}
    G(\omega) = \frac{1}{\sqrt{1 + (\omega/\gamma)^2}}.
\end{equation}
Eq.~\eqref{low_pass_G} is actually just one element of the family of \emph{Butterworth filters}, defined by the gain function 
\begin{equation}\label{butter_G}
    G(\omega) = \frac{1}{\sqrt{1 + (\omega/\gamma)^{2n}}},
\end{equation}
where $n$ is called the order of the filter, such that $n=1$ recovers Eq.~\eqref{low_pass_G}. 
Results for different values of $n$ are shown in Fig.~\ref{fig:gain}(a).
As anticipated, the filter passes frequencies $\omega < \gamma$, and attenuates frequencies $\omega > \gamma$.
Increasing $n$ gives rise to sharper cut-offs. 
For $n\rightarrow \infty$ the Gain  becomes rectangular at $\gamma$, leaving frequencies $\omega<\gamma$ unaffected, while entirely suppressing  $\omega>\gamma$.
There is a phase freedom in how to define the corresponding transfer function $h(\omega)$ which generates Eq.~\eqref{butter_G}. One possible choice is to write it as
\begin{equation}\label{butter_h}
    h(\omega) = \prod\limits_{k=1}^n \frac{(-1)^n}{(i\omega/\gamma) + e^{i (2k+n-1)\pi/2n}}.
\end{equation}

We can also have highpass filters, which have the opposite effect. 
For example, consider a function $h(t)$ similar to Eq.~\eqref{low_pass_basic}, but convolving with $dX/dt$ instead of $X(t)$: 
\begin{equation}
    Y(t) =\int\limits_{-\infty}^\infty e^{-\gamma (t-t')} \theta(t-t') \frac{dX}{dt'},
\end{equation}
with no factor of $\gamma$ in front in order to make $Y(t)$ have the same units as $X(t)$.
Applying the convolution theorem, and recalling that the Fourier transform of $dX/dt$ is $-i\omega X(\omega)$, we get 
\begin{equation}
    Y(\omega) =  \frac{-i \omega/\gamma}{1-i \omega/\gamma} X(\omega) := \tilde{h}(\omega) X(\omega).
\end{equation}
The corresponding gain is shown in Fig.~\ref{fig:gain}(b) and, as we can seen, behaves exactly opposite to the lowpass case, filtering out low frequencies but allowing high frequencies to pass. 
A more systematic way of creating highpass filters is to start with the lowpass one and apply the transformation $\omega \to 1/\omega$ (and, to make things dimensionless, we also set $\gamma \to 1/\gamma$). 
With this, one can then obtain a family of highpass Butterworth filters starting with Eq.~\eqref{butter_G}.
Results for this are shown in Fig.~\ref{fig:gain}(b) for different values of $n$. 

Finally, we can have a bandpass filter, which selects only within a certain window, specified by two frequencies $\omega_l$ and $\omega_h > \omega_l$. 
Bandpass filters can also be constructed from lowpass filters, by shifting the frequencies as 
\begin{equation}
    \frac{\omega}{\gamma} \rightarrow \frac{1}{\sqrt{\Delta \omega}}\left(\frac{\omega}{\omega_{0}} + \frac{\omega_{0}}{\omega}\right)\,,
\end{equation}
where $\Delta \omega = \omega_{h} - \omega_{l}$ and $\omega_{0} = \sqrt{\omega_{h} \omega_{l}}$.
Examples for the Butterworth case with different $n$ are shown in Fig.~\ref{fig:gain}(c).

\section{Adjoint Lindblad evolution}
\label{app:adjoint_diss}

The adjoint Liouvillian defined in Eq.~\eqref{adjoint_liouvillian} can be understood as the generator for a kind of ``Heisenberg-picture'' evolution for open quantum systems. In particular, the operator $A(t) = {\rm e}^{\mathcal{L}^\dagger t}A$ can be obtained as the solution of the adjoint QME
\begin{equation}
    \frac{d}{dt}A(t) = \mathcal{L}^\dagger A(t), \qquad A(0) = A.
\end{equation}
This solution can then be used to generate expectation values,
\begin{equation}
    \langle A(t)\rangle = \tr \left[A(t) \rho(0)\right] = \tr \left[A \rho(t)\right],
\end{equation}
in addition to two-time correlation functions as shown in Eq.~\eqref{two_time_function}. This is especially useful since one may only need to know the evolution of specific observables, rather than the full information contained in the density matrix. 

As a simple example, let us consider Example D: a cavity without driving or nonlinearity so that $G=U=0$. Moreover, let us allow for the presence of finite-temperature damping so that the Liouvillian becomes
\begin{equation}
    \mathcal{L}\bullet = -i[H,\bullet] + \kappa(\bar{N}+1)\mathcal{D}[a]\bullet + \kappa \bar{N}\mathcal{D}[a^\dagger]\bullet.
\end{equation}
The adjoint QME for the annihilation operator is 
\begin{equation}
    \frac{da}{dt} = -i \Delta a - \frac{1}{2}\kappa a,
\end{equation}
while the equation for the number operator $n=a^\dagger a$ is 
\begin{equation}
    \frac{dn}{dt} = \kappa (\bar{N}-n).
\end{equation}
The solutions of these equations are easily found to be
\begin{align}
    \label{Lindblad_heisenberg_solutions_a}
    & a(t) = e^{-i\Delta t - \kappa t/2} a(0), \\ \label{Lindblad_heisenberg_solutions_n} & n(t) = \bar{N} + e^{-\kappa t}\left[n(0) - \bar{N}\right],
\end{align}
which is far simpler than solving the master equation $d\rho/dt = \mathcal{L}\rho$, as the latter is an infinite system of coupled differential equations for the matrix elements of $\rho$.

It is crucial to note, however, that --- unlike for purely unitary dynamics generated by a Hamiltonian --- one cannot obtain the solution~\eqref{Lindblad_heisenberg_solutions_n} for $n(t)$ directly from Eq.~\eqref{Lindblad_heisenberg_solutions_a}, i.e. $n(t)\neq a^\dagger (t) a(t)$, or more precisely
\begin{equation}
    e^{\mathcal{L}^\dagger t}(a^\dagger a) \neq  e^{\mathcal{L}^\dagger t}(a^\dagger) e^{\mathcal{L}^\dagger t}(a).
\end{equation}
This shows that the evolution operator $e^{\mathcal{L}^\dagger t}$ is not distributive over the operator product. Equivalently, the adjoint Liouvillian does not obey the Leibniz product rule, since 
\begin{equation}
    \frac{d n}{dt} = \mathcal{L}^\dagger(a^\dagger a) \neq \mathcal{L}^\dagger(a^\dagger) a + a^\dagger \mathcal{L}^\dagger(a) .
\end{equation}
Therefore, care must be taken when working with the adjoint QME and its solutions: superoperators always act on \textit{everything} to their right. For example, it is a common misconception that the adjoint Liouvillian does not preserve the canonical commutation relations because $[a(t),a^\dagger(t)] \neq \mathbbm{1}$. However, this misunderstanding arises simply from an incorrect application of the evolution superoperator, which should act on the commutator as a whole:
\begin{equation}
    e^{\mathcal{L}^\dagger t}\left( [a,a^\dagger]\right) = e^{\mathcal{L}^\dagger t}\left( \mathbbm{1}\right) = \mathbbm{1}.
\end{equation}
The last equality follows because $\mathcal{L}^\dagger(\mathbbm{1})=0$ (this is a consequence of trace preservation, as shown explicitly below). Note that the same caution applies to all open quantum systems, not only to the specific example of a quantum oscillator. 

The only exception to this rule are the purely Hamiltonian superoperators of the form $\mathcal{L}\bullet = -i[H,\bullet]$, which generate the evolution superoperator $\mathcal{U} = e^{\mathcal{L}t}$. The corresponding adjoint superoperator acts as $\mathcal{U}^\dagger(A) = U^\dagger A U$, where $U = e^{-iHt}$ is a unitary operator. It is straightforward to check that $\mathcal{U}(\mathcal{U}^\dagger(A)) = A$ for any operator $A$, hence $\mathcal{U}$ itself is a \textit{unitary superoperator}: its adjoint is also its inverse. Such unitary superoperators are distributive because 
\begin{equation}
   \mathcal{U}(AB) = UABU^\dagger = UAU^\dagger UBU^\dagger = \mathcal{U}(A) \mathcal{U}(B).
\end{equation}
Equivalently, a Leibniz rule holds for a Hamiltonian generator $\mathcal{L}^\dagger = i[H,\bullet]$, since
\begin{equation}
 [H,AB] = [H,A]B + A[H,B],
\end{equation}
which is a well-known commutator identity. 

Finally, we note that the adjoint Liouvillian $\mathcal{L}^\dagger$ is defined by the relation
\begin{equation}
    \tr[A^\dagger \mathcal{L}(B)] = \tr[\mathcal{L}^\dagger(A^\dagger)B].
\end{equation}
In the vectorization language of Sec.~\ref{sec:vectorization}, this is equivalent to the standard definition of the adjoint with respect to the Hilbert-Schmidt inner product, $\llangle A|B\rrangle = \tr[A^\dagger B]$~\footnote{Note that $\mathcal{L}$ and its adjoint preserve hermiticity, i.e. $\mathcal{L}(A^\dagger) = [\mathcal{L}(A)]^\dagger$.}, i.e.
\begin{equation}
    \llangle A|\mathcal{L}B\rrangle = \llangle \mathcal{L}^\dagger A|B\rrangle.
\end{equation}
This is the underlying reason for the choice of notation $\mathcal{L}^\dagger$. Moreover, it implies that $\mathcal{L}^\dagger$ can be found simply by taking the conjugate transpose of the matrix $\mathcal{L}$, i.e.
\begin{equation}
    \mathcal{L} = \sum_{j\neq 0} \lambda_j |x_j\rrangle\llangle y_j| \quad \Rightarrow \quad \mathcal{L}^\dagger = \sum_{j\neq 0} \lambda^*_j |y_j\rrangle\llangle x_j|.
\end{equation}
Therefore, the right (left) eigenvectors of $\mathcal{L}$ are the left (right) eigenvectors of $\mathcal{L}^\dagger$, with the same eigenvalues up to complex conjugation. In particular, since the identity is always a left eigenvector of $\mathcal{L}$ with eigenvalue $0$, we have
\begin{equation}
    \mathcal{L}^\dagger |\mathbbm{1}\rrangle = 0 \quad \Rightarrow\quad \mathcal{L}^\dagger(\mathbbm{1}) = 0.
\end{equation}

\section{Full counting statistics for quantum diffusion}
\label{app:fcs_diffusive}

\subsection{As a limiting case of quantum jumps}

To derive the tilted Liouvillian for quantum diffusion, as in Sec.~\ref{sec:FCS_quantum_diffusion}, we start with Eq.~\eqref{tilted_liouvillian} and apply the transformation in Eq.~\eqref{gauge}, which results in
\begin{equation}
    \label{eq:tiltdiff1}
    \mathcal{L}_\chi\rho = \mathcal{L}\rho + \sum_{k=1}^\nops\left(e^{i\chi\nu_k}-1\right)(\alpha_k+L_k)\rho(\alpha_k^*+L_k^\dagger).
\end{equation}
The corresponding probability distribution stemming from FCS refers to the total charge $N(t)=\sum_k\nu_kN_k(t)$. 
Instead, we are interested in 
\begin{equation}
    \label{eq:ndiff}
    N_{\rm diff}(t) = \int_0^t dt' I_{\rm diff}(t') = \sum_{k=1}^\nops\nu_k \Bigg(\frac{N_k(t)}{|\alpha_k|}  - |\alpha_k|t\Bigg),
\end{equation}
where each counting variable is  re-scaled and there is an off-set that is independent of the system dynamics. The re-scaling is obtained by changing the weights in $\mathcal{L}_\chi$ as $\nu_k\rightarrow \nu_k/|\alpha_k|$. The off-set can be included by the replacement
\begin{equation}
    \label{eq:shifttiltliou}
    \mathcal{L}_\chi \rightarrow \mathcal{L}_\chi - i\chi\sum_{k=1}^\nops\nu_k|\alpha_k|.
\end{equation}
This can actually be anticipated from Eq.~\eqref{FCS_P} since the probability distribution $\tilde{P}$ for $\tilde{n}=n-ct$, where $ct$ is a constant offset in the total charge, can be written as
\begin{equation}
    \label{eq:shiftedpn}
    \tilde{P}(\tilde{n},t) = P(\tilde{n}+ct,t) =  \int\limits_{-\infty}^\infty \frac{d\chi}{2\pi} e^{-i \tilde{n} \chi}~\tr \{e^{(\mathcal{L}_\chi-i\chi c)t}\rho_0\big\},
\end{equation}
where $P$ denotes the probability distribution for $n$.

Applying these replacements to Eq.~\eqref{eq:tiltdiff1}, we obtain the tilted Liouvillian that describes the full counting statistics for $N_{\rm diff}(t)$
\begin{equation}
    \label{eq:tiltdiff2}
    \begin{aligned}
    \mathcal{L}_\chi\rho = &\mathcal{L}\rho + \sum_{k=1}^\nops\left(e^{i\chi\frac{\nu_k}{|\alpha_k|}}-1\right)(\alpha_k+L_k)\rho(\alpha_k^*+L_k^\dagger)\\&- i\chi\sum_{k=1}^\nops\nu_k|\alpha_k|.
    \end{aligned}
\end{equation}
Taking the limit $|\alpha_k|\rightarrow\infty$, this reduces to Eq.~\eqref{L_chi_diffusion} in the main text.

\subsection{For Gaussian POVMs}

We can also arrive at the same result for the case of Gaussian measurements in Eq.~\eqref{gaussian_POVM_kraus} (see Ref.~\cite{bednorz_2008} for a similar discussion).
As discussed in Appendix~\ref{app:povm}, the outcomes $z$ represent the output current in this case, corresponding to a measurement of the operator $x = 2L = 2 \sqrt{\lambda}Y$ with weight $\nu = 1/(2\sqrt{\lambda})$. 
As in Sec.~\ref{sec:FCS} and Eq.~\eqref{eq:nresolvedkraus}, the full counting statistics can be obtained by considering the $n$-resolved density matrix. Under a continuous measurement described by Eq.~\eqref{gaussian_POVM_kraus}, its time-evolution is given by
\begin{equation}
\rho_n(t+dt) =e^{\mathcal{L}dt}\int_{-\infty}^\infty dz{M}_z\rho_{n-z dt}(t)M_z,
\end{equation}
where $\mathcal{L}$ denotes the Liouvillian in the absence of the continuous measurement.
This equation implies that the measurement outcome $z$ (corresponding to the instantaneous diffusive current) increases the total charge by $zdt$. Fourier transforming this equation, we find
\begin{equation}
    \label{eq:povmfcs2}
    \rho_\chi(t+dt) = e^{\mathcal{L}dt}\int_{-\infty}^\infty dze^{i\chi zdt}{M}_z\rho_{\chi}(t)M_z.
\end{equation}
The integral over $z$ can be evaluated upon inserting identities $\sum_y |y\rangle\langle y|$, resolved in the eigenstates of $Y$, on the left and right of the density matrix. Expanding to linear order in $dt$, we find the generalized QME
\begin{equation}
    \label{eq:povmfcs3}
    \frac{\partial}{\partial t}\rho_\chi(t) = \mathcal{L}\rho_\chi(t)+\lambda \mathcal{D}[Y]\rho_\chi(t)-\frac{\chi^2}{8\lambda}\rho_\chi(t)+i\frac{\chi}{2}\{Y,\rho_\chi(t)\}.
\end{equation}
This equation nicely illustrates the trade-off between measurement backaction (term proportional to $\lambda$) and imprecision noise (term proportional to $1/\lambda$). We note that for $L = \sqrt{\lambda}Y$ and $\nu=1/(2\sqrt{\lambda})$, the last equation is equivalent to Eq.~\eqref{L_chi_diffusion}. Note that in the latter, the term $\mathcal{D}[Y]$ is included in $\mathcal{L}$, in contrast to Eq.~\eqref{eq:povmfcs3}.

\section{Moments from Full Counting Statistics}
\label{app:FCS_moments}

In this appendix we consider the following problem. 
We have a system described by a generic tilted Liouvillian $\mathcal{L}_\chi$, whose form we do not wish to specify. 
Our goal is to obtain the formulas for the first two cumulants.
The starting point is Eq.~\eqref{FCS_moments} which, written more explicitly, reads
\begin{equation}
    E\big[ N(t)^j \big] = (-i \partial_\chi)^j {\rm tr} \big\{ e^{\mathcal{L}_\chi t} \rho(0)\big\} \Bigg|_{\chi = 0}
\end{equation}
Thus, to use this formula we must be able to compute  $(-i\partial_\chi)^j e^{\mathcal{L}_\chi t}$. 
Derivatives of this kind, however, are quite difficult to handle because $\mathcal{L}_\chi$ generally does not commute with $\partial_\chi \mathcal{L}_\chi$. 
We can see this using the following general Baker-Campbell-Hausdorff formula: Let $G_\phi$ denote any operator (or superoperator) depending on some parameter $\phi$. 
Then 
\begin{align}
\label{BCH_derivative}
    \frac{d e^{G_\phi}}{d\phi} &= \Bigg( G' + \frac{1}{2} [G,G'] + \frac{1}{3!} [G,[G,G']] + \ldots\Bigg) e^{G_\phi}\\[0.2cm]
    &= e^{G_\phi}\Bigg( G' - \frac{1}{2} [G,G'] + \frac{1}{3!} [G,[G,G']] + \ldots\Bigg) ,
\label{BCH_derivative2}    
\end{align}
where  $G' = \partial_\phi G$.
The series is infinite because $G$ and $G'$ generally do not commute with each other. 
An alternative formula, also highlighting the non-commutativity, is 
\begin{equation}\label{BCH_integral}
    \partial_\phi e^{G_\phi} = \int\limits_0^1 dy~ e^{G_\phi y} G' e^{G_\phi(1-y)},
\end{equation}
which also shows that this will generally be a difficult computation. 

Because of these difficulties, we shall take another route. 
Start with the first moment:
\begin{equation}\label{app_FCS_moments_average_charge}
    E\big(N(t)\big) = -i \partial_\chi \tr\big\{ e^{\mathcal{L}_\chi t} \rho_0\big\}\Bigg|_{\chi=0}.
\end{equation}
Instead of computing it, we look instead at the current:
\begin{equation}\label{app_FCS_moments_tmp1}
    J(t) = \frac{d}{dt} E\big(N(t)\big) 
    = - i \frac{\partial}{\partial t} \frac{\partial}{\partial \chi} \tr\big\{ e^{\mathcal{L}_\chi t} \rho_0\big\}\Bigg|_{\chi=0}.
\end{equation}
If we differentiate first with respect to $t$, we do not encounter the difficulties  in Eq.~\eqref{BCH_derivative} because $G = \mathcal{L}_\chi t$ and $\dot{G} = \mathcal{L}_\chi$, so that $[G,\dot{G}]=0$.
In other words, $\partial_t e^{\mathcal{L}_\chi t} = \mathcal{L}_\chi e^{\mathcal{L}_\chi t}$.
In terms of the solution $\rho_\chi(t) = e^{\mathcal{L}_\chi t} \rho_0$ of the generalized QME~\eqref{gQME},
Eq.~\eqref{app_FCS_moments_tmp1} can thus be written as 
\begin{equation}
    J(t) = -i \frac{\partial}{\partial_\chi} \tr \big\{ \mathcal{L}_\chi \rho_\chi(t)\big\}\Bigg|_{\chi=0}.
\end{equation}
Finally, we take the derivative with respect to $\chi$, and  set $\chi=0$, leading to:
\begin{equation}
    J(t) = -i \tr\big\{ \mathcal{L}' \rho(t) + \mathcal{L} \rho'(t)\big\},
\end{equation}
where, here and henceforth, prime always denotes the derivative with respect to $\chi$, evaluated at $\chi=0$. That is, 
$\mathcal{L}' = \partial_\chi \mathcal{L}_\chi \big|_{\chi=0}$, and 
$\rho' = \partial_\chi \rho_\chi \big|_{\chi=0}$.
The second term in $J(t)$ vanishes, because $\mathcal{L} =\mathcal{L}_{\chi=0}$ is the true Liouvillian, which is traceless. 
We therefore finally arrive at 
\begin{equation}
    J(t) = -i ~\tr\big\{ \mathcal{L}' \rho(t)\big\} = \tr\big\{ \mathcal{J} \rho(t)\big\},
\end{equation}
with the identification $\mathcal{J} = -i \mathcal{L}'$, as in Eq.~\eqref{J_K_general_tilted}.

Next we do the same for the variance, or its time-derivative, which is the noise $D$  [Eq.~\eqref{D}]. 
Using Eq.~\eqref{FCS_moments} for the second moment we get 
\begin{align}
    D(t) &= (-i)^2 \frac{\partial}{\partial t} \frac{\partial^2}{\partial \chi^2} \tr \big\{ e^{\mathcal{L}_\chi t} \rho_0\big\}\Bigg|_{\chi=0} - 2 J(t) E\big(N(t) \big)
    \nonumber
    \\[0.2cm]
    &= 
    - \frac{\partial^2}{\partial \chi^2} \tr\big\{ \mathcal{L}_\chi \rho_\chi(t)\big\}\Bigg|_{\chi=0} - 2 J(t) E\big(N(t)\big)
    \nonumber
    \\[0.2cm]
    &= - \tr\big\{ \mathcal{L}'' \rho + 2 \mathcal{L}' \rho' + \mathcal{L} \rho''\big\} - 2 J(t) E\big(N(t)\big).
\end{align}
The term $\tr \big\{\mathcal{L} \rho'' \big\}$ vanishes because $\mathcal{L}$ is traceless. 
In the last term, we can use $J(t) = -i \tr\big\{ \mathcal{L}' \rho(t)\big\}$ and 
$E\big(N(t)\big) = -i \tr \big\{ \rho'(t)\big\}$ [which follows from Eq.~\eqref{app_FCS_moments_average_charge}].
We are then left with
\begin{equation}\label{app_FCS_moments_average_D_tmp}
    D(t) = -\tr \big\{ \mathcal{L}'' \rho(t)\big\} - 2 \tr\big\{ \mathcal{L}' \rho'\big\} + 2 \tr\big\{ \mathcal{L}' \rho\big\} \tr \big\{ \rho' \big\}.
\end{equation}
The complicated term here is $\tr(\rho') = \partial_\chi\tr\big\{ e^{\mathcal{L}_\chi t} \rho(0)\big\} \big|_{\chi=0}$. 
We can compute it using Eq.~\eqref{BCH_integral} [and changing variables to $\tau = t(1-y)$]:
\begin{equation}
    \frac{\partial e^{\mathcal{L}_\chi t}}{\partial \chi} \Bigg|_{\chi=0} 
    = \int\limits_0^t d\tau~ e^{\mathcal{L}(t-\tau)} \mathcal{L}' e^{\mathcal{L}\tau}.
\end{equation}
Multiplying by $\rho_0$ and noticing that $e^{\mathcal{L}\tau} \rho_0 = \rho(\tau)$ we then finally conclude that 
\begin{equation}\label{app_FCS_moments_rho_prime}
    \rho'(t) \equiv \partial_\chi \rho_\chi(t) \big|_{\chi=0} 
    = \int\limits_0^t d\tau~e^{\mathcal{L}(t-\tau)} \mathcal{L}' \rho(\tau).
\end{equation}
Inserting this in  Eq.~\eqref{app_FCS_moments_average_D_tmp} finally yields 
\begin{equation}
\begin{aligned}
    D = - \tr\big\{ \mathcal{L}'' \rho(t)\big\} -& 2\int\limits_0^t d\tau \Big\{
        \tr \Big[
            \mathcal{L}' e^{\mathcal{L}(t-\tau)} \mathcal{L}' \rho(\tau) 
        \Big]
        \\[0.1cm]
        &
        - \tr\big[ \mathcal{L}'\rho(t)\big ]
        \tr\big[ \mathcal{L}'\rho(\tau)\big ]
    \Big\}.
\end{aligned}
\end{equation}
This can now be compared with the general definition~\eqref{D_integral_time}, which allows us to identify the two-point correlation function $F(t,t-\tau)$. 
As a result, we find exactly Eq.~\eqref{F_jump}, provided we identify $\mathcal{J}$ and $K$ as per Eq.~\eqref{J_K_general_tilted}. 

This analysis, albeit somewhat involved, proves that for \emph{any} tilted Liouvillian, all formulas for $J,F,S,D$ developed in this tutorial continue to hold, provided we make the identification 
in Eq.~\eqref{J_K_general_tilted}.

\section{The long-time limit and the SCGF}
\label{app:scgf}
Here we show how the SCGF reduces to an eigenvalue of the tilted Liouvillian [cf.~Eqs.~\eqref{tilted_liouvillian} and \eqref{L_chi_diffusion}]. To this end, we re-write the moment generating function as
\begin{equation}
    \label{eq:momgenspec}
    M(\chi,t) =\tr \big\{ e^{\mathcal{L}_\chi t} \rho_0\big\} = \sum_j e^{\lambda_j(\chi) t}\llangle \id |x_j(\chi)\rrangle\llangle y_j(\chi)|\rho_0\rrangle,
\end{equation}
where we used the eigendecomposition of $\mathcal{L}_\chi$, in the same form as that for $\mathcal{L}$ shown in Eq.~\eqref{L_eigen_decomp}. 
In the long-time limit, the sum will be dominated by the eigenvalue with the largest real part, denoted $\lambda_0(\chi)$. Taking the logarithm, we find the cumulant generating function will therefore behave asymptotically as 
\begin{equation}
    \label{eq:cgflongtime}
    C(\chi,t) \simeq \lambda_0(\chi) t +\ln \llangle \id |x_0\rrangle\llangle y_0|\rho_0\rrangle,
\end{equation}
Since the second term is independent of time, it follows that the SCGF is given by
\begin{equation}
    \label{eq:scgfeig}
    C(\chi) = \lim\limits_{t\to \infty} \frac{d}{dt} C(\chi,t) = \lambda_0(\chi).
\end{equation}

\section{Drazin inverse and connection with Moore-Penrose}
\label{app:pseudoinverse}

The Drazin pseudo-inverse was defined in Eqs.~\eqref{Drazin_integral_def} or~\eqref{eigen_Drazin}.
\rev{The equivalence between the two definitions follows readily from Eq.~\eqref{exp_L}.}
We use pseudo-inverses because $\mathcal{L}$ itself is not invertible. 
In this appendix we explore in more detail the properties of the Drazin pseudo-inverse. 
For Hermitian matrices, the Drazin pseudo-inverse coincides with the Moore-Penrose pseudo-inverse (to be more properly defined below). 
But when it is not Hermitian, as is always the case with Liouvillians, the two are different.
In this appendix, we also clarify their relation [Eq.~\eqref{Drazin_MP_relation} below].

The basic intuition of any pseudo-inverse is that ``we invert what we can.'' This is clear from Eq.~\eqref{eigen_Drazin}, where we invert all eigenvalues that are non-zero: $\mathcal{L}^+ := \sum_{j\neq0} \frac{1}{\lambda_j} |x_j\rrangle \llangle y_j|$.
\rev{For example, pseudo-inverses appear in the context of resolvents and perturbation theory, although this is not always mentioned explicitly. Indeed, consider}
standard quantum mechanical perturbation theory. Let $H|n\rangle = E_n |n\rangle$ denote a Hamiltonian and $V$ a small perturbation. 
Assuming the $E_n$ are non-degenerate, the eigenvalues of $H' = H + V$ can be written up to second order as 
\begin{align}
    E_n' =& E_n + \langle n | V |n \rangle + \langle n| V  (E_n-H)^+ V |n\rangle
    \\[0.2cm]\nonumber
    &+ \langle n | V (E_n-H)^+ V (E_n-H)^+ V |n\rangle + \ldots.
\end{align}
Here $E_n - H = \sum_{j\neq n} (E_n-E_j) |j\rangle\langle j|$, which is not invertible. 
The pseudo-inverse hence reads $(E_n-H)^+ = \sum_{j\neq n} \tfrac{1}{E_n-E_j} |j\rangle\langle j|$; again, we invert what we can, and leave the singular parts untouched. 

In the vectorized notation of Sec.~\ref{sec:vectorization}, the Drazin pseudo-inverse~\eqref{Drazin_integral_def} can be written as 
\begin{equation}
    \mathcal{L}^+ = - \int\limits_0^\infty d\tau~e^{\mathcal{L} \tau} \mathcal{P},
\end{equation}
where 
\begin{equation}\label{projector}
    \mathcal{P} = \id - \rhossV \idV = \sum_k |x_k\rrangle\llangle y_k |.
\end{equation}
In fact, using  Eq.~\eqref{exp_L}, one may verify that this reduces exactly to Eq.~\eqref{eigen_Drazin}.
One may also verify that
\begin{equation}\label{Drazin_basic_prop}
    \mathcal{L}^+ \mathcal{L} = \mathcal{L} \mathcal{L}^+ = \id - \rhossV \idV. 
\end{equation}
That is, it returns the identity on the subspace complementary to that defined by $\rhossV \idV$. 
From this it also follows that $\mathcal{L} \mathcal{L}^+ \mathcal{L} = \mathcal{L}$ and $\mathcal{L}^+ \mathcal{L} \mathcal{L}^+ = \mathcal{L}^+$.

Technically speaking, the Drazin inverse we are using is actually called the ``Group inverse,'' which is a special case of Drazin inverses to matrices satisfying ${\rm rank}(A^2) = {\rm rank}(A)$. 
The Drazin/group inverse can be more generally defined as the unique matrix $\mathcal{L}^+$ satisfying 
\begin{align}
\label{Drazin_prop1}
    \mathcal{L} \mathcal{L}^+ \mathcal{L} &= \mathcal{L},
    \\[0.2cm]
\label{Drazin_prop2}    
    \mathcal{L}^+ \mathcal{L} \mathcal{L}^+ &= \mathcal{L}^+,
    \\[0.2cm]
\label{Drazin_prop3}    
    \mathcal{L}^+ \mathcal{L} &= \mathcal{L} \mathcal{L}^+.
\end{align}
On the other hand, the Moore-Penrose pseudo-inverse, which we will denote by 
$\mathcal{L}\MP$, is defined as the unique matrix satisfying 
\begin{align}
\label{MP_prop1}
    \mathcal{L} \mathcal{L}\MP \mathcal{L} &= \mathcal{L},
    \\[0.2cm]
\label{MP_prop2}    
    \mathcal{L}\MP \mathcal{L} \mathcal{L}\MP &= \mathcal{L}\MP,
    \\[0.2cm]
\label{MP_prop3}    
    (\mathcal{L}\MP \mathcal{L})^\dagger &= \mathcal{L}\MP \mathcal{L},
    \\[0.2cm]
\label{MP_prop4}
    (\mathcal{L} \mathcal{L}\MP)^\dagger &= \mathcal{L} \mathcal{L}\MP.
\end{align}
Eqs.~\eqref{MP_prop1} and~\eqref{MP_prop2} are the same as~\eqref{Drazin_prop1} and~\eqref{Drazin_prop2}.
But~\eqref{Drazin_prop3} is not equivalent to \eqref{MP_prop3}-\eqref{MP_prop4}, unless the matrix is Hermitian.

The relation between the Drazin and Moore-Penrose pseudo-inverses is given by 
\begin{equation}\label{Drazin_MP_relation}
    \mathcal{L}^+ = \mathcal{P} \mathcal{L}\MP \mathcal{P},
\end{equation}
which we now prove. 
The calculation of $\mathcal{L}\MP$ is a bit more complicated. If $\mathcal{L}^\dagger \mathcal{L}$ is invertible then $\mathcal{L}\MP = (\mathcal{L}^\dagger \mathcal{L})^{-1} \mathcal{L}^\dagger$. But for Liouvillians this is never the case, so this formula is not useful. 
One way to compute it is via the singular value decomposition. 
Another, which will be useful for our purposes, is through rank decomposition. 
Let $|i\rrangle$ denote a generic orthonormal basis of dimension $d^2-1$ where $d^2$ is the dimension of the Liouvillian (and $d$ is the dimension of the Hilbert space). 
We now define two matrices, 
\begin{equation}
    B = \sum_i |x_i\rrangle\llangle i|, 
    \qquad 
    C = \sum_i \lambda_i |i\rrangle\llangle y_i|,
\end{equation}
which are of dimensions $d^2 \times (d^2-1)$ and $(d^2-1)\times d^2$ respectively. 
They are such that $\mathcal{L} = BC$, as written in Eq.~\eqref{L_eigen_decomp}.

The Moore-Penrose pseudo-inverse is then given by 
\begin{equation}\label{MP_rank_formula}
    \mathcal{L}\MP = C^\dagger (C C^\dagger)^{-1} (B^\dagger B)^{-1} B^\dagger,
\end{equation}
which satisfy Eqs.~\eqref{MP_prop1}-\eqref{MP_prop4}, as one may verify.
The matrices $CC^\dagger$ and $B^\dagger B$ are both invertible and of dimension $d^2-1$.
From~\eqref{MP_rank_formula} it follows that
\begin{equation}\label{MP_action_xy}
    \mathcal{L} \mathcal{L}\MP |x_k\rrangle = |x_k\rrangle, 
    \qquad 
    \llangle y_k | \mathcal{L}\MP \mathcal{L} = \llangle y_k |. 
\end{equation}
Together with Eq.~\eqref{identity_matrix_decomposition} this implies 
\begin{equation}
    \mathcal{P}\mathcal{L}\MP\mathcal{L} = \mathcal{L} \mathcal{L}\MP \mathcal{P} = \mathcal{P}.
\end{equation}
With these relations, one may now verify that the Drazin inverse, as defined in Eq.~\eqref{Drazin_MP_relation},  will satisfy all  properties~\eqref{Drazin_prop1}-\eqref{Drazin_prop3}. 
And since the matrix satisfying said properties is unique, this proves that~\eqref{Drazin_MP_relation} is indeed the correct relation between the Drazin and Moore-Penrose pseudoinverses.

\section{Numerical computation of $S(\omega)$ and $D$}
\label{app:numerics}

In this appendix we discuss methods to compute $S(\omega)$ and $D$ efficiently. 
Crucial to these is the fact that the vectorized Liouvillian $\mathcal{L}$ in Eq.~\eqref{vec_vectorized_L} is generally very sparse. 
In such cases, methods involving the solution of linear systems of equations (as opposed to diagonalization) become particularly efficient. 
Conversely, if $\mathcal{L}$ is not large, it might be simpler to just diagonalize it.

\subsection{Power spectrum}

Suppose we choose to diagonalize $\mathcal{L}$. 
Then for each $\omega$, $S(\omega)$ can be efficiently reconstructed from Eq.~\eqref{eigen_S} by precomputing 
and storing the coefficients $\idV \mathcal{J} |x_j\rrangle\llangle y_j|\mathcal{J} \rhossV$.

When $\mathcal{L}$ is large, diagonalization becomes costly and it is more efficient to transform $S(\omega)$ into the solution of a linear system of equations. We use Eq.~\eqref{inverse_S} and start by precomputing $\mathcal{L}^2$. 
For each $\omega$, we define a new vector $|z\rrangle$ as the solution of 
\begin{equation}
    (\mathcal{L}^2 + \omega^2) |z\rrangle = \mathcal{J} \rhossV.
\end{equation}
Since $\mathcal{L}^2 + \omega^2$ is invertible (for $\omega \neq 0$), this solution is unique. 
The power spectrum is then written as 
\begin{equation}
    S(\omega) = K - 2 \idV \mathcal{J} \mathcal{L}|z\rrangle.
\end{equation}

\subsection{Drazin inverse and noise}

To compute the noise in Eq.~\eqref{eigen_D} we require the Drazin inverse. Again, if all eigenvalues and eigenvectors are known, we can compute it from Eq.~\eqref{eigen_Drazin}. However,  this is usually too costly if $\mathcal{L}$ is large.
Instead, we notice that we never require $\mathcal{L}^+$ itself, but only  $|z\rrangle := \mathcal{L}^+|\alpha\rrangle$, i.e., the Drazin acting on a certain vector $|\alpha\rrangle$ In the case of ~Eq.~\eqref{eigen_D} we have $|\alpha \rrangle = \mathcal{J} \rhossV$. 
But let us focus here on a general $|\alpha\rrangle$, and develop a general result. 

Multiplying $|z\rrangle$ by $\mathcal{L}$ and using Eq.~\eqref{Drazin_basic_prop} we see that $|z\rrangle$ is the solution of the linear system of equations
\begin{equation}\label{numerics_z_Drazin_equation}
    \mathcal{L}|z\rrangle = |\alpha \rrangle - \rhossV \idV \alpha\rrangle.
\end{equation}
But because $\mathcal{L}$ is not invertible, Eq.~\eqref{numerics_z_Drazin_equation} has an infinite number of solutions.
Since $|z\rrangle := \mathcal{L}^+|\alpha\rrangle$ and since $\idV \mathcal{L}^+ = 0$ [which follows from Eq.~\eqref{eigen_Drazin}], the solution we are interested in is the one satisfying $\idV z\rrangle = 0$. We can enforce the solver to produce this solution by adding an additional row to Eq.~\eqref{numerics_z_Drazin_equation}, as follows: 
\begin{equation}\label{numerics_z_Drazin_equation_2}
    \begin{pmatrix}
    ~~~~~~~~~~&\vdots& ~~~~~~~~~~ \\[0.2cm]
    &\mathcal{L}& \\[0.2cm]
    &\vdots& \\[0.2cm]
    & \idV &
    \end{pmatrix} |z\rrangle 
    = \begin{pmatrix}
    \vdots \\[0.2cm]
    |\alpha \rrangle - \rhossV \idV \alpha\rrangle \\[0.2cm]
    \vdots \\[0.2cm]
    0 
    \end{pmatrix}.
\end{equation}
The matrix on the left is now of dimension $(d^2+1)\times d$. 
The solution of this system will be exactly $|z\rrangle := \mathcal{L}^+|\alpha\rrangle$.

Returning now to the case of $D$, if we define $|z\rrangle = \mathcal{L}^+ \mathcal{J}|\alpha\rrangle$ then 
\begin{equation}
    D = K - 2 \idV \mathcal{J} |z\rrangle. 
\end{equation}
As an example, suppose $\mathcal{L}\rho =  \gamma \mathcal{D}[\sigma_-]\rho$.
In vectorized form this becomes 
\begin{equation}
    \mathcal{L} = \left(
\begin{array}{cccc}
 -\gamma  & 0 & 0 & 0 \\
 0 & -\frac{\gamma }{2} & 0 & 0 \\
 0 & 0 & -\frac{\gamma }{2} & 0 \\
 \gamma  & 0 & 0 & 0 \\
\end{array}
\right).
\end{equation}
Similarly, $\idV = (1~0~0~1)$. The matrix on the left-hand side of~\eqref{numerics_z_Drazin_equation_2} would then be
\begin{equation}
     \left(
\begin{array}{cccc}
 -\gamma  & 0 & 0 & 0 \\
 0 & -\frac{\gamma }{2} & 0 & 0 \\
 0 & 0 & -\frac{\gamma }{2} & 0 \\
 \gamma  & 0 & 0 & 0 \\
 1 & 0 & 0 & 1
\end{array}
\right).
\end{equation}

\section{Power spectrum for weak dissipation}
\label{app:weak_dissipation}

We analyze the power spectrum $S(\omega)$ in the limit where all dissipative terms in Eq.~\eqref{M} are very small.
Our starting point is Eq.~\eqref{eigen_S}. 
We split the Liouvillian as $\mathcal{L} = \mathcal{L}_H + \mathcal{L}_D$, where $\mathcal{L}_H = -i [H,\bullet]$ and $\mathcal{L}_D = \sum_k \mathcal{D}[L_k]$. 
To construct the power spectrum, we will use perturbation theory to build the eigenvalues and eigenvectors of $\mathcal{L}$ in the limit where $\mathcal{L}_D$ is small. 
In terms of $H|n\rangle = E_n |n\rangle$, the unitary part $\mathcal{L}_H$ can be written as 
\begin{equation}
    \mathcal{L}_H \rho = -i \sum_{n,m} \omega_{nm} |n\rangle\langle n | \rho |m\rangle\langle m|,
\end{equation}
where $\omega_{nm} = E_n - E_m$. 
The eigenvalues of $\mathcal{L}_H$ are thus $-i \omega_{nm}$, while the corresponding right and left eigenvectors are 
\begin{equation}
    |x_{nm} \rrangle = |m\rangle^* \otimes |n\rangle, 
    \qquad 
    \llangle y_{nm}| = \langle m|^* \otimes \langle n|.
\end{equation}
Before we use perturbation theory, however, we must be careful about the zero eigenvalues. 
For $\mathcal{L}_H$ there are at least $d$ of them, corresponding to $\omega_{nn}\equiv 0$. 
There might also be more if the $E_n$ are degenerate. 
Let $\mathsf{Z}$ denote the set of indices $(n,m)$ having $\omega_{nm} = 0$, and $\mathsf{nZ}$ the set having $\omega_{nm} \neq 0$.
All eigenvectors in $\mathsf{Z}$, which we will henceforth refer to as $|x_\alpha\rrangle$ and $\llangle y_\alpha|$, therefore have $\lambda_\alpha^0 = 0$. 
Those in $\mathsf{nZ}$, on the other hand, will be referred to as $x_{nm}\rrangle$ and $\llangle y_{nm}|$, and have $\lambda_{nm}^0 = -i \omega_{nm} \neq 0$. 

After we include the perturbation $\mathcal{L}_D$, the power spectrum~\eqref{eigen_S} will  change to 
\begin{align}
    S(\omega) &= K - 2 \sum_{\alpha \in \mathsf{Z}} \frac{\lambda_\alpha^1}{(\lambda_\alpha^1)^2 + \omega^2} g_\alpha 
    - 2 \sum_{n,m, \omega_{n,m} \neq 0} 
    \frac{\lambda_{nm}^1}{(\lambda_{nm}^1)^2 + \omega^2} g_{nm}.
\end{align}
where $\lambda_\alpha^1$ and $\lambda_{nm}^1$ are the perturbed eigenvalues (once $\mathcal{L}_D$ is included) and 
\begin{equation}
    g_\alpha = \idV \mathcal{J} |x_\alpha\rrangle\llangle y_\alpha | \mathcal{J} \rhossV, 
    \quad 
    g_{nm} = \idV \mathcal{J} |x_{nm}\rrangle\llangle y_{nm} | \mathcal{J} \rhossV.
\end{equation}
To lowest order in the perturbation, we can actually use the unperturbed eigenvectors; that is, include only the perturbation of the eigenvalues. 

For the set $\mathsf{Z}$, we need to use degenerate perturbation theory. This is much more difficult, and we will therefore not touch on this part. These terms, however, will only contribute to the peak/dip at $\omega=0$. 
This will be true even if the $\lambda_\alpha^1$ are complex, because by hypothesis the real parts are always small. 
The peaks at $\omega\neq 0$ will therefore be associated to the states in $\mathsf{nZ}$. 
These states might also be degenerate. However, for simplicity we will assume this is not the case. Therefore, all that remains to be done is to write down the perturbed eigenvalue, which will be 
\begin{equation}
    \lambda_{nm}^1 = -i \omega_{nm} + \llangle y_{nm} |\mathcal{L}_D | x_{nm}\rrangle := -i \omega_{nm} + \gamma_{nm}.
\end{equation}
The $\gamma_{nm}$ represent new effective damping rates, and are responsible for broadening the peaks/dips.
In general, they may have an imaginary part. 
But since they are small by hypothesis, this can just be absorbed into $\omega_{nm}$. 
We will therefore assume that $\gamma_{nm}$ is real. 

From this analysis, we see that the peaks/dips will be positioned at $\omega = \omega_{nm}$, and will have width equal to $\gamma_{nm}$. 
Finally, to understand whether they will be peaks or dips, we can write the power spectrum more explicitly as 
\begin{align}
&S(\omega) = K - 2\sum_{\alpha \in Z} \frac{\lambda_\alpha}{\lambda_\alpha^2 + \omega^2} g_\alpha 
\nonumber \\[0.2cm]
&- 2\sum_{\omega_{nm}>0} \frac{i\omega_{nm} (\omega_{nm}^2-\omega^2)(g_{nm}-g_{nm}^*) - \gamma_{nm} (\omega_{nm}^2 + \omega^2)(g_{nm} + g_{nm}^*)}{\omega_{nm}^4 + 2 \omega_{nm}^2 (\gamma_{nm}^2 - \omega^2) + (\gamma_{nm}^2 + \omega^2)^2}.
\end{align}
This shows that if $g_{nm}$ is real, we will have a peak when $g_{nm}>0$ or a dip if $g_{nm}<0$. Conversely, if it is complex, there will be peaks immediately followed by dips. 

\rev{To illustrate these ideas consider Example A, Eq.~\eqref{ExampleA_M} with $\Omega = 0$. 
The vectorized Liouvillian reads 
\begin{equation}
    \mathcal{L} = \begin{pmatrix}
        -\gamma (\bar{N}+1) & 0 & 0 & \gamma \bar{N} \\
        0 & i \Delta - \gamma(\bar{N}+1/2) & 0 & 0 \\
        0 & 0 & -i \Delta - \gamma(\bar{N}+1/2) & 0 \\ 
        \gamma (\bar{n}+1) & 0 & 0 & -\gamma \bar{N}
    \end{pmatrix}.
\end{equation}
The exact eigenvalues/eigenvectors of this matrix read: 
\begin{IEEEeqnarray*}{rRLl}
        \lambda_{00} &=& 0, \quad  &|x_{00}\rrangle = \bar{N} |00\rangle + (\bar{N}+1)|11\rangle, \\
        \lambda_{01} &=& -i \Delta - \gamma(\bar{N}+1/2), \quad  &|x_{01}\rrangle = |10\rangle, 
        \\
        \lambda_{10} &=& i\Delta - \gamma(\bar{N}+1/2), \quad  &|x_{10}\rrangle = |01\rangle, \\
        \lambda_{11} &=& -\gamma (2\bar{N}+1), \quad  &|x_{11}\rrangle = |00\rangle - |11\rangle ,
\end{IEEEeqnarray*}
where $|ij\rangle = |i\rangle\otimes |j\rangle$ and the eigenvectors are not normalized (the left eigenvectors $\llangle y_{nm}|$ are also not shown). 
These results can now be plugged in Eq.~\eqref{eigen_S}. 
In the limit of weak dissipation the elements linked to non-zero transition frequencies are $|x_{01}\rrangle$ and $|x_{10}\rrangle$. However, for these two states it follows that 
$\mathcal{J} |x_{01}\rrangle = 0$, so that they actually do not contribute. 
This is tantamount to the fact that $L_k^\dagger L_k = \{ \sigma_+\sigma_-, \sigma_-\sigma_+\}$ being diagonal in the computational basis. 
As a consequence, we find that the only term actually contributing to Eq.~\eqref{eigen_S} is $\lambda_{11}$.
}

\section{Extracting probabilities from a cavity  using FCS}
\label{app:cavity_probabilities}

In this appendix we demonstrate the results discussed in Sec.~\ref{sec:cavity_statistics}. 
Our system is an optical cavity prepared in a generic state $\rho_0$ and evolving according to the Liouvillian $\mathcal{L}\rho = \kappa \mathcal{D}[a]\rho$. 
Our goal will be to solve for the generalized master equation~\eqref{gQME}, with the appropriate tilted Liouvillian $\mathcal{L}_\chi$ for each type of measurement.

\subsection{Direct photo-detection}

For direct photo-detection the tilted Liouvillian is given by~\eqref{tilted_liouvillian}, with $L = \sqrt{\kappa} a$ and $\nu = 1$: 
\begin{equation}
    \mathcal{L}_\chi \rho = \kappa \Big[ e^{i \chi} a \rho a^\dagger - \frac{1}{2} \{a^\dagger a, \rho\}\Big]
\end{equation}
We want to solve for the moment generating function $M(\chi,t) = \tr\big\{\rho_\chi(t)\big\} = \tr\big\{e^{\mathcal{L}_\chi t} \rho_0\big\}$ and then take the limit $t\to\infty$. 
The tilted Liouvillian $\mathcal{L}_\chi$ turns out to have exactly one zero eigenvalue, all others having negative real parts. 
The corresponding right eigenvector, as one may verify, is the vacuum state $|{\rm vac}\rrangle = |0\rangle\langle 0 |$, which is the same as the right eigenvector of $\mathcal{L}$.
The left eigenvector is different, however, and acts as 
$\llangle \chi | \bullet \rrangle = \sum_n \langle n| \bullet |n\rangle e^{i n \chi}$. 
That is, it yields a tilted version of the trace. 
In fact, one can verify that $\tr\big\{ e^{i \chi a^\dagger a} \mathcal{L}_\chi(\bullet)\big\} = 0$, which therefore proves that this is a left eigenvector with zero eigenvalue. 
One also sees that if $\chi=0$ we recover the usual trace, $\llangle \chi | \bullet \rrangle = \tr(\bullet)$.

The evolution under the tilted master equation will therefore have the form 
\begin{equation}
    e^{\mathcal{L}_\chi t} |\rho_0\rrangle = |{\rm vac}\rrangle\llangle \chi | \rho_0\rrangle + \sum_j e^{\lambda_{j,\chi} t} |x_{j,\chi}\rrangle\llangle y_{j,\chi}|,
\end{equation}
where the sum contains eigenvalues $\lambda_{j,\chi}$ with strictly negative real part. 
In the long-time limit these terms will all vanish and we are therefore left with $|\rho_\chi(t\to\infty)\rrangle = |{\rm vac}\rrangle \llangle \chi | \rho_0\rrangle$.
Hence, taking the trace we find 
\begin{equation}
   M(\chi,\infty) =  \tr\big\{ \rho_\chi(\infty)\big\} = \llangle \chi | \rho_0\rrangle = \sum_n e^{i n \chi} \langle n|\rho_0|n\rangle.
\end{equation}
Plugging this into Eq.~\eqref{FCS_P2} and carrying out the now trivial Fourier transform, we arrive at Eq.~\eqref{empty_cavity_photo_detection}: viz., $P(n,\infty) = \langle n |\rho_0|n\rangle$.

\subsection{Homodyne detection}

For homodyne detection, the tilted Liouvillian is given by Eq.~\eqref{L_chi_diffusion} (see also Table~\ref{tab:diffusion_experiments}).
Unlike standard homodyne detection, however, in this case we choose a time-dependent weight factor $\nu(t)=\nu_0\exp(-\kappa t/2)$, with $\nu_0 = \sqrt{\kappa/2}$.
The tilted Liouvillian will thus be 
\begin{equation}
    \label{eq:liouchihom}
    \mathcal{L}_\chi(t) = \kappa\mathcal{D}[a]+i\chi\mathcal{H}(t)-\frac{\chi^2}{2}\nu_0^2e^{-\kappa t},
\end{equation}
where 
\begin{equation}
    \label{eq:hsupop}
    \mathcal{H}(t)\rho = \nu(t)\sqrt{\kappa}(a\rho+\rho a^\dagger).
\end{equation}
The moment generating function is again the solution of the generalized QME~\eqref{gQME}. However, since the dynamics is time-dependent, it will now be given by 
\begin{equation}
    \label{eq:momgenfcshom}
    M(\chi,\infty) = \tr\{\mathcal{T}e^{\int_0^\infty dt \mathcal{L}_\chi(t)}\rho_0\},
\end{equation}
where $\mathcal{T}$ denotes the time-ordering operator.
Using the identity
\begin{equation}
    \label{eq:timord}
    \mathcal{T}e^{\int_0^t d\tau [\mathcal{A}(\tau)+\mathcal{B}]}=e^{\mathcal{B}\tau}\mathcal{T}e^{\int_0^t d\tau e^{-\mathcal{B}\tau}\mathcal{A}(\tau)e^{\mathcal{B}\tau}},
\end{equation}
we may write 
\begin{equation}
    \label{eq:momgenfcshom2}
    M(\chi,\infty) = e^{-\frac{\nu_0^2}{2\kappa}\chi^2}\tr\{\mathcal{T}e^{i\chi\int_0^\infty dt \tilde{\mathcal{H}}(t)}\rho_0\},
\end{equation}
with
\begin{equation}
    \label{eq:htilap}
    \tilde{\mathcal{H}}(t) = e^{-\kappa t\mathcal{D}[a]}\mathcal{H}(t)e^{\kappa t\mathcal{D}[a]}.
\end{equation}
To obtain Eq.~\eqref{eq:momgenfcshom2}, we made use of $\tr\{e^{\tau\kappa\mathcal{D}[a]}\rho\}=\tr\{\rho\}$,
for any $\tau$ and $\rho$.

From the definitions of $\mathcal{H}(t)$ and $\mathcal{D}[a]$, one may show
\begin{equation}
    \label{eq:commhd}
    [\mathcal{H}(t),\mathcal{D}[a]]=-\frac{1}{2}\mathcal{H}(t).
\end{equation}
Through the Baker-Campbell-Hausdorff formula, this implies that
\begin{equation}
    \label{Htilap}
    \tilde{\mathcal{H}}(t) = e^{-\frac{\kappa}{2}t}\mathcal{H}(t)=e^{-\kappa t}\mathcal{H}(0).
\end{equation}
Equation~\eqref{eq:momgenfcshom2} then reduces to
\begin{equation}
    \label{eq:momgenfcshom3}
    M(\chi,\infty) = e^{-\frac{\nu_0^2}{2\kappa}\chi^2}\tr\{e^{i\frac{\chi}{\kappa}\mathcal{H}(0)}\rho_0\} = e^{-\frac{\nu_0^2}{2\kappa}\chi^2}\tr\{\rho_0e^{i\frac{\chi}{\kappa}\mathcal{H}^\dagger(0)}\id\},
\end{equation}
where 
\begin{equation}
    \label{eq:Had}
    \mathcal{H}^\dagger(0)X = \nu_0\sqrt{\kappa}(Xa+a^\dagger X),
\end{equation}
is the adjoint of $\mathcal{H}(0)$. It is straightforward to show that
\begin{equation}
    \label{eq:Hadid}
    [\mathcal{H}^\dagger(0)]^n\id = (\nu_0\sqrt{\kappa})^n:\!(a+a^\dagger)^n\!:,
\end{equation}
where $:\!\!:$ denotes normal ordering. This allows us to write
\begin{equation}
    \label{eq:momgenfcshom4}
    \begin{aligned}
    M(\chi,\infty) =&e^{-\frac{\nu_0^2}{2\kappa}\chi^2}\tr\{\rho_0:\!e^{i\chi\frac{\nu_0}{\sqrt{\kappa}}(a+a^\dagger)}\!:\} 
    \\[0.2cm]=& e^{-\frac{\nu_0^2}{2\kappa}\chi^2}\tr\{\rho_0e^{i\chi\frac{\nu_0}{\sqrt{\kappa}}a^\dagger}e^{i\chi\frac{\nu_0}{\sqrt{\kappa}}a}\}
    \\[0.2cm]=&
    \tr\{\rho_0e^{i\chi\frac{\nu_0}{\sqrt{\kappa}}(a^\dagger+a)}\}
    \\[0.2cm]=&
    \tr\{\rho_0e^{i\chi x}\}
    \\[0.2cm]=&
    \int dx \langle x|\rho_0|x\rangle e^{i\chi x},
    \end{aligned}
\end{equation}
where for the last equality, we used $\nu_0 = \sqrt{\kappa/2}$ and $x=(a^\dagger+a)/\sqrt{2}$. 
Plugging this in Eq.~\eqref{FCS_P} yields us the probability $P(N_{\rm diff}(t)=x)$. 
The remaining Fourier transform is trivial, and we therefore arrive at Eq.~\eqref{empty_cavity_homodyne}; viz., $P(x,\infty) = \langle x |\rho_0|x\rangle$.

\subsection{Heterodyne detection}
For heterodyne detection, we include two counting fields $\chi_x$ and $\chi_y$. The moment generating function is still given by Eq.~\eqref{eq:momgenfcshom}, with the Liouvillian
\begin{equation}
    \label{eq:liouchihet}
    \mathcal{L}_\chi(t) = \kappa\mathcal{D}[a]+i\chi_x\mathcal{H}_x(t)+i\chi_y\mathcal{H}_y(t)-\frac{\chi_x^2+\chi_y^2}{2}\nu_0^2e^{-\kappa t},
\end{equation}
where
\begin{equation}
    \label{eq:hsupophet}
    \begin{aligned}
    \mathcal{H}_x(t)\rho &= \nu_0e^{-\frac{\kappa}{2}t}\sqrt{\frac{\kappa}{2}}(a\rho+\rho a^\dagger),\\\mathcal{H}_x(t)\rho &= -i\nu_0e^{-\frac{\kappa}{2}t}\sqrt{\frac{\kappa}{2}}(a\rho-\rho a^\dagger).
    \end{aligned}
\end{equation}
Note that $\mathcal{H}_x(t)=\mathcal{H}(t)/\sqrt{2}$ [cf.~Eq.~\eqref{eq:hsupop}]. This is because for heterodyne detection, we need to split the outgoing beam in two before performing homodyne detection on each output.

A calculation completely analogous to the last subsection then results in
\begin{equation}
    \label{eq:momgenhet1}
    \begin{aligned}
    M(\chi_x,\chi_y)=&e^{-(\chi_x^2+\chi_y^2)\frac{\nu_0^2}{2\kappa}}\tr\{\rho_0:\!e^{i\chi_x\frac{\nu_0}{\sqrt{2\kappa}}(a+a^\dagger)-\chi_y\frac{\nu_0}{\sqrt{2\kappa}}(a^\dagger-a)}\!:\}
    \\[0.2cm]=&
    \tr\{e^{i\frac{\nu_0}{\sqrt{2\kappa}}(\chi_x+i\chi_y)a^\dagger}\rho_0e^{i\frac{\nu_0}{\sqrt{2\kappa}}(\chi_x-i\chi_y)a}\}
    \\[0.2cm]=&
    \int dx dy e^{i\chi_x x+i\chi_y y}\frac{1}{\pi}\left\langle\frac{x+iy}{\sqrt{2}}\right|\rho_0\left|\frac{x+iy}{\sqrt{2}}\right\rangle,
    \end{aligned}
\end{equation}
where in the last equality, we expressed the trace in terms of the coherent states $a|\alpha+i\beta\rangle=(\alpha+i\beta)|\alpha+i\beta\rangle$ and we chose $\nu_0=\sqrt{\kappa}$. The probability distribution $P(N_x(t)=x,N_y(t)=y)$ in the limit $t\rightarrow\infty$ is thus given by the Husimi $Q$-function
\begin{equation}
    \label{eq:husimiq}
    Q(x,y) = \frac{1}{\pi}\left\langle\frac{x+iy}{\sqrt{2}}\right|\rho_0\left|\frac{x+iy}{\sqrt{2}}\right\rangle.
\end{equation}

 \section{Calculations in the Gaussian case}
\label{app:Gaussian}

In this appendix we prove Eq.~\eqref{Gaussian_two_point_correlations}. 
The proof is based on the Quantum Regression Theorem described in Sec.~\ref{sec:Coherence}, according to which 
\begin{equation}
    \begin{aligned}
        \langle R_j(t) R_i(t+\tau)\rangle &= \tr\Big\{ R_i e^{\mathcal{L}\tau} \big( \rho(t) R_j\big) \Big\},
        \\[0.2cm]
        \langle R_j(t+\tau) R_i(t)\rangle &= \tr\Big\{ R_j e^{\mathcal{L}\tau} \big(  R_i\rho(t)\big) \Big\}.
    \end{aligned}
\end{equation}
In terms of the adjoint Liouvillian $\mathcal{L}^\dagger$ [defined as per Eq.~\eqref{adjoint_liouvillian}], we can also write this as 
\begin{equation}\label{Gaussian_appendix_correlation_functions_2}
    \begin{aligned}
        \langle R_j(t) R_i(t+\tau)\rangle &= \langle R_j \hat{R}_i(\tau) \rangle_t        
        \\[0.2cm]
        \langle R_j(t+\tau) R_i(t)\rangle &= \langle \hat{R}_j(\tau) R_i \rangle_t,
    \end{aligned}
\end{equation}
where $\hat{\mathcal{O}}(\tau) := e^{\mathcal{L}^\dagger \tau} \mathcal{O}$ and where $\langle \ldots \rangle_t$ is an expectation value over $\rho(t)$.

The key to proving Eqs.~\eqref{Gaussian_two_point_correlations} is to now find a closed evolution equation for $\hat{R}_i(\tau)$. 
For Gaussian processes, as one can verify starting from the general Liouvillian~\eqref{Gaussian_Liouvillian}, this will evolve in the same way as the mean vector in Eq.~\eqref{Gaussian_mean_evolution_2N}. 
That is, 
\begin{equation}
    \frac{d \hat{\bm{R}}(\tau)}{d\tau} = - \mathcal{W} \hat{\bm{R}}_i(\tau) + \Omega \bm{f}. 
\end{equation}
This equation is linear and the solution is simply 
\begin{equation}\label{Gaussian_appendix_591839812789}
    \hat{\bm{R}}(\tau) = G(\tau) \bm{R}  + \int\limits_0^\tau ds G(\tau-s) \Omega \bm{f},
\end{equation}
where $\bm{R} = \hat{\bm{R}}(0)$ and $G(\tau) = e^{-\mathcal{W}\tau}$.
\rev{
The last term can be simplified by noting that, even though this is an operator equation for $\hat{\bm{R}}$, the last term is proportional to the identity matrix. And, in fact, it is the same term that appears in the equation for the mean vector $\bm{r}$. 
In fact, solving 
Eq.~\eqref{Gaussian_mean_evolution_2N} for $\bm{r}$, between two arbitrary times $t$ and $t+\tau$, yields
\begin{equation}\label{Gaussian_appendix_938174938917}
    \bm{r}(t+\tau) = G(\tau) \bm{r}(t) + \int\limits_0^\tau ds~G(\tau-s) \Omega\bm{f}. 
\end{equation}
This shows that the 2nd term in Eq.~\eqref{Gaussian_appendix_591839812789} is nothing but $r(t+\tau) - G(\tau) r(t)$. 
This statement is actually true for any time $t$, since Eq.~\eqref{Gaussian_appendix_938174938917} only depends on the time difference $\tau$. However, with the last term in Eq.~\eqref{Gaussian_appendix_correlation_functions_2} in mind, we actually want quantities evaluated at times $t$ and $t+\tau$, it is convenient to stick with $r(t+\tau) - G(\tau) r(t)$.  }
Hence 
\begin{equation}
    \hat{\bm{R}}(\tau) = G(\tau) \bm{R} + \bm{r}(t+\tau) - G(\tau) \bm{r}(t).
\end{equation}
This is a general formula for $e^{\mathcal{L}^\dagger \tau}\bm{R}$, with the right-hand side written in terms only of $G(\tau)$ and the mean vector.

Plugging this in Eq.~\eqref{Gaussian_appendix_correlation_functions_2} then yields, e.g. for the first correlation function 
\begin{equation}
    \langle R_j(t) R_i(t+\tau)\rangle = \sum_k G_{ik}(\tau) \langle R_jR_k \rangle_t + \langle R_j\rangle_t \Big[ \bm{r}(t+\tau) - G(\tau) \bm{r}(t) 
    \Big]_i,
\end{equation}
where, of course, $\langle R_i \rangle_t = r_j(t)$. 
To finish, we write this in terms of 
\begin{equation}
    \tilde{\Theta}_{ij} = (\Theta - i \Omega/2)_{ij} = \pm \langle R_j R_i \rangle - \langle R_i \rangle\langle R_j \rangle. 
\end{equation}
We then arrive exactly at Eq.~\eqref{Gaussian_two_point_correlations}.

\section{Relation between $\mathcal{L}^+$ and $\mathcal{L}_0^{-1}$}
\label{app:drazin_L0_relation}

\rev{
The Liouvillian relates to the no-jump superoperator $\mathcal{L}_0$ as 
$\mathcal{L} = \mathcal{L}_0 + \mathcal{J}$, where $\mathcal{J} = \sum_k \mathcal{L}_k$. 
We prove here an identity between $\mathcal{L}_0^{-1}$ and the Drazin inverse $\mathcal{L}^+$ [Eq.~\eqref{eigen_Drazin}]. 
We begin by noting the identity ($\mathcal{P} = \rhossV \idV$)
\begin{equation}\label{app_drazin_alternative_definition}
    \mathcal{L}^+ \equiv \lim\limits_{\epsilon\to0} (\mathcal{L}+\epsilon)^{-1} - \frac{1}{\epsilon} \mathcal{P},
\end{equation}
which readily follows from Eq.~\eqref{eigen_Drazin} if we write
\begin{equation}
    (\mathcal{L} + \epsilon)^{-1} = \frac{1}{\epsilon} \rhossV \idV + \sum_{j\neq 0} \frac{1}{\lambda_j + \epsilon} |x_k\rrangle\llangle y_k|,
\end{equation}

We now consider the matrix identity 
\begin{equation}
    (A+B)^{-1} = A^{-1} - A^{-1} B (A+B)^{-1},
\end{equation}
and set $A = \mathcal{L}_0$ and $B = \mathcal{J}+\epsilon$, leading to 
\begin{align}\nonumber
(\mathcal{L}+\epsilon)^{-1} &= \mathcal{L}_0^{-1} - \mathcal{L}_0^{-1} (\mathcal{J}+\epsilon) (\mathcal{L}+\epsilon)^{-1}
\\[0.2cm]\nonumber
&= \mathcal{L}_0^{-1} - \mathcal{L}_0^{-1} (\mathcal{J}+\epsilon) \bigg( \frac{\mathcal{P}}{\epsilon} + \mathcal{L}^+\bigg)
\\[0.2cm]
&= \mathcal{L}_0^{-1} - \frac{1}{\epsilon}\mathcal{L}_0^{-1} \mathcal{J} \mathcal{P}  - \mathcal{L}_0^{-1} \mathcal{J} \mathcal{L}^+ - \mathcal{L}_0^{-1} \mathcal{P} - \epsilon\mathcal{L}_0^{-1} \mathcal{L}^+.
\label{app_tmp30298130928039810298310928}
\end{align}
The steady-state and normalization conditions imply $\mathcal{L} \mathcal{P} = 0$. Together with $\mathcal{L} = \mathcal{L}_0 + \mathcal{J}$ we then see that $\mathcal{L}_0^{-1} \mathcal{J} \mathcal{P} = - \mathcal{P}$. 
Plugging this in Eq.~\eqref{app_tmp30298130928039810298310928} and using Eq.~\eqref{app_drazin_alternative_definition} we then find
\begin{equation}
\begin{aligned}
    \mathcal{L}^+ &= \mathcal{L}_0^{-1} \mathcal{Q} - \mathcal{L}_0^{-1} \mathcal{J} \mathcal{L}^+
    \\
    &= \mathcal{Q} \mathcal{L}_0^{-1} \mathcal{Q} - \mathcal{Q} \mathcal{L}_0^{-1} \mathcal{J} \mathcal{L}^+,
\end{aligned}
\end{equation}
where $\mathcal{Q} = 1- \mathcal{P}$ and, in the last line, we used the fact that $\mathcal{L}^+ = \mathcal{Q} \mathcal{L}^+ \mathcal{Q}$. 
Isolating $\mathcal{L}^+$ we then finally arrive at 
\begin{equation}\label{drazin_L0_relation}
    \mathcal{L}^+  = (1+\mathcal{Q}\mathcal{L}_0^{-1} \mathcal{J})^{-1} \mathcal{Q} \mathcal{L}_0^{-1} \mathcal{Q},
\end{equation}
which is the desired relation.
}

\section{Proof of Eq.~(\ref{renewal_D_sigma})}
\label{app:WTD}

We consider a single renewal jump channel which, in vectorized notation, reads $\mathcal{J} = |\sigma\rrangle\llangle \xi|$ with $\idV \sigma\rrangle = 1$. 
The average current $J$ and the dynamical activity $K$ coincide, $J = K = \tr(\mathcal{J}\rhoss) = \llangle \xi  \rhossV$. 
The noise is given by Eq.~\eqref{eigen_D}, which now becomes 
\begin{equation}
    D = 
    K - K \llangle \xi | \mathcal{L}^+ |\sigma\rrangle. 
\end{equation}
The WTD, on the other hand, reads $W(t) = \llangle \xi | e^{\mathcal{L}_0 t} |\sigma\rrangle$, where $\mathcal{L}_0 = \mathcal{L}- \mathcal{J}$. 
The mean waiting time is $\mu = 1/K$ [Eq.~\eqref{Mean_waiting_time}] and the variance of the waiting time, according to Eq.~\eqref{moments_WTD}, is 
\begin{equation}
    \sigma^2 = - \frac{2}{K} \idV \mathcal{L}_0^{-1} \rhossV - \mu^2. 
\end{equation}
Equation~\eqref{renewal_D_sigma}, which we wish to prove, therefore reduces to the identity 
\begin{equation}\label{app_WTD_identity_to_prove}
    \llangle \xi |\mathcal{L}^+ |\sigma\rrangle = 1 + K \idV \mathcal{L}_0^{-1} \rhossV. 
\end{equation}
To prove it, we first establish the following relations:
\begin{equation}\label{app_WTD_identities_L0}
    \llangle \xi |\mathcal{L}_0^{-1} = - \idV, 
    \qquad 
    \mathcal{L}_0^{-1} |\sigma\rrangle = - \frac{\rhossV}{K}.
\end{equation} 
The first, for example, can be derived by using the fact that  $\mathcal{J} = \mathcal{L}- \mathcal{L}_0$ and  $\idV \mathcal{L} = 0$ to write $\idV \mathcal{J} \mathcal{L}_0^{-1} = - \idV$. 
But since $\idV \sigma\rrangle = 1$, it is also true that $\idV \mathcal{J} \mathcal{L}_0^{-1} = \llangle \xi | \mathcal{L}_0^{-1}$. 
The second relation follows from a similar reasoning and the fact that $\mathcal{L} \rhossV = 0$.

The proof of Eq.~\eqref{app_WTD_identity_to_prove} now follows from \rev{Eq.~\eqref{drazin_L0_relation}. 
Because of~\eqref{app_WTD_identities_L0} we have that $\mathcal{Q} \mathcal{L}_0^{-1} \mathcal{J} = 0$, a result which is special for renewal processes with only a single jump channel. 
As a consequence Eq.~\eqref{drazin_L0_relation} reduces to  $\mathcal{L}^+ = \mathcal{Q} \mathcal{L}_0^{-1} \mathcal{Q}$. 
Or, more explicitly, 
}
\begin{equation}\label{app_WTD_identities_inverse}
    \begin{aligned}
        \mathcal{L}^+ &= \mathcal{L}_0^{-1} - \mathcal{L}_0^{-1} \rhossV \idV - \rhossV \idV \mathcal{L}_0^{-1}
        \\[0.2cm]
        &+\idV \mathcal{L}_0^{-1} \rhossV ~\rhossV \idV.
    \end{aligned}
\end{equation}
One may verify that, together with~\eqref{app_WTD_identities_L0}, this relation establishes Eq.~\eqref{app_WTD_identity_to_prove}.


\section{Fluctuation Theorem for $\Theta^2=-1$}
\label{app:ftthetam1}
\rev{
In this Appendix, we show that the symmetry in the SCGF in Eq.~\eqref{eq:symmscgf} follows from Eq.~\eqref{vec_vectorized_Ltilt} under the assumption of time-reversal symmetry, even for $\Theta^2=-1$, where $H$ and $L_k$ may in general not be real-valued. To this end, we consider the Liovillian
\begin{IEEEeqnarray}{rCl}
\label{vec_vectorized_Ltiltapp}
\bar{\mathcal{L}}_{\bm \chi}  &=& -i ( H \otimes \id-  \id\otimes H\trans) 
\\[0.2cm]&&
+ \sum\limits_{k=1}^\nops  \Bigg[e^{i\nu_{\alpha k}\chi_{\alpha}}L_k \otimes L_k^* - \frac{1}{2} L_k^\dagger L_k \otimes \id- \frac{1}{2} \id \otimes (L_k^\dagger L_k)\trans \Bigg]\,,
\nonumber
\end{IEEEeqnarray}
which is obtained from $\mathcal{L}_{\bm \chi}$ in Eq.~\eqref{vec_vectorized_Ltilt} by exchanging the matrices before and after the Kronecker products. The Liouvillians $\bar{\mathcal{L}}_{\bm \chi}$ and $\mathcal{L}_{\bm \chi}$ have the same eigenvalues. Indeed, $\bar{\mathcal{L}}_{\bm \chi}$ corresponds to a different but completely legitimate vectorization prescription, where we first transpose the matrix before stacking its columns, and thus describes the same physics. For time-reversal symmetric $H$ and $L_k$ [c.f.~\eqref{eq:timerev}], and using the local detailed-balance condition in Eq.~\eqref{eq:local_db}, we find the symmetry
\begin{equation}
    \label{eq:symmlioutheta}
    \bar{\mathcal{L}}_{\bm \chi} = \Theta \mathcal{L}_{\bm \chi+i\bm\sigma}\trans\Theta^{-1}.
\end{equation}
Due to the anti-linearity of $\Theta$, this implies the following symmetry for the SCGF
\begin{equation}
    \label{eq:scgfsymapp}
    C(\bm \chi) = C^*(\bm \chi+i\bm\sigma)= C(-\bm \chi+i\bm\sigma),
\end{equation}
where the last equality follows from Eq.~\eqref{SCGF_reconstructing_P}. Equation \eqref{eq:scgfsymapp} is identical to Eq.~\eqref{eq:scgfsymlt} in the main text. The remaining results thus follow analogously.
}

\bibliography{library}
\end{document}